\documentclass[preprint]{aastex62}

\usepackage{rotating}
\usepackage{float}
\usepackage{subfigure}
\usepackage{graphicx,amsmath}

\usepackage{xcolor}
\usepackage{tikz}
\usetikzlibrary{shapes,arrows,backgrounds,calc,positioning}
\usetikzlibrary{calc,trees,positioning,arrows,chains,shapes.geometric,
decorations.pathreplacing,decorations.pathmorphing,shapes,
matrix,shapes.symbols}
\usetikzlibrary{calc}
\usepackage{array}
\usepackage{amsmath}
\usepackage{amsxtra}
\usepackage{amsfonts}
\usepackage{amssymb}
\usepackage{hyperref}
\usepackage{xspace}
\usepackage{multirow}
\usepackage{booktabs}
\let\new=\newcommand 
\new{\diff}{{\rm d}}
\new{\be}{\begin{equation}}
\new{\ee}{\end{equation}}

\received{2019 September 5}
\revised{2020 March 17}
\accepted{2020 April 8}

\submitjournal{ApJ}
\shorttitle{Evolution of spin and mass of black holes}
\shortauthors{Bhattacharyya \& Mangalam}

\begin{document}

\title{Cosmic spin and mass evolution of black holes and its impact}

\correspondingauthor{A. Mangalam}
\email{mangalam@iiap.res.in}

\author[0000-0002-6904-3952]{Dipanweeta Bhattacharyya}
\affiliation{Indian Institute of Astrophysics, Sarjapur road, Koramangala 2nd Block, Bangalore-560034, India}
\email{dipanweeta@iiap.res.in}

\author[0000-0001-9282-0011]{A. Mangalam}
\affiliation{Indian Institute of Astrophysics, Sarjapur road, Koramangala 2nd Block, Bangalore-560034, India}
\bibliographystyle{aasjournal}

\begin{abstract}
 We build an evolution model of the central black hole that depends on the processes of gas accretion, the capture of stars, mergers, and electromagnetic torque. In the case of gas accretion in the presence of cooling sources, the flow is momentum driven, after which the black hole reaches a saturated mass; subsequently, it grows only by stellar capture and mergers. We model the evolution of the mass and spin with the initial seed mass and spin in $\Lambda$CDM cosmology. For stellar capture, we have assumed a power-law density profile for the stellar cusp in a framework of relativistic loss cone theory that includes the effects of black hole spin, Carter's constant, loss cone angular momentum, and capture radius. Based on this, the predicted capture rates of $10^{-5}$ to $10^{-6}$ yr$^{-1}$ are closer to the observed range. We have considered the merger activity to be effective for $z \lesssim 4$, and we self-consistently include the Blandford--Znajek torque. We calculate these effects on the black hole growth individually and in combination, for deriving the evolution. Before saturation, accretion dominates the black hole growth ($\sim 95\%$ of the final mass), and subsequently stellar capture and mergers take over with roughly equal contributions. The simulations of the evolution of the $M_{\bullet}$--$\sigma$ relation using these effects are consistent with available observations. We run our model backward in time and retrodict the parameters at formation. Our model will provide useful inputs for building demographics of the black holes and in formation scenarios involving stellar capture.
\end{abstract}

\keywords{Black hole physics (159); Accretion (14); Stellar dynamics (1596);
Cosmological evolution (336); Galaxy nuclei (609)}

\section{Introduction} \label{1} 
It is now widely accepted that all massive galaxies have supermassive black holes (SMBHs) at their centers \citep{1995ARA&A..33..581K}. At distances close to the center of these galaxies, stellar or gas motions are nearly completely dominated by the gravitational potential of the SMBH; this region is described as the sphere of influence. Black holes grow their mass and spin by accretion of gas, by the capture of stars, and by the merger activity of galaxies [that include the central black holes; \cite{2010PhRvD..82l4045K}]. Here we discuss these processes to motivate our model of spin and mass evolution and its impact on various diagnostics like stellar capture rate, the $M_\bullet$--$\sigma$ evolution, and retrodicting the properties of the seed black hole.

\cite{2017NatAs...1E.147A} have studied the evolution of the mass of the black holes by star capture, as well as accretion with and without merger activity. Nonrelativistic loss cone theory is used for their analysis, but the spin evolution of the black hole has not been considered. The loss cone is a region in velocity space where a star is captured by the black hole if it is within this region. The rates of tidal disruption events (TDEs) for a single black hole in steady state have been derived by different authors. \cite{1999MNRAS.306...35S} found the rate of capture to be $10^{-6}$ to $10^{-4}$ yr$^{-1}$ gal$^{-1}$ for main-sequence stars in the galaxies following the Nuker profile. \cite{1999MNRAS.309..447M} find the rate to be $10^{-9}$ to $10^{-4}$ yr$^{-1}$ gal$^{-1}$ using a two-integral model for nonspherical galaxies (triaxial) assuming that all the stars have centrophobic loop orbits and the refilling of loss cone occurs by the two-body relaxation process. \cite{1996NewA....1..149R} found an enhancement in the rate of tidal disruption due to resonant relaxation processes for stars bound to the black hole, but \cite{1998MNRAS.299.1231R} find that in the presence of relativistic precession of black hole masses $\geq 10^{8} M_{\odot}$, this effect is quenched. By assuming a single mass star distribution and solving the steady-state Fokker--Planck equation for 51 galaxies following the Nuker profile, \cite{2004ApJ...600..149W}, derived the rate of disruption to be $10^{-9}$ to $10^{-4}$ yr$^{-1}$ with a revised $M_{\bullet}$--$\sigma$ relation. \cite{2011MNRAS.418.1308B}, using Aarseth's NBODY 6 code, found the rate to be $10^{-6}$ to $10^{-4}$ yr$^{-1}$ gal$^{-1}$ assuming the Sersic profile with $n = 4$ for initial stellar distribution around the black hole. \cite{2012PhRvD..85b4037K} derived the capture rate in the presence of the spin of the back hole to be $\sim 10^{-5}$to $10^{-6}$ yr$^{-1}$. \cite{2015ApJ...814..141M} derived the rate of $10^{-4}$--$10^{-5}$ yr$^{-1}$ for $M_{8} =  10^{-2}$ to $10^{2}$ in a nonrelativistic steady-state loss cone regime. \cite{2015JHEAp...7..148K}, \cite{2002AJ....124.1308D} (ROSAT surveys), and \cite{2009ApJ...698.1367G} (in UV band) have provided observed values of TDEs for different wavelength bands to be about $10^{-5}$ yr$^{-1}$.

The connection of the SMBHs to their host galaxies is evidenced by the strong correlation between the mass of SMBH and velocity dispersion, $\sigma$, of the stars in the rest of the galaxy. This is somewhat surprising because the velocity dispersion is measured for the stars that are too far from the SMBH to be affected by its gravitational field. Its origin is still a topic of debate. This relation is important since the mass of SMBH, which is very difficult to measure directly, can be calculated with relatively better precision using the quantity $\sigma$ (the velocity dispersion of stars far from the SMBH), which is easier to measure for nearby systems. The cosmological $M_{\bullet}- \sigma$ relation is given by the equation
\be 
\displaystyle M_\bullet (z) = K_{0} (z) \sigma^{p (z)}; ~~M_{7} = k_{0} (z) \sigma_{100}^{p (z)}, \label{msigma}
\ee
where $\displaystyle M_{7} = M_{\bullet} /10^{7} M_{\odot}$ and $\displaystyle\sigma_{100} = \sigma /(100~ {\rm km ~sec}^{-1}$). The value of $k_{0}(0)$ is typically $\simeq 1$ [see Table \ref{msigmat}]. \citet{2000ApJ...539L...9F} first reported the index $p$ of the relation to be 4.8 $\pm$ 0.5, which can be explained, for example, by the gas feedback argument of \cite{1998A&A...331L...1S} based on energy-driven flow. \cite{2000ApJ...539L..13G} reported $p$ to be 3.75 $\pm$ 0.3, which is close to the prediction of $p$ = 4 given by a feedback argument by \cite{2003ApJ...596L..27K} based on momentum-driven flow. The exact explanation of the origin of this relation is still not understood properly, but various models give $p$ between 4 and 5, which is in rough agreement with observations as summarized in \cite{2018JApA...39....4B}. \cite{2009ApJ...694..867S} renormalized Equation (\ref{msigma}) using $k_{0} \rightarrow k_{0} (1 + z)^{\alpha}$ and found $\alpha$ = 0.33. \cite{2002ApJ...578...90F} finds the $M_{\bullet}$--$\sigma$ relation to be valid until $z \simeq 1$ (the limit of the survey) and expects that the relation was likely to hold beyond $z=1$.

 \cite{2001ApJ...547..140M} have analyzed a sample of 32 galaxies to determine the ratio of bulge mass and black hole mass using the $M_{\bullet}$--$\sigma$ relation. The mass density of the black hole in the local universe is consistent with the observations. The local black hole scaling relations (with $\sigma$ and bulge mass, $M_{b}$), given by \cite{1999MNRAS.307..637S} were used to derive a black hole mass function \citep{2013CQGra..30x4001S}. The redshift variation of the scaling relations has been taken to be $\propto (1+z)^{\alpha}$ and the value of $\alpha$ has been determined for $M_{\bullet}$--$M_{b}$, as well as the $M_{\bullet}$--$\sigma$ relation, and thus the black hole mass density has been determined. \cite{2010IAUS..267..213N} has derived the evolution of $M_{\bullet}$ and $L / L_{{\rm Edd}}$ with redshift for type 1 radio-quiet active galactic nuclei (AGNs). 

The mass evolution equation for black holes (ignoring mergers) can be approximated by
\be
\displaystyle
M_\bullet(t)= f(t) {\cal M} = K_{0} \sigma^p \simeq M_s +\int {\rm d}t (\dot{M}_* + \dot{M}_g) \simeq K_{0}(t) \sigma^{p(t)} \label{paradigm}
\ee

where $\dot{M}_* = <m_*> \dot{N} =  k_{2} \sigma^{p_1}$, where $p_1=4.3$ for nonrelativistic loss cone theory (eg. \cite{2015ApJ...814..141M}), and $\dot{M_g}= k_{1} \sigma^{p_2}$ where $p_2 \simeq 4$ is the gas accretion rate from the momentum-driven flow. The seed black hole mass is derived
from black hole formation models and is roughly given by $M_s \propto {\cal M}$ or $M_{s} = k_3 \sigma^{p_3}$ (Faber--Jackson law by a fiducial argument here gives $p_3 \simeq 5$), although $p_{3}$ is quickly irrelevant as $M_{\bullet} >> M_{s}$ during the evolution. The similarity of $p_{1} \simeq p_{2} \simeq p_{3}$ is why we think that the form of the $M_{\bullet}$--$\sigma$ relation approximately holds at all epochs; hence, Equation (\ref{paradigm}) is the basic paradigm of the paper with a model to predict $p$, given an evolutionary model of mass and spin. While nonrelativistic simulations \citep{2009ApJ...694..867S, 2015MNRAS.452..575S} obtained the evolution of the $M_{\bullet}$--$\sigma$ relation previously, we have derived here the {\em joint} evolution of the mass and spin of SMBH using a semianalytic model that takes into account the relativistic effects on the critical radii like the horizon, the capture radius, and the cross section that is incorporated into our steady loss cone formalism. Further, our calculations consider all the possible factors contributing to the growth of black holes like accretion, stellar capture, mergers, and Blandford--Znajek (BZ) torque simultaneously; previously, there were models for determination of the evolution for different factors separately. We have built a model for the evolution of measured spin of black holes and estimated its impact on the $M_{\bullet}$--$\sigma$ relation in $\Lambda$CDM cosmology, which is predicated on the physics of gas accretion, star capture, and mergers. We have self-consistently solved coupled equations to get a more complete picture of the evolution of the spin and mass of the SMBH. Our results are shown to agree well with a preliminary analysis of observational data of different galaxies. 

From the analysis of \cite{1972ApJ...178..347B}, without thermodynamic effects, the nonrotating black holes can attain a maximum spin of $\approx$ 1 by the accretion of gas from the innermost stable circular orbit (ISCO). The black hole spin is limited by an upper limit of the spin of 0.998 based on radiation torque due to a difference in cross section for counter and corotating photons near this limit, which is responsible for the saturation \citep{1974ApJ...191..507T}. From Seyfert 1.2 galaxy MCG-06-20-15, XXM-Newton observations have analyzed the upper limit to be 0.989$^{+0.009}_{-0.002}$ with 90\% confidence \citep{2006ApJ...652.1028B}. Later on, \cite{2004ApJ...602..312G} have found the maximum value to be around 0.9, less than 0.998 for relativistic MHD disks. This may not be applicable for thin disc cases; this suggests that the black holes that have grown through MHD accretion are not maximally rotating. For the thick-disk cases, the saturated value of spin was found to be 0.93. The thin-disk analysis by \cite{2004ApJ...602..312G} indicates that through sub-Eddington accretion, the spin can be very close to the maximal rotation. \cite{2005ApJ...620...69V} and \cite{2005ApJ...633..624V} have argued that the effect of accretion torque always results in spinning up of the black hole. In the former paper, the spin-up process of black holes is assumed to be caused by the accretion and binary coalescence, where the SMBH spins up even if the direction of the spin axis varies with time as accretion dominates over coalescences. But if the accretion disks become self-gravitating, their angular momentum per unit mass will be less than that of the black hole. Therefore, in such cases, a black hole having sufficient spin will be spun down \citep{2008MNRAS.385.1621K}. If the black hole is growing by the merger process, the upper limits can be different. After extrapolation of data, \cite{2008PhRvD..77f4010M} have suggested that, for merging two similar-mass black holes with maximum initial spin and aligned with their orbital angular momentum, the upper value can be 0.951 $\pm$ 0.004. \cite{2005ApJ...633..624V} have calculated the growth of black holes taking the range of the formation redshift as $z_{f} \sim$ 10--20, using an accretion rate given by the Bondi--Hoyle formula which is $\propto M_\bullet^2$.

Our goal in this paper is to consider all these processes: gas accretion, stellar capture, mergers, and black hole electrodynamical spin-down to build a self-consistent model of spin and mass evolution. The main motivation of this study is to construct a detailed evolution model of the black hole, that can be a useful tool to study the coevolution of the black hole and the galaxy. We take a comprehensive approach by including all the growth channels semianalytically, with an aim to isolate the important effects. The relativistic treatment is important, as all the channels depend on spin and hence would modulate the black hole growth. The mass evolution equation taking into account gas accretion, stellar capture, and mergers is given by
\be
\frac{\diff M_{\bullet}}{\diff t} = \epsilon_{I}(j) \dot{M}_{\bullet g}+ \epsilon(j) \dot{M}_{\bullet *} + \dot{M}_{\bullet m},\label{mtdimintro}
\ee
and the evolution equation for the spin parameter, $j$, is given by
\be \displaystyle
\frac{\diff j}{\diff t} = \frac{\dot{M}_{\bullet g}}{M_{\bullet}} \bigg(l_{I} (j) - 2 \epsilon_{I}(j)j\bigg) + \frac{\dot{M}_{\bullet *}}{M_{\bullet}} \bigg(l_{*} (j) - 2 \epsilon(j)j\bigg) + \dot{M}_{\bullet m} \cdot \frac{j}{M_{\bullet}}\bigg(-\frac{7}{3} + \frac{9q}{\sqrt{2}j^{2}}\bigg) + x_{+}^{3}(j)j\frac{\mathcal{G}_{0}}{\mathcal{J}_{0}}, \label{jtdimintro}
\ee
where the first three terms in Equations (\ref{mtdimintro}) and (\ref{jtdimintro}) are for accretion, stellar capture, and mergers, respectively, while the last term in Equation (\ref{jtdimintro}) arises as a result of the BZ torque. The mass accretion efficiency, $\epsilon (j)$, is  given by
\be 
        \epsilon(j) = \left\{\begin{array}{lr}
        \epsilon_{I}(j) & \text{\rm for~} M_{\bullet} < M_{c} \\
        1 & \text{\rm for~} M_{\bullet} \geq M_{c},
        \end{array}\right. \label{acceffintro}
\ee 
\\where $M_{c}$ is the critical mass defined in Section \ref{2.2}, and
\be
\epsilon_{I}(j) = \frac{z_{m}^{2}(j)-2z_{m}(j)+j\sqrt{z_{m}(j)}}{z_{m}(j)(z_{m}^{2}(j)-3z_{m}(j)+2j\sqrt{z_{m}(j)})^{1/2}},
\ee
is the efficiency of gas accretion through ISCO, where the ISCO radius is given by \citep{1972ApJ...178..347B}
\begin{eqnarray}
\nonumber z_{m}(j)=  \frac{r_{ms}}{M_{\bullet}} = 3+Z_2-k\sqrt{(3-Z_1)(3+Z_1+2Z_2)}~; \\
Z_1=1+(1-j^2)^{1/3}((1+j)^{1/3}+(1-j)^{1/3});~ Z_2=(3j^2+Z^2_1)^{1/2},
\end{eqnarray}
where $k = 0$ for $j = 0$, +1 for prograde and -1 for retrograde cases, and $j \in \{0, 1\}$. We will derive and discuss each term in Equations (\ref{mtdimintro}) and (\ref{jtdimintro}) in the next section. The symbols used in these equations are defined in Table \ref{symbols}. Figure \ref{diagram} and Table \ref{effects} in Section 2 show the domains of operation and the strengths of the terms (calculated in Appendix \ref{timescales}), respectively. The solution of these two equations will provide the joint mass and spin evolution of black holes. This self-consistent evolution model would be handy in comparing results with simulations and in retrodictions of the formation parameters, thus constraining models of black hole formation. In the future, detailed demographic studies can be carried out to evolve black hole mass and spin distributions. The three applications that we consider are as follows: The first is the capture rates of stars taking into account relativistic corrections to the tidal radius and the capture radius \citep{2019CQGra..36d5009R} to the loss cone model given by \cite{2015ApJ...814..141M}. Second, using the model, we predict the evolution of the $M_\bullet$--$\sigma$ relation. The third application is to retrodict the seed mass and spin of the black hole and the formation redshift under various assumptions, given the recent observations at the epoch near $z \simeq 7$ \citep{2019A&A...625A..23C}. 

In Table \ref{symbols}, a glossary of the symbols used is given. In Section \ref{2} we discuss the growth of the SMBH by gas accretion, star capture, mergers, and the effects of BZ torque on the spin evolution of the black hole individually and discuss the individual evolution equations. In Section 3, we build the required collective evolution equations for spin and mass of SMBH and discuss the cumulative effects. In Section 4, we discuss the impact on the $M_{\bullet}$--$\sigma$ relation, and we also retrodict from the known parameters of mass and spin of the quasars ULASJ134208.10+092838.61 ($z$ = 7.54), ULASJ112001.48+064124.3 ($z$ = 7.08) and DELSJ003836.10-152723.6 ($z$ = 7.02) \citep{2019A&A...625A..23C}, the seed spin, and the mass of the black holes. In Section \ref{5} we summarize the results. The discussions are presented in Section \ref{6} and the conclusions in Section \ref{7}.

\begin{table}[H]
\large
%\hspace{-2.7 cm}
\begin{center}
%\vspace{-2cm}

\scalebox{0.46}{
\begin{tabular}{|c l  c l|}\hline\hline
Common Parameters & & & \\ \hline\hline
$M_{\bullet}$ & Mass of the black hole  & $j$ &   Spin parameter of the black hole  \\ 
$M_{s}$ & Seed black hole mass &  $t_{0}$ & Unit of time ( 1 Gyr)  \\ 
$j_{0}$ & Seed black hole spin parameter & $p$ & Index of the $M_{\bullet}$--$\sigma$ relation  \\ 
$M_{b}$ & Bulge mass & $\sigma$  & Velocity dispersion of stars  \\ 
$f_{b}$ & $\displaystyle M_{\bullet} / M_{b}$  & $M_{\odot}$ & Solar mass \\ 
$M_{x}$ & $M_{\bullet}$ / $10^{x} M_{\odot}$ & $\sigma_{x}$ & $\displaystyle \sigma / (x ~ {\rm{km~ sec^{-1}}})$\\ 
$G, c$ & Gravitational constant, speed of light & $r_{g}$ &  $GM_{\bullet} / c^{2}$\\ 
$t$, $z$ & Look-back time and redshift & $z_{f}$ & Formation redshift \\
$\Omega_{m}$, $a$ & Cosmological parameter, scale factor & $H_{0}$ & Hubble constant \\
$\mu_{\bullet}$ & $M_{\bullet} / M_{s}$ & $t_{0}$ & 1 Gyr \\
\hline
Gas dynamical parameters & & & \\ \hline\hline
$\dot{M}_{\bullet g}$ & Rate of mass growth by accretion & $\eta$ & Efficiency factor of Eddington accretion = $\dot{M}_{\bullet} / \dot{M}_{E}$ \\ 
$\epsilon_{M}$ & Radiation efficiency=$L/\dot{M}_{0}c^{2}$, where $L$ is luminosity, $\dot{M}_{0}$ is rest mass accretion rate & & \\\hline
Stellar dynamical parameters & & & \\ \hline\hline
$x_{c}, x_{\ell}$ & Capture and loss cone radius in units of $r_{g}$ & $l_{\ell}$ & Angular momentum in units of $\displaystyle \frac{GM}{c}$ at $x_{l}$ \\ 
$l_{\ell p}$, $l_{\ell r}$ & Angular momentum in units of $\displaystyle \frac{GM}{c}$ at $x_{c}$ & $x_{t}$ & Tidal radius in units of $r_{g}$\\
$\eta_{t}$ & Strength of tidal encounter &  $r_{p}$ & Radius of pericenter \\ 
$x_{+}$ & Horizon in units of $r_{g}$ & $V_{{\rm eff}}$ & Effective potential in Kerr metric \\
$m_{*}$ & Mass of a star & $R_{*}$ & Radius of a star \\
$l_{\ell}$ & Loss cone angular momentum & $\rho$ & Stellar mass density \\
$\gamma$ & Power-law index of mass density & $P$ & Radial orbital period \\
$f_{s}(E)$ & Distribution function of stars & $\dot{N}_{f}$, $\dot{N}_{s}$ & Rate of capture of stars in full and steady loss cone \\
$q_{s}$ & Diffusion parameter & $J$ & Orbital angular momentum \\
$J_{c}$ & Angular momentum of circular orbit & $E$ & Orbital energy \\
$r_{h}$ & Influence radius of black hole & $s_{t}$ & $\displaystyle r_{t} / r_{h}$ \\
$\epsilon$ & $\displaystyle E / \sigma^{2}$, normalized energy & $\dot{M}_{\bullet *}$ & Rate of mass growth by stellar capture \\ \hline
Mergers and BZ torque & & & \\ \hline\hline
$q$ & Merger mass ratio & $\dot{M}_{\bullet m}$ & Rate of growth of mass by mergers \\
$\dot{N}_{m}$ & Rate of mergers & $\mathcal{J}_{0}$ & Angular momentum budget of black hole \\
$\mathcal{G}_{0}$ & BZ torque & $B_{4}$ & Magnetic field in units of $10^{4}$ G \\
$f_{h}$ & Ratio of black hole mass to the halo mass &  & \\
\hline
Evolution & & & \\ \hline\hline
$M_{\bullet t}$ & Saturation mass & $t_{s}$ & Saturation time \\
$z_{s}$ & Saturation redshift & $\mu_{M}$ & Ratio of mass gained and seed mass \\ 
$M_{*c}$ & Critical mass below which $\dot{M}_{\bullet *}>\dot{M}_{\bullet g}$ & & \\
\hline
\end{tabular}
}
\end{center}
\caption{Glossary of Symbols Used.}
\label{symbols}
\end{table}

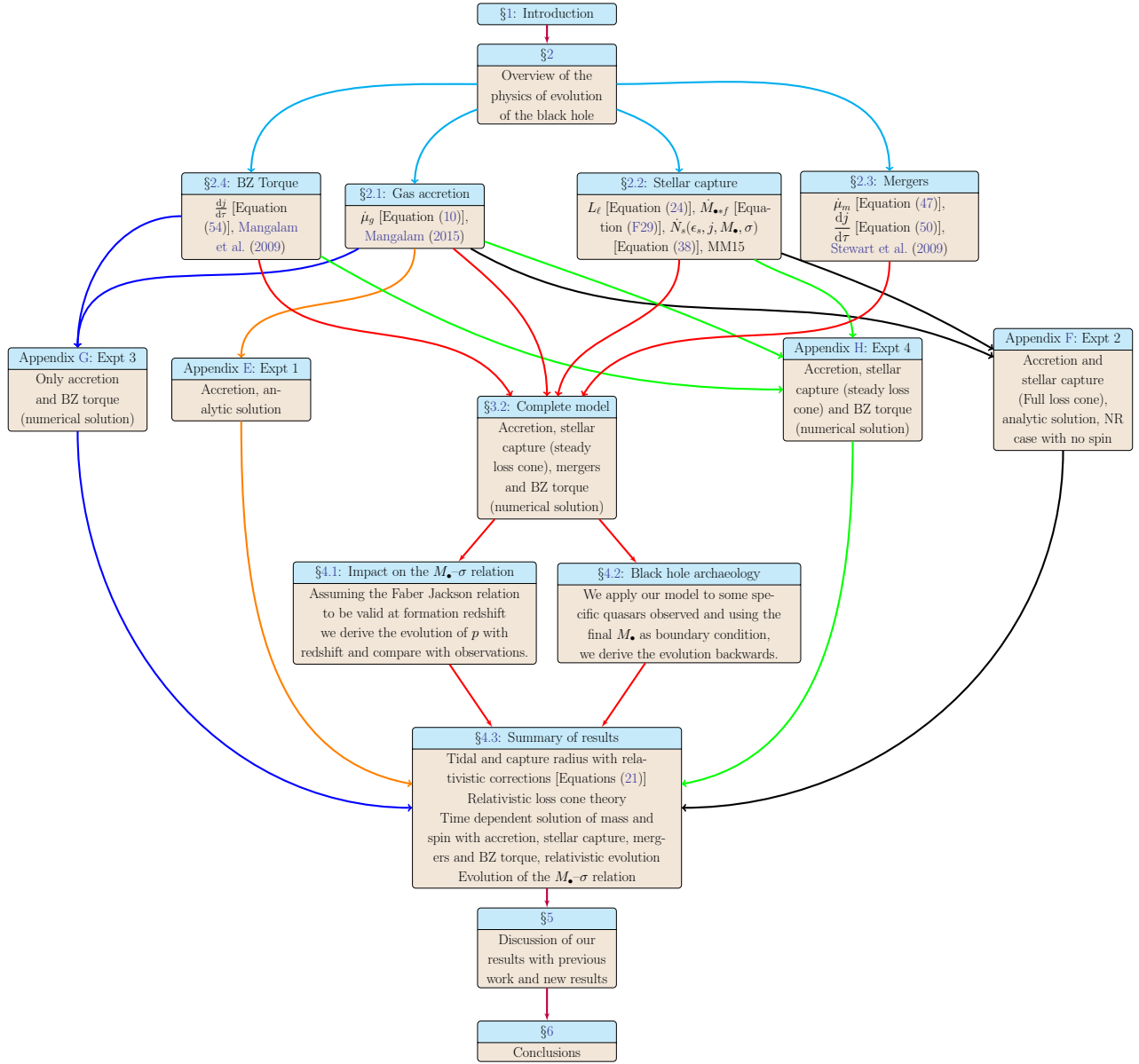
\begin{figure}[H]
\Large
\begin{center}
\scalebox{0.35}{
\begin{tikzpicture}[node distance = 3cm, auto]
\tikzstyle{decision} = [diamond, draw,  
    text width=5em, text badly centered, node distance=3cm, inner sep=0pt]
\tikzstyle{block} = [rectangle, draw, 
    text width=10em, text centered, rounded corners, minimum height=0.7em]
\tikzstyle{line} = [draw, -latex']
\tikzstyle{cloud} = [draw, ellipse,node distance=1cm,
    minimum height=4 em]
\tikzstyle{arrow} = [thick,->,>=stealth]
\tikzstyle{acro}     = [draw,->]
    \node [block,fill=cyan!20] (para19) {\S \ref{1}: Introduction};
    %\node [block,fill=cyan!20, below of= paras, node distance=3cm] (para19) {\S \ref{1}: Introduction};
\node [block, below of = para19, node distance=3cm, rectangle split,rectangle split parts=2,rectangle split part fill={cyan!20,brown!20}] (para) {\S \ref{2} \\ \nodepart{second} Overview of the physics of evolution of the black hole};
\node [block, below left of= para, node distance=8cm, rectangle split,rectangle split parts=2,rectangle split part fill={cyan!20,brown!20}] (para1) {\color{black}\S \ref{2.1}: Gas accretion \\ \nodepart{second} $\dot{\mu}_{g}$ [Equation (\ref{acc_mt})], \cite{2015ASInC..12...51M}};

\node [block,left of= para1,node distance=7cm, rectangle split,rectangle split parts=2,rectangle split part fill={cyan!20,brown!20}] (para5) {\color{black}\S \ref{2.4}: BZ Torque \\ \nodepart{second} $\frac{\diff j}{\diff \tau}$ [Equation (\ref{bz_jt})], \cite{2009MNRAS.397.2216M}};

\tikzstyle{block} = [rectangle, draw, 
    text width=15em, text centered, rounded corners, minimum height=0.7em]
\node [block, below right of= para,node distance=8cm, rectangle split,rectangle split parts=2,rectangle split part fill={cyan!20,brown!20}] (para2) {\color{black} \S \ref{2.2}: Stellar capture \\ \nodepart{second} $L_{\ell}$ [Equation (\ref{llcEquation})], $\dot{M}_{\bullet *f}$ [Equation (\ref{flceq})], $\dot{N}_{s}(\epsilon_{s}, j, M_{\bullet}, \sigma)$ [Equation (\ref{ndtsd_mt})], MM15};

\tikzstyle{block} = [rectangle, draw, 
    text width=13em, text centered, rounded corners, minimum height=0.7em]
\node [block,right of= para2,node distance=9cm, rectangle split,rectangle split parts=2,rectangle split part fill={cyan!20,brown!20}] (para4) {\color{black}\S \ref{2.3}: Mergers \\ \nodepart{second} $\dot{\mu}_{m}$ [Equation (\ref{mg_mt})], $\displaystyle \frac{\diff j}{\diff \tau} $ [Equation (\ref{mg_jt})], \cite{2009ApJ...702.1005S}};

\tikzstyle{block} = [rectangle, draw, 
    text width=10em, text centered, rounded corners, minimum height=0.7em]
\node [block,below left of= para5,node distance=10.5cm, rectangle split,rectangle split parts=2,rectangle split part fill={cyan!20,brown!20}] (para9) {\color{black}Appendix \ref{3.1}: Expt 3\\ \nodepart{second}Only accretion and BZ torque (numerical solution)};

\node [block, below left of= para1,node distance=10.5cm, rectangle split,rectangle split parts=2,rectangle split part fill={cyan!20,brown!20}] (para10) {\color{black}Appendix \ref{AppendixA}: Expt 1 \\ \nodepart{second}  Accretion, analytic solution};

\node [block, below right of= para2,node distance=10.5cm, rectangle split,rectangle split parts=2,rectangle split part fill={cyan!20,brown!20}] (para11) {\color{black}Appendix \ref{3.2}: Expt 4\\ \nodepart{second} Accretion, stellar capture (steady loss cone) and BZ torque (numerical solution) };

\node [block, below right of= para4,node distance=10.5cm, rectangle split,rectangle split parts=2,rectangle split part fill={cyan!20,brown!20}] (para12) {\color{black}Appendix \ref{AppendixB}: Expt 2\\ \nodepart{second} Accretion and stellar capture (Full loss cone), analytic solution, NR case with no spin};

\node [block, below of= para,node distance= 16 cm, rectangle split,rectangle split parts=2,rectangle split part fill={cyan!20,brown!20}] (para13) {\color{black}\S \ref{3.3}: Complete model\\ \nodepart{second}  Accretion, stellar capture (steady loss cone), mergers and BZ torque (numerical solution)};

\tikzstyle{block} = [rectangle, draw, 
    text width=18em, text centered, rounded corners, minimum height=0.7em]
\node [block, below of= para1, node distance=17 cm, rectangle split,rectangle split parts=2,rectangle split part fill={cyan!20,brown!20}] (para14) {\color{black}\S \ref{4.1}: Impact on the $M_{\bullet}$--$\sigma$ relation \\ \nodepart{second} Assuming the Faber Jackson relation to be valid at formation redshift we derive the evolution of $p$ with redshift and compare with observations.};

\tikzstyle{block} = [rectangle, draw, 
    text width=18em, text centered, rounded corners, minimum height=0.7em]
\node [block, below of= para2, node distance=17 cm, rectangle split,rectangle split parts=2,rectangle split part fill={cyan!20,brown!20}] (para20) {\color{black}\S \ref{4.2}: Black hole archaeology \\ \nodepart{second} We apply our model to some specific quasars observed and using the final $M_{\bullet}$ as boundary condition, we derive the evolution backwards.};

\tikzstyle{block} = [rectangle, draw, 
    text width=20em, text centered, rounded corners, minimum height=0.7em]
\node [block, below of= para13, node distance=15 cm,rectangle split,rectangle split parts=2,rectangle split part fill={cyan!20,brown!20}] (para16) {\color{black}\S \ref{5}: Summary of results \\ \nodepart{second}  Tidal and capture radius with relativistic corrections [Equations (\ref{rtEquation})]\\  Relativistic loss cone theory \\ Time dependent solution of mass and spin with accretion, stellar capture, mergers and BZ torque, relativistic evolution \\ Evolution of the $M_{\bullet}$--$\sigma$ relation};

\tikzstyle{block} = [rectangle, draw, 
    text width=10em, text centered, rounded corners, minimum height=0.7em]
\node [block, below of= para16, node distance=6 cm, rectangle split,rectangle split parts=2,rectangle split part fill={cyan!20,brown!20}] (para17) {\color{black}\S \ref{6} \\ \nodepart{second} Discussion of our results with previous work and new results};

\node [block, below of= para17, node distance=4 cm, rectangle split,rectangle split parts=2,rectangle split part fill={cyan!20,brown!20}] (para18) {\color{black}\S \ref{7} \\ \nodepart{second} Conclusions};

\draw [very thick,->, cyan, line width=0.8 mm](para)to [out=-160,in=90](para1);
\draw [very thick,->, cyan, line width=0.8 mm] (para) to [out=-20,in=90] (para2);
\draw [very thick,->, cyan, line width=0.8 mm] (para) to [out=0,in=90] (para4);
\draw [very thick,->, cyan, line width=0.8 mm] (para) to [out=-180,in=90] (para5);
\draw [very thick,->, blue, line width=0.8 mm](para1)to [out=-150,in=90](para9);
\draw [very thick,->, blue, line width=0.8 mm](para5)to [out=-180,in=90](para9);
\draw [very thick,->, orange, line width=0.8 mm](para1)to [out=-90,in=90](para10);
\draw [very thick,->, black, line width=0.8 mm](para2)to [out=-20,in=150](para12);
\draw [very thick,->, black, line width=0.8 mm](para1)to [out=-30,in=155](para12);

\draw [very thick,->, green, line width=0.8 mm](para1)to [out=-20,in=155](para11);
\draw [very thick,->, green, line width=0.8 mm](para5)to [out=-30,in=-180](para11);
\draw [very thick,->, green, line width=0.8 mm](para2)to [out=-30,in=90](para11);

\draw [very thick,->, red, line width=0.8 mm](para5)to [out=-80,in=120](para13);
\draw [very thick,->, red, line width=0.8 mm](para1)to [out=-40,in=90](para13);
\draw [very thick,->, red, line width=0.8 mm](para2)to [out=-90,in=80](para13);
\draw [very thick,->, red, line width=0.8 mm](para4)to [out=-90,in=60](para13);
\path [line, red, line width=0.8 mm] (para13)--node{}(para14);
\path [line, red, line width=0.8 mm] (para14)--node{}(para16);
\path [line, red, line width=0.8 mm] (para13)--node{}(para20);
\path [line, red, line width=0.8 mm] (para20)--node{}(para16);
\draw [very thick,->, blue, line width=0.8 mm](para9)to [out=-90,in=-180](para16);
\draw [very thick,->, orange, line width=0.8 mm](para10)to [out=-90,in=170](para16);
\draw [very thick,->, green, line width=0.8 mm](para11)to [out=-90,in=10](para16);
\draw [very thick,->, black, line width=0.8 mm](para12)to [out=-90,in=0](para16);
\path [line, purple, line width=0.8 mm] (para16)--node{}(para17);
\path [line, purple, line width=0.8 mm] (para17)--node{}(para18);
%\path [line, purple, line width=0.8 mm] (paras)--node{}(para19);
\path [line, purple, line width=0.8 mm] (para19)--node{}(para);

 \end{tikzpicture}

}
\end{center}
\caption{Flowchart of concepts in the paper discussing the input physics of relativistic stellar capture, gas accretion, electromagnetic torque and mergers}.
\label{flowchartsec}
\end{figure}

\section{Overview of the Physics of Evolution of the Black Hole}\label{2}
The seed black holes grow their mass mainly through three processes: accretion, stellar capture, and mergers. Though the stellar capture presumably does not contribute to the evolution of its spin, the other two do contribute to spin evolution. The BZ mechanism \citep{1977MNRAS.179..433B} of electromagnetic braking of the black hole contributes to spinning, but it does not have any effect on mass growth. In this section, we discuss all these processes individually, derive Equations (\ref{mtdimintro}) and (\ref{jtdimintro}) and state the assumptions of our model.

%In this section, we present all these processes one by one and discuss the assumptions of our model.

\subsection{\it{Growth of the Black Hole by Gas Accretion}}\label{2.1}
The black hole mainly grows by accretion flow of the gas. In the case of an energy-driven flow \citep{1998A&A...331L...1S} it is assumed that all the energy from the accretion is used for unbinding the mass of the bulge and the maximum possible mass the SMBH can attain from this accretion process is
\be M_{\bullet} \simeq 8 \times 10^{8} \bigg(\frac{\sigma}{300 \rm km~sec^{-1}}\bigg)^{5} M_{\odot}. \ee
 \cite{2003ApJ...596L..27K} proposed that black hole growth occurs by gas flow until it reaches a saturated mass $M_{\bullet t}$, which is a different approach than that of \cite{1998A&A...331L...1S}, who propose an energy-driven flow by assuming that the energy from accretion is completely used in unbinding the mass of the galactic bulge, while there is no loss of energy due to radiation. \cite{2003ApJ...596L..27K} considers Compton cooling, for which some energy is lost to radiation and the remaining energy is available for unbinding the mass of the bulge. From the analysis of \cite{2003ApJ...596L..27K} after saturation, the outflow velocity exceeds the escape velocity of the medium and the gas is driven away, causing the accretion process to stop. The saturated mass is found to be $M_{\bullet t} = 9.375\times 10^{6} \sigma_{100}^{4}M_{\odot}$ derived for a spherical geometry of the ambient gas. However, the infalling matter must possess some amount of angular momentum so that an accretion disk forms, and thus there is a small solid angle where only inflow occurs. If most of the gas lies in the plane of the galaxy, the momentum-driven outflow would not halt the inflow; this also implies that accretion from this point adds little mass to the hole. In our model, we ignore accretion after saturation and consider that only stellar capture and mergers contribute to the growth of the black hole and also at this point $p \approx 4$.
We take sub-Eddington accretion throughout so that
\be
\displaystyle
\dot{M}_{\bullet g} = k_{1}M_{\bullet}, \label{acceq}
\ee
where $\displaystyle k_{1} = \frac{4\pi G m_{p}\eta }{\sigma_{e}c}$ and the factor $\eta$ = $\dot{M}_{\bullet} / \dot{M}_{E}$, where $\dot{M}_{E}$ is the Eddington accretion rate.

Therefore, in units of $\displaystyle \mu_{\bullet} = \frac{M_{\bullet}}{M_{s}}$, where $M_{s}$ is the seed mass, $\displaystyle \tau = \frac{t}{t_{0}}$, where $t_{0}$ = 1 Gyr, we can express
\be
\displaystyle
\dot{\mu}_{g} = \frac{\dot{M}_{\bullet g} t_{0}}{M_{s}}. \label{acc_mt}
\ee
\cite{2015ASInC..12...51M} used a theoretical model for mass and spin evolution of the black hole taking into account the angular momentum torque caused by the electrodynamical jet, where it was shown that the spin evolution with the accretion rate is taken to be a given fraction of the Eddington rate for different cases such as the thin disk, Bondi accretion, and also MHD disk. The mass evolution equation is given by
\be
\displaystyle
\frac{\diff M_{\bullet}}{\diff t} =  \epsilon_{I}(j) \dot{M}_{\bullet g}, \label{macc}
\ee 
where $\displaystyle \epsilon_{I}(j)$ [see Equation (2)] is the efficiency of energy conversion with the innermost radius of the disk to be taken typically at ISCO, and $\dot{M}_{\bullet g}$ is the rate of accretion. The spin evolution equation is given by [see \cite{2015ASInC..12...51M}, \cite{2009MNRAS.397.2216M}, \cite{2005ApJ...620...59S}]
\be
\displaystyle
\frac{\diff j}{\diff t} = \frac{\dot{M}_{\bullet g}}{M_{\bullet}} \bigg(l_{I} (j) - 2 \epsilon_{I}(j)j\bigg),\label{jacc}
\ee
where $l_{I} (j)$ [see Equation (5)] is the angular momentum per unit mass at ISCO. This can be seen by the following arguments. The first term is due to the accretion of angular momentum at ISCO, while the second represents the spin-down due to an increase in the black hole inertia; these arguments give
\begin{eqnarray}
\displaystyle
\nonumber J_{\bullet} = \frac{G M_{\bullet}^{2} j}{c},
\end{eqnarray}
\begin{eqnarray}
\dot{j} = \frac{c}{G}\frac{\diff}{\diff t} \bigg[\frac{J_{\bullet}}{M_{\bullet}^{2}} \bigg] =  \frac{\dot{M}_{\bullet g}}{M_{\bullet}} \bigg(l_{I} (j) - 2 \epsilon_{I}(j)j\bigg).
\end{eqnarray}

\subsection{\it{Growth of the Black Hole by stellar capture}}\label{2.2}
The SMBHs can also grow by the capture of stars in two ways. Those stars that fall into the event horizon without disruption are directly captured, and the indirect capture occurs when the stars are tidally disrupted outside the horizon. For SMBHs more massive than $10^{8} M_{\odot}$, the direct capture of solar-type stars is possible if the angular momentum of the star is smaller than some critical value \citep{1976MNRAS.176..633F} for the nonrelativistic case. We proceed to calculate the limiting value for the relativistic case. For a Kerr black hole, the standard effective potential is written as [\cite{PhysRev.174.1559, 1973grav.book.....M, 1998bhp..book.....F, 2019CQGra..36d5009R}; \cite{RM2019} (RM19)]
 \be 
V_{eff} (x, l, j, Q) = -\frac{1}{x} + \frac{l^{2} + Q}{2 x^{2}} - \frac{[(l - j)^{2} + Q]}{x^{3}} + \frac{j^{2}Q}{2 x^{4}}. \label{kerrpot}
\ee 
Here $l \equiv L / (GM / c)$, where $L$ is the angular momentum, $x = r / r_{g}$, $r_{g} = GM_{\bullet} / c^{2}$, $j$ is the spin parameter, and $Q$ is the Carter's constant. The solution of $V_{{\rm eff}} (x_{p})$ = 0 and ${V}^{\prime}_{{\rm eff}} (x_{p})$ = 0 gives the equation of separatrix orbit where $x_{p}$ is the pericenter, as shown in Fig 2(b) of RM19. From the two conditions $V_{{\rm eff}} (x_{p}) = 0$ and ${V}^{\prime}_{{\rm eff}}(x_{p})$ = 0, we find the equation for the separatrix $x_{p}(Q, l, j)$ to be given by
\be 
x_{p}^{3} - [(l - j)^{2} + Q] x_{p} + j^{2} Q = 0,\label{rc}
\ee 
which represents a turning point condition for an orbit that is just bound or just unbound. This also represents a marginally bound spherical orbit (MBSO). The innermost stable spherical orbit (ISSO) and MBSO are the end points of the separatrix curve from ($e = 0$, ISSO) to ($e =1$, MBSO). The star is captured at MBSO (as $r \rightarrow \infty$ in Figure 2(b) of RM19) and $r_{s}$ is the pericenter. This capture radius (MBSO) $x_{c} (Q, j)$ in units of $r_{g}$, is found by reducing Equation (\ref{rc}) to (by substituting for ($l$--$j$) from the translation formulae [see Equation (7) in RM19]; see Appendix D of RM19 for details)
\begin{eqnarray}
x_{c}^8 -8 x_{c}^7 - 2j^2 x_{c}^6 + 16 x_{c}^6 + 2 j^2 Q x_{c}^5 -8 j^2 x_{c}^5 -6 j^2 Q x_{c}^4 + j^4 x_{c}^4 -2 j^4 Q x_{c}^3 + \nonumber \\
8 j^2 Q x_{c}^3 +  j^4 Q^2 x_{c}^2 -2j^4 Q x_{c}^2 -2 j^4 Q^2 x_{c} + j^4 Q^2 =0. \label{MBSOEquation}
\end{eqnarray}
Solving Equation (\ref{MBSOEquation}) for real roots (numerically for $Q \neq 0$) that are higher than the light radius \citep{1973blho.conf..215B}, we find the capture radii for both prograde and retrograde cases. If $Q$ = 0, Equation (\ref{MBSOEquation}) reduces to
\be 
x_{c}^{4} (x_{c}^{2} - 2jx_{c} + j^{2} - 4x_{c}) (x_{c}^{2} - 4x_{c} + j^{2} + 2 j x_{c}) = 0,
\ee 
which leads to the known result \citep{2002NewA....7..385Z}
\be 
x_{c} (j) = \left\{\begin{array}{lr}
        -j + 2(1 + \sqrt{1 - j}) & \text{\rm for~ prograde} \\
       j + 2(1 + \sqrt{1 + j}) & \text{\rm for~ retrograde}
        \end{array}\right.\label{xcQ0}
\ee 
The angular momentum at $x_{c}$, from Equation (\ref{rc}), is found to be
\be 
\displaystyle l_{c} (M_{8}, j, Q) = j + k \sqrt{x_{c}^{2} - Q + \frac{j^{2}Q}{x_{c}}}. \label{lcEquation}
\ee 
The value of $l_{c}$ will be positive for the prograde case ($k = 1$) and negative for the retrograde case ($k$ = --1) owing to the direction of spin of the black hole. The capture of a star can occur in two ways, either by tidal disruption of the stars or by direct capture by the black hole. For direct capture, we determine $x_{c}$ using Equation (\ref{MBSOEquation}) for $Q \neq 0$ or by using Equation (\ref{xcQ0}) for $Q = 0$. Below a certain critical mass, the stars are tidally disrupted \citep{2013CQGra..30x4005M} and above this the stars are swallowed whole. Therefore, $r_{t}$ is defined as the radius below which the star gets disrupted by the black hole. We calculate the tidal radius in the presence of black hole spin applying Poisson's equation
\be \frac{\partial^{2}V_{eff}}{\partial r^{2}}\bigg|_{r = r_{t}} = - 4 \pi G\rho, \label{rteq}\ee 
where $\rho$ is the stellar mass density. Using the generalized form of the effective Kerr potential in natural units [Equation (\ref{kerrpot})], the tidal radius equation (Equation (\ref{rteq})), finally leads to
\be \bigg[-\frac{2}{x^{3}} + \frac{3(l^{2} + Q)}{x^{4}} - 12\frac{[(j - l)^{2} + Q]}{x^{5}} - \frac{10 j^{3}Q}{x^{6}} \bigg]_{x = x_{t}} = - 4 \pi \tilde{\rho},\label{rtEquation} \ee 
where 
\begin{eqnarray} 
\tilde{\rho} & = & \frac{\rho}{ M_{\bullet}} \cdot r_{g}^{3} \simeq 0.3 M_{8}^{2} \bigg(\frac{\rho_{*}}{\rho_{\odot}}\bigg) ,\\ \nonumber 
x & = & r / r_{g},\\ \nonumber
l & = &  L / ~\frac{GM_{\bullet}}{c}.
\end{eqnarray}
where $\rho_{*}$ is the density of the star. We solve Equation (\ref{rtEquation}) numerically for $x_{t}(j, k, M_{8}, Q)$ by considering $l = l_{c}$, the angular momentum of capture taking the above approximation for $\tilde{\rho} \simeq 0.3 M_{8}^{2}$ (assuming the star to be of solar type). An analytic approximation to $r_{t}$ has been calculated in Appendix \ref{Appendix}.

The loss cone radius, $x_{\ell} \equiv {\rm Max}[x_{t}, x_{c}]$ is given by
\be 
\displaystyle x_{\ell} (M_{8}, j, k, Q) = r_{\ell} / r_{g} = {\rm Max}[r_{t} (M_{8}, j, k, Q), r_{c} (j, Q)] / r_{g}.
\ee 
The angular momentum at $x_{\ell}$ is found by putting $V_{{\rm eff}} (x, l, j,Q) = 0$ (see Equation (\ref{kerrpot})) to be

\be 
\displaystyle l_{\ell}( M_{8}, j, k, Q) = 2j + k\sqrt{\frac{2x_{\ell}j^{2}}{(x_{\ell}-2)^{2}} - \frac{Qj^{2}}{x_{\ell} (x_{\ell} - 2)} + \frac{2x_{\ell}^{2}}{(x_{\ell} - 2)} - Q}. \label{llcEquation}
\ee 
{\normalfont The value of $l_{\ell}$ is positive for the prograde case ($k = 1$) and negative for the retrograde case ($k$ = --1) owing to the direction of spin of the black hole. We show both the cases in Figure \ref{figllc}. For $Q$ = 0 and $j=0$, in the nonrelativistic limit (for high values of $x_{\ell}$), using Equation (\ref{llcEquation}), we obtain
\be
L_{\ell}^{2} = \bigg(\frac{GM_{\bullet}}{c}\bigg)^{2} l_{\ell}^{2} = \bigg(\frac{GM_{\bullet}}{c}\bigg)^{2} 2 x_{\ell} = 2GM_{\bullet} r_{\ell}, \label{nonrell}
\ee 
which is the well-known nonrelativistic result.

We show the variation of $x_{\ell} (M_{8}, j, k, Q)$ in Figure \ref{figllca} (which is an input to the loss cone theory) and that of $x_{t} (M_{8}, j, k, Q)$ in Appendix \ref{Appendix}.

\begin{figure}[H]
\centering
\subfigure[]{\includegraphics[scale=0.2]{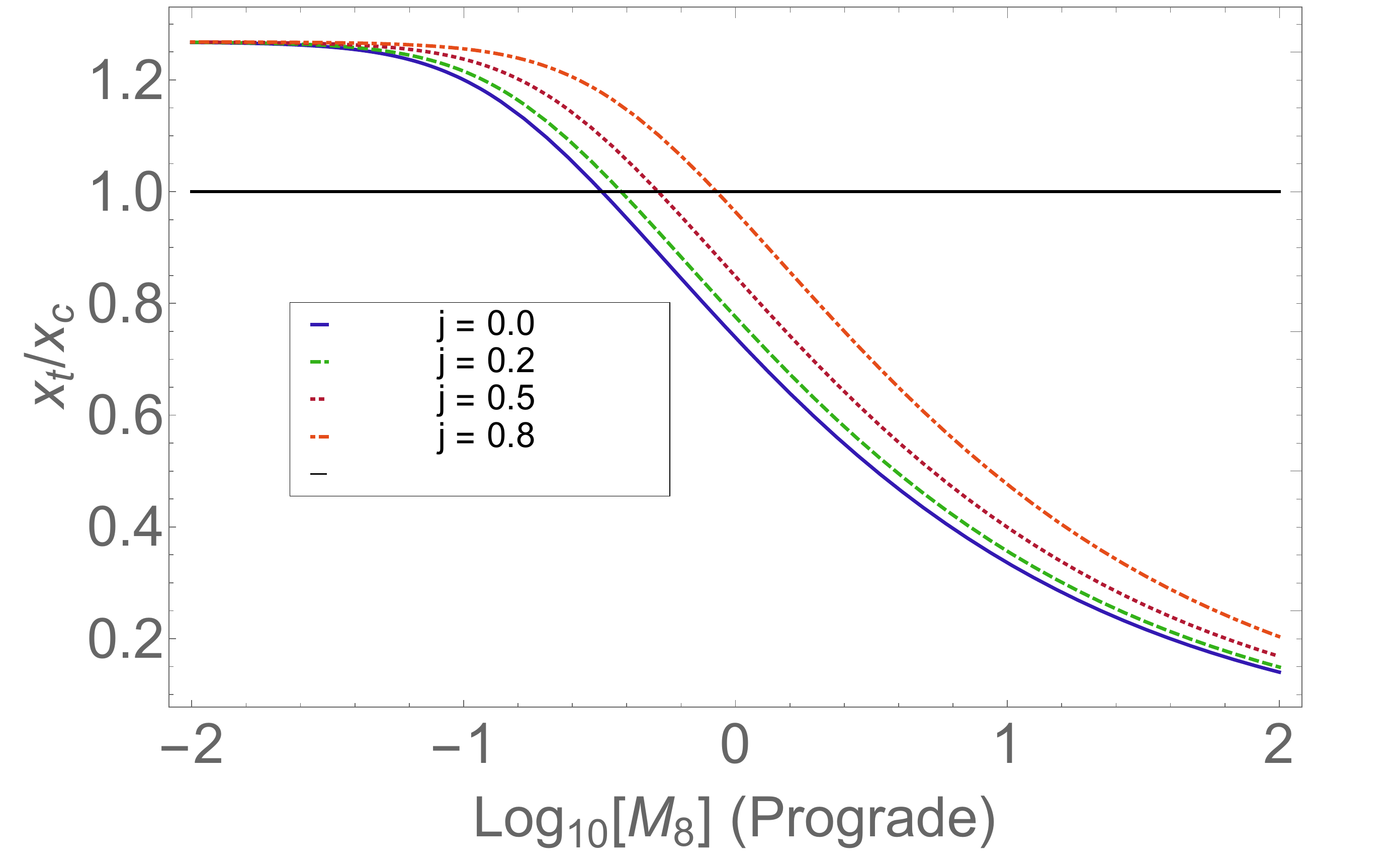}} \hspace{0.3 cm}
\subfigure[]{\includegraphics[scale=0.2]{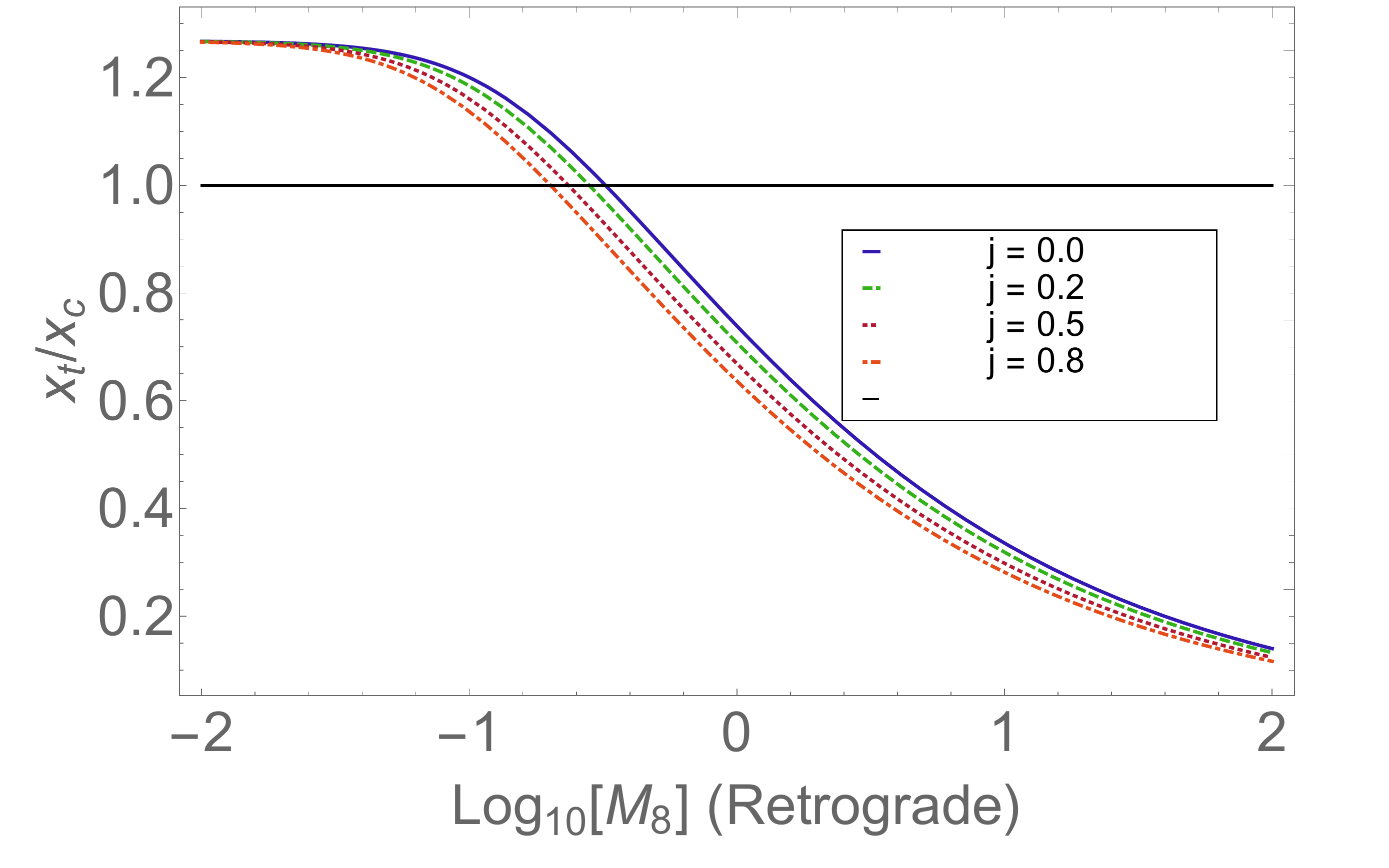}} \hspace{0.3 cm}\\
\subfigure[]{\includegraphics[scale=0.2]{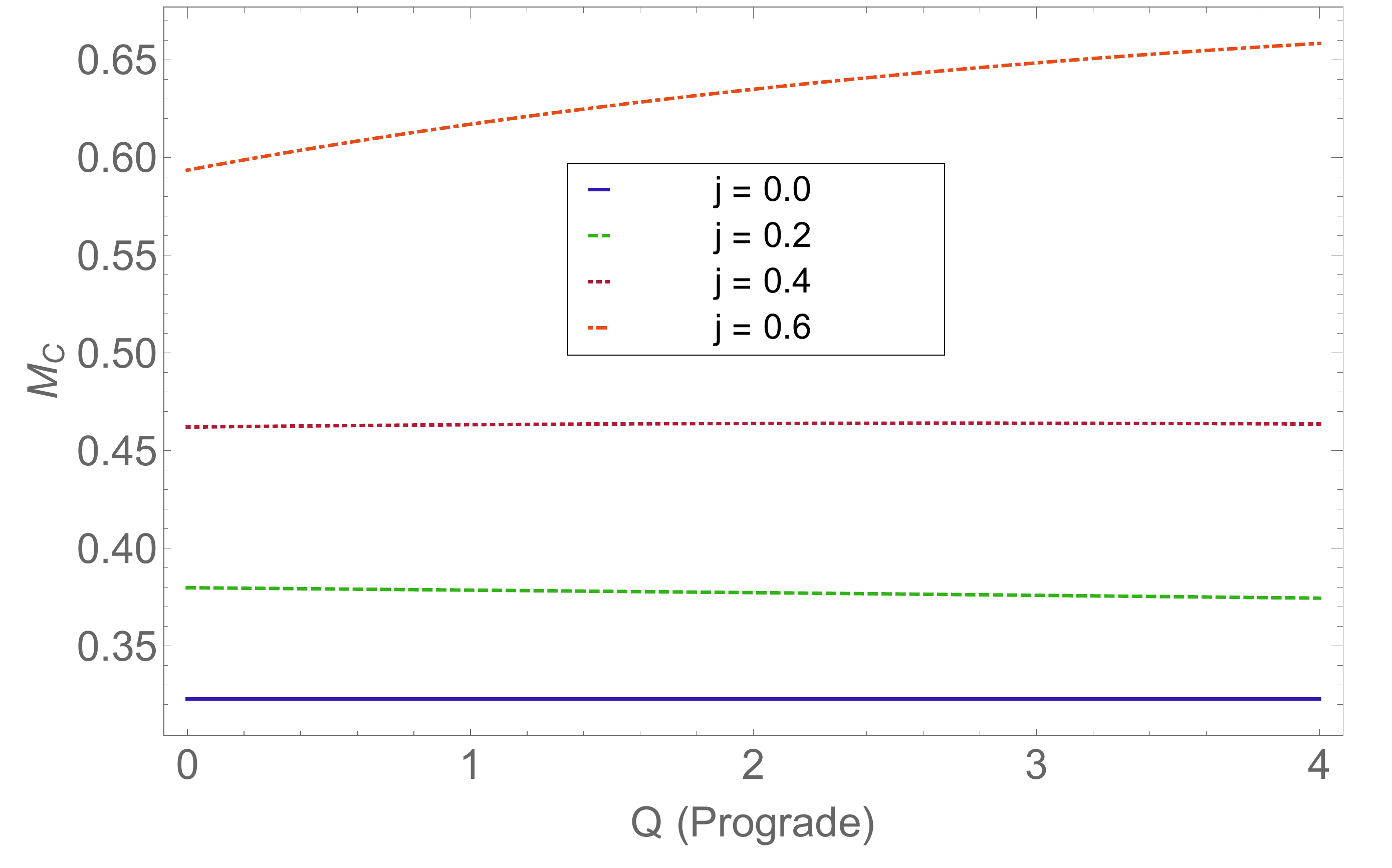}} \hspace{0.3 cm}
\subfigure[]{\includegraphics[scale=0.2]{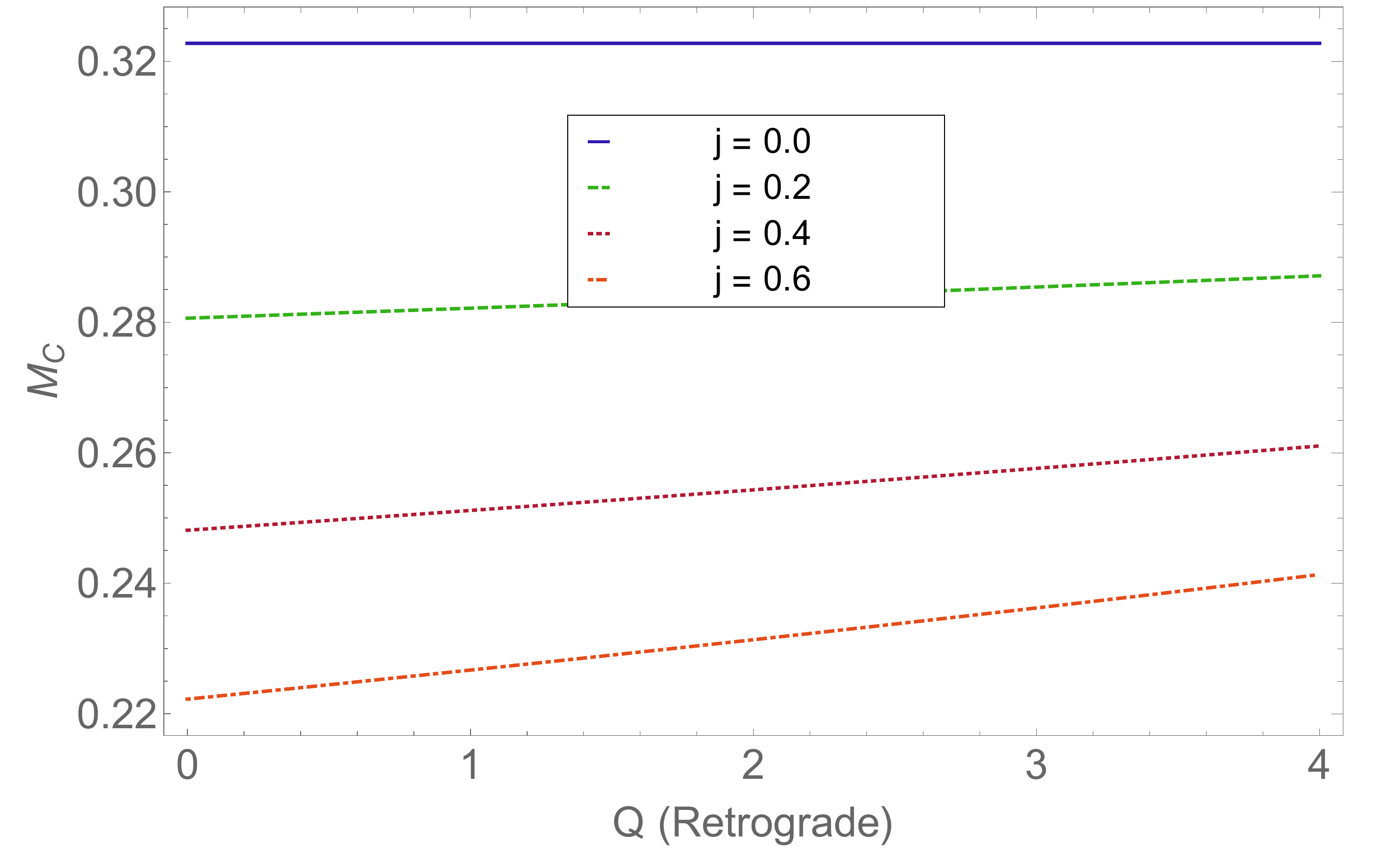}} 
\caption{Ratio of tidal radius to the capture radius [$r_{t}(M_{8}, j, k, Q) / r_{c}(M_{8}, j, Q)$ = $x_{t}(M_{8}, j, k, Q) / x_{c}(j, Q)$] is shown as a function of $M_{8}$ for $Q$ = 0 (a, b) and the locus of the critical mass, $M_{c}(j, Q)$ for different $j$ as a function of $Q$ (c, d). The critical mass of the black hole is determined from the plots when $x_{t} / x_{c}$ = 1; this critical mass is represented by the black line in (a) and (b).}
\label{figrtc}
\end{figure}

\begin{figure}[H]
\centering
\subfigure[]{\includegraphics[scale=0.2]{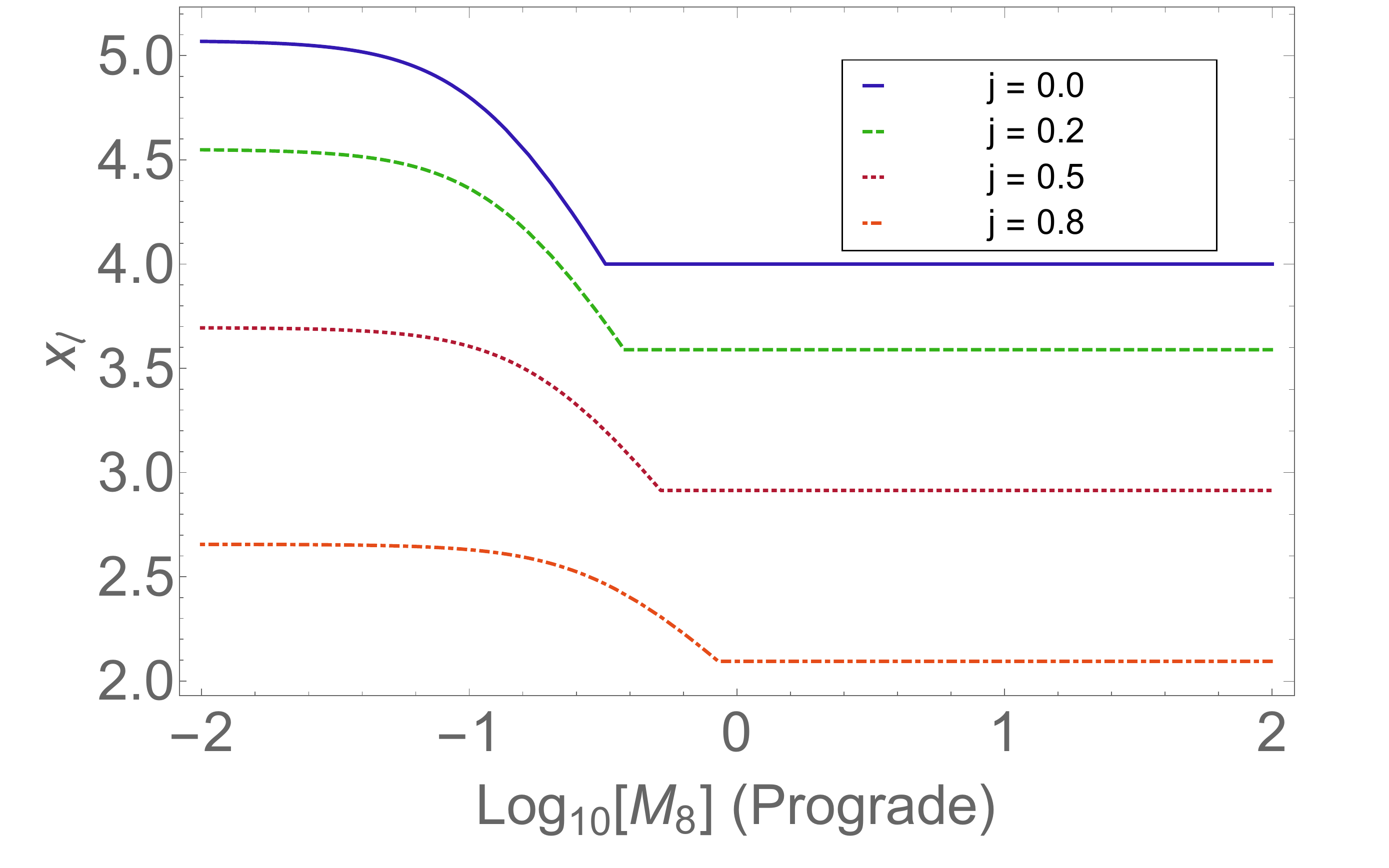}} \hspace{0.3 cm}
\subfigure[]{\includegraphics[scale=0.2]{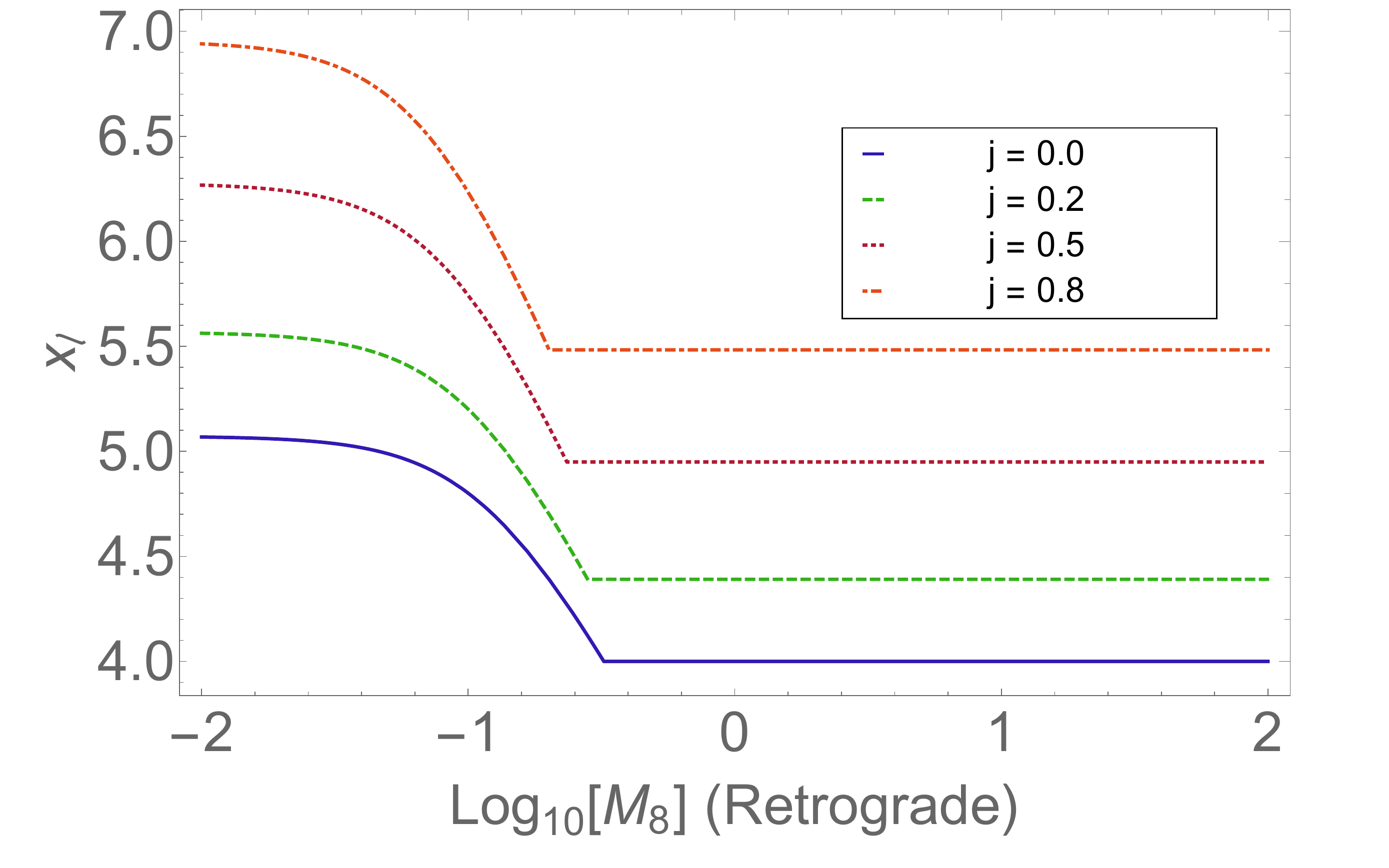}} \hspace{0.3 cm}\\
\subfigure[]{\includegraphics[scale=0.2]{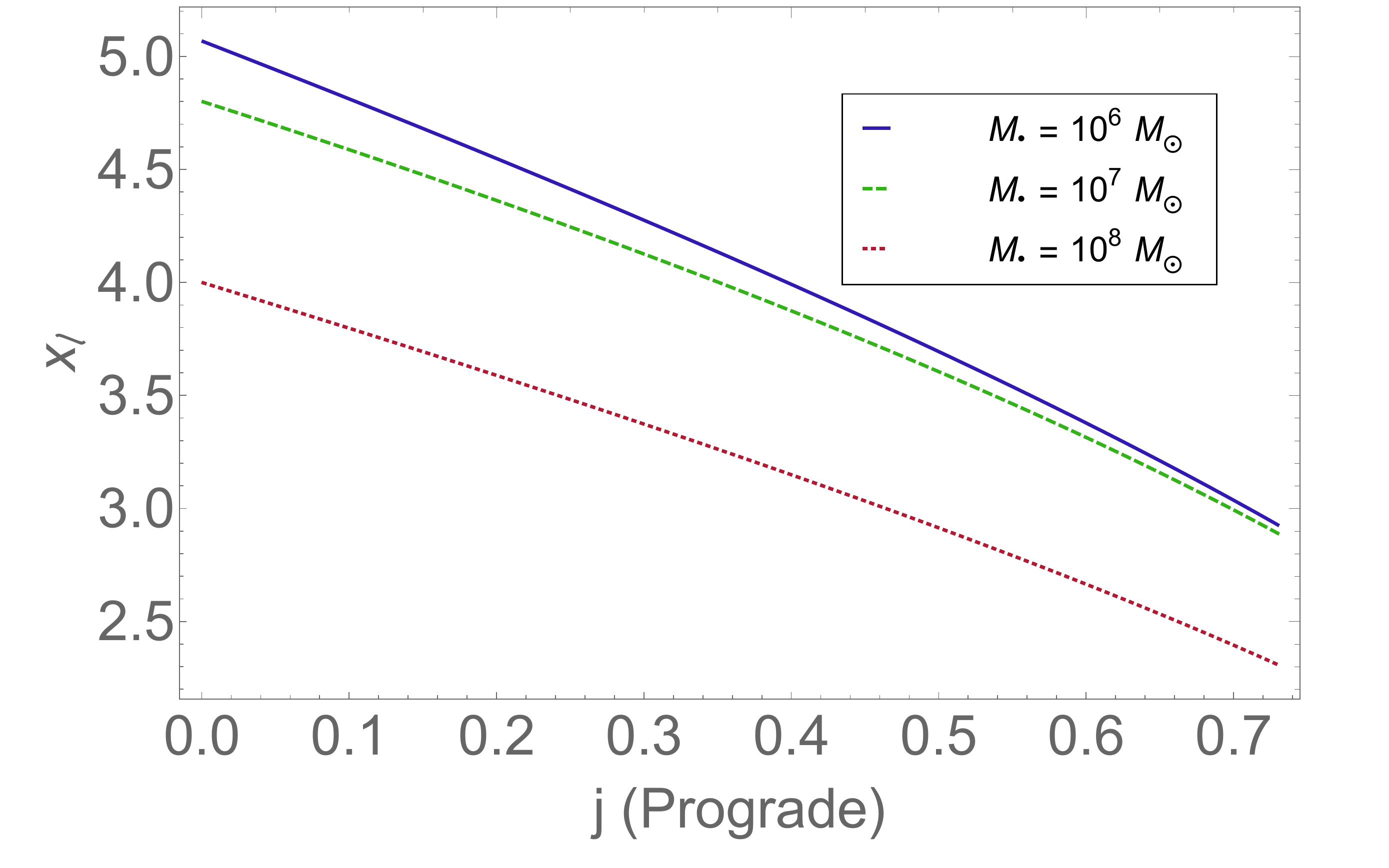}} \hspace{0.3 cm}
\subfigure[]{\includegraphics[scale=0.2]{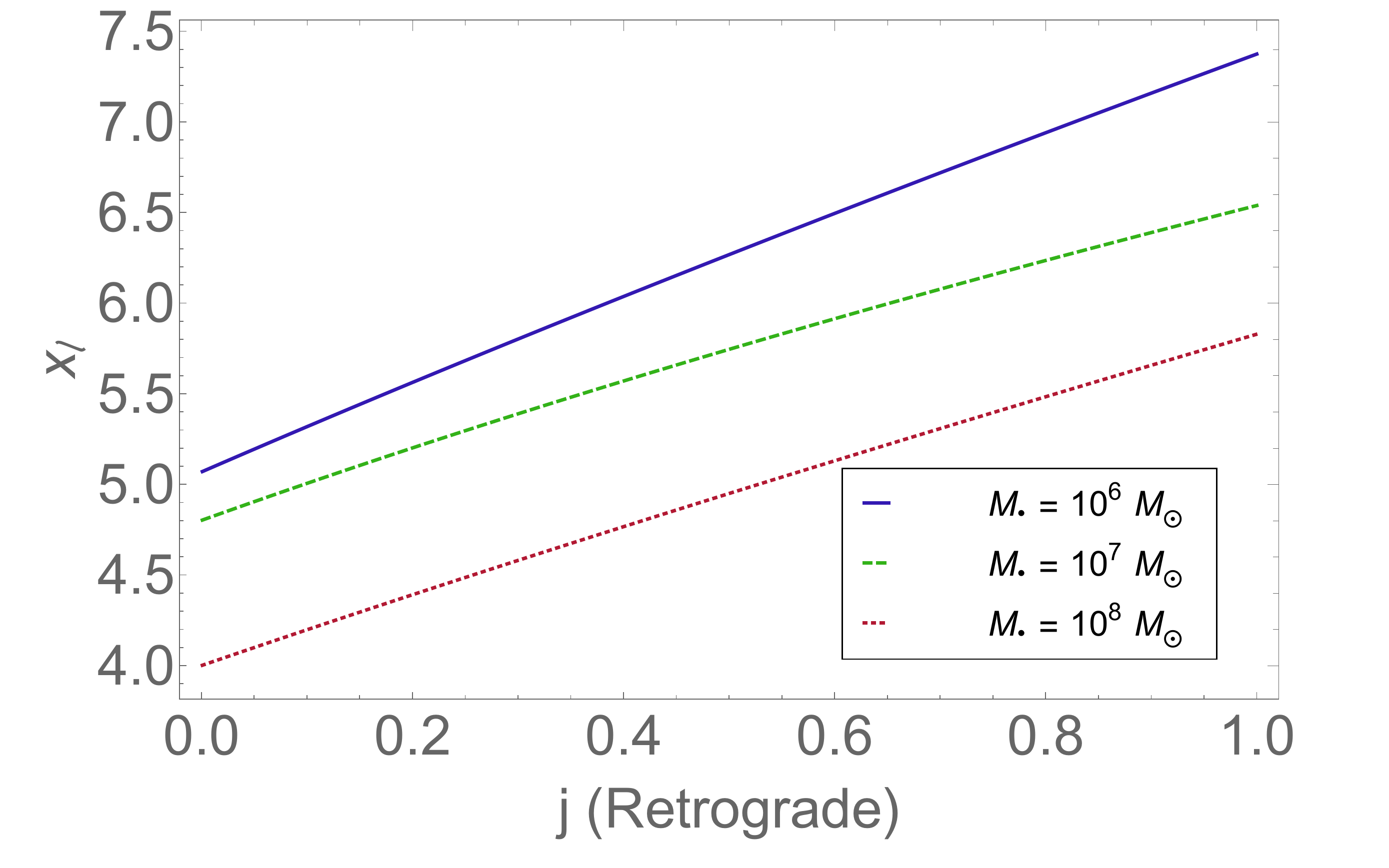}} 
\caption{Loss cone radius ($\displaystyle x_{\ell} (M_{8}, j, k, Q) \equiv r_{\ell} (M_{8}, j, k, Q) / r_{g}$) = Max[$x_{t}(M_{8}, j, k, Q)$, $x_{c}(j, Q)$] is shown as a function of $M_{8}$ (a, b) and $j$ (c, d) for $Q$ = 0.}
\label{figllca}
\end{figure}

\begin{figure}[H]
\centering
\subfigure[]{\includegraphics[scale=0.2]{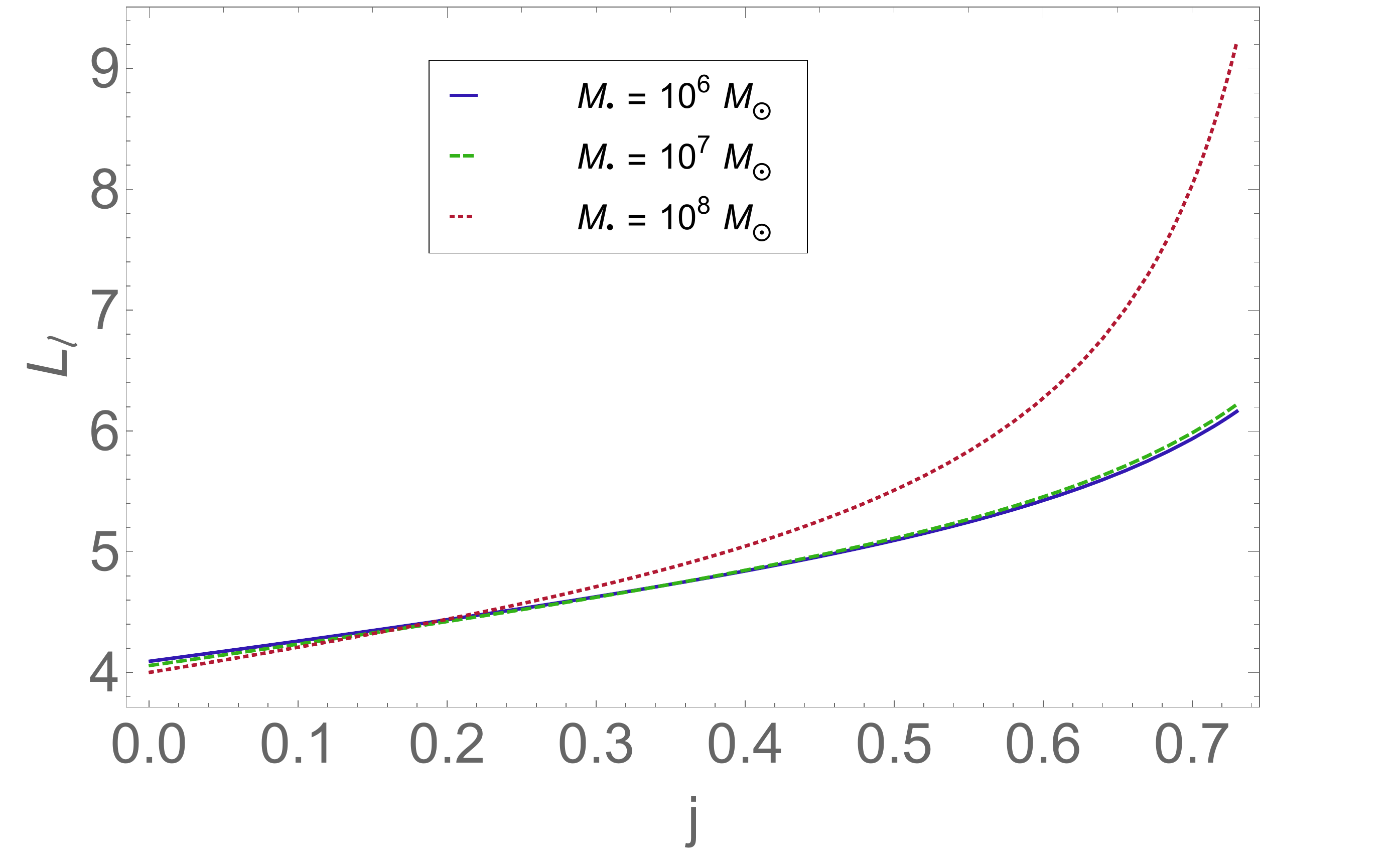}} \hspace{0.3 cm}
\subfigure[]{\includegraphics[scale=0.2]{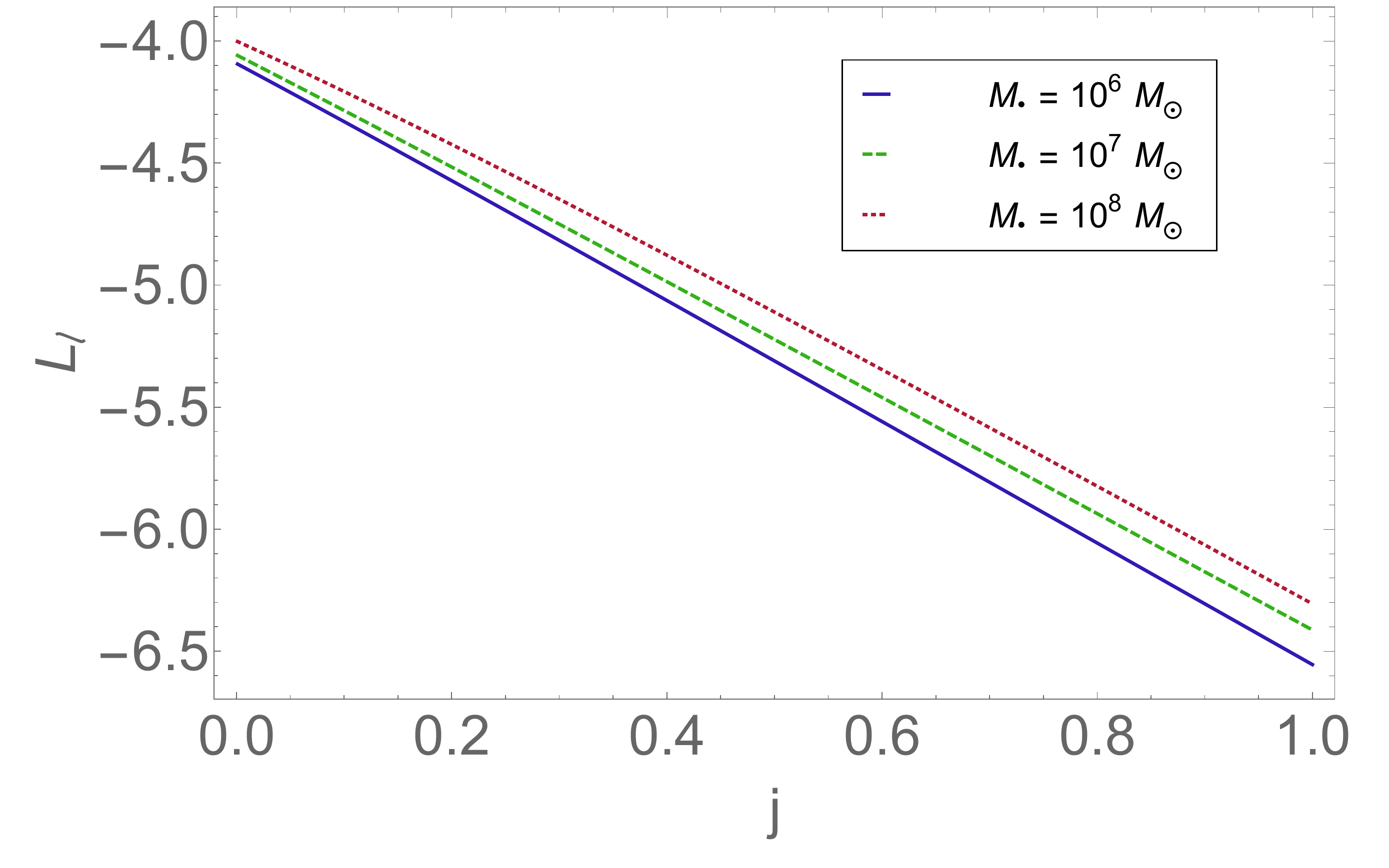}} \hspace{0.3 cm}\\
\subfigure[]{\includegraphics[scale=0.2]{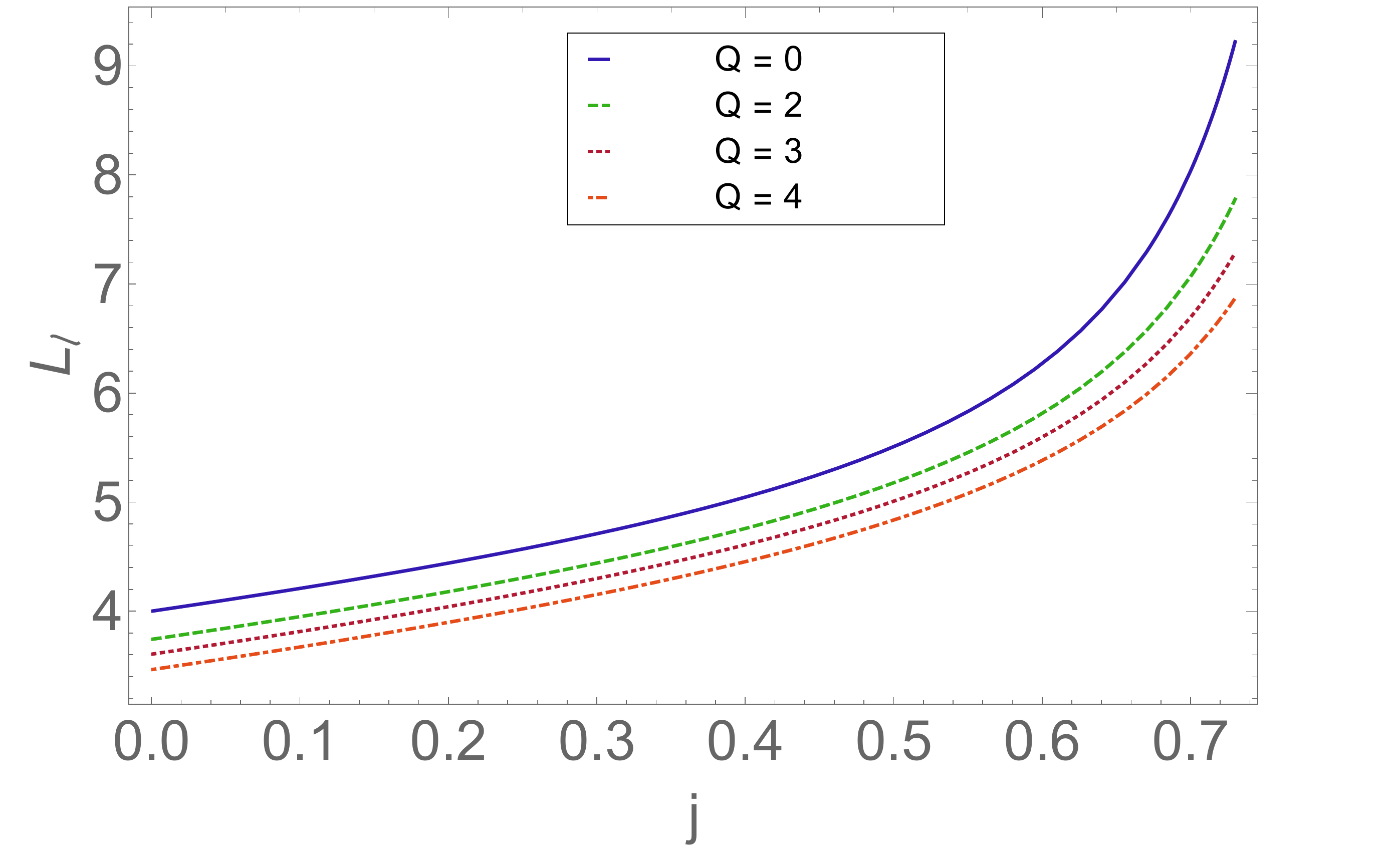}} \hspace{0.3 cm}
\subfigure[]{\includegraphics[scale=0.2]{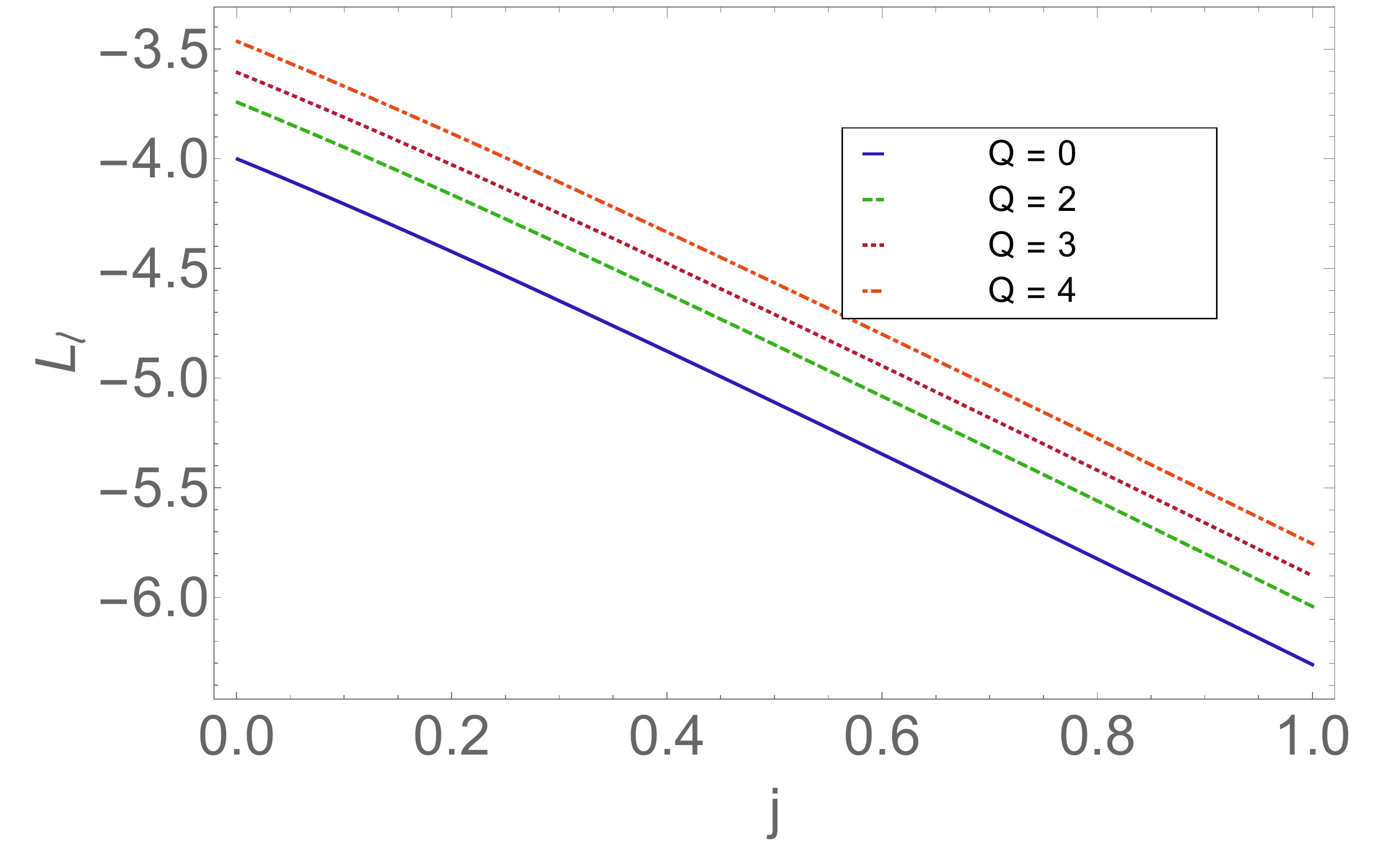}} \hspace{0.3 cm}
\caption{Loss cone angular momentum $l_{\ell} (M_{8}, j, k, Q)$ is shown as a function of $j$ for $Q$ = 0 (panels (a) and (b)) and different $Q$ values with $M_{8}$ =1 (panels (c) and (d)) for prograde (panels (a) and (c)) and retrograde (panels (b) and (d)) cases.}
\label{figllc}
\end{figure}

\normalfont
We explore the dependence of $M_{c}(j, Q)$, $r_{t} (M_{8},j, k, Q)$ and $r_{\ell}(M_{8},j, k, Q)$ in Figures \ref{figrtc}, \ref{figrta}, \ref{figllca} and \ref{figrcaQ} respectively. We observe the following:
\begin{enumerate}
\item Figure \ref{figrtc} shows the ratio $r_{t}(M_{8}, j, k, Q)$/$r_{c}(M_{8}, j, Q)$ as a function of both $j$ and $M_{\bullet}$ for $Q$ = 0. $M_{c}(j, Q) = M_{\bullet} (r_{t}$/$r_{c} = 1)$ is the critical mass, which has a dependence on $j$ and matches with the previous simulation results of \cite{2012PhRvD..85b4037K} who shows that the critical mass changes as a function of $j$, in the relativistic limit. For black holes more massive than $10^{7} M_{\odot}$, but below the critical mass, the tidal disruption occurs very close to the horizon and Newtonian treatment of the tidal interactions cannot be applied. \cite{2012PhRvD..85b4037K} calculates equatorial ($Q$ = 0) stellar orbits in the Kerr metric to evaluate the relativistic tidal tensor at the pericenter for the stars that are not directly captured by the black hole and also combine their relativistic treatment with previous calculations of the population of these orbits in order to determine tidal disruption rates for spinning black holes. They found a strong dependence of tidal disruption rates on black hole spin for $M_{8} > 1$. Our calculation of $r_{t}$ shows an increase at the high-mass end ($M_{8} > 1$) as suggested by \cite{2012PhRvD..85b4037K}. The bottom panels of Figure \ref{figrtc} show that $M_{c}(j, Q)$ is nearly flat in $Q$. 
\item Figure \ref{figllca} shows $x_{\ell}(M_{8}, j, k, Q)$ for $Q$ = 0. When the value of $x_{c} (j, Q)$ exceeds $x_{t}(M_{8}, j, k, Q)$, the stars will be directly captured instead of getting tidally disrupted; hence, $x_{\ell}(M_{8}, j, Q)$ flattens out after $M_{\bullet}> M_{c}$. $x_{\ell}(M_{8}, j, k, Q)$ for different $Q$ values are shown in Appendix \ref{Appendix} for both prograde and retrograde cases. For fixed $j$, in the retrograde case we see that $x_{\ell}(M_{\bullet}, j, k, Q)$ increases with $Q$ and decreases with $Q$ for the prograde case. $x_{\ell} (M_{8}, j, k, Q)$ for fixed $M_{\bullet}$ is nearly the same for different $Q$. This is true because as $Q$ increases, $L - L_{z}$ decreases, causing the pericenter to shrink in the prograde case, and the opposite occurs in the retrograde case.
%\item 
\item The dimensionless angular momentum at $r_{\ell}(M_{8}, j, k, Q)$ defined as $l_{\ell}(M_{8}, j, k, Q)$ [Equation (\ref{llcEquation})] is the loss cone angular momentum in the relativistic regime. Fig \ref{figllc} shows $l_{\ell}(M_{8}, j, k, Q$) for different $Q$ values, and it increases with $M_{8}$ for both prograde and retrograde cases for $Q$ = 0. For fixed $M_{8}$, $l_{\ell} (M_{8}, j, k, Q$) decreases with an increase in $Q$ for the prograde case, while it increases for the retrograde case. This can be understood from the fact that $Q$ is a measure of $L - L_{z}$; so that $L_{z}$ increases when $Q$ decreases for $L_{z} > 0$ (prograde) and $|L_{z}|$ decreases when $Q$ decreases for $L_{z} < 0$ (retrograde).
\end{enumerate}
\normalfont

{\it{Steady loss cone theory:}}
For typical masses $M_{\bullet} \gtrsim 10^{5} M_{\odot}$, the more practical case is the steady-state theory of \cite{1978ApJ...226.1087C}. By using direct numerical integration of the Fokker--Planck equation in angular momentum and energy space, they derived the stellar distribution in the presence of a black hole in a steady state. The distribution of orbital energies near the black hole can never reach a steady state because no black hole is old enough \citep{2013degn.book.....M}, as expected for the distribution of orbital angular momenta near $l_{\ell}$ because $\tau_{M, *} / t_{r} << 1$ (see Appendix \ref{timescales}, Table \ref{effects}). Therefore, a hybrid approach should be used for the calculation of event rates based on the observed distribution of energies where the angular momentum distribution at each energy has reached an approximate steady state under the influence of gravitational encounters. We use the expression of the capture rate for full loss cone theory, $\dot{N}_{f}$, given by \cite{2013degn.book.....M}, and also derive the steady loss cone theory rate, $\dot{N}_{s}$. We discuss the conditions to determine which one is more appropriate.
To see this, we examine the stellar capture in two situations, one where the loss cone gets filled quickly and another where it is dominated by diffusion. For the typical black hole mass under consideration (as $\displaystyle\frac{M_{\bullet}}{M_{\odot}}$ evolves from $10^{4}$ to $10^{8}$), the diffusive regime operates, and hence it is more appropriate to use the steady loss cone theory (see Appendix \ref{flcslc} for a detailed justification).

\cite{2015ApJ...814..141M} (hereafter MM15) have constructed a detailed model of the tidal disruption events using stellar dynamical and gas dynamical inputs like black hole mass, specific orbital energy and also angular momentum, the mass of a single star, its radius, and the pericenter of the star orbit. Using the Cohn--Kulsrud boundary layer theory, they calculated the differential rate of number of stars falling in the steady loss cone to be (MM15)
\be \frac{\diff^{2}\dot{N}_{s}}{\diff \bar{e} \diff l^{2} \diff m} = 4\pi^{2}s_{t}^{-1} \sigma^{2} \xi(m) f_{*} (\bar{e}, M_{\bullet}, m) L_{\ell}^{2} (\bar{e}) F(\chi = 1, l),\label{stdeq} \ee
where $s_{t} = r_{t} (M_{\bullet}, j)/ r_{h}$, $\bar{e} = E / (GM_{\bullet}/r_{t})$, $E$ is the energy, $f_{*}$ is the probability that a star of mass $m$ is tidally captured as a main sequence, and $\xi(m)$ is the stellar mass function where $m = m_{*} / M_{\odot}$, $F = X(y_{lc})\zeta(q_{s})$, and
\be X(y_{lc}) = \frac{f_{s}(E)}{1 + q_{s}^{-1} \zeta(q_{s}) \log(1 / y_{\ell})}, \ee
with $\displaystyle q_{s} = \frac{<D(E)>}{y_{\ell}}$ and $\displaystyle y_{\ell} = \frac{L_{\ell}^{2}}{J_{c}^{2}}$ where $f_{s}(E)$ is the distribution function of stars in the galaxy. $<D(E)>$ is the orbit-averaged angular momentum diffusion coefficient, and $J_{c}$ is the angular momentum of circular orbit. MM15 have used the $M_{\bullet}$--$\sigma$ relation (taking $p = 4.86$) to get the expression for $\dot{N}_{s}$. By applying the steady-state Fokker-Planck equation while using a power-law stellar density profile (having power-law index $\gamma$), they obtained the rate of capture of stars $\dot{N} \propto M_{\bullet}^{\beta}$ , where $\beta$ = -- 0.3 $\pm 0.01$ for $M_{6} > 10$ and the value of $\dot{N}_{s}$ is $\sim 6.8 \times 10^{-5}$ ${\rm yr}^{-1}$ for $\gamma$ = 0.7. We apply the same technique to calculate $\dot{N}_{s}$, including the relativistic forms of $r_{t}$ and $r_{c}$, but do not assume the $M_{\bullet}$--$\sigma$ relation a priori and consider $\sigma$ as an independent parameter in our model.
We start from the basic equation for $N_{s}$ given by
\be
N_{s} = 4\pi^{2} \int  P(E) \diff E \int f_{s} (E, J) \diff J^{2}, 
\ee
where $P(E)$ is the orbital period. We use the same expressions and parameters given in MM15 with the following assumptions: \be \int \xi(m)\diff m = 1, ~ ~ f_{*} = 1, \ee  and 
\be <D(\epsilon_{s})> = \frac{32\sqrt{2}}{3} \frac{\pi^{2}G^{2}<m_{f}^{2}>\log \Lambda}{J_{c}^{2}}\frac{M_{\bullet}}{<m_{*}>}\frac{1}{\sigma^{2}}[2h_{1}(\epsilon_{s}) + 3h_{2}(\epsilon_{s}) - h_{3} (\epsilon_{s})], \ee
where $m_{f}$ is the mass of the field star, with the maximum mass taken to be 150 $M_{\odot}$, $\Lambda \approx M_{\bullet} / m_{*}$ $\epsilon_{s} = E / \sigma^{2}$, and $h_{1}$, $h_{2}$ and $h_{3}$ are defined in MM15. Now,
\be J_{c}^{2} = \sigma^{2} r_{h}^{2} [s_{c} (\epsilon_{s}) + 2 s_{c}^{4 - \gamma}(\epsilon_{s})], \ee
where $s_{c}$ is the ratio of the radius of circular orbit and the horizon radius, the $\epsilon_{s}$-dependent part is called as $\beta(\epsilon_{s})$, and
\be L_{\ell} (M_{\bullet}, j, k, Q)= \frac{GM_{\bullet}}{c} l_{\ell} (M_{\bullet}, j, k, Q). \ee
Therefore,
\be y_{\ell} (M_{\bullet},j, k, Q, \epsilon_{s}) = \displaystyle \frac{L_{\ell}^{2}}{J_{c}^{2}} = \frac{L_{\ell}^{2}(M_{\bullet}, j, k, Q)}{\sigma^{2}r_{h}^{2}\beta(\epsilon_{s})}, \ee
where $\beta(\epsilon) = [s_{c} (\epsilon_{s}) + 2 s_{c}^{4 - \gamma}(\epsilon_{s})]$. From the definition, $q_{s}(\epsilon_{s})$ is written as
\be q_{s}(\epsilon_{s}) = \frac{<D(\epsilon_{s})>}{y_{\ell}}, \ee
which can be simplified to
\be q_{s}(M_{\bullet},j, k, Q, \epsilon_{s}, \sigma) = \frac{32\sqrt{2}}{3} \frac{\pi^{2} <m_{f}^{2}>}{M_{\bullet} <m_{*}>}\frac{\log \Lambda ~\sigma^{2}r_{h}^{2}}{L_{\ell}^{2} (M_{\bullet}, j, k, Q)} [2h_{1}(\epsilon_{s}) + 3h_{2}(\epsilon_{s}) - h_{3} (\epsilon_{s})].\ee
The expression $\zeta(q_{s})$, as given by MM15 is\\
\be
\displaystyle
\zeta(q_{s}) = \left\{\begin{array}{lr}
        1 & \text{\rm for~} q_{s} \geq 4 \\
        q_{s} / (0.86 q_{s}^{0.5} + 0.384 q_{s} - 0379 q_{s}^{1.5} + 0.427 q_{s}^{2} - 0.095 q_{s}^{2.5}) & \text{\rm otherwise}.
        \end{array}\right.
\ee

By integrating Equation (\ref{stdeq}) assuming $f_{*}$ =1, we finally arrive at
\be
\frac{\diff N_{s}}{\diff \epsilon_{s}} (M_{\bullet},j, k, Q, \epsilon_{s}, \sigma) = \frac{\sqrt{2}\pi^{3} L_{\ell}^{2} (M_{\bullet}, j, k, Q) \sigma^{2} \epsilon_{s}^{-\frac{3}{2}}}{G^{2}M_{\bullet}<m_{*}>} g(\epsilon_{s}) \frac{\zeta(q_{s})}{1 + q_{s}^{-1} \zeta(q_{s}) \log(1 / y_{\ell})}.\label{ne}
\ee
Then, dividing Equation (\ref{ne}) by the orbital period $P(\epsilon_{s})$, we find an expression of $\displaystyle \frac{\diff \dot{N}_{s}}{\diff \epsilon_{s}}$ as
\be 
\frac{\diff \dot{N}_{s}}{\diff \epsilon_{s}} (M_{\bullet},j, k, Q, \epsilon_{s}, \sigma) = \frac{4\pi^{2} L_{\ell}^{2} (M_{\bullet}, j, k, Q) \sigma^{5}}{G^{3}M_{\bullet}^{2}<m_{*}>} g(\epsilon_{s}) \frac{\zeta(q_{s})}{1 + q_{s}^{-1} \zeta(q_{s}) \log(1 / y_{\ell})}., \label{ndtsd_mt}
\ee
where we have used the relativistic approximation to $r_{\ell} (M_{\bullet}, j, Q)$ to obtain $\dot{N}_{s}(M_{\bullet},j, k, Q, \epsilon_{s}, \sigma)$ in the relativistic limit as a function of the black hole spin. Since the diffusion occurs at very large radius, only the first term of the effective potential dominates. 
Integrating this expression numerically, we finally find the rate of capture of stars for the case of the steady loss cone, and Figure \ref{stdlc} shows the resulting variation of $\dot{N}_{s} (M_{\bullet},j, k, Q, \epsilon_{s}, \sigma)$ with the $M_{\bullet}$. $\dot{N}_{s}(M_{\bullet},j, k, Q, \epsilon_{s}, \sigma)$ has very little dependence on $Q$; our results for $Q = 4$ are similar to Figure \ref{stdlc} for $Q = 0$. Therefore, $\dot{N}_{s}(M_{\bullet},j, k, Q, \epsilon_{s}, \sigma)$ is nearly independent of the value of $Q$.
\begin{figure}[H]
\begin{center}
\subfigure[Prograde]{\includegraphics[scale=0.25]{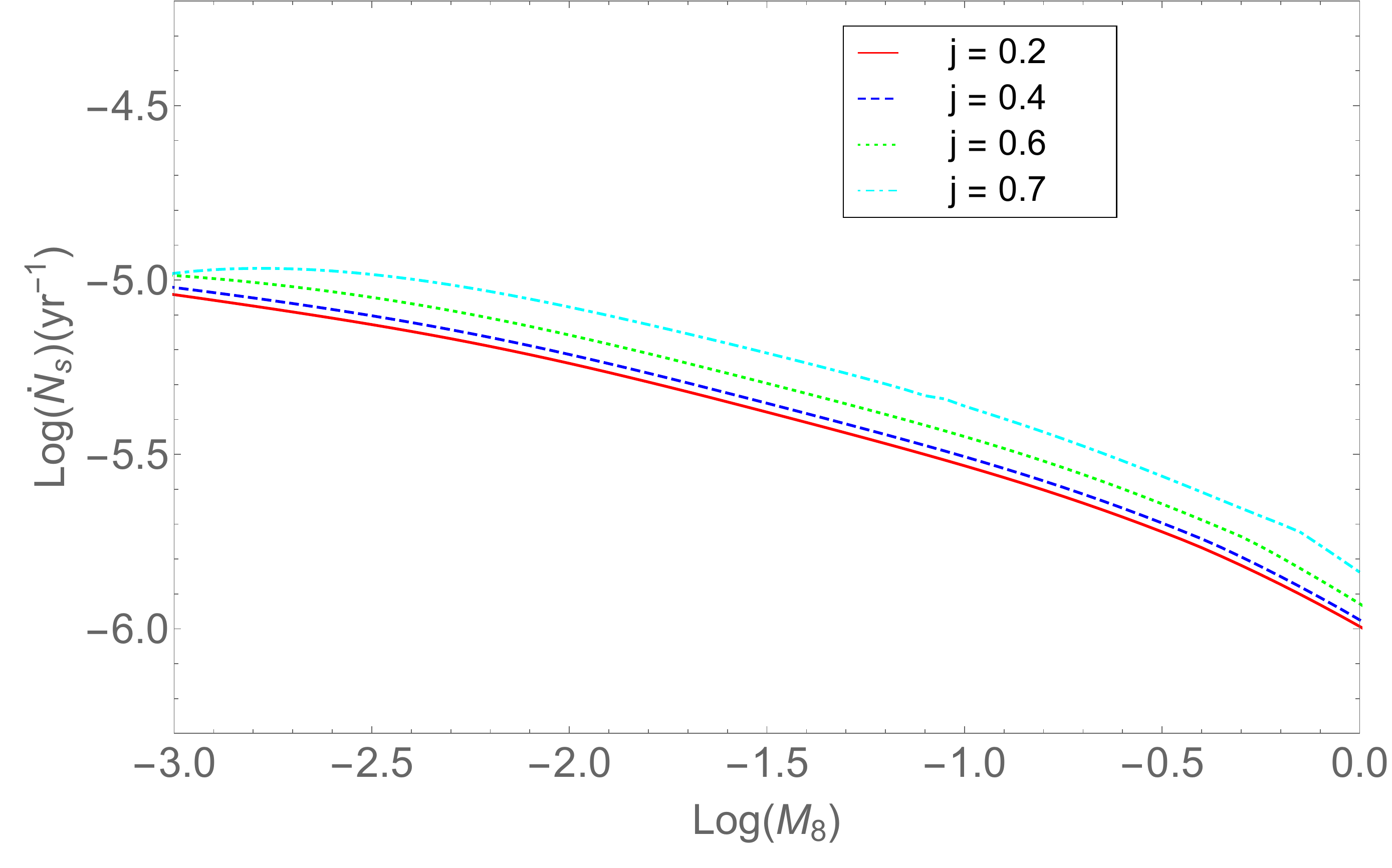}}
\subfigure[Retrograde]{\includegraphics[scale=0.25]{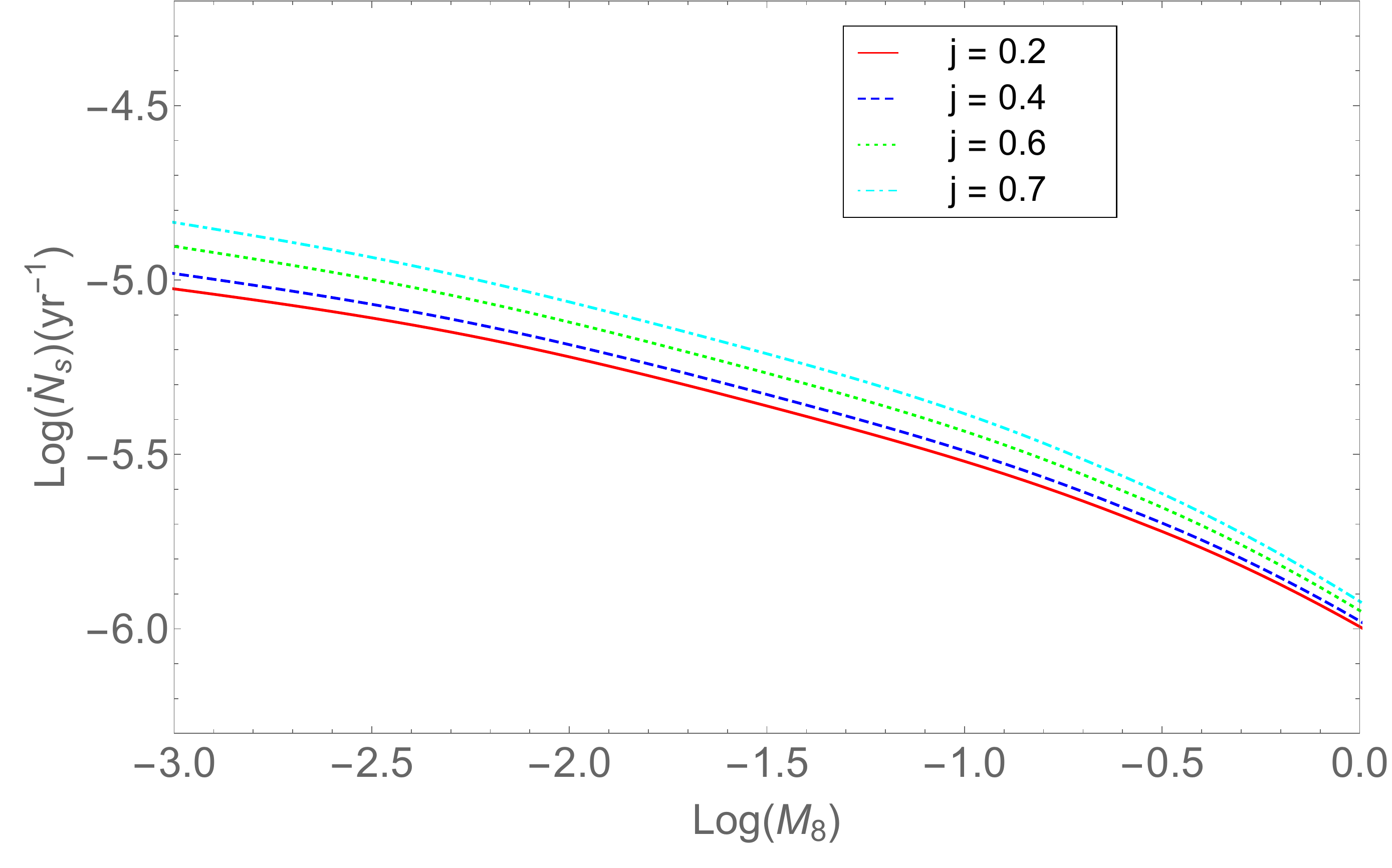}}\\
%\subfigure[Prograde]{\includegraphics[scale=0.25]{stdlcndk1Q.pdf}}
%\subfigure[Retrograde]{\includegraphics[scale=0.25]{stdlcndk0Q.pdf}}
\caption{Capture rate, $\dot{N}_{s}(M_{\bullet},j, k, Q, \epsilon_{s}, \sigma)$, is shown which reduces monotonically with $M_{8}$ and increases slightly with $j$ when $k$ = 1 (panel (a)) and --1 (panel (b)) for $Q$ = 0, %(top) and (bottom) the same plots for $Q$ = 4 
where the lower limit of the $\epsilon_{s}$ integration is taken to be $\epsilon_{m}$ = -10, $\gamma$ = 1.1, and $\sigma$ = 200 km s$^{-1}$.}
\label{stdlc}
\end{center}
\end{figure}
The monotonic decrease of $\dot{N}_{s}$ can be explained by the decrease of $l_{\ell}$ with $M_{8}$. From Figure \ref{stdlc}, we see that $\dot{N}_{s}(M_{\bullet},j, k, Q, \epsilon_{s}, \sigma)$ increases with an increase in $j$. When we apply the $M_{\bullet}$--$\sigma$ relation taking $p$ = 4.86, we find that the capture rate, $\dot{N}_{s}(M_{\bullet},j, k, Q, \epsilon_{s}, \sigma)$ (see Figure \ref{stdlcmsigma}), follows a similar trend to that of \cite{2012PhRvD..85b4037K} (see Figures 3, 4), where $\dot{N_{s}}$($j$) increases with $j$ with the assumption of the universal $M_{\bullet}$--$\sigma$ relation. For higher $\gamma$, $\dot{N}_{s}$ increases for both the prograde and retrograde cases, which is similar to the result of MM15. The difference between the full and steady loss cone capture rates is discussed in Appendix \ref{capturerate}. We derive the mass evolution only in the presence of stellar capture, and the result obtained is in rough agreement with that of \cite{2017NatAs...1E.147A}.
\begin{figure}[H]
    \centering
    \includegraphics[scale=0.25]{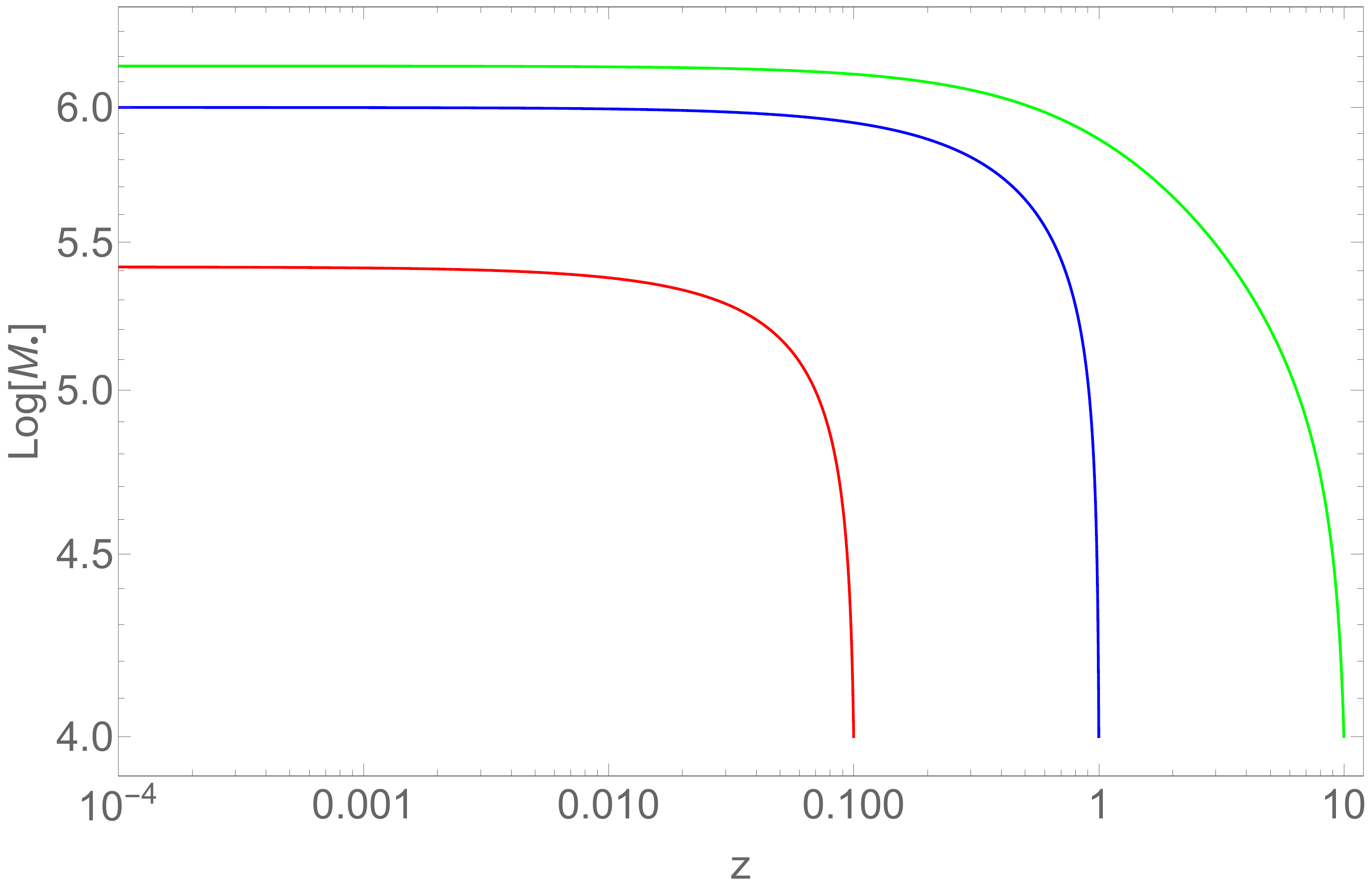}
    \caption{Mass evolution only in the presence of stellar capture is shown for a seed mass of $10^{4} M_{\odot}$, $\gamma$ = 1.1, and at different formation redshifts (red is for $z_{f}$ = 0.1, blue is for $z_{f}$ = 1, green is for $z_{f}$ = 10).}
    \label{tal_ev}
\end{figure}
Figure \ref{tal_ev} is similar to the result of \cite{2017NatAs...1E.147A} [see Figure 2] under the same conditions.

\subsection{\it{Growth of the Black Hole by Mergers}}\label{2.3}
The black holes can grow their mass also by the merger process, though the rate is generally much smaller compared to accretion, while minor mergers are more probable than the major mergers. When the accretion process stops owing to saturation, the dominant contribution to mass growth of the black hole comes from the effect of mergers. We compute the mass growth rate by merger activity by integrating the merger rate over the mass of the smaller black hole. \cite{2009ApJ...702.1005S} use high-resolution $\Lambda {\rm CDM}$ $N$-body simulations for predicting merger rates in dark matter halos and investigate the scaling of common merger-related observables with luminosity, stellar mass, merger mass ratio, and redshift $z$ = 4 $\rightarrow$ 0. They derive the expression for merger rate (infall) valid for $0\leq z \lesssim 4$ considering the peak of merger activity; the dependence on different parameters has been determined using simple fitting functions. They developed simulations that contained 512 particles of mass $3.16 \times 10^{8} h^{-1} M_{\odot}$ which was evolved within a comoving volume of $80 h^{-1} $ Mpc on a side by the Adaptive Refinement Tree (ART) $N$-body code developed by \cite{1997ApJS..111...73K, 2004ApJ...609...35K}. We use the rate of mergers given in \cite{2009ApJ...702.1005S} and integrate it over the mass of the smaller black hole to find $\dot{M}_{\bullet m}$. Following their assumptions, we also consider the merger activity to be valid in the range $z = 4 \rightarrow 0$.
In \cite{2009ApJ...702.1005S} the empirical expression for merger rate (infall)  is
\be 
\frac{\diff N_{m}}{\diff t} (m/M \in (0.1, 0.7)) = A_{t} (z, M) F(m / M),
\ee
where $m$ and $M$ are the masses of the smaller and larger merging galaxies and $N_{m}$ is the number of mergers,
\be 
A_{t} (z, M) = 0.02 {\rm Gyr ^{-1}} (1 + z)^{2.2} M_{12}^{b},
\ee
with $b = 0.15$ and $M_{12}$ = $M$ / $10^{12} h^{-1} M_{\odot}$ with $h = 0.7$ that is valid for $0\leq z\lesssim 4$. Adopting this, the rate of mass growth due to mergers is given as
\be 
\frac{\diff M}{\diff t} = A_{t} M \int_{q}^{1} F(q) dq = A_{t} M n(q), \label{nq}
\ee
where $q = m / M$, and $F(q)$ is given as
\be 
F(q) = q^{-c} (1 - q)^{d},
\ee
where $c$ = 0.5 and $d$ = 1.3 and $n(q)$ can be written as a combination of complete and incomplete Beta functions, where the complete and incomplete Beta functions are defined, respectively, as
\be
B (x, y) \equiv \int_{0}^{1} t^{x - 1} (1 - t)^{y - 1} \diff t,
\ee
and
\be
B_{z} (x, y) = \int_{0}^{z} t^{x - 1} (1 - t)^{y - 1} \diff t \equiv \frac{z^{x}}{x} {}_{2}F_{1}(x,1 - y; x + 1; z)
\ee
As a result, we can express
\be
n (q) = B (1 - c, 1 + d) - B_{q} (1 - c, 1 + d),
\ee
so that the merger mass rate becomes
\be 
\dot{M}_{\bullet m} = 8.058 \times 10^{-3} (1 + z)^{2.2} \bigg[\frac{M_{5}}{f_{h}}\bigg]^{1.15} n(q) M_{5}~ 10^{5}M_{\odot} / {\rm Gyr}; 
\ee
In units where $\displaystyle \mu_{\bullet} = \frac{M_{\bullet}}{M_{s}}$, where $M_{s}$ is the seed mass, $\displaystyle \tau = \frac{t}{t_{0}}$, where $t_{0}$ = 1 Gyr, this can further be expressed as
\be 
\displaystyle \dot{\mu}_{m} (q, M_{s}, z, z_{f}) = \frac{\dot{M}_{\bullet m} t_{0}}{M_{s}} = \frac{8.058 \times 10^{-3} (1 + z)^{2.2} \bigg[\frac{M_{5}}{f_{h}}\bigg]^{1.15} n(q) M_{5}} {M_{s5}}, \label{mg_mt}
\ee 
where $f_{h} = M_{\bullet} / M$ and $M_{5}$ is the mass of the SMBH in units of $10^{5} M_{\odot}$ which simplifies to
\be 
M_{\bullet 5} (q, M_{s}, z, z_{f}) = \bigg[ M_{s5}^{0.15} - 1.21 \times 10^{-3} \int_{z_{f}}^{z} (1 + z)^{2.2} n(q) \frac{\diff t}{\diff z} (z) \bigg]^{-\frac{20}{3}},
\ee
where we have used $f_{h}$ = 3 $\times 10^{-5}$, $\displaystyle \frac{\diff t}{\diff z}$ is given by Equation (\ref{tzcm}), and $z_{f}$ is the formation redshift. For simplicity, we assume a proportionality relation, $M_{\bullet} = f_{h}M$, while \cite{2002ApJ...578...90F} and \cite{2011ApJ...734...92J} have assumed the relation to be slightly nonlinear, with the index of the relation dependent on the choice of the dark matter profile. Furthermore, $f_{h}$ increases to $2 \times 10^{-4}$ for halo masses of $\sim 10^{14} M_{\odot}$ and decreases to $10^{-5}$ for halo masses of $\sim 10^{12} M_{\odot}$ \citep{2002ApJ...578...90F}. Therefore, as a reasonable approximation, we assume a mean value of $f_{h}$ in our model. The frequency of major mergers is much less than the frequency of minor ones \citep{2009ApJ...702.1005S}. \cite{2004ApJ...602..312G} consider the collapse of stars, accretion, and major and minor mergers that contribute to the spin of the astrophysical black holes. Major mergers contribute to spinning up the hole, whereas minor mergers contribute to spinning it down \citep{2004ApJ...602..312G}. Since accretion is dominant for spinning up of the black hole, we consider only the contribution of minor mergers for spin-down and neglect the major mergers for the spin evolution of the hole in our model. Different models suggest that black holes produced by the collapse of a supermassive star are likely to have $j \sim$ 0.7. Though the result of major mergers is not yet known, \cite{2004ApJ...602..312G} provide some current estimates and analytic bounds on $j$ for these processes. They apply the formalism of \cite{2003ApJ...585L.101H} to minor mergers assuming an isotropic distribution of orbital angular momentum and find that the spin-down occurs with $j \sim M^{7 /3} $ and evaluate a power law for spin decay for the limit of the small value of $j$, by expanding the radius and specific energy of ISCO as a function of $j$. Their simulations for accretion process from fully relativistic MHD flow indicates a spin equilibrium at $j \sim 0.9$, much less than the canonical value 0.998 of \cite{1974ApJ...191..507T} that was derived excluding the MHD effects. This suggests the possibility that the black holes that grow mainly by the accretion process are not maximally rotating. We use the spin-down term by minor mergers given by \cite{2004ApJ...602..312G} in our evolution model to be valid in the range $z = 4 \rightarrow 0$, which causes a significant decrease in the final spin value. \cite{2004ApJ...602..312G}, by taking the effect of minor mergers on spin evolution of the black hole, find
\be
\frac{\diff \log j}{\diff \log M_{\bullet}} = -\frac{7}{3} + \frac{9q}{\sqrt{2}j^{2}},
\ee
which can be written as
\be
\frac{\diff j}{\diff \tau} = \dot{\mu}_{m} \cdot \frac{j}{\mu_{\bullet}}\bigg(-\frac{7}{3} + \frac{9q}{\sqrt{2}j^{2}}\bigg).\label{mg_jt}
\ee
%Figure \ref{mergerm} shows the mass evolution of the black hole in the presence of only merger activities for different $z_{f}$. 
The merger term dominates after the accretion stops, which happens after the black hole reaches saturation. We have used $q$ = 0.1, as $q>0.1$ implies major mergers. But the frequency of major mergers is much less than the minor ones, and the growth rate by major mergers is almost of the same order for different $q$ values. We see that the mass growth due to mergers is significantly smaller in this case compared to the gas accretion, and hence we consider only the minor mergers, as they are more frequent.

\subsection{\it{Recipe for the Electromagnetic Spin-down of the Black Hole}}\label{2.4}

If magnetic field lines are present in a rotating black hole supported by external currents that are flowing in an equatorial disk, there will be an induced electrical potential difference. For large field strengths, the vacuum will be unstable to the cascade production of an electron--positron pair creating a force-free magnetosphere, leading to an electromagnetic extraction of energy and angular momentum. \cite{1977MNRAS.179..433B} have derived an approximate solution for spinning black holes to provide a model of the central engine of the AGN. The advantage of this model is that the relativistic electrons can be accelerated efficiently compared to other models. We include the BZ effect for causing the spin-down of the hole in our model. The spin-down due to BZ torque is implemented by Equation (14) in \cite{2009MNRAS.397.2216M}, where they study the case of rapid loss of cold gas due to AGN feedback, which may cause expansion in the effective radii of massive elliptical galaxies from $z\simeq 2$ to 0; they quantify the extent of the expansion in terms of the star formation parameters and time of the expulsion of the cold gas; and they show that cosmological changes are expected to have a major influence on the gas accretion mode, which at high redshifts can be dominantly cold thin disk accretion and at low redshifts could be dominantly hot Bondi-fed ADAF accretion. They calculate the spin-down to be $\tau_{j} \sim M_{9}^{2} 0.2$ Gyr, which explains the cosmological evolution of the luminosity function from powerful to weak radio galaxies. We use the expression of spin evolution caused by BZ torque as implemented in \cite{2009MNRAS.397.2216M} in our model. We calculate the spin evolution by the BZ effect for different initial and final spin values \citep{2009MNRAS.397.2216M} from
\be \frac{\diff j}{\diff t} = x_{+}^{3}(j)j\frac{\mathcal{G}_{0}}{\mathcal{J}_{0}},\ee 
where $x_{+}(j) = 1 + \sqrt{1 - j^{2}}$ and the  BZ torque, $\mathcal{G}_{0}$, is given by \be \mathcal{G}_{0} = \frac{m^{3}}{8} B_{\perp}^{2} f_{BZ} = 4 \times 10^{46} f_{BZ} B_{4} M_{8}^{3} ({\rm erg}),\ee 
and the angular momentum budget, $\mathcal{J}_{0}$ is \be \mathcal{J}_{0} = cM_{\bullet}m = 9 \times 10^{64} M_{8}^{2} ({\rm g~cm^{2}~s^{-1}}),\ee
and where $\displaystyle B_{4} = B / 10^{4} \rm Gauss$, $f_{{\rm BZ}}$ is a geometric factor that comes from the averaging of the angle over the horizon of magnetic flux and the spin of the magnetic field \citep{2009MNRAS.397.2216M}. Therefore, in dimensionless form
\be
\frac{\diff j}{\diff \tau} = \frac{4}{9}\times 10^{-5} f_{BZ} B_{4} \mu_{\bullet} M_{s5}x_{+}^{3}(j)j, \label{bz_jt}
\ee
whose analytic solution of spin-down time $\tau_{j, BZ}(j)$ is given by \citep{2009MNRAS.397.2216M}
\be
\displaystyle
\tau_{j, BZ} = \frac{\mathcal{J}_{0}}{\mathcal{G}_{0}} \int_{j_{f}}^{j_{i}} \frac{\diff j}{r^{3}(j) j} = 7.0 \times 10^{8} {\rm yr} \frac{(\kappa (j_{i}, j_{f}) / 0.1)}{B_{4}^{2} M_{9} f_{BZ}},
\ee
where $M_{9} = M_{\bullet} / (10^{9} M_{\odot})$ and 
\be
\kappa (j_{i}, j_{f}) = \bigg[\bigg(\frac{1}{16}\bigg) \log \bigg(\frac{2 - w}{w}\bigg) + \bigg(\frac{3 w^{2} + 3 w -4}{24 w^{3}} \bigg)\bigg]_{w_{f}}^{w_{i}},
\ee
with $w_{i} = x_{+}(j_{i})$, $w_{f} = x_{+} (j_{f})$. %Fig \ref{BZ_j} shows the spin down due to BZ effect for a certain set of parameters. 
This is equivalent to the study of the spin-down for the Bondi case with zero accretion in \cite{2015ASInC..12...51M}.

\begin{figure}[H]
    \centering
    \includegraphics[scale = 0.65]{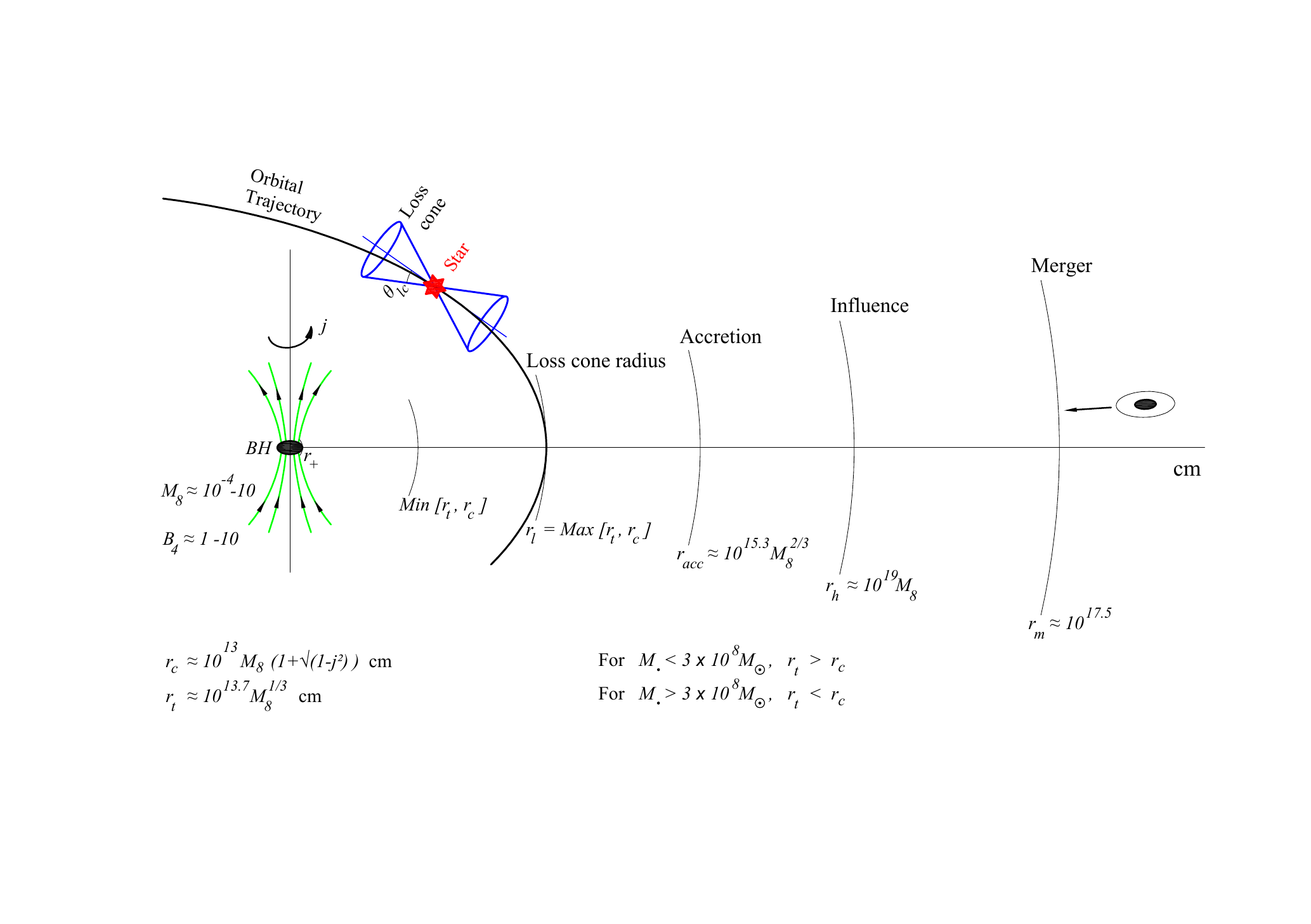}
    \caption{Important radii corresponding to all the processes contributing to the growth of the black hole is shown for $M_{\bullet} = 10^{4} - 10^{6} M_{\odot}$.}
    \label{diagram}
\end{figure}

\begin{table}[H]
\begin{center}
\begin{tabular}{|c |c |c |c |}\hline
Effects & Region & $\tau_{j}$ & $\tau_{M}$  \\ \hline\hline
Gas accretion & $r_{I} - r_{d}$ & 1 Gyr & 1 Gyr\\ \hline
Stellar capture & $r_{t} - r_{h}$  & - & $10$ Gyr\\ \hline
Mergers & $r_{M}$  & 10 Gyr & $\sim$ 10 Gyr\\ \hline
BZ torque & $r_{+} - r_{I}$  & 1 Gyr & - \\ \hline\hline 
\end{tabular}
\end{center}
\caption{The Domain and Timescales for Different Physical Effects (Shown in Figure \ref{diagram}) Contributing to the Growth of the Black Hole.}
\label{effects}
\end{table}
We indicate operative time scales in different physical regimes of gas accretion, stellar capture, mergers, and electromagnetic torque in Table \ref{effects}, where the evolution timescales for mass, $\tau_{M}$, and spin, $\tau_{j}$, are calculated in Appendix \ref{timescales}. It is clear that the evolution timescales of both mass and spin are of order 1--10 Gyr. This motivates us to use $t_{0}$ = 1 Gyr as the unit of time in our model. 

\section{Black hole evolution model in $\Lambda {\rm CDM}$ cosmology} \label{3}
The black hole growth can occur by gas flow and by capture of stars and mergers until it reaches a saturated mass $M_{\bullet t}$ at a time $t = t_{s}$ when the gas flow stops and it grows only by the capture of stars and mergers. This happens because the outflow velocity exceeds the escape velocity of the medium and the gas is driven away, causing the accretion process to stop. The saturated mass is given by \cite{2003ApJ...596L..27K} as 
\be M_{\bullet t} = 9.375\times 10^{6} \sigma_{100}^{4}M_{\odot}\label{satm}. \ee 
For the rate of growth of mass by mergers, we use Equation (\ref{nq}), which is valid from $z$ = 4 to the present time, given that the merger activity peaks at $z$ = 5--0.5 \citep{2009MNRAS.395.1376W}. We perform five experiments that we discuss in this section. For the mass evolution, we consider the contribution from both major mergers and the minor ones, and for the spin evolution, we consider only the contribution from the minor mergers to spinning down the black hole \citep{2004ApJ...602..312G}, as the contribution of the accretion process in spinning up the black hole is much higher than the contributions from mergers. But due to a smaller frequency of the major mergers, the final mass attained by the merger process does not vary significantly with the value of $q$. We have dealt with two scenarios: (i) for $z_{f} \lesssim 4$ the contribution of the mergers will be present throughout; and (ii) for $z_{f} \gtrsim 4$, initially, there will be only accretion and stellar capture, and mergers will come into play later than $z$ = 4; from then on until $t_{s}$, all the three terms will contribute, after which the accretion stops.  
\\
To summarize, our model is based on some assumptions and conditions:
\begin{enumerate}
\item Black hole seeds are formed at look-back times of the order of the Hubble time.
%\item The value $\sigma$ is constant from formation redshift till present time. {\bf We have assumed $\sigma$ to be a constant throughout since its variation is relatively small and reduces over Hubble time by a factor $\sim$ 15$\%$ (\cite{2009ApJ...694..867S}, Figure 8). We have discussed this in \S 6.}
\item At the saturation time, the mass reaches $M_{\bullet t} = 9.375\times 10^{6} \sigma_{100}^{4}M_{\odot}$ and $p \rightarrow$ 4.
%\item The value $\sigma$ is constant from formation redshift till present time. {\bf We have assumed $\sigma$ to be a constant throughout since its variation is relatively small and reduces over Hubble time by a factor $\sim$ 15$\%$ (\cite{2009ApJ...694..867S}, Figure 8). We have discussed this in \S 6.}
\item The merger activity exists only for $z \lesssim 4$ \citep{2009ApJ...702.1005S}.
%\item For the evolution of the $M_{\bullet}$--$\sigma$ relation, we assume that at formation the Faber - Jackson relation which implies $M_{\bullet} \propto \sigma^{5}$ holds and then we derive its subsequent evolution.
\end{enumerate}

As before, we have normalized mass by $\displaystyle \mu_{\bullet} = \displaystyle \frac{M_{\bullet}}{M_{s}}$, where $M_{s}$ is the seed mass, and time by $\displaystyle \tau = \frac{t}{t_{0}}$, where $t_{0}$ = 1 Gyr. The mass evolution equation is given by Equation (\ref{mtdimintro}) as
\be
\frac{\diff M_{\bullet}}{\diff t} = \epsilon_{I}(j) \dot{M}_{\bullet g}+ \epsilon(j) \dot{M}_{\bullet *} + \dot{M}_{\bullet m},\label{mtdim}
\ee

where $\epsilon (j)$ is the mass accretion efficiency given by
\be 
        \epsilon(j) = \left\{\begin{array}{lr}
        \epsilon_{I}(j) & \text{\rm for~} M_{\bullet} < M_{c} \\
        1 & \text{\rm for~} M_{\bullet} \geq M_{c},
        \end{array}\right. \label{acceff}
\ee 
\\where \citep{1972ApJ...178..347B}
\be
\epsilon_{I}(j) = \frac{z_{m}^{2}(j)-2z_{m}(j)+j\sqrt{z_{m}(j)}}{z_{m}(j)(z_{m}^{2}(j)-3z_{m}(j)+2j\sqrt{z_{m}(j)})^{1/2}},
\ee
and 
\be
z_{m}(j)=  \frac{r_{ms}}{M_{\bullet}} = 3+Z_2-k\sqrt{(3-Z_1)(3+Z_1+2Z_2)},
\ee
with $Z_1=1+(1-j^2)^{1/3}((1+j)^{1/3}+(1-j)^{1/3})$ and $Z_2=(3j^2+Z^2_1)^{1/2}$ \citep{1972ApJ...178..347B}. The dimensionless equation becomes 
\be
\frac{\diff \mu_{\bullet}}{\diff \tau} = \epsilon_{I}(j) \dot{\mu}_{g}+ \epsilon(j) \dot{\mu}_{*} + \dot{\mu}_{m}.\label{mt}
\ee
The first term on the right-hand side of Equation (\ref{mt}) represents the gas accretion and stems from Equation (\ref{macc}), the second term due to the stellar capture is calculated from Equation (\ref{ndtsd_mt}) and represented below by Equation (\ref{mdotst}), and the third term comes from the contribution of mergers provided by Equation (\ref{mg_mt}). In Equation (\ref{acceff}), we see that $\epsilon(j)$ is given by efficiency at ISCO for $M_{\bullet} < M_{c}$ and 1 for $M_{\bullet} > M_{c}$. This is because, beyond the critical mass, the stars are directly captured with the efficiency of 1, while for $M_{\bullet} < M_{c}$, the gas enters through ISCO (by accretion of tidally disrupted stars) with efficiency, $\epsilon_{I}$. The gas accretion is through ISCO, but, the stars can disrupt and enter by gas accretion through ISCO, as well as by direct capture.

The spin evolution equation of black holes taking into account gas accretion, stellar capture, mergers, and BZ torque is given by Equation (\ref{jtdimintro}) as (see \S \ref{2} for the various terms)
\be \displaystyle
\frac{\diff j}{\diff t} = \frac{\dot{M}_{\bullet g}}{M_{\bullet}} \bigg(l_{I} (j) - 2 \epsilon_{I}(j)j\bigg) + \frac{\dot{M}_{\bullet *}}{M_{\bullet}} \bigg(l_{*} (j) - 2 \epsilon(j)j\bigg) + \dot{M}_{\bullet m} \cdot \frac{j}{M_{\bullet}}\bigg(-\frac{7}{3} + \frac{9q}{\sqrt{2}j^{2}}\bigg) + x_{+}^{3}(j)j\frac{\mathcal{G}_{0}}{\mathcal{J}_{0}}. \label{jtdim}
\ee

The dimensionless version of Equation (\ref{jtdim}) is 
\be \displaystyle
\frac{\diff j}{\diff \tau} = \frac{\dot{\mu}_{g}}{\mu_{\bullet}} \bigg(l_{I} (j) - 2 \epsilon_{I}(j)j\bigg) + \frac{\dot{\mu}_{*}}{\mu_{\bullet}} \bigg(l_{*} (j) - 2 \epsilon(j)j\bigg) + \dot{\mu}_{m} \cdot \frac{j}{\mu_{\bullet}}\bigg(-\frac{7}{3} + \frac{9q}{\sqrt{2}j^{2}}\bigg) + \frac{4}{9}\times 10^{-5} f_{BZ} B_{4} \mu_{\bullet} M_{s5}x_{+}^{3}(j)j.\label{jt}
\ee
where \be 
    \displaystyle
   \dot{\mu}_{g} (M_{\bullet}, M_{s})= \left\{\begin{array}{lr} \displaystyle
       \frac{\dot{M}_{\bullet g} t_{0}}{M_{s}} =  \frac{k_{1}M_{\bullet} t_{0}}{M_{s}} & \text{\rm for~} M_{\bullet} \leq M_{\bullet t}\\
       0 & \text{\rm for~} M_{\bullet} > M_{\bullet t},\label{acceq2}
        \end{array};~ \displaystyle k_{1} = \frac{4\pi G m_{p}\eta}{\sigma_{e}c}\right.
\ee 
\be 
\displaystyle \dot{\mu}_{*} (M_{\bullet},j, k, Q, \epsilon_{s}, \sigma, M_{s})= \frac{\dot{M}_{\bullet *} t_{0}}{M_{s}} = \left\{\begin{array}{lr} m_{\star}\dot{N}_{f} t_{0}/ M_{s} & \text{\rm for full loss cone}\\
                   m_{\star}\dot{N}_{s} t_{0}/ M_{s} & \text{\rm for steady loss cone}\end{array}\right. \label{mdotst},
\ee 
where $\dot{N}_{f}$ and $\dot{N}_{s}$ are the stellar capture rates derived for full or steady loss cone theories. We define
\be 
    \displaystyle
   \dot{\mu}_{m} (M_{\bullet}, q, M_{s}, z, z_{f})= \left\{\begin{array}{lr} \displaystyle
       \frac{\dot{M}_{\bullet m} t_{0}}{M_{s}} & \text{\rm for~} z \leq 4\\
       0 & \text{\rm for~} z > 4\label{meq2}
        \end{array},\right.
\ee 
where $M_{s5}$ is the mass of the seed black hole in units of $10^{5} M_{\odot}$; for our calculations we have used $f_{{\rm BZ}}$ = 1. The first term on the right-hand side of Equation (\ref{jt}) for gas accretion stems from Equation (\ref{jacc}) [which shuts off after saturation, as implemented in Equation (\ref{acceq2})]. The second term represents the stellar capture, which can happen in two ways: by tidal disruption (for $M_{\bullet} < M_{c}$) when the gas has to pass through ISCO with an angular momentum and efficiency at ISCO, or by a direct capture (for $M_{\bullet} > M_{c}$), when it will retain its original angular momentum and efficiency, $\epsilon(j)$ = 1 as given by Equations (\ref{acceff}) and (\ref{lstar}). The third term represents mergers and stems from Equation (\ref{mg_jt}) (effective during $z = 4 \rightarrow 0$; see (4) in our assumptions as implemented in Equation (\ref{meq2})), and the last term represents the contribution of BZ torque (see Equation (\ref{bz_jt})). The angular momentum of the stellar component is given by

\be 
    \displaystyle
    l_{*} (M_{\bullet}, j, k, Q)= \left\{\begin{array}{lr} \displaystyle
       l_{I} (j) =  \frac{z_{m}^2(j)-2j\sqrt{z_{m}(j)}+j^2}{z_{m}^{1/2}(j)[z_{m}^2(j)+2j\sqrt{z_{m}(j)}-3z_{m}(j)]^{1/2}} & \text{\rm for~} M_{\bullet} < M_{c} (j)\\
       l_{\ell} (M_{\bullet}, j, k, Q) = \displaystyle 2j + k\sqrt{\frac{2x_{\ell}j^{2}}{(x_{\ell}-2)^{2}} - \frac{Qj^{2}}{x_{\ell} (x_{\ell} - 2)} + \frac{2x_{\ell}^{2}}{(x_{\ell} - 2)} - Q} & \text{\rm for~} M_{\bullet} \geq M_{c}(j),\label{lstar}
        \end{array}\right.
\ee 
where $l_{I} (j)$ is given by Equation (\ref{bardeen72}) \citep{1972ApJ...178..347B} and $l_{\ell} (M_{\bullet}, j, k, Q)$ is given by Equation (\ref{llcEquation}). %Also, $\displaystyle L_{\ell} = l_{\ell} \frac{GM_{\bullet}}{c}$ is given by 
which depends on $x_{\ell}(M_{\bullet}, j, k, Q)$. If we take \{$j = 0, Q$ = 0\}, we obtain the nonrelativistic result from the expression of $\displaystyle L_{\ell} (M_{\bullet}, j, k, Q) = l_{\ell} \frac{GM_{\bullet}}{c}$, as shown in Equation (\ref{nonrell}). For $\Lambda {\rm CDM}$ cosmology, we take $\Omega_{r}$= 0, $\Omega_{m} = 0.3$, $\Omega_{\Lambda} = 0.7$, and the look-back time as a function of redshift can be written as
\be
t(z) = \frac{1}{H_{0}}\int_{1/(1+z_{f})}^{1/(1+z)}da \frac{1}{\sqrt{\Omega_{m} a^{-1} + \Omega_{\Lambda} a^{2}}} = t_{z}(z) - t_{z}(z_{f}), \label{tz}
\ee
where $z_{f}$ is the formation redshift and $H_{0}$ is the present-day Hubble constant ($H_{0}$ = 70 km s$^{-1}$ Mpc$^{-1}$), and where we find by direct integration that
\be
t_{z}(z) = \frac{1}{H_{0}} \frac{2}{3} \frac{1}{\sqrt{1 - \Omega_{m}}}\log\bigg[\sqrt{1 - \Omega_{m}}\sqrt{\Omega_{m} + \frac{1 - \Omega_{m}}{(1 + z)^{3}}} + (1 - \Omega_{m}) \bigg(\frac{1}{1 + z}\bigg)^{\frac{3}{2}}\bigg]. \label{tzcm}
\ee

which matches with the result of \cite{2010gfe..book.....M} for $z_{f}$ = $\infty$. The boundary conditions are \\
\begin{enumerate}
\item 
At $t = 0$, $M_{\bullet} = M_{s}$, $z = z_{f}$ and $j = j_{0}$.\\\\
\item 
At $t = t_{s}$, $M_{\bullet} = M_{\bullet t}$, $z = z_{s}$.\\\\
\end{enumerate}

Equations (\ref{mtdim}) and (\ref{jtdim}) are the basic evolution equations of the black hole mass and spin that we solve along with all the auxiliary equations (Equations (\ref{acceff}--\ref{mt}), (\ref{jt}--\ref{tzcm})).

\begin{figure}[H]
\Large
\begin{center}
\scalebox{0.5}{
\begin{tikzpicture}[node distance = 3cm, auto]
\tikzstyle{decision} = [diamond, draw,  
    text width=5em, text badly centered, node distance=3cm, inner sep=0pt]
\tikzstyle{block} = [rectangle, draw, 
    text width=10em, text centered, rounded corners, minimum height=0.7em]
\tikzstyle{line} = [draw, -latex']
\tikzstyle{cloud} = [draw, ellipse,node distance=3cm,
    minimum height=4em]
\node [cloud,fill=black] (para) {{\bf \color{white}Growth of Black Hole}};
\tikzstyle{block} = [rectangle, draw, 
    text width=8em, text centered, rounded corners, minimum height=0.7em]
\node [block,fill=cyan!20, below left of= para, node distance=4cm] (para3) {{\color{black}Stellar capture}};
\node [block,fill=cyan!20, left of= para3, node distance=8cm] (para1) {{\color{black}Accretion}};
\node [block,fill=cyan!20, below right of= para,node distance=4cm] (para6) {{\color{black}BZ Torque}};
\node [block,fill=cyan!20, right of= para6,node distance=8cm] (para2) {{\color{black}Mergers}};
\tikzstyle{block} = [rectangle, draw, 
    text width=18em, text centered, rounded corners, minimum height=1em]
\node [block, above of= para1,node distance=9cm, rectangle split,rectangle split parts=2,rectangle split part fill={cyan!20, white!20}] (para4) {\underline{Case 1 : $\displaystyle z_{s} < 4$} \\ \nodepart{second} \[\displaystyle z : z_{f} \rightarrow  4 : \left\{\begin{array}{lr}
          \dot{M}_{\bullet}(\dot{M}_{\bullet g},  \dot{M}_{ \bullet *})\\
         \dot{j}(\dot{j}_{g}, \dot{j}_{BZ}).
        \end{array}\right.\] \\  
        \[\displaystyle  z: 4 \rightarrow z_{s} : \left\{\begin{array}{lr}
          \dot{M}_{\bullet}(\dot{M}_{\bullet g}, \dot{M}_{\bullet *}, \dot{M}_{\bullet m})\\
         \dot{j}(\dot{j}_{g}, \dot{j}_{BZ}, \dot{j}_{m}).
        \end{array}\right.\]
         \\ 
         \[\displaystyle  z: z_{s} \rightarrow 0 : \left\{\begin{array}{lr}
          \dot{M}_{\bullet}(\dot{M}_{\bullet *}, \dot{M}_{\bullet m})\\
        \dot{j}(\dot{j}_{BZ}, \dot{j}_{m}).
        \end{array}\right.\]
         };

\node [block, above of= para2,node distance=9cm, rectangle split,rectangle split parts=2,rectangle split part fill={cyan!20, white!20}] (para5) {\underline{Case 2 : $\displaystyle z_{s} > 4$} \\ \nodepart{second}\[\displaystyle z: z_{f} \rightarrow z_{s} : \left\{\begin{array}{lr}
       \dot{M}_{\bullet}(\dot{M}_{\bullet g}, \dot{M}_{\bullet *})\\
         \dot{j}(\dot{j}_{g}, \dot{j}_{BZ}).
        \end{array}\right.\] \\
        \[\displaystyle z: z_{s} \rightarrow 4 : \left\{\begin{array}{lr}
         \dot{M}_{\bullet}(\dot{M}_{\bullet *})\\
        \dot{j}(\dot{j}_{BZ}).
        \end{array}\right.\] \\  
        \[\displaystyle z: 4 \rightarrow 0 : \left\{\begin{array}{lr}
         \dot{M}_{\bullet}(\dot{M}_{\bullet *}, \dot{M}_{\bullet m})\\
        \dot{j}(\dot{j}_{BZ}, \dot{j}_{m}).
        \end{array}\right.\]};

\node [block, above of= para,node distance=6cm, rectangle split,rectangle split parts=2,rectangle split part fill={cyan!20, white!20}] (para5) {\underline{Evolution equations} \\ \nodepart{second}\[ \dot{M}_{\bullet g} \rightarrow {\rm Equation (\ref{acc_mt})}\] \[ \dot{M}_{\bullet *} \rightarrow {\rm Equation (\ref{mdotst})}\] \[ \dot{M}_{\bullet m} \rightarrow {\rm Equation (\ref{mg_mt})}\] \[ \dot{j}_{g} \rightarrow {\rm Equation (\ref{jacc})}\]\[ \dot{j}_{BZ} \rightarrow {\rm Equation (\ref{bz_jt})}\]\[ \dot{j}_{m} \rightarrow {\rm Equation (\ref{mg_jt})}\]};

\tikzstyle{block} = [rectangle, draw, 
    text width=12em, text centered, rounded corners, minimum height=1em]
\tikzstyle{block} = [rectangle, draw, 
    text width=35em, text centered, rounded corners, minimum height=0.5em]
\node [block, below of=para, node distance=8cm, rectangle split,rectangle split parts=2,rectangle split part fill={cyan!20, white!20}] (para7) {$\Lambda CDM$ Model of Cosmology \\ \nodepart{second}
\[\displaystyle
t(z) = \frac{1}{H_{0}}\int_{1/(1+z_{f})}^{1/(1+z)}da \frac{1}{\sqrt{\Omega_{m} a^{-1} + \Omega_{\Lambda} a^{2}}},
= t_{z}(z) - t_{z}(z_{f}),
\] \\ \[\displaystyle t_{z}(z) = \frac{1}{H_{0}} \frac{2}{3} \frac{1}{\sqrt{1 - \Omega_{m}}}\log\bigg[\sqrt{1 - \Omega_{m}}\sqrt{\Omega_{m} - \frac{\Omega_{m}-1}{(1 + z)^{3}}} - (\Omega_{m}-1) \bigg(\frac{1}{1 + z}\bigg)^{\frac{3}{2}}\bigg]\]};
\path [line, line width=0.8 mm] (para3)--node{}(para);
\path [line, line width=0.8 mm] (para6)--node{}(para);
\draw [->, draw=black, line width=0.8 mm](para1)to [out=90,in=-180](para);
\draw [->, black,line width=0.8 mm](para2)to [out=90,in=0](para);

 \end{tikzpicture}

}
\end{center}
\caption{A schematic for our model of evolution of the mass and the spin of black hole in $\Lambda CDM$ cosmology.}
\label{flowchart}
\end{figure}
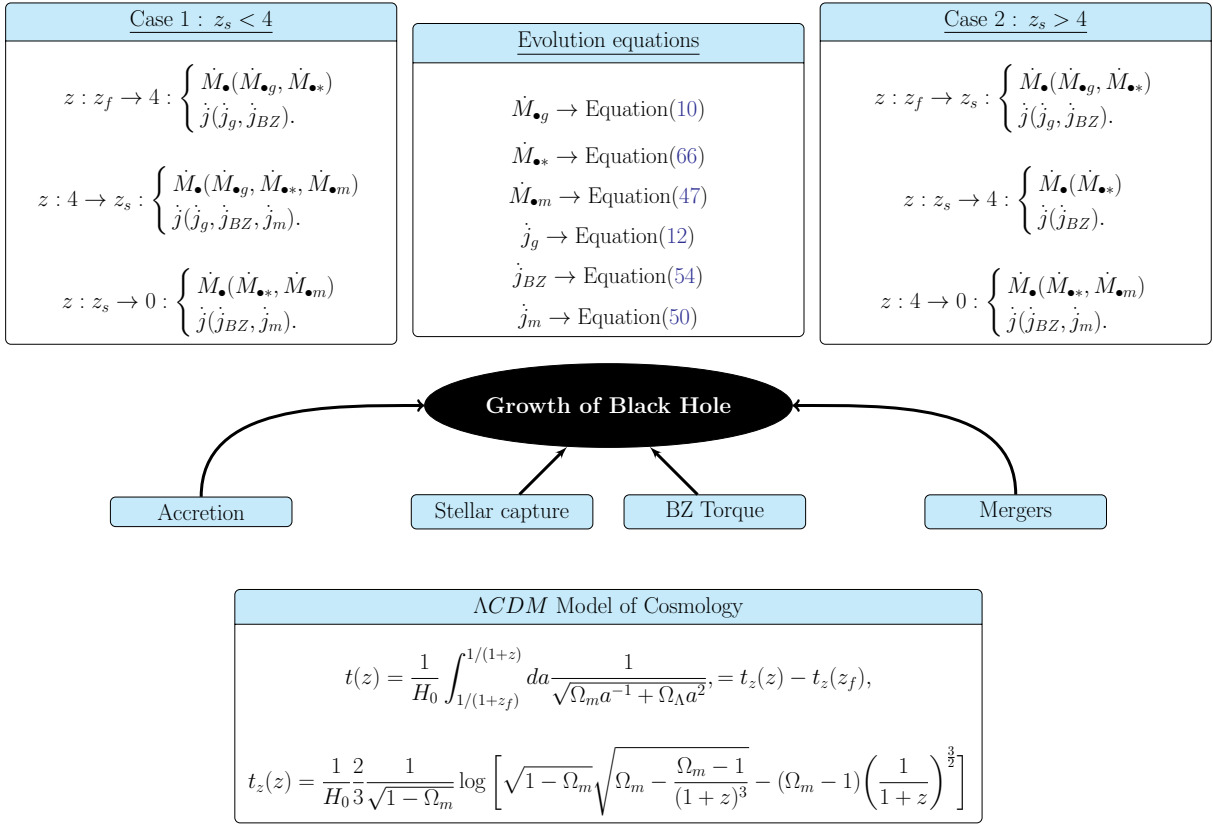

We perform the following experiments, which we tabulate in Table \ref{expts}. Next, we discuss the parameter range.\\

\begin{table}[H]
\begin{center}
\begin{tabular}{|c |c |c |c |c |c |}\hline\hline
Models  & Accretion & BZ Torque & Stellar Capture & Mergers & Parameter Sets\\ \hline
Expt 1 & $\surd$ & & & & ${M_{s}, \eta, j_{0}, z_{f}}$\\ \hline
Expt 2 & $\surd$ & & (FLC)$\surd$ & & ${M_{s}, \eta, j_{0}, z_{f}, \sigma_{100}}$\\ \hline
Expt 3 & $\surd$ & $\surd$ &  & & ${M_{s}, \eta, j_{0}, z_{f}, B_{4}}$\\ \hline
Expt 4 & $\surd$ & $\surd$ & (SLC)$\surd$ & & ${M_{s}, \eta, j_{0}, z_{f}, \sigma_{100}}$\\ \hline
Complete model & $\surd$ & $\surd$ & (SLC)$\surd$ & $\surd$ & ${M_{s}, \eta, j_{0}, z_{f}, \sigma_{100}, q}$\\ \hline 
\end{tabular}
\caption{Description of the Five Different Experiments Performed for Various Combinations of Astrophysical Components Included, along with the Parameter Sets.}
\label{expts}
\end{center}
\end{table}

{\it{Justification of the chosen parameter ranges: }}
To perform the experiments given in Table \ref{expts}, we choose the ranges of the input parameters, which are justified by observational values. The minimum $\sigma$ measured to date is around 30 - 40 km s$^{-1}$ \citep{2011ApJ...739...28X}. For this low $\sigma$, the saturated mass is of the order of around 10$^{5} M_{\odot}$. Therefore, the seed masses considered should be $\lesssim$ $10^{5} M_{\odot}$. The ranges we consider for seed mass and formation redshift are consistent with the values considered by \cite{2017NatAs...1E.147A}. The values of $z_{f}$ are taken to be in the range $z_{f}$ = 5-8. Average observed values of $\sigma$ are within the range of 100--200 km s$^{-1}$. Values of $\gamma \simeq$ 1.1-1.5 are consistent with the observed values \citep{2013CQGra..30x4005M}. The $B_{4}$ values are taken to be in the typical range of 1-10 \citep{1990agn..conf.....B}, for a black hole of mass $10^{8}$--$10^{10} M_{\odot}$; \cite{1977MNRAS.179..433B} show that the field strength should be more than $10^{5}$ G for supplying electromagnetic power equal to or more than the Eddington power. The $\eta$ values are typically sub-Eddington ($\eta \gtrsim 0.07$), and below that it will not be possible to attain the high mass of the present-day black holes. We have used $\eta$ in the range [0.07, 0.09] as given in \cite{2009ApJ...690...20S}, including the effect of duty cycles. We have also illustrated the case of $\eta = 0.01$, which clearly indicates a very slow mass growth.\\

%\begin{center}
%{\underline{Parameter sets used}}
%\end{center}
\begin{table}[H]
\centering
\scalebox{1}{
\begin{tabular}{|c |c |c |}\hline\hline
Parameters & Ranges & References\\ \hline\hline
$M_{\bullet s}$ & $10^{3} - 10^{5}$ $M_{\odot}$ & \cite{2017NatAs...1E.147A}\\ \hline
$j_{0}$ & 0.001 - 0.4 & \cite{2015ASInC..12...51M}\\ \hline
$z_{f}$ & 5 - 8 & \cite{2017NatAs...1E.147A}\\ \hline
$\eta$ & 0.07 - 0.09 & \cite{2009ApJ...690...20S}\\ \hline
$\sigma_{100}$ & 1 - 2.5 & \citep{2011ApJ...739...28X}, \cite{2018JApA...39....4B}\\ \hline
$\gamma$ & 1.1 - 1.5 & \cite{2013CQGra..30x4005M}\\ \hline
$B_{4}$ & 1 - 10 & \cite{1990agn..conf.....B} \\ \hline
\end{tabular}
}
\caption{The Ranges of the Parameters Used in Our Model Are Shown and Are Based on the Papers Cited.}
\label{prmtr}
\end{table}

%\begin{center}
%{\underline{Table of the parameter sets for runs}}
%\end{center}
\begin{table}[H]
\centering
\scalebox{0.75}{
\begin{tabular}{|c |c |c |c |c |c |c |c |c |}\hline\hline
Run \# & $M_{s} (10^{5} M_{\odot})$ & $B_{4}$  & $\sigma_{100}$ & $z_{f}$ & $\eta$  & $j_{0}$ & Varying Parameter & Comments\\ \hline\hline
1.1 & 1 & 5 & & 4 & 0.09 & 0.001 &  $j_{0} = 0$ & Expt 1\\
1.2 & 1 & 10 & & 4 & 0.09 & 0.2 &  &\\ \hline \hline
2.1. & 0.1 &  &  1 & 10 & 0.09 &   & $M_{s}$ & Expt 2\\ 
2.2 & 0.5 &  &  1 & 10 & 0.09 & & &  \\ 
2.3 & 1 & &  1 & 10 & 0.09 &  & & \\ \hline\hline
3.1.1 & 1 & 1 & & 4 & 0.09 & 0.2 & $B_{4}$ & Expt 3.1 \\ 
3.1.2 & 1 & 5 & & 4 & 0.09 & 0.2 & & \\ 
3.1.3 & 1 & 10 & & 4 & 0.09 & 0.2 & & \\ \hline
3.2.1 & 1 & 5 & & 4 & 0.01 & 0.2 & Lower limit of $\eta$ & Expt 3.2\\ 
3.2.2 & 1 & 5 & & 4 & 0.05 & 0.2 & & \\ \hline\hline
4.1.1 & 0.5 & 5 &  1 & 6 & 0.07 &  0.2 & $M_{s}$ & Expt 4.1\\ 
4.1.2 & 0.6 & 5 &  1 & 6 & 0.07 & 0.2 & & /Expt 5.1* \\ 
4.1.3 & 0.7 & 5 &  1 & 6 & 0.07 & 0.2 & & \\ 
4.1.4 & 1 & 5 & 1 & 6 & 0.07  & 0.2 & &\\ \hline
4.2.1 & 1 & 5  & 1 & 6 & 0.07  & 0.2 & $B_{4}$ & Expt 4.2\\
4.2.2 & 1 & 6  & 1 & 6 & 0.07  & 0.2 & & /Expt 5.2*\\ 
4.2.3 & 1 & 8  & 1 & 6 & 0.07  & 0.2 & &\\ 
4.2.4 & 1 & 10 & 1 & 6 & 0.07  & 0.2 & &\\ \hline
4.3.1 & 1 & 5  & 1 & 6 & 0.07  & 0.2 & $\sigma_{100}$ & Expt 4.3\\ 
4.3.2 & 1 & 5  & 1.5 & 6 & 0.07  & 0.2 & & /Expt 5.3*\\ 
4.3.3 & 1 & 5  & 2 & 6 & 0.07  & 0.2 & & \\ 
4.3.4 & 1 & 5  & 2.5 & 6 & 0.07 & 0.2 & &\\ \hline
4.4.1 & 1  & 5 & 1 & 5 & 0.07  & 0.2 & $z_{f}$ & Expt 4.4\\ 
4.4.2 & 1  & 5 & 1 & 6 & 0.07  & 0.2 & & /Expt 5.4*\\ 
4.4.3 & 1  & 5 & 1 & 7 & 0.07  & 0.2 & & \\ 
4.4.4 & 1 & 5 & 1 & 8 & 0.07  & 0.2 & &\\ \hline
4.5.1 & 1 & 5 & 1 & 6 & 0.07  & 0.2 & $\eta$ & Expt 4.5\\
4.5.2 & 1 & 5 & 1 & 6 & 0.075  & 0.2 & & /Expt 5.5*\\ 
4.5.3 & 1 & 5 & 1 & 6 & 0.08  & 0.2 & & \\ 
4.5.4 & 1 & 5 & 1 & 6 & 0.09  & 0.2 & &\\ \hline
4.6.1 & 1 & 5 & 1 & 6 & 0.07  & 0.0 & $j_{0}$ & Expt 4.6\\ 
4.6.2 & 1 & 5 & 1 & 6 & 0.07  & 0.2 & & /Expt 5.6*\\ 
4.6.3 & 1 & 5 & 1 & 6 & 0.07  & 0.3 & & \\ 
4.6.4 & 1 & 5  & 1 & 6 & 0.07  & 0.4 & &\\ \hline\hline
\end{tabular}
}
\caption{Sets of the Parameters Used for the runs with \{$k$ = 1, $\gamma$ = 1.1\} for the experiments given in Table \ref{expts}.}
\label{table_acc_bz_st_mg}
  {\raggedright Note: For each experiment we specify the parameter sets used. An asterisk indicates that, in addition to the parameter set for Expt 4, we have one more parameter $q$ = 0.1 for Expt 5, which prescribes the complete model. \par}
\end{table}

\subsection{\it{Summary of Experiments 1-4}}
Experiment 1 (only accretion) has been discussed in Appendix \ref{AppendixA}, experiment 2 (accretion and stellar capture in full loss cone theory) in Appendix \ref{AppendixB}, experiment 3 (only accretion and BZ effect) in Appendix \ref{3.1} and experiment 4 (accretion, stellar capture in steady loss cone and BZ effect; see Table \ref{expts}) in Appendix \ref{3.2} along with their results. Here we present a summary and salient points of these experiments:

\begin{enumerate}
\item  {\it Experiment 1} (Appendix \ref{AppendixA}):
    In experiment 1, we recover the well-known result of \cite{1970Natur.226...64B}, where we see that, in the presence of only accretion, the black hole spin saturates very fast, and subsequently only the mass increases, leaving the spin parameter unchanged at the saturated value of 1. This demonstrates that the effects of mergers and BZ are important components required to spin down the black hole.
    
\item {\it Experiment 2} (Appendix \ref{AppendixB}):
In experiment 2, we model the mass evolution of the black hole in the presence of accretion (and feedback) and stellar capture in the full loss cone limit. The main usefulness of this result is that we can derive a completely analytic solution in this case, to obtain fiducial timescales, which can be compared with other nearby models. This is a nonrelativistic treatment (sans spin), and it is useful to obtain the time (or redshift) of mass saturation given by Equation (\ref{satm}). Taking into account the saturation, we present the results for the more realistic evolution experiments (3 \& 4 in Table \ref{expts}), where we include the effects one at a time.
    
\item  {\it Experiment 3} (Appendix \ref{3.1}):
    In experiment 3, we model the mass and spin evolution of the black hole in the presence of accretion and the BZ effect \citep{2015ASInC..12...51M}. It can be seen that the BZ torque causes the spin-down of the black hole, reducing it from the highest saturated spin value. As the $B_{4}$ value is increased, the spin-down is more effective, while the accretion is enhanced. Also, it can be seen that mass growth by accretion with an efficiency of $\eta$ = 0.01 or 0.05 is very small, which cannot generate high-mass black holes in the universe. Therefore, $\eta \geq$ 0.05.

\item {\it Experiment 4} (Appendix \ref{3.2}):
    In experiment 4, we consider the mass and spin evolution of black holes in the presence of accretion, stellar capture, the steady loss cone regime, and the BZ process. The chosen input parameters are given in Table \ref{table_acc_bz_st_mg}. It is seen that BZ moderates the spin evolution. It is seen that spin buildup is not as rapid, but the mass accretion proceeds to saturation similar to experiment 3.
    \end{enumerate}

We have discussed how the complete model differs from experiment 4 in \S \ref{3.3}.

\subsection{\it{Complete Model with Accretion, Stellar Capture, Mergers, and BZ Torque}}\label{3.3}
Here we add the contribution of mergers to the spin and mass evolution and retain all the terms in Equations (\ref{jt}) and (\ref{mt}) for our calculations (see Figure \ref{wtmwm}).

\begin{figure}[H]
\centering
%\subfigure[Mass]{\includegraphics[scale=0.24]{com_m2.pdf}} \hspace{0.1 cm}
%\subfigure[Spin]{\includegraphics[scale=0.26]{com_j2.pdf}} \hspace{0.1 cm}\\
\subfigure[Mass]{\includegraphics[scale=0.29]{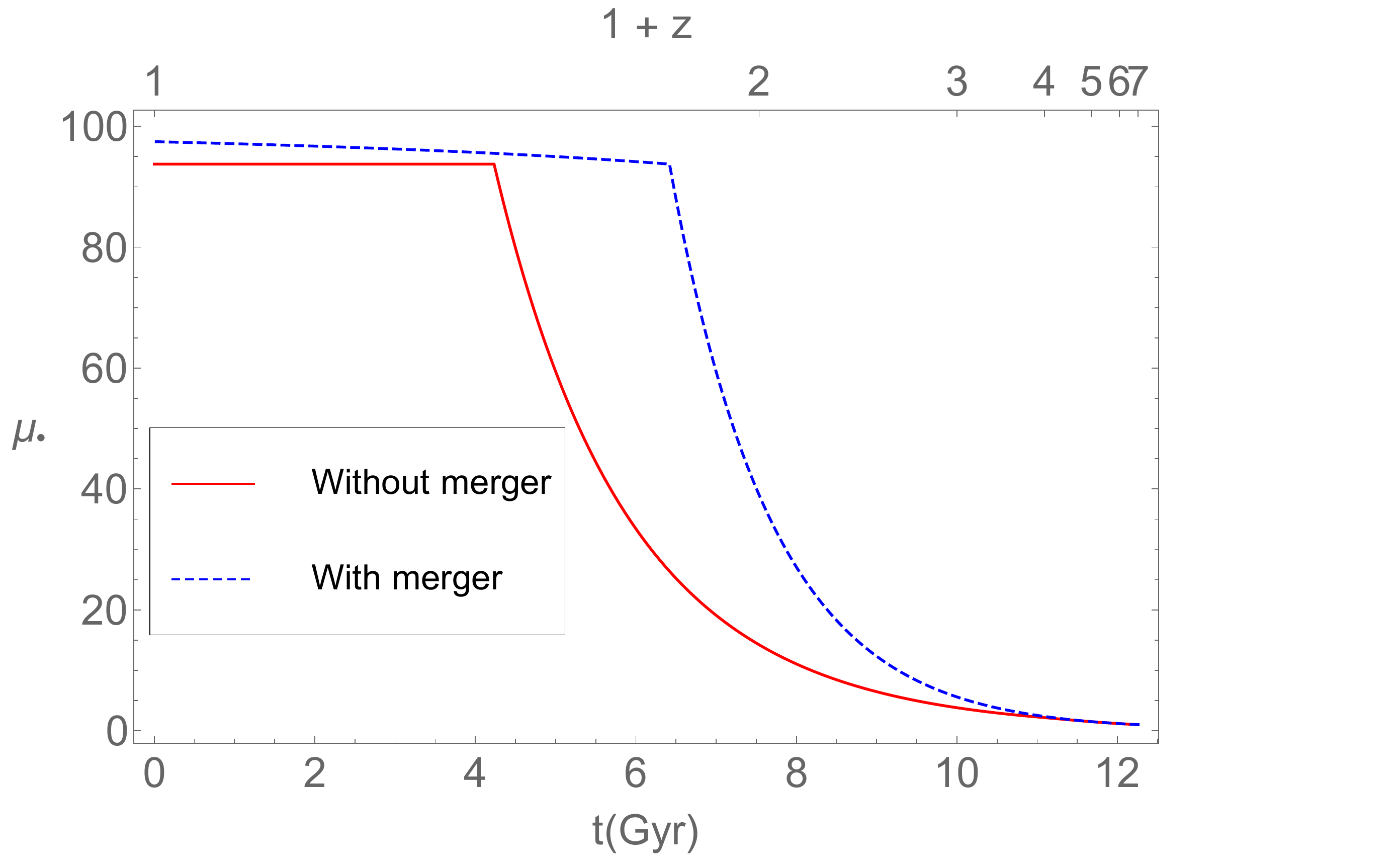}} \hspace{0.1 cm}
\subfigure[Spin]{\includegraphics[scale=0.29]{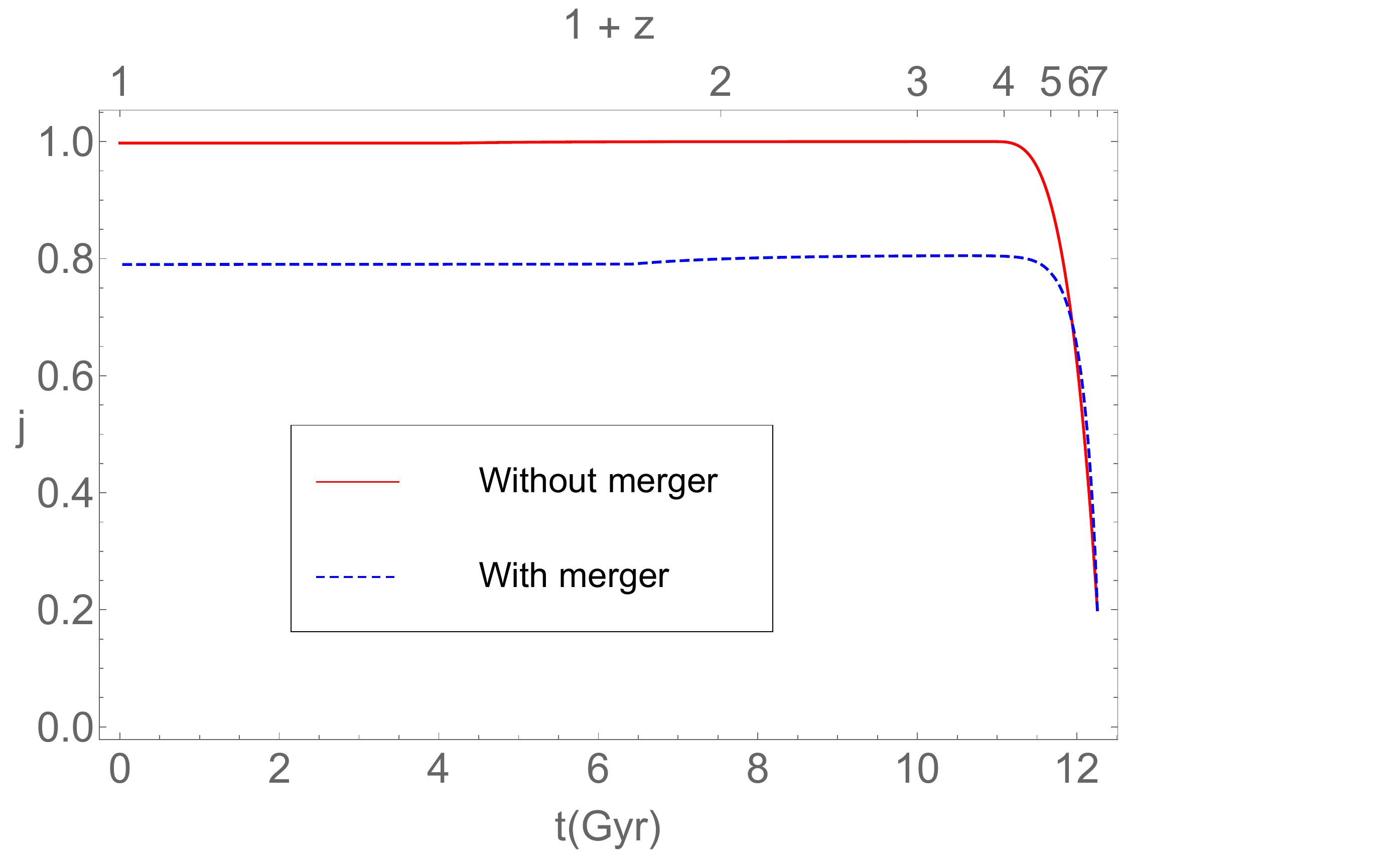}} \hspace{0.1 cm}
\caption{Evolution  of $\mu_{\bullet}(t)$ (a) and (b) $j(t)$ of the black hole are shown for the canonical case of the complete model with mergers (run \# 5.5.4) and without the effect of mergers with the other parameters being the same. }
\label{wtmwm}
\end{figure}
Our results are the following:
\begin{enumerate}
\item We see a change in the slope of the mass evolution in Figure \ref{wtmwm}(a). This is due to the saturation of black hole mass, where the dominant term, accretion, stops contributing and stellar capture and mergers take over. Since the mass growth rate by accretion is more than the latter two, the slope changes dramatically.
\item From Figure \ref{wtmwm} (run \# 5.5.4 / 4.5.4) we observe a difference in evolution in the presence and absence of mergers. It is clearly seen that in the presence of the mergers the black hole reaches the saturation mass earlier owing to the higher mass growth rate and that the final mass attained is higher because of the contribution of mergers. 
\item As we consider the merger activity to be effective from $z \lesssim 4$, we see that the two curves start deviating from each other after $z \gtrsim 4$ owing to an overall increase in the mass growth rate.
\item We observe from the spin evolution (see Figure \ref{wtmwm}(b)), that the saturated or the final spins are different for the two cases. This is due to minor mergers that cause the spin-down of the black holes; again, the evolution changes after $z \simeq 4$. This emphasizes the importance of the contribution of mergers and the BZ effect in the spin evolution; otherwise, the black holes will be maximally spinning.
\end{enumerate}

Next, we discuss the evolution in the presence of all the effects and its dependence on the input parameters. Figure \ref{wtmwm} represents the evolution for the canonical case (run \# 5.5.4). For the runs (run \# 5.1 to \# 5.6 in Table \ref{table_acc_bz_st_mg}), we discuss our results obtained in Figure \ref{acc_bz_st_mg_m} for the mass evolution and in Figure \ref{acc_bz_st_mg_j} for the spin evolution. 

\begin{enumerate}
\item We found that the mass evolution has a small dependence on the parameters $\{k$, $\gamma$, $j_{0}, B_{4}\}$ in the input range.
\item If the $\sigma$ is the same, then the final mass will be almost the same, irrespective of their initial masses [see Figure \ref{acc_bz_st_mg_ma}]. 
\item Change in $z_{f}$ (run \# 5.4) has little impact on the evolution and does not affect the final mass much [see Figure \ref{acc_bz_st_mg_me}]. 
\item Variation of $\sigma$ (run \# 5.3) shifts the saturation point owing to the dependence on $\sigma$ [see Equation (\ref{satm})]. Higher $\sigma$ implies higher saturation mass and larger time taken to reach the saturation point [see Figure \ref{acc_bz_st_mg_md}]. 
\item Increase of $\eta$ (run \# 5.5) increases the accretion rate, which is the main source of mass growth. Hence, for higher $\eta$, the system reaches the saturation point earlier [see Figure \ref{acc_bz_st_mg_mf}].
\item The difference between the complete model (see Figure \ref{acc_bz_st_mg_m}) and Expt 4 (see Figure \ref{acc_bz_st_m}) is that the mass evolution is faster after saturation because of the presence of the merger term, as this contributes along with the stellar capture when the gas accretion stops. 
%\end{itemize}

\begin{figure}[H]
\centering
\centering
\subfigure[\label{acc_bz_st_mg_ma}]{\includegraphics[scale=0.2]{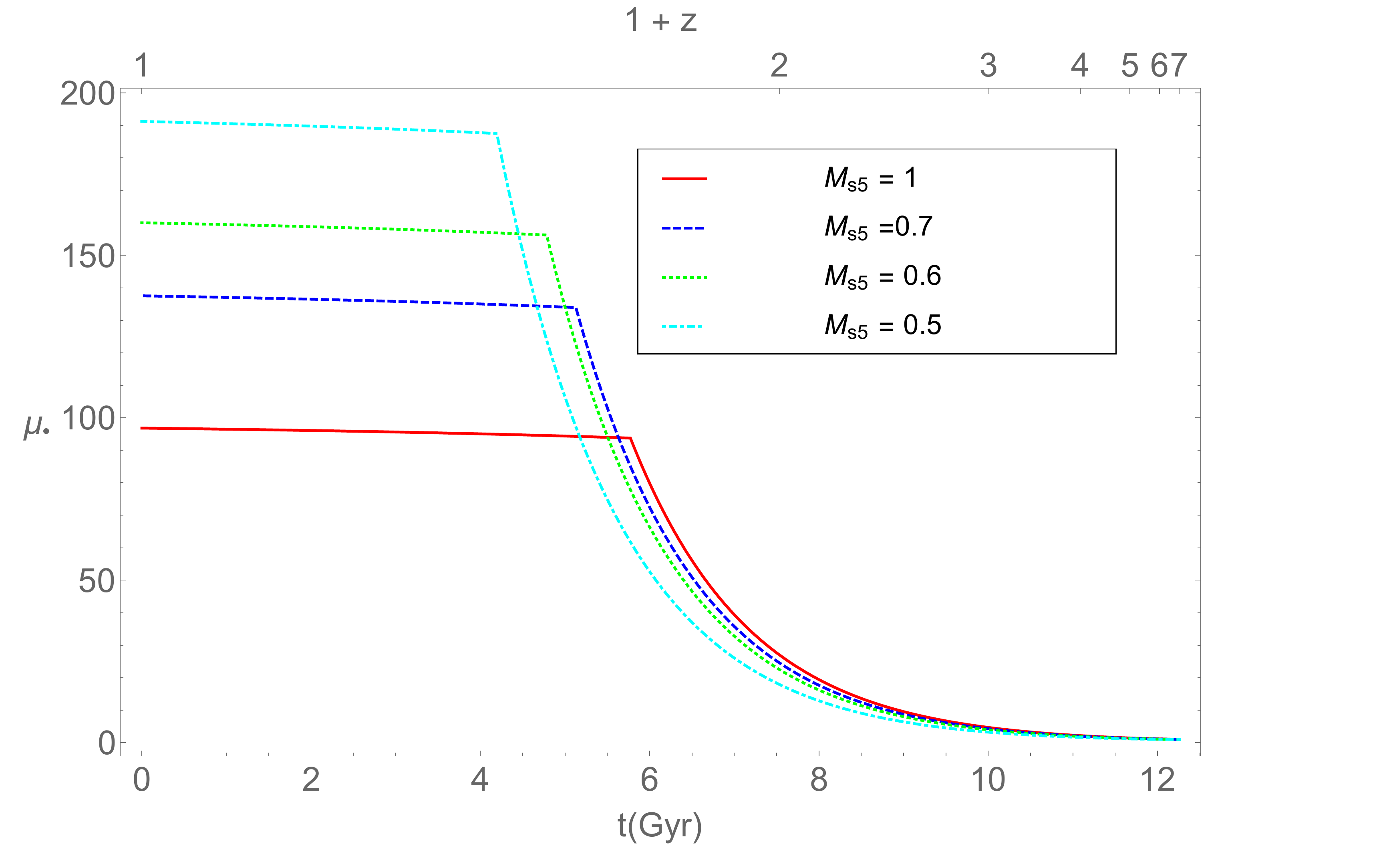}} \hspace{0.1 cm}
%\subfigure[\label{acc_bz_st_mg_mb}]{\includegraphics[scale=0.25]{acc_bz_st_mg_m_bz10_rel.pdf}} 
%\\
\subfigure[\label{acc_bz_st_mg_md}]{\includegraphics[scale=0.2]{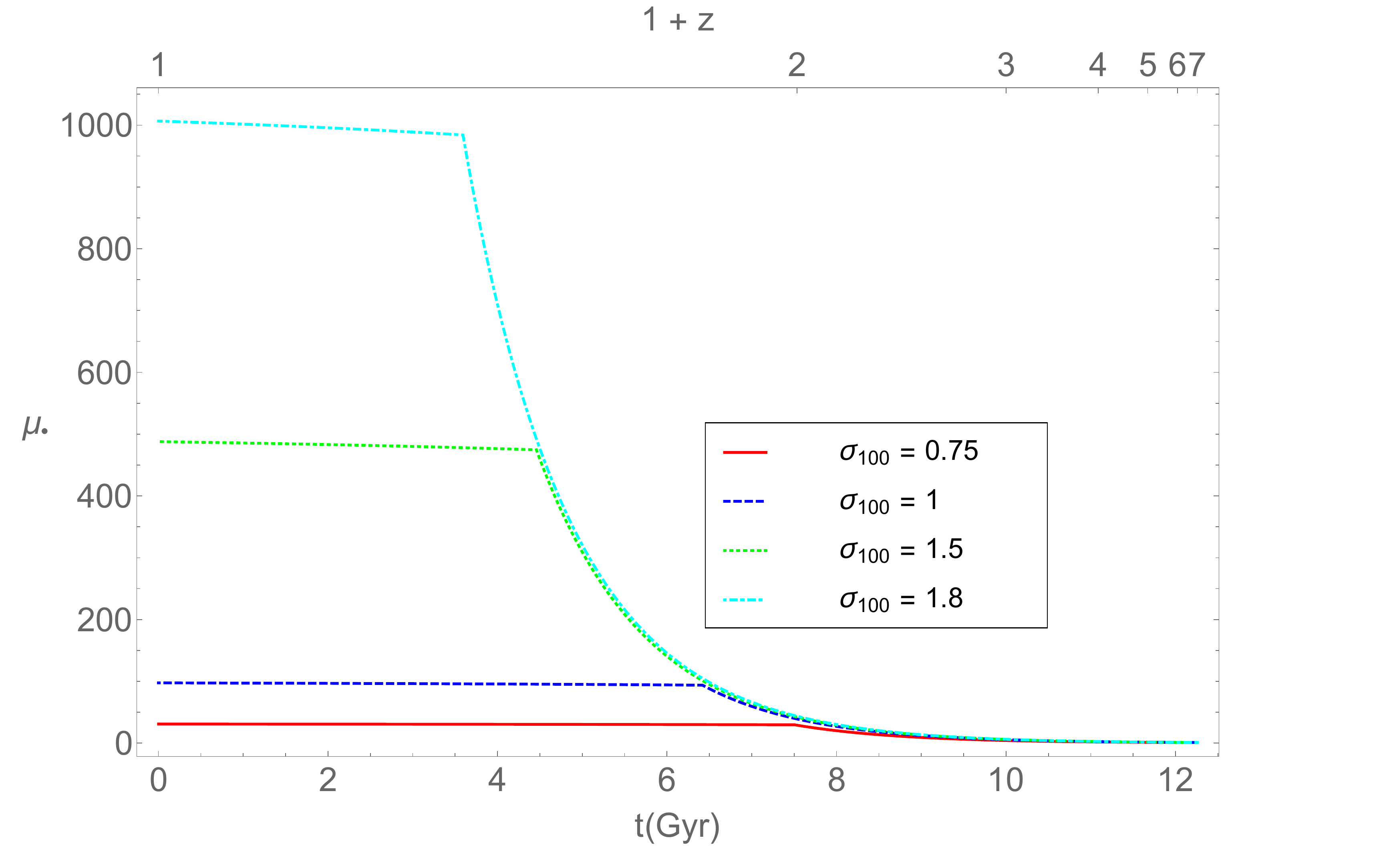}} \\
\subfigure[\label{acc_bz_st_mg_me}]{\includegraphics[scale=0.2]{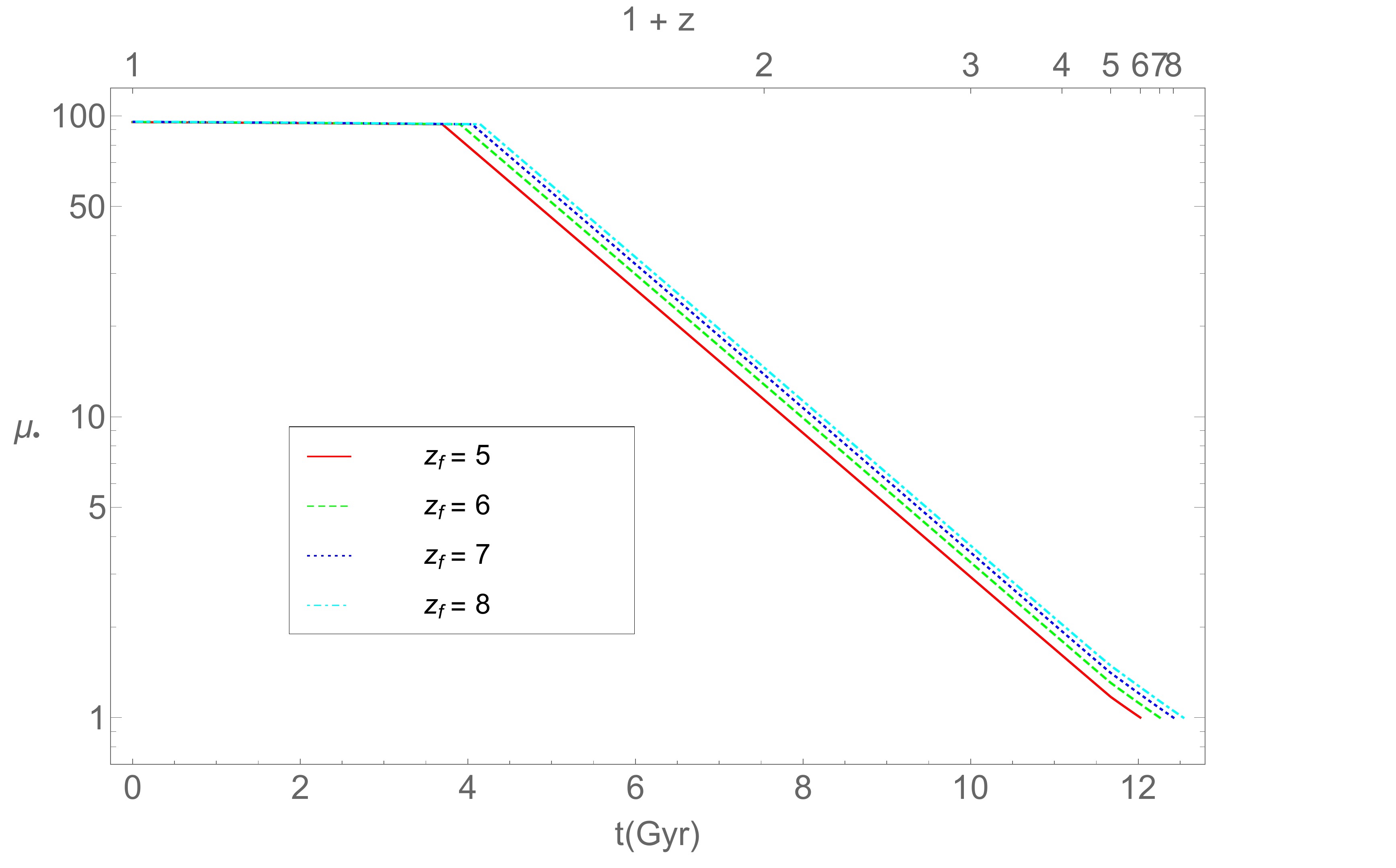}} \hspace{0.1 cm}
\subfigure[\label{acc_bz_st_mg_mf}]{\includegraphics[scale=0.20]{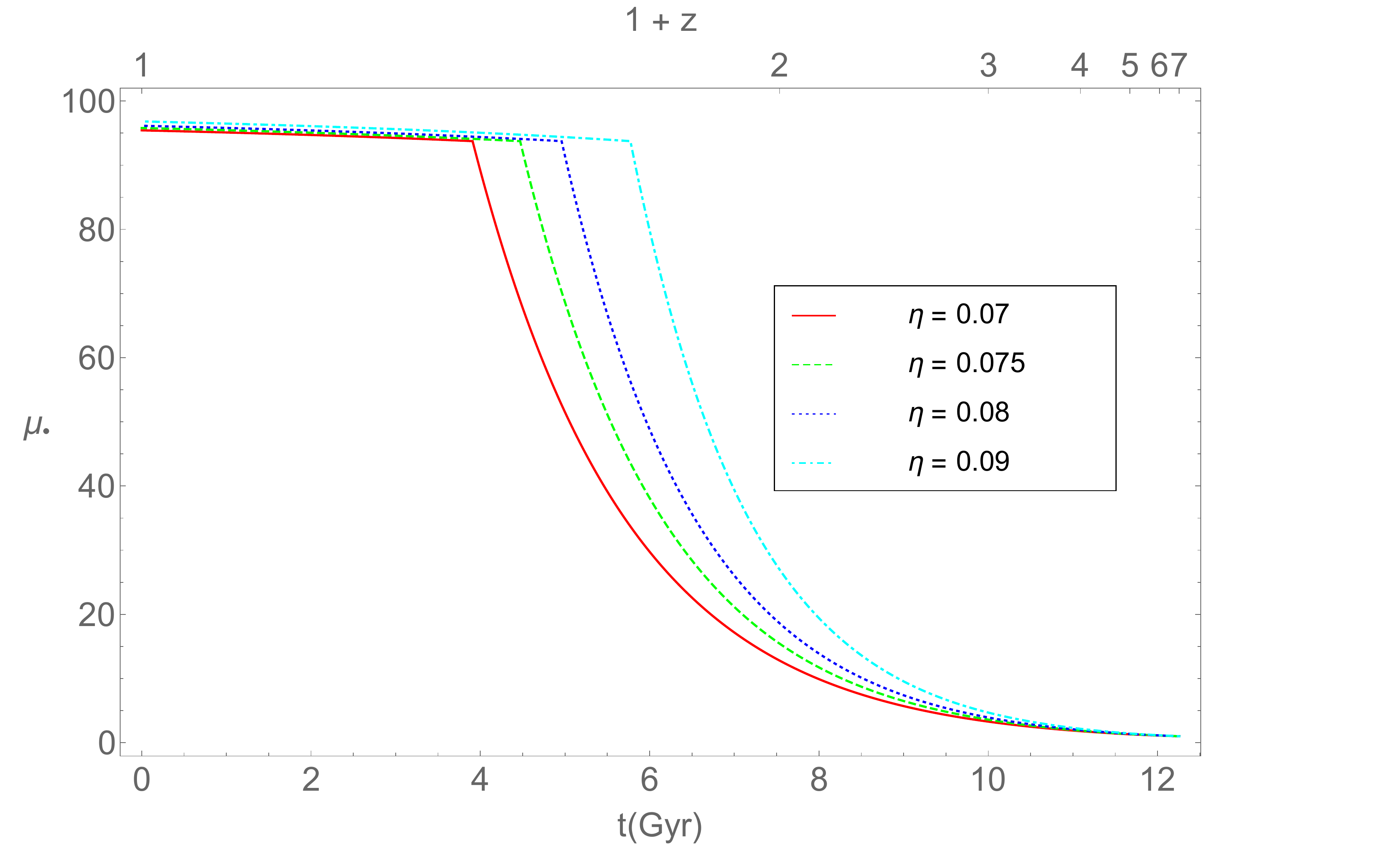}} \\
\caption{The mass evolution, $\displaystyle\mu_{\bullet}$($t$), for run \# 5.1, \# 5.3, \# 5.4, \# 5.5 [see Table \ref{table_acc_bz_st_mg}] (a -- d) are shown, when there is accretion, stellar capture, merger and BZ torque present for deviation of various parameters from their values in the canonical set.}
\label{acc_bz_st_mg_m}
\end{figure}

%\begin{itemize}
\item Again, we observe from run \# 5.1 to \# 5.6 (see Figure \ref{acc_bz_st_mg_j}) that there is little variation of $j$ for changes in parameters $\{k$, $\gamma$ , $M_{s}\}$. 
\item Figures \ref{acc_bz_st_mg_ja}, \ref{acc_bz_st_mg_je}, \ref{acc_bz_st_mg_jf}, and \ref{acc_bz_st_mg_jh} show variation at the starting points due to different initial values, but the final values are nearly the same. Therefore, $M_{s}$, $z_{f}$, $\eta$, and $j_{0}$ do not affect the final spin value of the black hole.
\item The decrease in $j$ occurs at the high-mass end because of the BZ effect. It is also seen that an increase in $B_{4}$ value (run \# 5.2) decreases the final spin, as expected.
%\item Figure \ref{acc_bz_st_jb} shows that an increase in $B_{4}$ value (run \# 4.2), decreases the final spin, as expected.
\item A higher $\sigma$ (run \# 5.3) causes a higher final mass of the black hole; hence, the final spin value decreases with an increase in $\sigma$ (see Figure \ref{acc_bz_st_mg_jd}), while keeping $M_{s}$ constant.

\item The difference of the complete model with the experiments (3 and 4) is due to the presence of the mergers; the final value of spin acquired is lesser since the minor merger contributes to spinning down the hole (see Figure \ref{acc_bz_st_mg_j}).
\end{enumerate}
%By calculating $M_{\bullet}(\sigma, z$), we obtain the evolution of the $M_{\bullet}$--$\sigma$ relation. 
Our evolution model is summarized schematically in a flowchart (Figure \ref{flowchart}). The motivation is to isolate the contribution of different effects to the evolution of the black hole individually, and also together from $z = z_{f} \rightarrow 0$. This, in turn, can give us information about for the coevolution of the black hole and the galaxy.\\
%Next, we discuss some applications of our model.
We discuss two applications of the evolution model: the impact on the $M_{\bullet}-\sigma$ relation and black hole archaeology.

\begin{figure}[H]
\centering
\subfigure[\label{acc_bz_st_mg_ja}]{\includegraphics[scale=0.2]{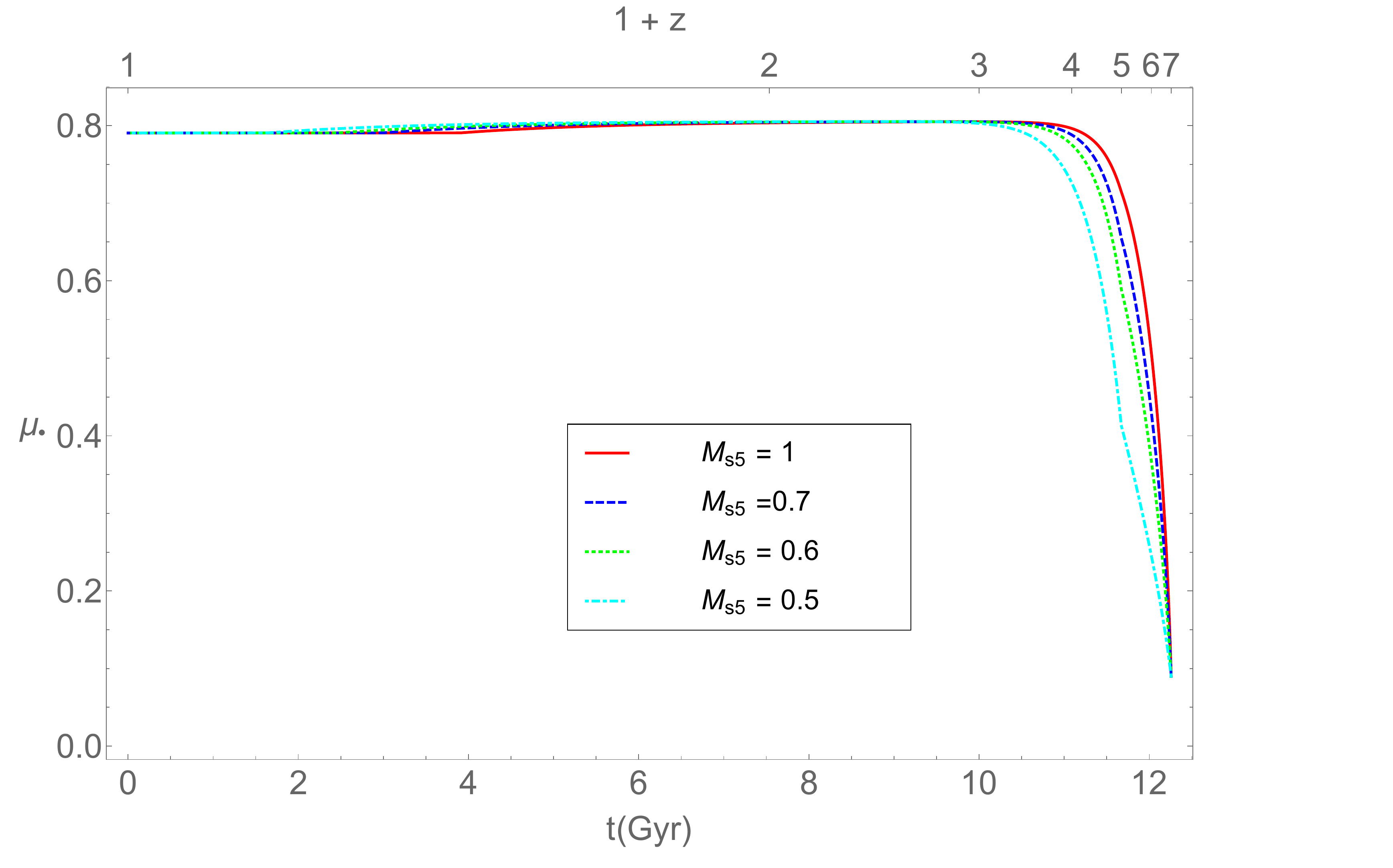}} \hspace{0.1 cm}
\subfigure[\label{acc_bz_st_mg_jb}]{\includegraphics[scale=0.2]{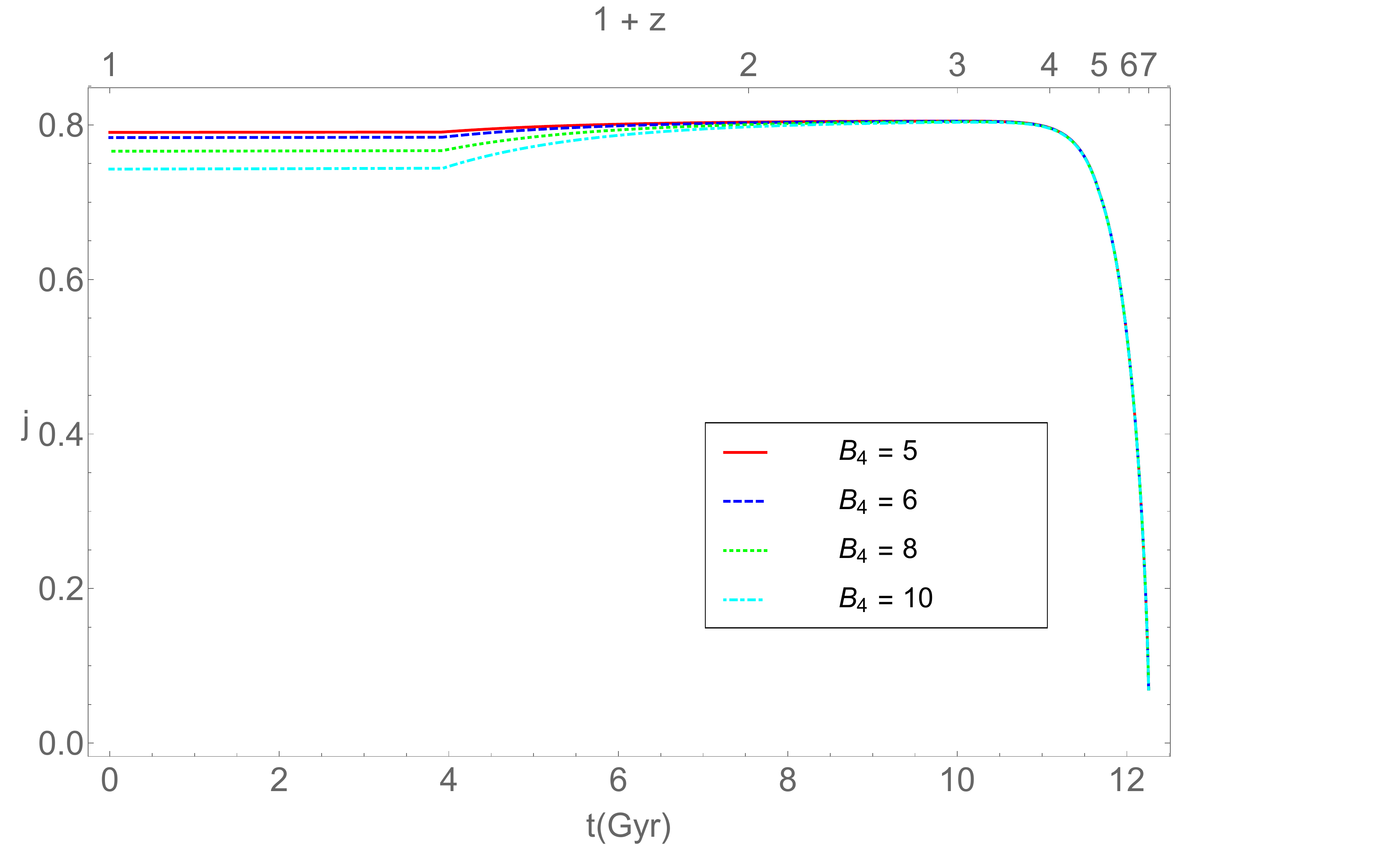}} 
\\
\subfigure[\label{acc_bz_st_mg_jd}]{\includegraphics[scale=0.2]{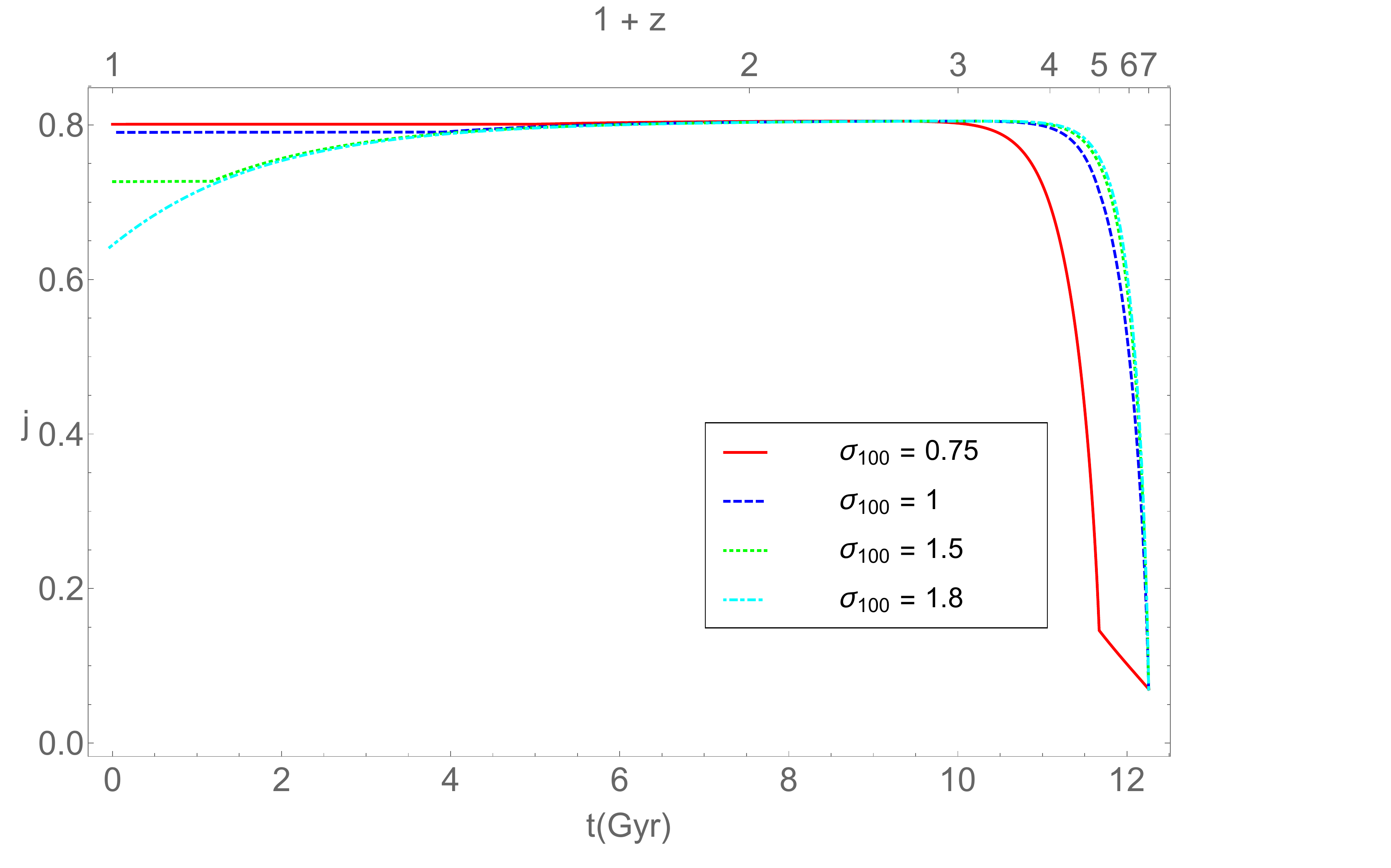}} 
\subfigure[\label{acc_bz_st_mg_je}]{\includegraphics[scale=0.2]{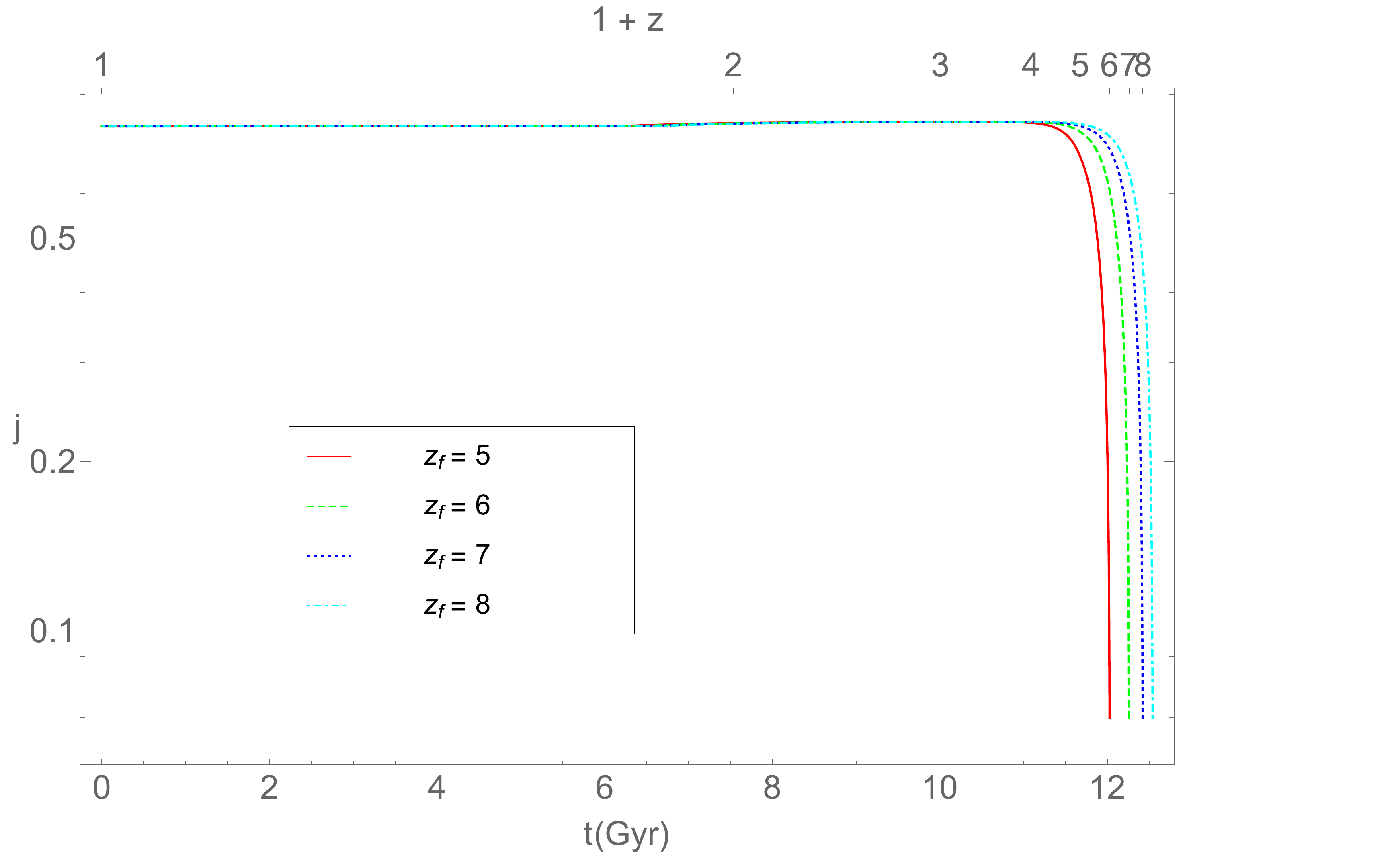}} \hspace{0.1 cm}\\
\subfigure[\label{acc_bz_st_mg_jf}]{\includegraphics[scale=0.2]{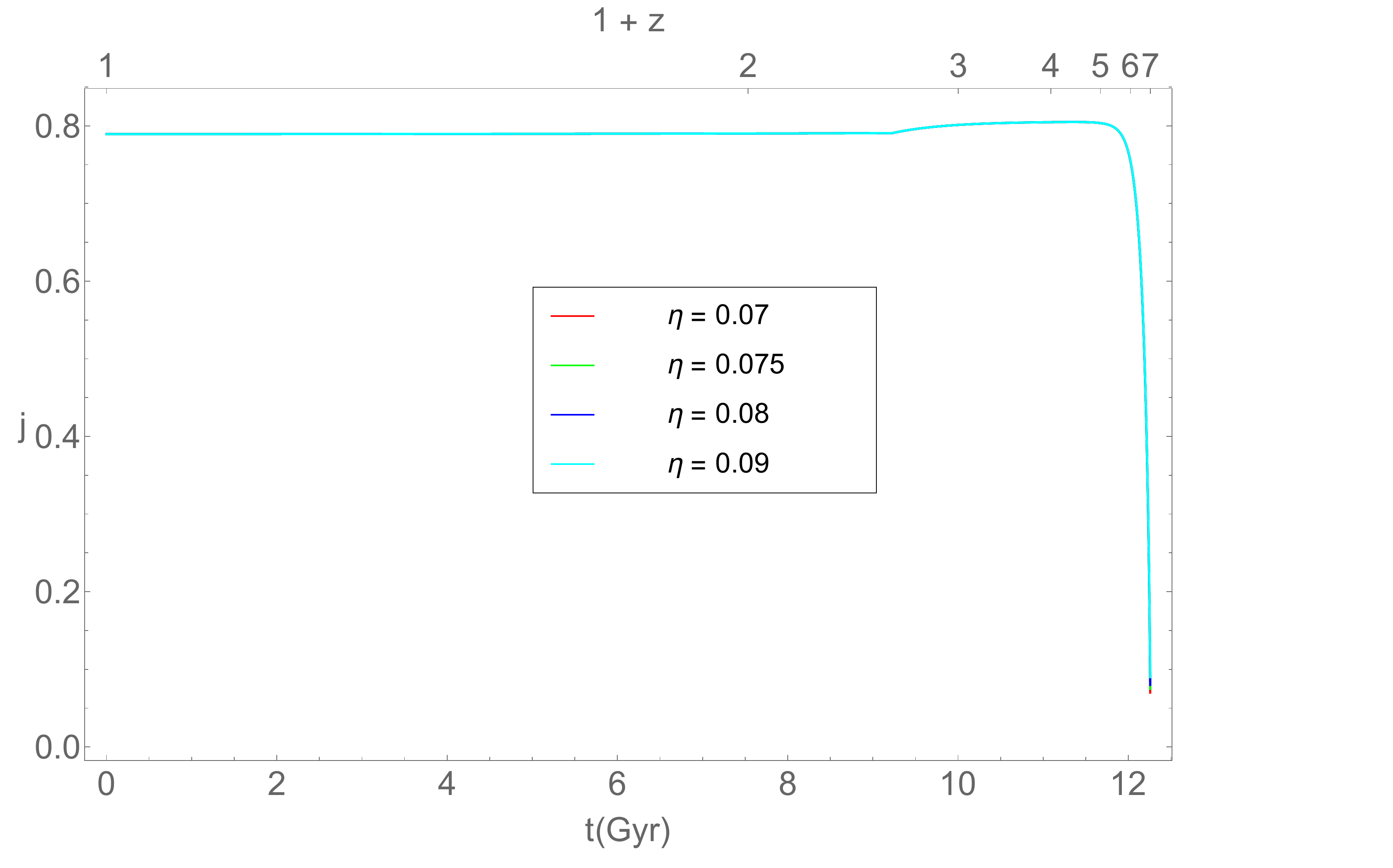}} 
\subfigure[\label{acc_bz_st_mg_jh}]{\includegraphics[scale=0.2]{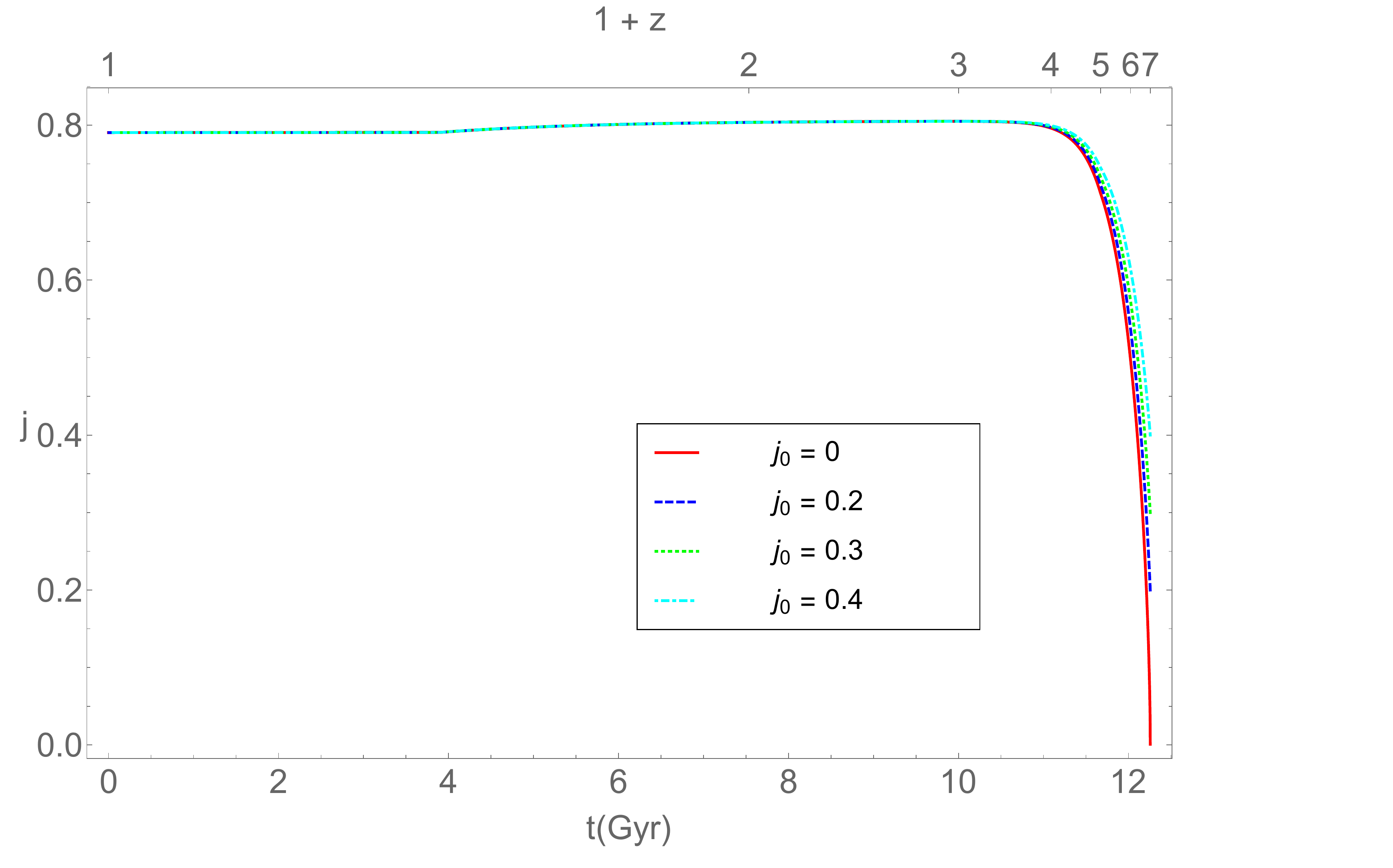}}  
\caption{The spin evolution, $j$($t$), for run \# 5.1 -- \# 5.6 [see Table \ref{table_acc_bz_st_mg}] (a -- f) are shown, when there is accretion, stellar capture, merger and BZ torque present for deviation of various parameters from their values in the canonical set.}
\label{acc_bz_st_mg_j}
\end{figure}

\section{Applications of Our Model}\label{4}

\subsection{{\it Impact on the $M_{\bullet}$--$\sigma$ relation}}\label{4.1}

All the solutions of $M_{\bullet}$ are dependent on the value of $\sigma$, which fixes the value of $t_{s}$ and $z_{s}$ for different galaxies given the same $M_{s}$. By calculating $M_{\bullet}(\sigma, z$), we obtain the evolution of the $M_{\bullet}$--$\sigma$ relation. We have assumed that the value $\sigma$ is constant from formation redshift until the present time since its variation is relatively small and reduces over Hubble time by a factor of $\sim$ 15$\%$ [see \cite{2009ApJ...694..867S}, Figure 8]. We have discussed this in \S 6. In our future models, we plan to include the time variation of $\sigma$ using an empirical form motivated by  \cite{2009ApJ...694..867S}, which assumes a small variation $\sigma(z) = \sigma_{0} (1 + z)^{-\gamma}$, where $\sigma_{0}$ is the present-day value of $\sigma$. But for now, the focus is to isolate all the other effects first. We have calculated $M_{\bullet}(\sigma, z)$ and derived $p(z)$, the index of the $M_{\bullet}$--$\sigma$ relation [see Fig \ref{msigmarl}]. In deriving $p(z)$, we have considered the observed range of $\sigma (z)$, to derive the corresponding range of $M_{\bullet} (z)$ using our evolution model. To start with, we assumed that, at the formation redshift, $p=5$, which is set by the Faber--Jackson relation. It is a reasonable assumption, given that black hole formation models produce masses proportional to the bulge mass \citep{2001A&A...379.1138M}. Even if this were not true for the small initial seed mass, the power-law index $p(z)$ would eventually be dominated by the gas and star accretion that inflates the final mass by a factor $M_\bullet/M_{s} \simeq 10^{3} - 10^{4}$. This is an initial fiducial value to derive the evolution that clearly does not change the long-term or near-term value of $p$. At the saturation time, the value of $p$ = 4, as predicted by the \cite{2003ApJ...596L..27K} model. Thereafter, the black holes grow by stellar capture and mergers alone. Since the growth rate reduces, the slope almost remains near 4 after the saturation.
\begin{figure}[H]
\centering
\subfigure[\label{slope_drop}]{\includegraphics[scale=0.2]{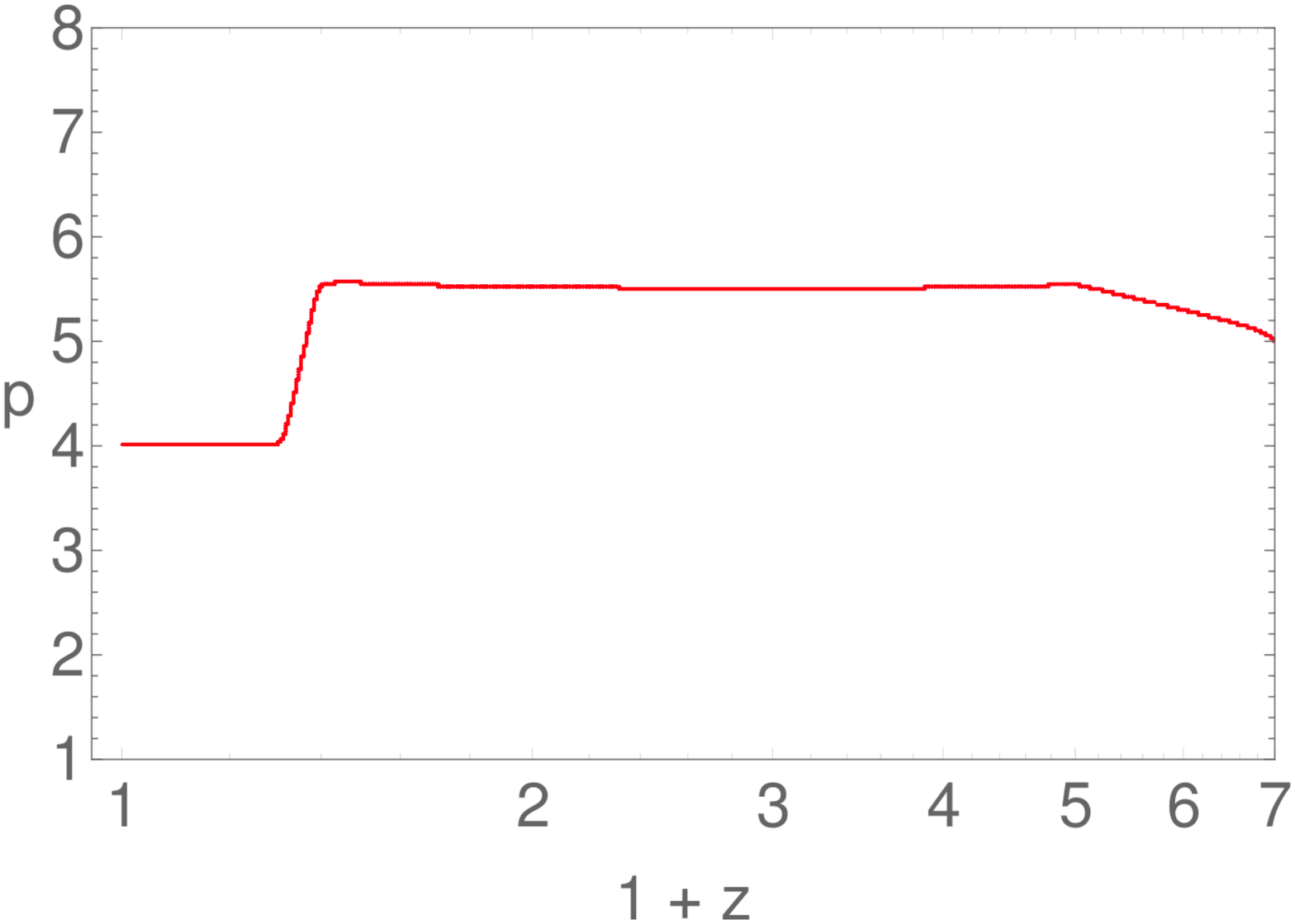}}\hspace{0.1 cm}
\subfigure[\label{msigmarla}]{\includegraphics[scale=0.2]{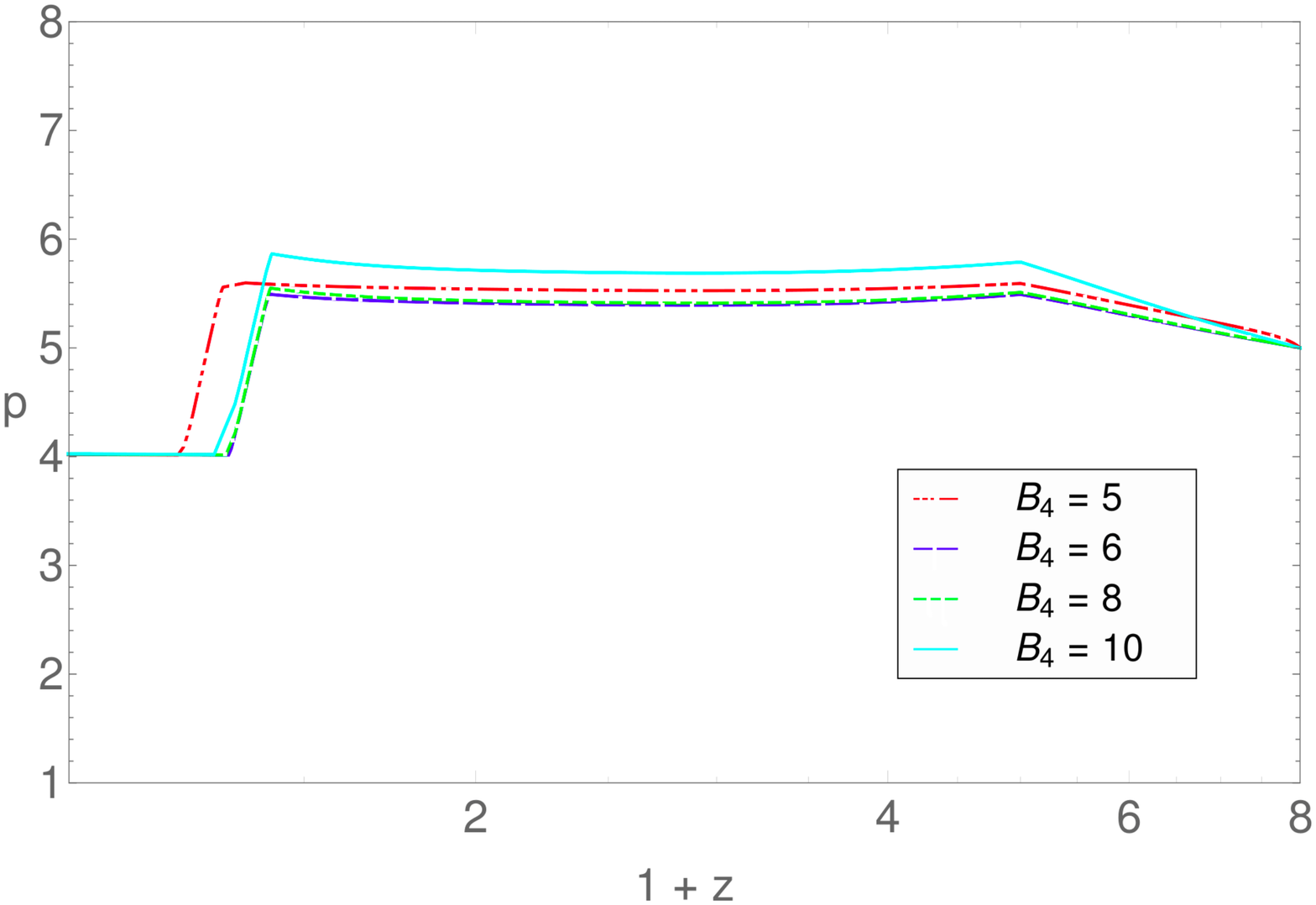}}  \\
\subfigure[\label{msigmarlb}]{\includegraphics[scale=0.2]{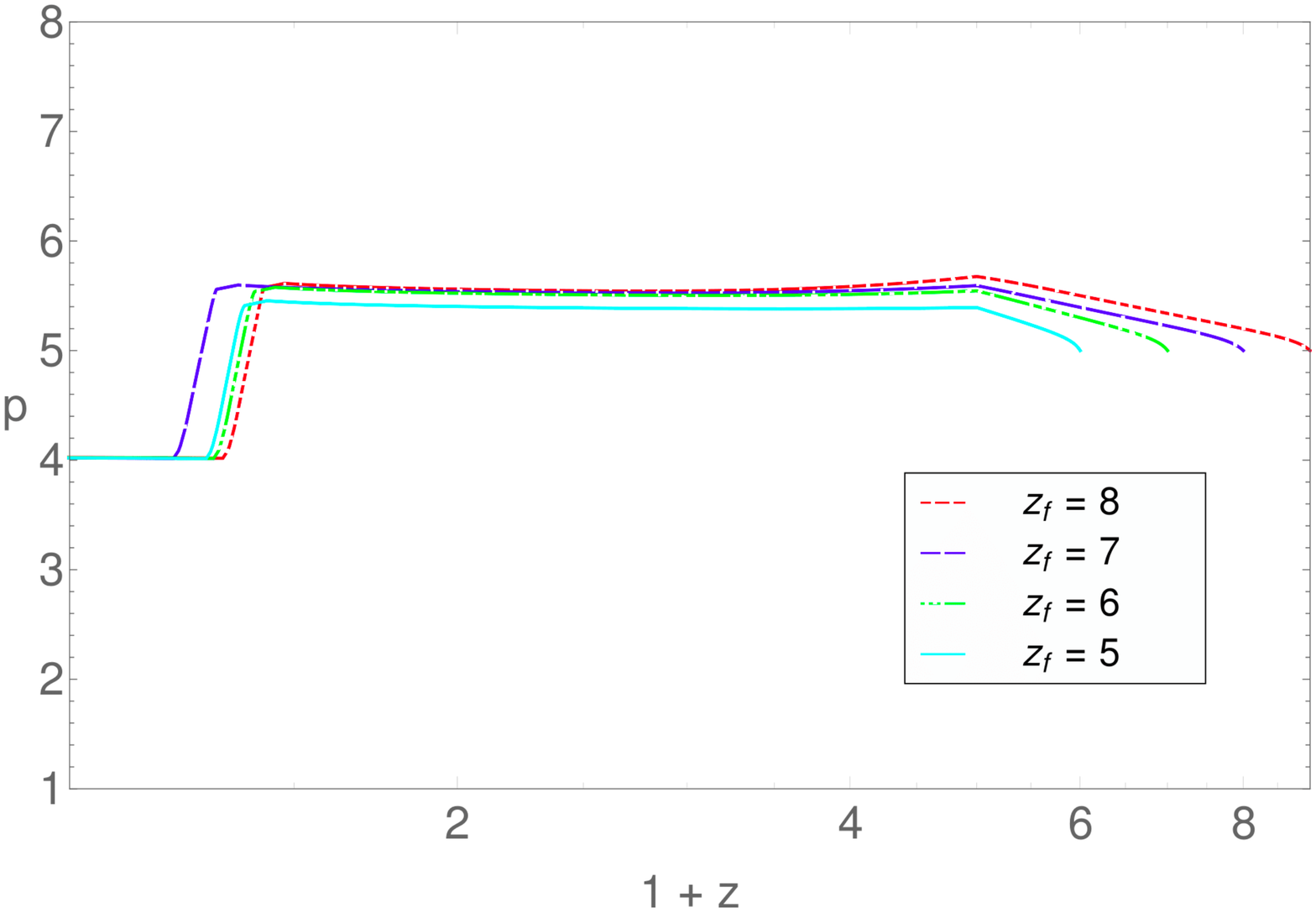}}\hspace{0.1 cm}
\subfigure[\label{msigmarlc}]{\includegraphics[scale=0.2]{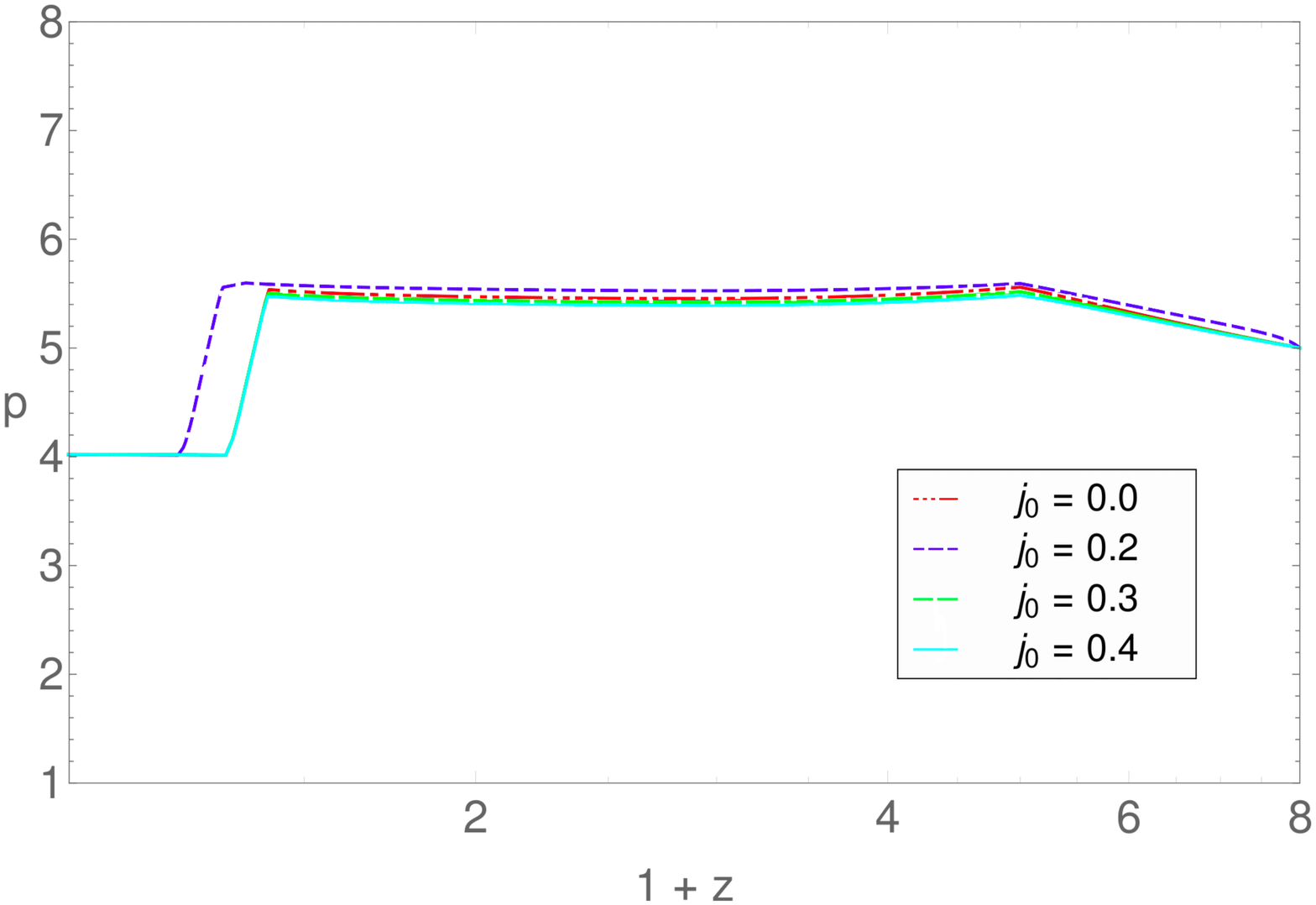}}
\caption{The evolution of $p(z)$ for $\gamma$ = 1.1, $M_{s}$ = $10^{4} M_{\odot}$ is presented above [Figure \ref{slope_drop} shows the canonical case, the variation with $B_{4}$ is seen in Figure \ref{msigmarla} for $\{z_{f}$ =7, $j_{0}$ = 0.2$\}$, the variation with $z_{f}$ in Figure \ref{msigmarlb} for $\{B_{4}$ = 5, $j_{0}$ = 0.2$\}$, and the variation with $j_{0}$ in Figure \ref{msigmarlc} with $\{z_{f}$ = 7, $B_{4}$ = 5$\}$].}
\label{msigmarl}
\end{figure}

Now, we discuss the dependence of $p(z)$ on the parameters $\{{B_{4}, z_{f}, j_{0}}\}$. Figure \ref{slope_drop} in the upper panel of Figure \ref{msigmarl} shows $p(z)$ for the canonical case, and Figures \ref{msigmarl}(b)--(d) show its deviation in the parameter space of $\{B_{4}$, $z_{f}$, $j_{0}\}$. We see a change of slope to $p = 4$ near the saturation point as expected, following the dependence for momentum-driven flow [see Equation (\ref{satm})]. Before $z_{s}$, the $p$ value is almost constant, which agrees with the previous work that finds little evolution of the $M_{\bullet}$--$\sigma$ relation. A more accurate evolution can be carried out by considering the mass and redshift distribution function of the black holes to carry out a population synthesis to derive $p(z)$ \citep{2015MNRAS.452..575S}. In Figure \ref{msigmarl}, we observe for all the cases that there is a little variation with changes in $\{B_{4}, z_{f}, j_{0}\}$ in the considered range. Thus, we conclude that this relation is expected to be within the observed range of 4--5 as $z_{f} \rightarrow 0$. Next, we compare our results with data obtained in \cite{2018JApA...39....4B} from the observed intensity profiles of these galaxies listed in \cite{2004ApJ...600..149W}. These galaxies are within the redshift range 0.004 -- 0.002 (see Table \ref{data}).

\begin{table}[H]
\begin{center}
\scalebox{0.7}{
\begin{tabular}{|c |c |c |c |c |}\hline
$\#$ & Galaxy & $M_{\bullet}$ (in $10^{7} M_{\odot}$) & $\sigma$ (km s$^{-1}$) & $z$ \\ \hline\hline
1 & NGC 3379 & 13.6 & 230 & 0.00304 $\pm$ 0.00001\\ \hline
2 & NGC 3377 & 2.60  & 217 & 0.00222 $\pm$ 0.00001\\ \hline
3 & NGC 4486 & 188  & 433 & 0.00428 $\pm$ 0.00002\\ \hline
4 & NGC 4551 & 3.77  & 218 & 0.00392 $\pm$ 0.00002\\ \hline
5 & NGC 4472 & 117  & 542 & 0.00327 $\pm$ 0.00002\\ \hline
6 & NGC 3115 & 17.0  & 230 & 0.00221 $\pm$ 0.00001\\ \hline
7 & NGC 4467 & 0.493  & 77 & 0.00475 $\pm$ 0.00004\\ \hline
8 & NGC 4365 & 67.7  & 453 & 0.00415 $\pm$ 0.00002\\ \hline
9 & NGC 4636 & 58.0  & 251 & 0.00313 $\pm$ 0.00001\\ \hline
10 & NGC 4889 & 299  & 467 & 0.02167 $\pm$ 0.00004\\ \hline
11 & NGC 4464 & 1.12  & 112 & 0.00415 $\pm$ 0.00001\\ \hline
12 & NGC 4697 & 20.76  & 215 & 0.00414 $\pm$ 0.00001\\ \hline\hline 
\end{tabular}
}
\end{center}
\caption{Data from BM18 [Based on \cite{2004ApJ...600..149W}] for 12 galaxies Used for matching Our results with observations are given above.}
\label{data}
\end{table}

\begin{table}[H]
\begin{center}
\scalebox{1}{
\begin{tabular}{|c |c |c |c |}\hline
 \# & References & $p$  &  $k_{0}$\\ \hline\hline
1 &\cite{2000ApJ...539L...9F} & 4.8  & 0.5 \\ \hline
2 & \cite{2000ApJ...539L..13G} & 3.75 & 0.9 \\ \hline
3 & \cite{2001ApJ...547..140M} & 4.72 & 0.5\\ \hline
4 & \cite{2002ApJ...578...90F} & 4.58 & 0.7\\ \hline
5 & \cite{2002ApJ...574..740T} & 4.02 & 0.83\\ \hline
6 & \cite{2005SSRv..116..523F} & 4.86 & 0.57\\ \hline
7 & \cite{2009ApJ...698..198G} & 4.24  & 0.7 \\ \hline
%8 & {\em ibid} & 3.96 $\pm$ 0.42 & 25 Elliptical galaxies \\ \hline
8 & \cite{2013ARAA..51..511K} & 4.38 & 1.48 \\ \hline
9 & \cite{2013ApJ...764..184M} & 5.64 & 0.42\\ \hline
%10 & \cite{2013ApJ...764..151G} & 5.53 $\pm$ 0.34 & 51 non - barred galaxies\\ \hline
10 & \cite{2013ApJ...765...23D} & 4.06 & 0.97 \\ \hline
11 & \cite{2017ApJ...838L..10B} & 4.76 & 1.69 \\ \hline
%14 & {\em ibid} & 3.90 $\pm$ 0.93 & 16 AGN host galaxies \\ \hline
12 & \cite{2019ApJ...887...10S} & 6.10 & 0.27 \\ \hline
\hline \hline
\end{tabular}
}
\caption{Survey of the $M_{\bullet}-\sigma$ relation [see Equation (\ref{msigma})] Giving the Historical Determinations of the Slopes and Constant When $M_{\bullet}$ Is in units of $10^{7} M_{\odot}$ and $\sigma$ is in units of 100 km s$^{-1}$.}
\label{msigmat}
\end{center}
\end{table}

In Figure \ref{obs}, the red curve corresponds to $z = 0.003$ and the green curve corresponds to $z =  0.23$. We see that the red curve gives the better fit to the data presented in Table \ref{data}, which is similar to observed values, as the range of redshifts of these galaxies are in the range of 0.001 -- 0.004. We provide a scatter plot of \{$p, k_{0}$\} pairs (see Equation (\ref{msigma})) from the literature (shown in Table \ref{msigmat} and Figure \ref{scatter_plot}) below, along with the values obtained from our model, which is within the observed range. The value of $p$, in our model is nearly 4 owing to the saturation mass used in our evolution model [see Equation (\ref{satm})], based on the prescription of the momentum-driven flow \citep{2003ApJ...596L..27K}. If the energy-driven flow \citep{1998A&A...331L...1S} dominates, we expect $p\simeq 5$ in the local universe. The \cite{2003ApJ...596L..27K} model invokes the presence of cooling sources, whereas the energy-driven flow assumes that there are no cooling processes involved in the medium. Similarly for full loss theory, $\dot{N}_{f} \propto \sigma^{5}$. With one or more of the effects of substantial merger rates, full loss cone stellar capture rates, absence of cooling sources, and heavy seeds, the values of \{$p, k_{0}$\} can deviate from \{4, 1\} [see Equation (\ref{paradigm}) in our paper, which represents our basic paradigm]. We hold that the assumptions of the momentum-driven flow and the steady loss cone theory are more appropriate.

\cite{2009ApJ...694..867S} analyzed the data of over 40,000 early-type galaxies from the  Sloan Digital Sky Survey (SDSS) and they determined $k_{0} (z)$ $\propto$ $(1 + z)^{0.33}$ [see Equation (\ref{msigma})]. According to their analysis, this relation almost holds throughout the age of the universe. Our model also predicts an almost constant $p$ throughout the entire redshift range as expected from Equation (\ref{paradigm}). We have also shown $k_{0}(z)$ starting from an approximate value (considering the Faber--Jackson relation with the seed mass within the considered range and $\sigma$). Since we have considered a constant $\sigma$, $k_{0}(z)$ is predicted to decrease. According to \cite{2009ApJ...694..867S}, the $M_{\bullet}$--$\sigma$ relation is given by
\be 
\log[M_{\bullet} / M_{\odot}] = 8.21 + 3.83 \log[\sigma_{200}] + \alpha\log[1 + z]. \label{shankareq}
\ee 
If we consider the relation above using $\alpha$ = 0.33 and assume the empirical relation $\sigma (z)= \sigma_{0} (1 + z)^{-0.25}$, where $\sigma_{0}$ is the present-day velocity dispersion to calculate $M_{\bullet} (t)$, we find that the final $M_{\bullet}$ is similar to the prediction from our evolution model; see Figure \ref{shankar_mass}, which compares $M_{\bullet 5}(t)$ from Equation (\ref{shankareq}) with our model prediction for ($M_{s} = 10^{5} M_{\odot}$, $\sigma_{100}$ = 1). We see that if the $\sigma$ stays constant, the final mass attained is also nearly same, which was also concluded by \cite{2017NatAs...1E.147A}.

\begin{figure}[H]
\centering
\subfigure[\label{obs}]{\includegraphics[scale=0.2]{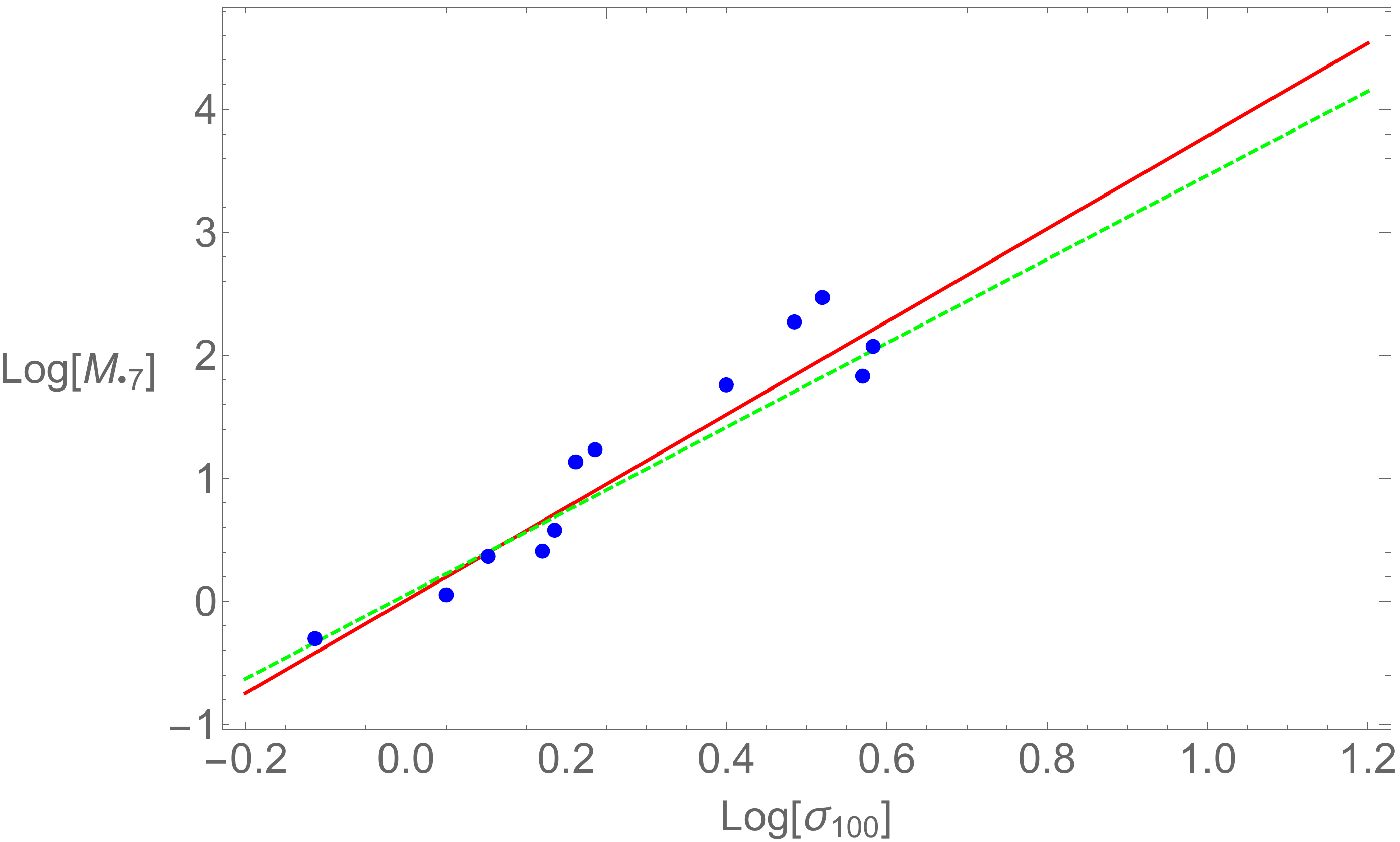}}\hspace{0.4 cm}
\subfigure[\label{scatter_plot}]{\includegraphics[scale=0.31]{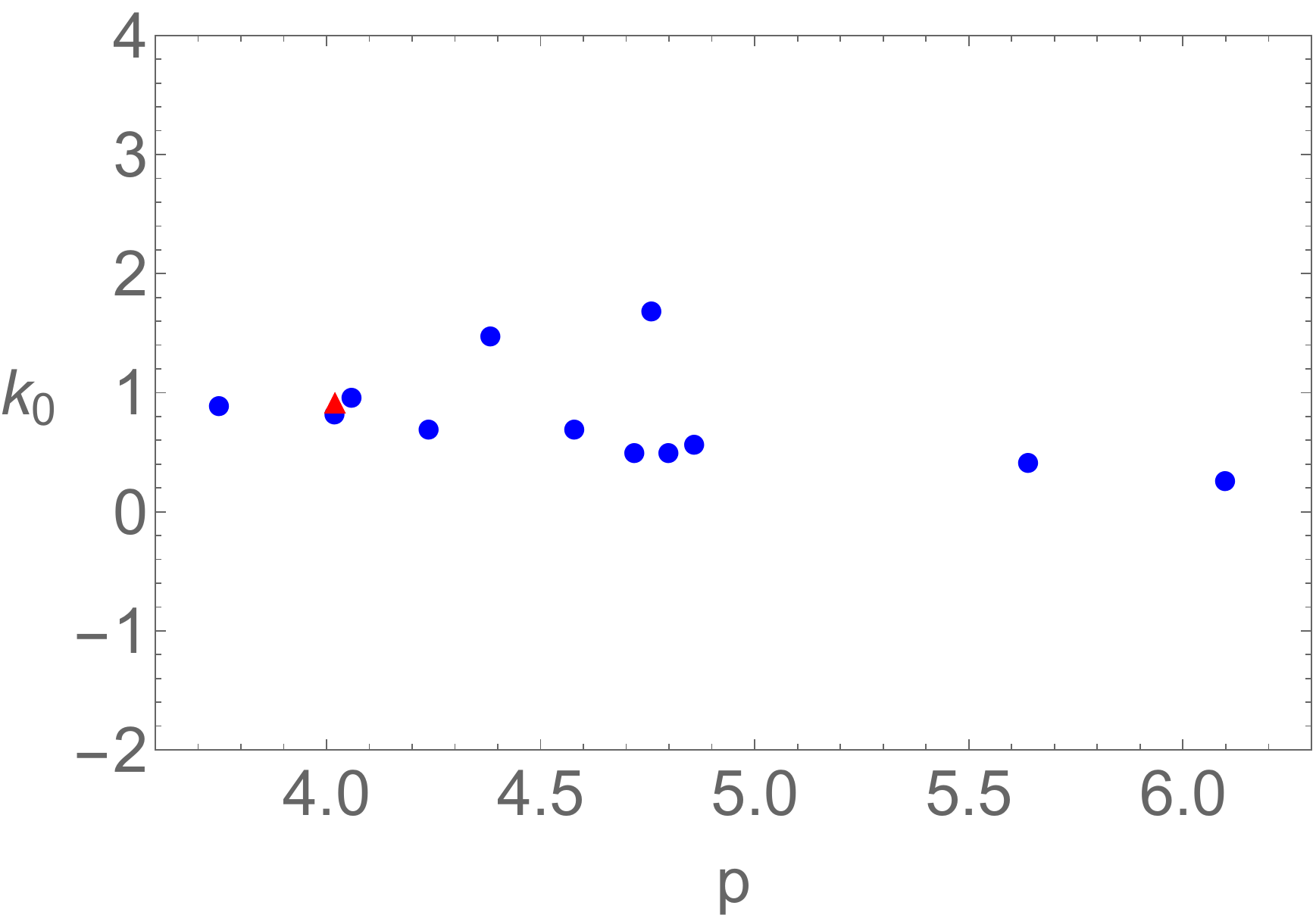}}
\caption{(a) A plot of $\log(M_{\bullet {7}})$ vs $\log(\sigma_{100})$ for two different redshifts [$z = 0.003$, (red) and $z = 0.23$, (green)] calculated from our evolution model is shown and compared with the data obtained from our model in BM18 for the 12 elliptical galaxies (whose) redshift lies in the range 0.004 -- 0.002) and (b) scatter plot of the values of $k_{0}$ and $p$ [see Equation (\ref{msigma})] available in the literature (represented by the blue dots) and the red point shows the values obtained using our model.}
%\label{obs}
\end{figure}

\begin{figure}[H]
    \centering
    \subfigure[\label{intcpt_drop}]{\includegraphics[scale=0.2]{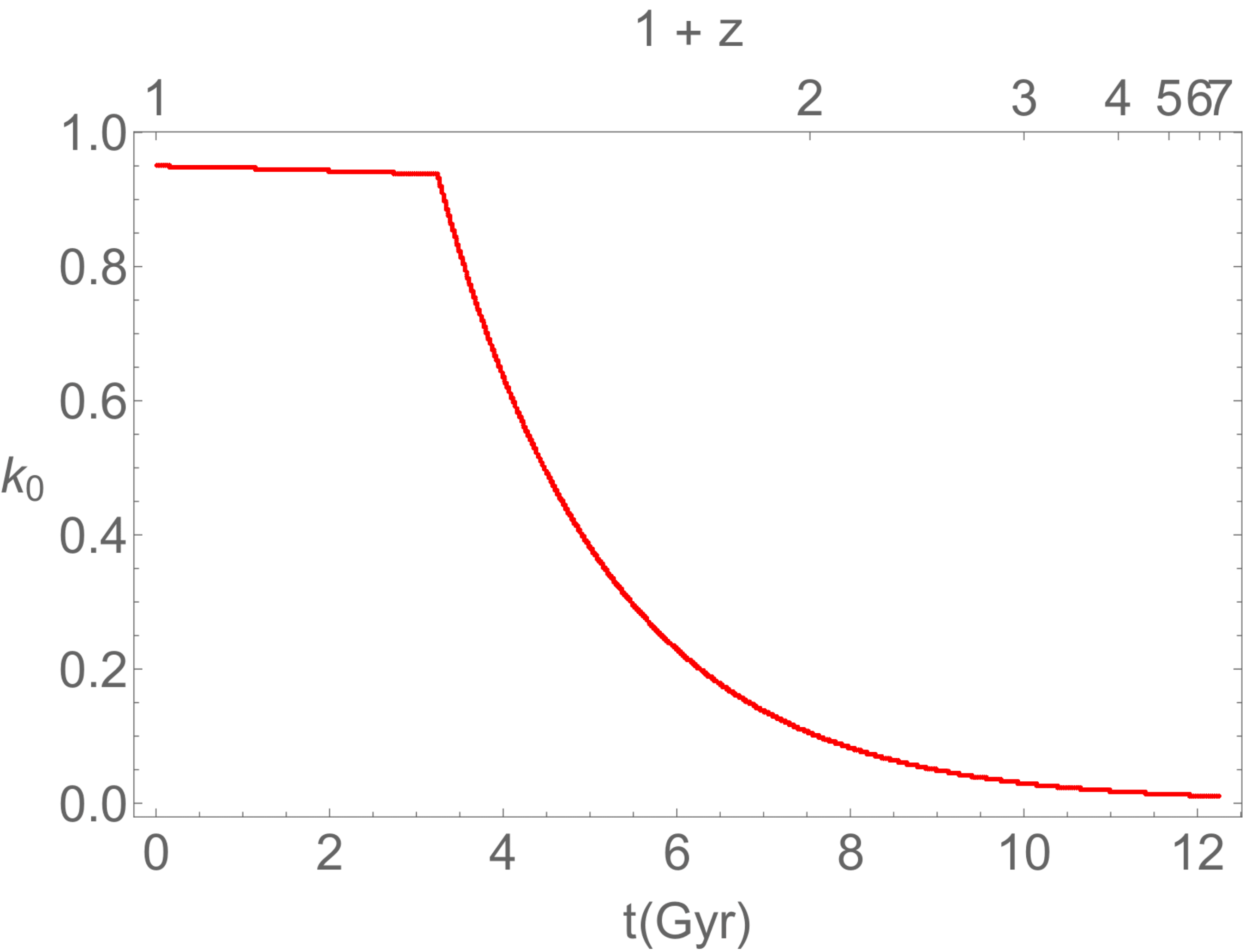}}\hspace{0.1 cm}
\subfigure[\label{shankar_mass}]{\includegraphics[scale=0.21]{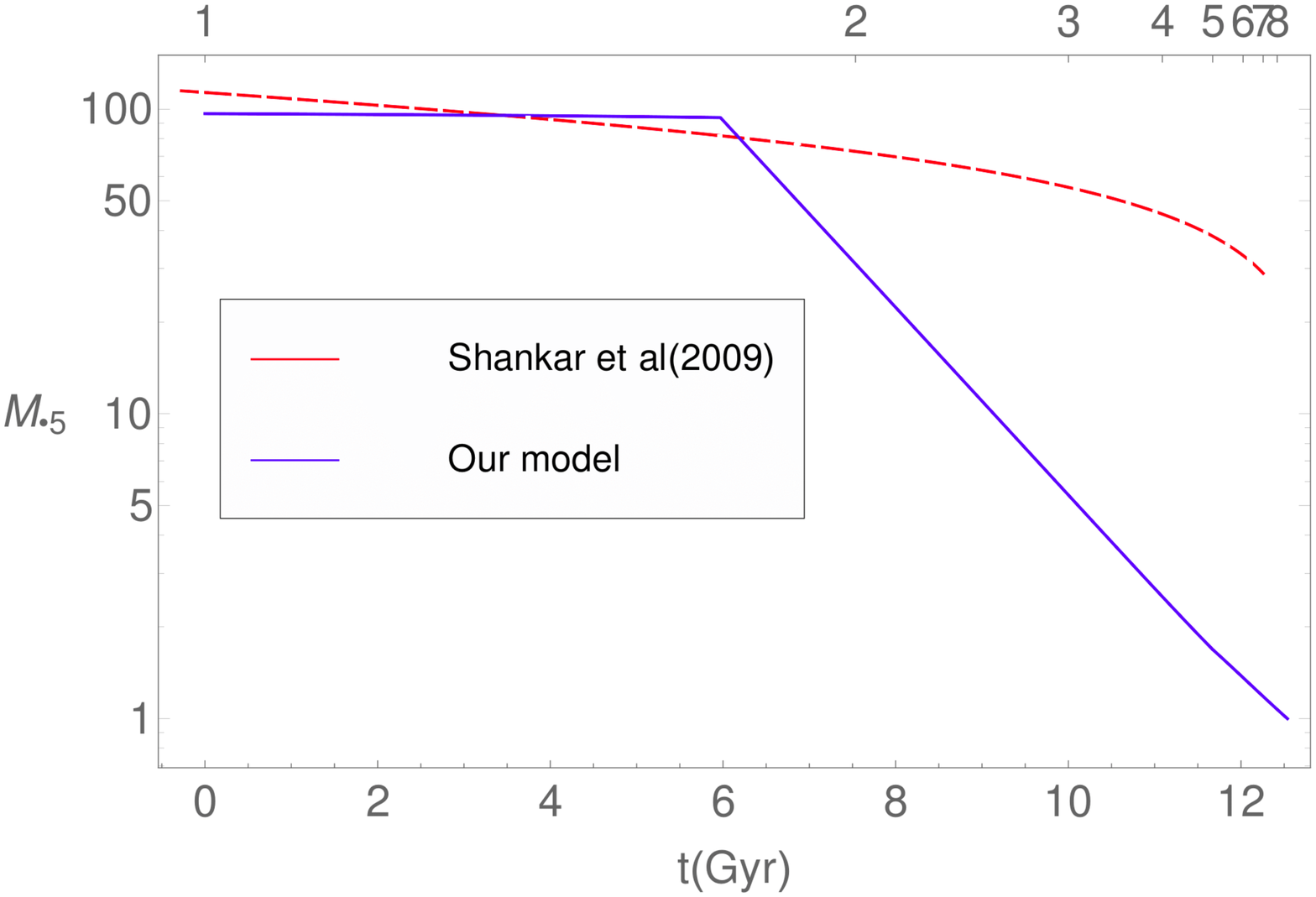}}
    \caption{The evolution of the index $k_{0}(z)$ for $\gamma$ = 1.1 is shown for the canonical case (a) and (b) $M_{\bullet 5} (t)$ from prescription of \cite{2009ApJ...694..867S} and our model for $M_{s} = 10^{5} M_{\odot}$.}
\end{figure}

\subsection{\it{Black hole archaeology}}\label{4.2}
If we use the final mass and spin as boundary conditions of the mass evolution, we can evolve our model backward in time, a process that we call {\it black hole archaeology}. \cite{2019A&A...625A..23C} analyzed the Optical--UV emission of distant quasars ULASJ134208.10+092838.61 ($z$ = 7.54), ULASJ112001.48+064124.3 ($z$ = 7.08) and DELSJ003836.10-152723.6 ($z$ = 7.02) to study their properties and found the presence of an accretion disk. They used the relativistic disk models KERRBB and SLIMBH to model the emission with approximations to describe the emission as a function of {$M_{\bullet}$, $\eta$, $j$} and the viewing angle $\theta_{\nu}$. They found that the accretion rate for all sources is sub-Eddington and thus conclude that all three have reached the last stages of their evolution. 

\begin{table}[H]
\begin{center}
\scalebox{0.8}{
\begin{tabular}{|c |c |c |c |c |c |}\hline
 & \multicolumn{2}{|c|}{Input Parameters} & \multicolumn{3}{|c|}{Combinations of \{$M_{\bullet s}$ (in $10^{9} M_{\odot}$), $j_{0}$\}} \\ \hline
\# & $\eta$ & $j_{f}$ &  $z =10$ &  $z =15$ & $z =20$\\ \hline\hline
1 & 1 & 0.7 & \{0.12, 0.76\} & \{0.06, 0.8\} & \{0.02, 0.5\}\\ \hline
2 & 0.1 & 0.7 & \{0.75, 0.65\}  & \{0.63, 0.55\} & \{0.6, 0.27\} \\ \hline
3 & 1 & 0.45 & \{0.1, 0.7\}  & \{0.05, 0.75\} & \{0.02, 0.5\}\\ \hline
4 & 0.1 & 0.45 & \{0.75, 0.45\} & \{0.63, 0.38\} & \{0.6, 0.25\}\\ \hline\hline 
\end{tabular}
}
\end{center}
\caption{Combinations of Seed Mass and Spin, \{$M_{\bullet s}$, $j_{0}$\}, at $z_{f} = \{10, 15, 20\}$ for Quasars with Mass $\simeq$ $10^{9} M_{\odot}$ at $z \simeq 7$ for Different Sets of Input Parameters, $\{\eta, j_{f}\}$, Where, $j_{f}$ Is the final spin at $z = 7$.}
\label{arch_table}
\end{table}

\begin{figure}[H]
    \centering
    \subfigure[\label{sm_m_t}]{\includegraphics[scale=0.25]{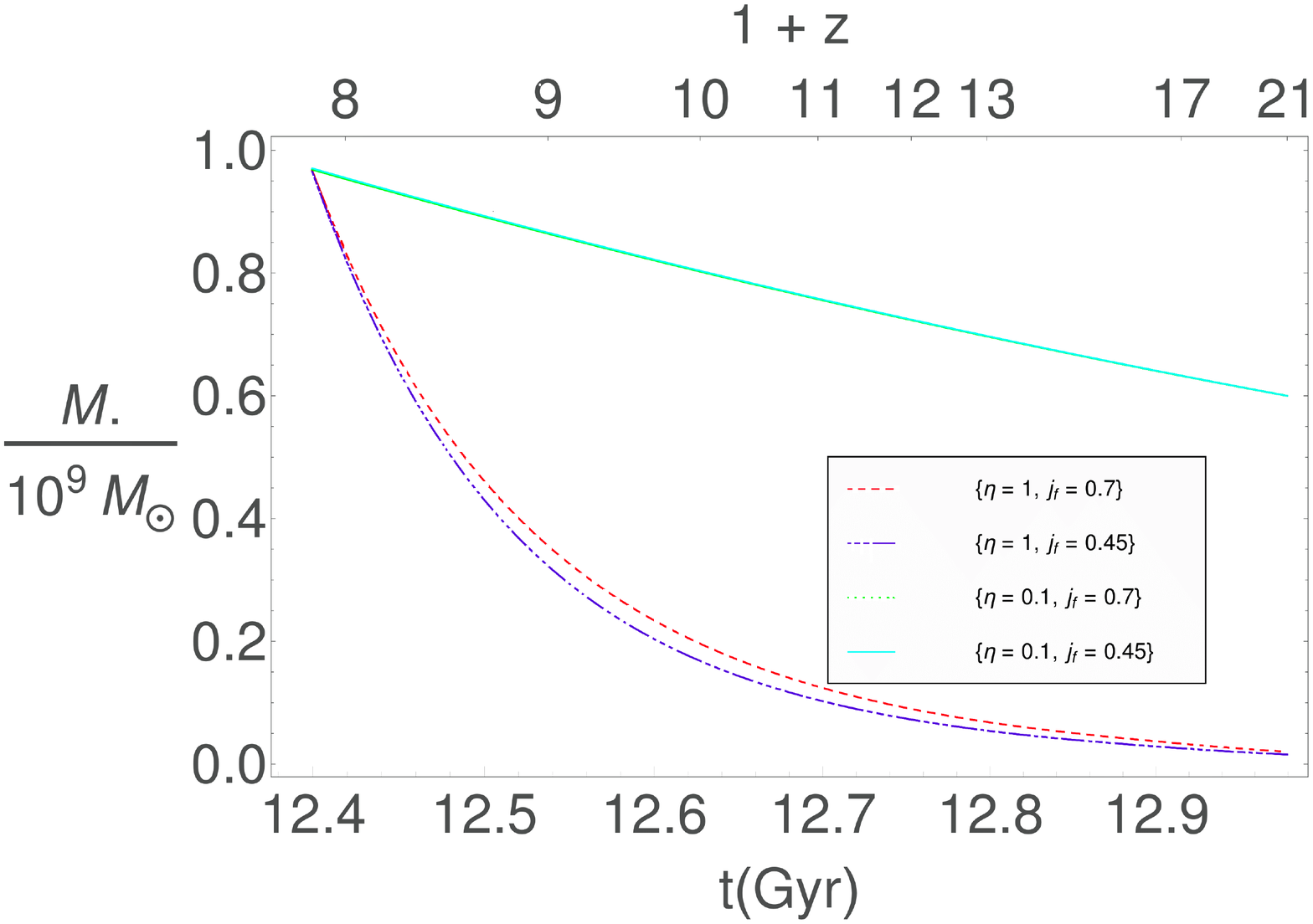}} \hspace{0.1 cm}
     \subfigure[\label{sm_j_t}]{\includegraphics[scale=0.25]{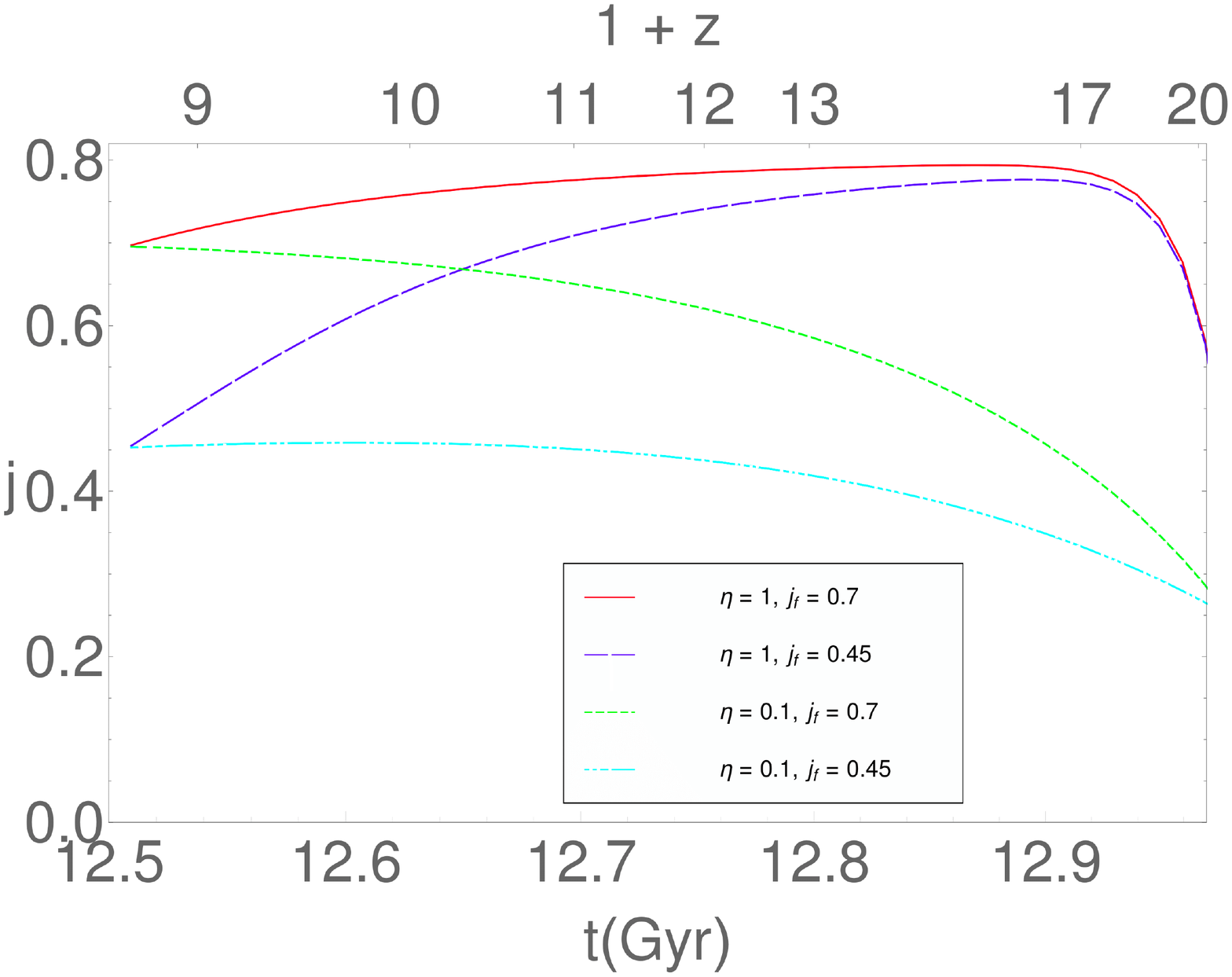}} 
    \caption{(a) $M_{\bullet}(t)$ and (b) $j(t)$ evolution for different combinations of $\eta$ and final spin at $z \simeq 7$, $j_{f}$ for $z_{f} = 20$ are shown for final mass at $z \simeq 7$, $M_{f} \simeq 10^{9} M_{\odot}$.}
    \label{samuele_mass}
\end{figure}

 We observe from the mass and spin values for the quasars listed in Table 1 of \cite{2019A&A...625A..23C}, (as determined through KERRBB and SLIMBH models), that the following input sets of $\{\eta, j_{f}\} =\{ \{1, 0.7\},\{0.1, 0.7\},\{1, 0.45\},\{0.1, 0.45\} \}$ are suggested. They have also calculated $M_{\bullet}$ for $j_{f} = \{0, 1\}$, the extreme ends of the spin values. We have taken the final mass to be $M_{\bullet} \simeq 10^{9} M_{\odot}$ at $z \simeq 7$ (as suggested by their models) and evolved our model backward the for the different sets of $\{\eta, j_{f}\}$ given above to find the initial seed masses at $z_{f} = \{10, 15, 20\}$. We see that when $\eta = 1$, the seed mass is also quite small as compared to the case of $\eta = 0.1$ (see Table \ref{arch_table}); this is expected owing to the difference in accretion rate [see Figure \ref{sm_m_t}]. The $j_{f}$ values does not make much difference to $M_{\bullet}(t)$ when $\eta$ is fixed. For the case of spin evolution, when $\eta = 1$, $j$ increases and then decreases, but for $\eta = 0.1$, it continues to decrease [see Figure \ref{sm_j_t}]. This is because, for high accretion rate, the spin reaches its maximum value rapidly and then it reduces owing to BZ torque and minor mergers to $j_{f}$; however, when $\eta = 0.1$, the mass growth is slower, so it does not reach the maximum spin within a gigayear, as both BZ and merger terms are mass dependent and hence not as effective. It seems that a heavy seed of nearly $M_s =10^7 M_\odot$ is required at $z=20$ even if $\eta=1$ (see Table \ref{arch_table}). This poses difficulties for black hole formation models [eg. \cite{2018JApA...39....9P}] or for the mass suggested by \cite{2019A&A...625A..23C}. One possible resolution can be a two-phase accretion for the growth of quasars, with a short super-Eddington phase in the beginning without considering feedback, under very favorable conditions, in an environment where there is a lot of cold gas around the black hole, followed by a long sub-Eddington phase with feedback effects, as suggested by \cite{2012MNRAS.424.1461L}. The existence of a short super-Eddington phase is also suggested by \cite{2019A&A...625A..23C} and \cite{2014ApJ...782...69L}. \cite{2019A&A...625A..23C} find that if the seed black holes in these sources, with masses in the range $10^{2}$--$10^{4} M_{\odot}$, grow during $z_{f}$ = 20--10 at 15--30 times of the Eddington accretion rate with a low radiative efficiency ($\sim 10\%$), then they can reach the present-day mass within 0.7 Gyr.

\begin{figure}[H]
\centering
\subfigure[\label{m_t_opp}]{\includegraphics[scale=0.27]{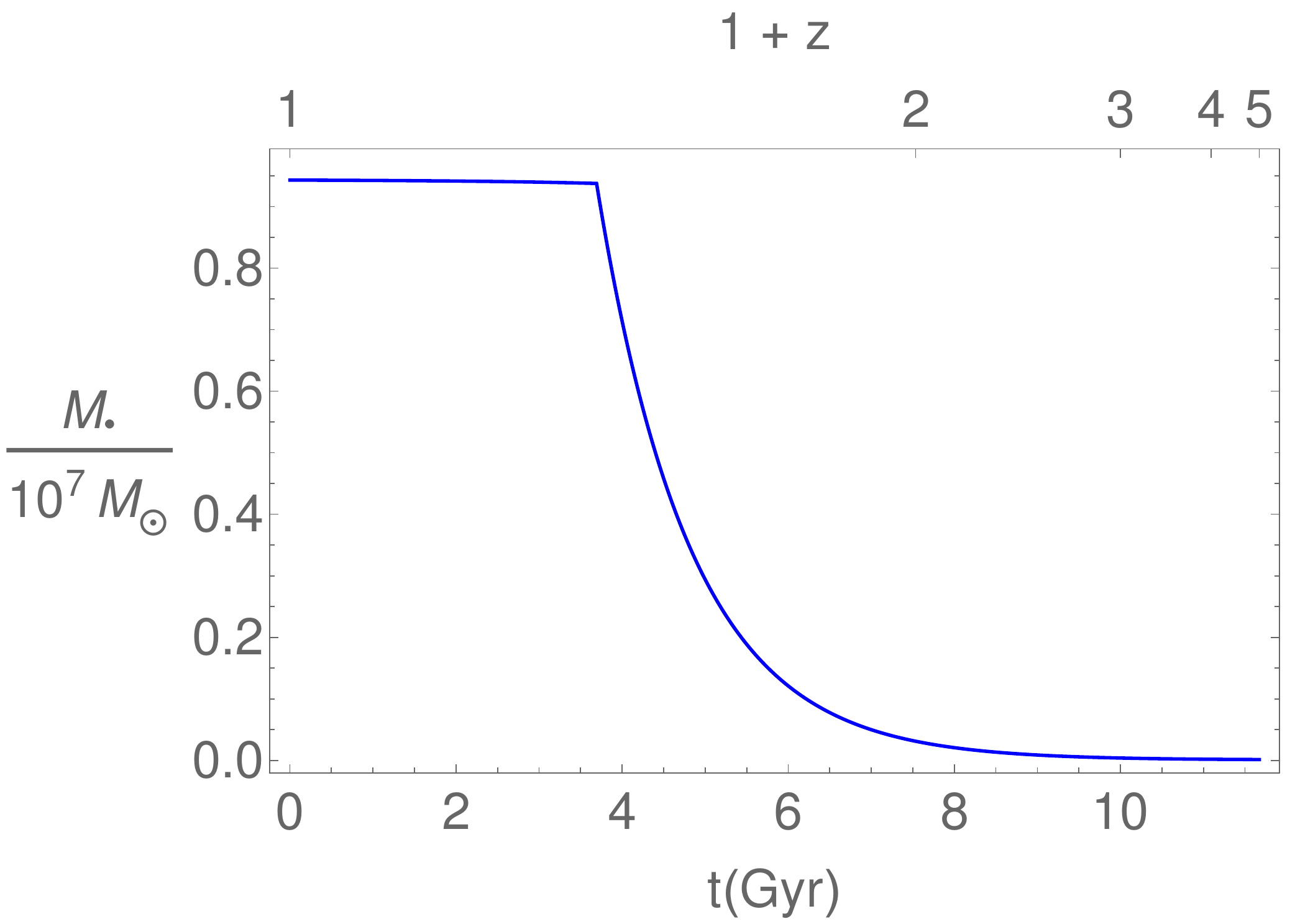}} \hspace{0.1 cm}
\subfigure[\label{j_t_opp}]{\includegraphics[scale=0.25]{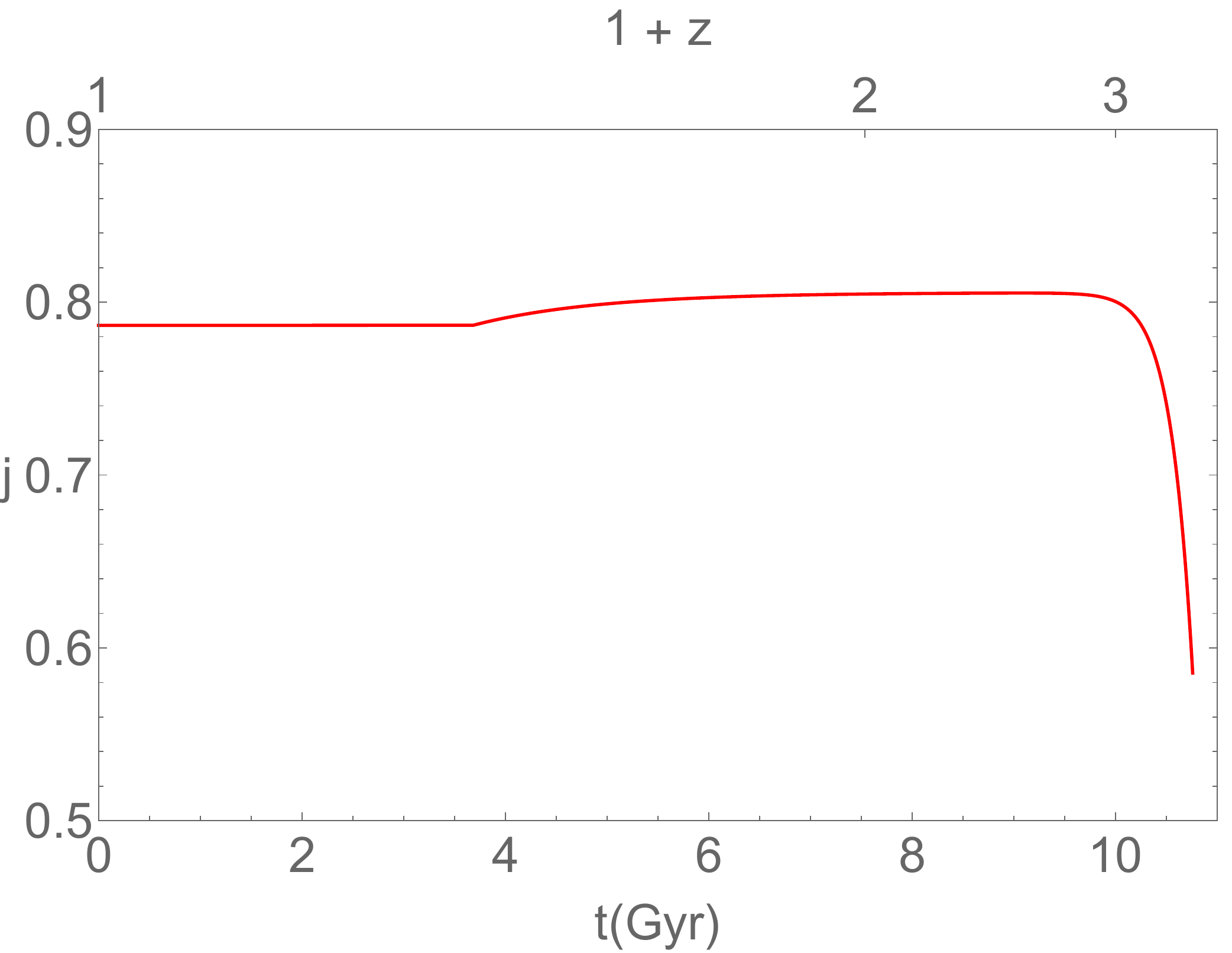}} 
\caption{(a) $M_{\bullet} (t)$ and (b) $j(t)$ evolution for the complete model, are shown, starting from final mass $\mu_{\bullet 5}$ = 94.3 with other parameters the same as that of run \# 5.1.4.}
\label{ev_opp}
\end{figure}

For comparison, we also evolve the final configuration for SMBHs, $\{ M_\bullet =10^7 M_\odot, j_f =0.8, z_f=0\}$, which is shown in Figure \ref{ev_opp}. We see from Fig \ref{m_t_opp} that the mass reaches a seed value of $3.5 \times 10^{4} M_{\odot}$, which is typical, and Figure \ref{j_t_opp} indicates a seed spin of $j_s =0.58$. With these illustrations, it is clear that our model is a useful tool for black hole archaeology.

\subsection{\it{Summary of the Results and Caveats}}\label{5}
We summarize our results here.

The key novel aspects of the paper are the relativistic inputs of the capture radius and tidal radius to the loss cone formalism, determining the applicable range of steady loss cone theory, and including all known contributions of gas, stellar, electromagnetic torque, and mergers through detailed formulae as recipes for calculating the joint spin and mass evolution relativistically while taking into account the effects of saturation, merger regimes, and the mode of stellar ingestion. We have applied this elaborate model to make predictions for the capture rate of stars, $\dot{N}_{s}$, for the evolution of the
$M_\bullet -\sigma$ relation and in retrodicting the initial black hole configurations from their more recent inferred ones. The detailed findings are summarized as follows:

\begin{enumerate}
\item We calculate $r_{t}$ using the effective Kerr potential to include the effect of the spin parameter and find $x_{t}(M_{8}, j, k, Q)$ (see Figure \ref{figrta} for both prograde and retrograde cases). We see from Figure \ref{figrta}, that a higher $j$ reduces $x_{t}$ owing to the relativistic potential. We see that $x_{t}(M_{8}, j, k, Q)$ is important in deriving $r_{t}$ and $l_{\ell} (M_{8}, j, k, Q)$, which has an impact on $\dot{N}$. Even a small change in $x_{t} (M_{8}, j, k, Q)$ has an impact on $l_{\ell}(M_{8}, j, k, Q)$.
\item We calculate the loss cone radius $x_{\ell}$ = Max[$x_{t}$, $x_{c}$] (Figure \ref{figllca}). For higher-mass black holes, the prograde capture radius $x_{c}$ goes down dramatically, so it reduces the capture rate. $x_{\ell}$ (see Figure \ref{figllca}) has an impact on $\dot{N}_{s}(M_{\bullet},j, k, Q, \epsilon_{s}, \sigma)$ which reduces with mass but increases with spin for both the prograde and the retrograde cases. This can be further explored with axisymmetric distributions $f(E, L_{z})$, as it is known that the Carter's constant is a function of $L^{2} - L_{z}^{2}$. A critical mass value of $M_{c} (j, Q) \simeq 3\times 10^{8} M_{\odot}$ is found; for higher masses $r_{\ell}$ is set by $r_{c}$ instead of the tidal radius (the black lines in Figure \ref{figrtc}a and Figure \ref{figrtc}b show the critical point, when $r_{t}/r_{c}$ = 1); $M_{c}(j, Q)$ changes significantly with spin, and this has implications for cosmic evolution and its impact on $\dot{N}$ and black hole growth that need to be further explored (see Figure \ref{figrtc}).
\item We also calculate a relativistic correction to the tidal radius given by Equation (\ref{rtEquation}).
\item We calculate the effects of stellar capture for both full and steady loss cone theory. For most cases, we find that the steady loss cone model is appropriate. We have calculated the $\dot{N}_{s}$ using the prescription given by MM15 (but by not assuming an $M_{\bullet}$--$\sigma$ relation) while adding the relativistic corrections to $r_{\ell}$ (as shown in Figure \ref{stdlc}) to obtain $\dot{N}_{s}(M_{\bullet},j, k, Q, \epsilon_{s}, \sigma)$. This is smaller typically by a factor of 10 than the nonrelativistic model of MM15. Our predicted capture rates (see Figure \ref{stdlc}) of $10^{-5}$ to $10^{-6}$ yr$^{-1}$ can explain the observed rates \citep{2002AJ....124.1308D, 2009ApJ...698.1367G, 2015JHEAp...7..148K} of around $10^{-5}$ yr$^{-1}$ (dominated by black holes of $M_8 \lesssim 0.01$) and are a key result.
\item We calculate the impact of the evolution on the spin and mass of the SMBH [Figures \ref{acc_bz_st_mg_j}, \ref{acc_bz_st_mg_m}], and $M_{\bullet}$--$\sigma$ relation (Figure \ref{msigmarl}) as a function of redshift in a $\Lambda$CDM cosmology. We performed five experiments by adding the contributions of gas accretion, stellar capture, BZ effect, and mergers one by one and studied how it impacts the evolution. These are useful illustrations of the individual effects.
  
\item In Appendix \ref{3.2}, we derived the mass evolution in the nonrelativistic case assuming full loss cone theory by analytical expressions considering only accretion and stellar capture. We present the evolution of $M_{\bullet} (z)$ for different cases in Figure \ref{nrwtsp}. In Appendix \ref{3.1}, we have also considered the BZ torque, which contributes to spinning down the black hole with a strong poloidal magnetic field that extracts the spin energy, causing a spin-down of the black hole. Next, we studied the evolution of the spin and mass including all the effects one at a time. Figure \ref{acc_bz_j} shows the evolution in the presence of only accretion and BZ torque, while Figures \ref{acc_bz_st_j} and \ref{acc_bz_st_m} show the evolution in the presence of accretion, stellar capture, and BZ torque. All the effects of accretion, stellar capture, mergers, and BZ torque have been included for different parameter sets in Figures \ref{acc_bz_st_mg_m} and \ref{acc_bz_st_mg_j}. We see that the accretion term dominates over the other terms until saturation. This is because the stellar capture rate decreases with mass, and at the same time the mass growth rate by accretion increases. The mergers and stellar capture contribute significantly to the mass growth after the halt of accretion. The merger activity drops off after $z\gtrsim  4$. Therefore, in the presence of the merger term, the mass and spin evolution start to deviate from those of the case for evolution without mergers near $z\simeq 4$; due to an overall increase in the mass growth rate, the saturation occurs earlier, and the final mass of the black hole is also higher than that of the case for evolution without mergers.
\item We compare our results of $p(z)$ with available observations (see Figure \ref{obs} and Table \ref{data}) given in \cite{2018JApA...39....4B}, where $\sigma$ was calculated from observed intensity profiles for a set of galaxies given in \cite{2004ApJ...600..149W}.
\item We model $p(z)$ in the range $z = z_{f} \rightarrow 0$ with a constant $\sigma$, assuming that minor mergers do not change it substantially. This is seemingly consistent with observed $p$ in the nearby redshift range. But our predictions need to be tested by simulations and data available from future missions like the Thirty Meter Telescope (TMT), Very Large Telescope (VLT), and Extremely Large Telescope (ELT).
\item We have assumed that the seed mass $M_{s} \propto \sigma^{5}$, as suggested by the Faber--Jackson relation for deriving the evolution of the $M_{\bullet}$--$\sigma$ relation, as an application of our evolution model. Subsequently, the black hole grows, impacting $p(z)$.
\item We conclude that $p(z)$ changes gradually with redshift. Therefore, we expect that the late-type galaxies will have a higher $p$ compared to the early-type galaxies as suggested by \cite{2013ApJ...764..184M}. 
\item Our model is useful for carrying out black hole archaeology. Figure \ref{ev_opp} shows the evolution obtained when we run our model backward using the present-day initial conditions of $\{M_{\bullet}, j\}$ and we find that $M_{s} \simeq 3.5 \times 10^{4} M_{\odot}$, which is within the range of seed masses considered in the literature.
\end{enumerate}
We discuss these results in the next section.

\section{Discussion}\label{6}
{\it{Stellar capture rate of black holes: }}
The rates of TDEs for a single black hole in steady state have been derived by different authors as already mentioned in \S \ref{1}, with various physical effects included such as the Nuker profiles \citep{1999MNRAS.306...35S}, nonspherical galaxies \citep{1999MNRAS.309..447M}, resonant relaxation \citep{1996NewA....1..149R} and its quenching by relativistic precession \citep{1998MNRAS.299.1231R}, and black hole spin \citep{2012PhRvD..85b4037K}. The theoretical estimates range from $10^{-6}$ to $10^{-4}$ yr$^{-1}$ for the most part, while the observational results of \cite{2015JHEAp...7..148K}, \cite{2002AJ....124.1308D} (ROSAT surveys), and \citep{2009ApJ...698.1367G} (in UV band) have provided rates of TDEs for different wavelength bands to be about $\lesssim 10^{-5}$ yr$^{-1}$. MM15 model the nonrelativistic steady-state loss cone regime, taking into account the angular momentum dependence. We have expanded the theory to include relativistic effects in a Kerr potential to calculate the tidal and capture radius, which in turn, is an input to the loss cone theory that determines the rate of stellar capture. MM15 considered nonrelativistic theory and used $L_{\ell}(\sigma, r_{t}) = \sqrt{2r_{t}^{2} \phi_(r_{t}) - E}$. In our relativistic model,  the loss cone angular momentum $L_{\ell}(j, k, x_\ell, Q)$ is given by Equation (\ref{llcEquation}), where the loss cone radius $x_\ell$ is used instead of $x_t$ used in MM15; this causes a decrease in the value of $\dot{N}_{s}$ by a factor of a few, due to the decrease of the loss cone radius, bringing it more in line with observed estimates.

\cite{2017NatAs...1E.147A} determined the minimal mass of the present-day black holes by including the stellar capture process. They conclude irrespective of the seed masses that if $\sigma$ of the galaxies are nearly equal, then all the black holes reach almost the same mass, assuming that the $M_{\bullet}$--$\sigma$ relation holds throughout. All the black holes grow over the age of the universe to the present-day mass scale of $M_{6} \gtrsim 0.2$ (with 5\% lower confidence level), independent of their initial seed mass and the formation process. They conclude that the present-day $M_{\bullet}$ is nearly independent of the uncertainties in $z_{f}$, and provide a universal minimal mass estimate for the black holes that grow by gas accretion or mergers. This can explain the reason for not finding any intermediate-mass black holes with $M_{6} \lesssim 0.2$, which in turn implies that present-day galaxies that have $\sigma \lesssim$ 35 km s$^{-1}$ (at 5\% lower confidence level) do not contain a central black hole. We derive the evolution without any a priori assumption of $M_{\bullet}$--$\sigma$ relation throughout and also take into account all major effects causing the growth of the black hole including relativistic effects of stellar capture and spin evolution, which were not considered by \cite{2017NatAs...1E.147A}. Our result agrees with their finding that the final mass attained by the hole is nearly independent of the formation time. Figure \ref{tal_ev} shows the mass evolution in the presence of only stellar capture, which matches the result of \cite{2017NatAs...1E.147A} [see Figure \ref{tal_ev}]. The black holes with higher seed masses will reach the saturation point earlier, as they will grow lesser by accretion and more by mergers and stellar capture as compared to the lower-mass seed black holes.

The stellar capture process can indeed be important for the formation of the SMBH seeds. Recently, \cite{2020arXiv200203645B} have suggested that the mergers of stellar mass black holes and neutron stars via gas dynamical friction in $\sim 10^{7}$ yr and in a dense cluster, whose size is $\sim$ kpc and contains very high gas mass of $\sim$ 10$^{10} M_{\odot}$ (leading to a high density of $10 M_{\odot}$ pc$^{-3}$), form $10^{4}$--$10^{6} M_{\odot}$ seeds. Our proposal is different: using our relativistic steady loss cone theory (in a not so dense cluster and hence neglecting dynamical friction), the mass growth rate due to stellar capture alone can be approximated from our numerical model [see Figure \ref{stdlc}] to be
\begin{equation}
\dot{M}_{\bullet *} = 5 \times 10^{-6} M_{6}^{-0.33} M_{\odot} {\rm yr^{-1}}, \label{st_mdot_app}
\end{equation}
%\newpage
for typical values of $\sigma$ = 200 km s$^{-1}$ and $\gamma$ = 1.1. The rate of mass growth by accretion process is given by
\begin{equation}
\dot{M}_{\bullet g}(\eta) \simeq 10^{-2} \eta M_{6} M_{\odot} {\rm yr^{-1}}.\label{mdaccap}
\end{equation}
From Equations (\ref{st_mdot_app}) and (\ref{mdaccap}), we find that the critical mass below which stellar capture dominates over accretion is given by $M_{*c}(\eta)\simeq 5\times 10^{3} \eta^{-0.75} M_{\odot}$. Therefore, stellar capture can be an important process for forming SMBH seeds with $M_{\bullet s} \lesssim M_{*c}$.
\begin{center}
\begin{figure}[h!]
\centering
\includegraphics[scale=0.3]{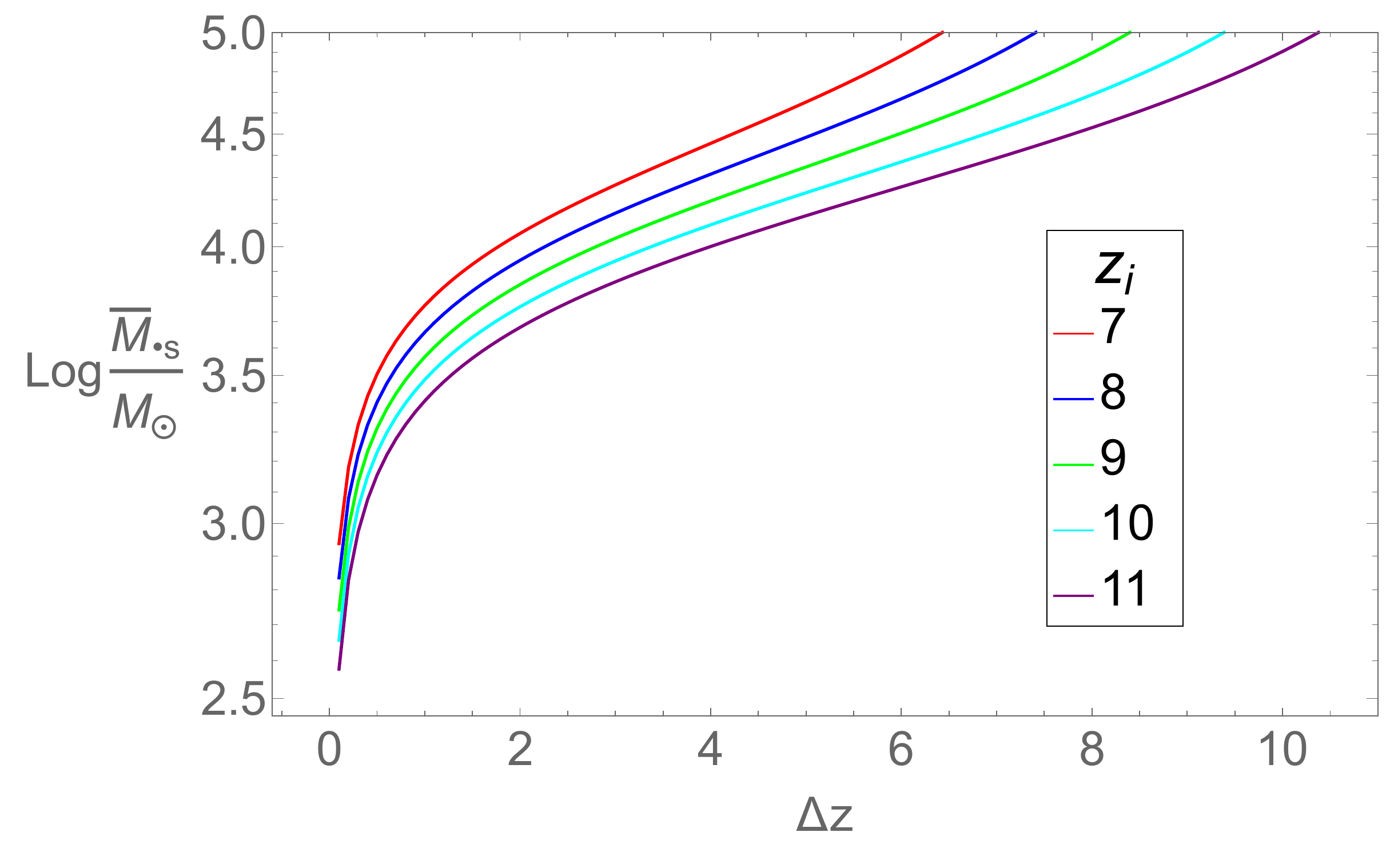}
\caption{The mass growth as a function of change in redshift is shown above where it is seen that $10^{4} M_{\odot}$ seed is obtained for $\{z_{i}, z_{f}\} = \{\{11, 7.01\}, \{10, 6.63\}, \{9, 6.21\}, \{8, 5.74\}, \{7, 5.23\}\}$.}
\label{deltam}
\end{figure}
\end{center}

Solving Equation (\ref{st_mdot_app}), for $\sigma$ = 200 km sec$^{-1}$ and $\gamma$ = 1.1, we arrive at
\begin{equation}
\Delta \bar{M}_{\bullet} = \bar{M}_{\bullet s} - \bar{M}_{\bullet *};~\bar{M}_{\bullet s}^{1.33} - \bar{M}_{\bullet *}^{1.33} \simeq \bar{M}_{\bullet s}^{1.33}= 6.35 \times 10^{5} \Delta t;~ \Delta t = t(z_{f}) - t(z_{i}); ~ \Delta z = z_{i} - z_{f},
\end{equation}
where the masses are in units of $M_{\odot}$, and $\Delta t = t(z_{f}) - t(z_{i})$ [see Equation (\ref{tz}) for $t(z)$] is in units of Gyr. $\bar{M}_{\bullet s}$ is the seed mass formed at $z_{f}$, and $\bar{M}_{\bullet *}$ is the mass of the stellar mass black hole at an initial redshift of $z_{i}$. $\Delta t$ can be expressed as a function of $\Delta z = z_{i} - z_{f}$ and $z_{i}$ using Equation (\ref{tz}) in our paper, so that $\Delta \bar{M}_{\bullet} (z_{i}, \Delta z)\simeq \bar{M}_{\bullet s} (z_{i}, \Delta z)$. From the Figure \ref{deltam}, we see that the seed mass reaches $\sim 10^{4} M_{\odot}$ for the following combinations of initial and final redshifts:  $\{z_{i}, z_{f}\} = \{\{11, 7.01\}, \{10, 6.63\}, \{9, 6.21\}, \{8, 5.74\}, \{7, 5.23\}\}$. Therefore, stellar capture can be considered as a viable process for formation of SMBH seeds where it is seen that SMBH seeds of $10^{4} M_{\odot}$ can be formed in $10^{7}$--$10^{8}$ years depending on the initial redshift range, $z_{i}$ = 7--10. For binary systems, the rate can be higher by an order of magnitude compared to our case, as shown by \cite{2019MNRAS.488.4042T}, which is comparable to MM15.

{\it{Mass and spin evolution of the black hole: }}
The spin and the mass evolution of an SMBH are mainly dependent on three processes: gas accretion, the capture of stars, and mergers. We built a formation for relativistic loss cone theory and included it in the mass and spin evolution of black holes. For accretion, we have used a constant sub-Eddington accretion efficiency throughout the process, taking into account duty cycles. In the case of gas accretion with cooling sources, the flow is momentum driven \citep{2003ApJ...596L..27K}.  
The stellar capture rate has been carried out in both full and steady loss cone theory frameworks. We have incorporated the prescription of saturated mass by \cite{2003ApJ...596L..27K}, which causes a halt in accretion, leaving the stellar capture and mergers as the only channels to contribute to the growth of the black hole. For the mass growth of SMBHs by mergers, we have considered both the contributions of major and minor mergers \citep{2009ApJ...702.1005S}. The rate of minor mergers is more frequent compared to the major ones. However, major mergers contribute to spinning up the black hole, while minor mergers spin it down \citep{2004ApJ...602..312G}. We neglect the contribution of the major mergers in spinning up the hole and consider only the effect of the minor mergers in spinning down the hole. We have considered the mergers to be effective for $z \lesssim 4$. 

We now compare our results with previous studies. \cite{2015ASInC..12...51M} used a theoretical model for mass and spin evolution of the accreting black hole taking into account the spin-down torque caused by the electrodynamical jet. The evolution in the presence and absence of accretion was studied for different cases such as the thin disk, Bondi accretion, and the MHD disk. When accretion stops, the jet power shows an increase before a gradual decrease if the initial spin, $j > \sqrt{3} /2 $, because of the increase in the size of the black hole. The results indicate that the black hole achieves the maximum spin value in the absence of a jet. We expanded these evolution equations to include terms representing stellar capture and mergers. \cite{2014MNRAS.440.1590D} have derived the mass evolution from simulations caused by accretion and mergers and applied semianalytic methods for spin evolution. Our results agree with their findings that the low-mass black holes grow their mass mainly by the accretion process, whereas high-mass black holes ($> 10^{8} M_{\odot}$) grow their mass mostly by mergers. This is because accretion halts owing to saturation beyond $M_{\bullet} > M_{\bullet t}$ while the low-mass black holes grow slowly by stellar capture. We have considered a constant rate of Eddington accretion, while they have considered it to be reducing over time owing to gas rarefaction in galaxies. The justification for considering our parameter ranges is given in \S \ref{3}. From Figure \ref{wtmwm}, we see a change in the slope of mass evolution near the saturation time. This is expected since the accretion of gas stops and stellar capture and merger activity take over for subsequent growth. The rate of growth for mass accretion is much greater than the other two channels, so the halt of accretion causes the slope change. \cite{2019ApJ...873..101Z} have studied the spin evolution via two-phase accretion and have found that higher-mass black holes have intermediate spin ($\sim 0.5$), while the low-mass black holes have higher spin ($\gtrsim 0.8$). In our paper, we have studied accretion only in the the thin-disk mode \citep{2008ApJ...680..169S, 2010A&A...516A..87S, 2015ApJ...815..129S}. The low value of spin for higher masses can be explained as follows: when the BZ effect dominates, it causes the spin-down of the black hole, while it is possible that the low-mass black holes are a result of gas accretion alone and without mergers; hence, the spins are higher. \cite{2019ApJ...873..101Z} have also used a power-law dependence of the radiative efficiency with the black hole mass where it decreases with an increase in mass, though they have found the dependence to be weak. This is in contradiction with the model of \cite{2011ApJ...728...98D}, who claim an increase with mass with a power-law index of 0.5. Here, we aim to capture a complete picture of black hole growth using all the factors contributing to it, and for simplicity, we consider the radiative efficiency, $\epsilon_{M}$, to be constant with respect to mass (but varying as a function of $j$), with a goal to study and compare all the other contributing factors. We intend to include the mass variation of radiative efficiency in the future. However, the dependence is weak, and it is not likely to make a significant difference to our results. Recently, \cite{2020MNRAS.493.1500S} suggested a higher mean radiative efficiency, $\epsilon_{M} \sim $ 0.15, which is defined as $\epsilon_{M} = L/\dot{M}_{0}c^{2}$, where $L$ is the luminosity and $\dot{M}_{0}$ is the rest mass accretion rate. We can write $\epsilon_{M} (j)= 1 - \epsilon_{I}(j)$  following the standard prescription of \cite{1972ApJ...178..347B}, \cite{2005ApJ...620...59S}, and \cite{2015ASInC..12...51M} for the thin-disk case, where $\epsilon_{I}(j)$ is given by Equation (6) in our paper. This factor $\epsilon_{M}(j)$ is shown in Figure 7 of \cite{2020MNRAS.493.1500S}; they conclude from observational values, using their de-biased relation of black hole mass and star mass, that a steady-state value of $\epsilon_{M}$  $\sim 0.15$ is expected. In our theoretical evolution model, we use the standard relativistic form of $\epsilon_{M}(j)$ which varies between $\displaystyle (1 - \frac{2\sqrt{2}}{3}) \simeq $ 0.06 (for $j$ = 0) and $\displaystyle (1 - \frac{1}{\sqrt{3}}) \simeq $ 0.42 (for $j$ = 1). The mean value is approximately 0.24, which, after incorporating the duty cycle, can reduce further [to values near 0.15 as suggested by \cite{2020MNRAS.493.1500S}].

From Figure \ref{wtmwm}, we see that the spin of the black hole initially increases because of accretion, after which there is a spin-down due to the BZ torque. Since we have considered a thin-disk accretion, the spin value very quickly reaches the maximum spin as mentioned by \cite{2012MNRAS.424.1461L}. \cite{2004ApJ...602..312G} showed how accretion, major mergers, and minor mergers contribute to the spin evolution of the black hole. Using the prescription given by \cite{2004ApJ...602..312G} for minor mergers, we see that the value of the maximum spin attained is much less than those where the contribution of mergers is included. We have incorporated minor mergers only for spin-down of the hole, as the spin-up process is already dominated by the accretion process.

Our model is useful for retrodicting the initial black hole configuration when we run our model backward from the observed $\{M_{\bullet}, j\}$ as the initial conditions as shown in \S \ref{4.2}. More observations and models that provide the final \{$M_{\bullet}, j$\} state will provide useful clues for such exercises in black hole demographics.

{\it{Evolution of the $M_{\bullet}$--$\sigma$ relation: }}
We have combined all the known effects contributing to the mass and spin evolution of the black hole and thus derived the evolution of the $M_{\bullet} = k_{0}(z) \sigma^{p(z)}$ relation by semianalytic methods; some preliminary results were shown in \cite{bhattacharyya_mangalam_2018}. \cite{2015ApJ...805...96S} and \cite{2013ApJ...764...80S} have studied the evolution of the relation from SDSS data for quasars and have found no evolution of the $M_{\bullet}$--$\sigma$ relation up to $z \simeq 1$. Numerical simulations of the large-scale structure of the universe by \cite{2015MNRAS.452..575S} and \cite{2016MNRAS.463.2465T} show that this relation holds almost up to $ z \simeq 4$. \cite{2006ApJ...641...90R} have studied the evolution of this relation until $z$ = 6 for merging disk galaxies through hydrodynamic simulations while taking into account the effects of accretion and supernovae. They have found almost no change in $p(z)$ and a very small change of $k_{0}(z)$ similarly as suggested similarly by \cite{2009ApJ...694..867S}. From their analysis, $k_{0}(z)$ $\propto$ $(1 + z)^{\alpha}$, with $\alpha$ = 0.33. In our analysis, since we consider $\sigma$ to be a constant throughout, so that the value of $k_{0}$ is expected to decrease at higher redshift as seen in Fig \ref{intcpt_drop}. At the saturation point, the value of $p$ $\simeq$ 4 from our model following \cite{2003ApJ...596L..27K}. We have considered a range of values of $\sigma_{100}$ = $\{1$ -- $2\}$, which is the average observed range of $\sigma$ for different galaxies, and assumed $M_{\bullet} (z = z_{f})$ $\propto$ $\sigma^{5}$ as set by the Faber--Jackson relation. From Figure \ref{msigmarl}, we see that $p (z)$ remains in the range of 4 -- 5 throughout, which roughly agrees with the empirical result of \cite{2009ApJ...694..867S}. Figure \ref{msigmarl} shows higher values of $p(z)$ at higher redshifts. Therefore, we conclude that $p(z)$ will be higher for late-type galaxies as suggested by \cite{2013ApJ...764..184M}. One possibility is that the $\sigma$ varies with redshift owing to major mergers, but this is outside the scope of this paper. For minor mergers, a constant $\sigma$ is a reasonable assumption that is based on the work of several authors [\cite{2009ApJ...697.1290B}, \cite{2012ApJ...744...63O}, \cite{2009MNRAS.398..898H},  and the dissipative model described by \cite{2009ApJ...694..867S}], where they have found that $\sigma$ changes little with redshift  [$\sigma$ reduces over Hubble time by a factor of $\sim$ 15$\%$, \cite{2009ApJ...694..867S}]. The elliptical galaxies that obey the $M_{\bullet}$--$\sigma$ relation are within this specified redshift. We conclude from our simulations that although $p(z)$ varies with $\{j_{0}$, $B_{4}$, $z_{f}\}$, it stays within the predicted range of 4 -- 5.

\section{Conclusions}\label{7}
Our model of deriving the joint evolution of black hole mass, spin and $M_{\bullet}$--$\sigma$ relation throws light on the coevolution of the black hole and its environment from the time of formation. We have incorporated all the factors contributing to the growth of the black hole to build a comprehensive evolution model of the black hole.
\begin{enumerate}
    \item We have included relativistic effects in the process of tidal and direct capture. A key consequence is that the capture rate reduces to the range $10^{-5} - 10^{-6}$ yr$^{-1}$, which is more in line with observations.
    \item We have built a semianalytic self-consistent evolution model of the black hole.
    \item We have explored the roles and phases of importance of each of the growth channels. Though the contributions from stellar capture ($\sim$ 3\%) and mergers ($\sim$ 2\%) in mass growth of the black hole are small compared to accretion ($\sim$ 95\%), irrespective of the parameters before saturation, these two factors play a major role after the saturation when the accretion process stops or contributes negligibly \citep{2003ApJ...596L..27K}. The estimates of the contribution of the effects mentioned here are computed for the canonical case, and this can vary up to 5\% within the context of our model assumptions and the chosen parameter ranges. The stellar capture contributes to the mass growth while not changing the spin substantially, whereas the mergers contribute to both. Minor mergers reduce the maximum spin value achieved by accretion by $\sim$20\%. BZ torque does not contribute to mass growth, but only to the spin-down of the black hole (in the presence of all the effects, for $B_{4}$ = 10, the spin-down is $\sim$3\% from the maximum value attained owing to accretion) as discussed in \S 2.4. Mergers and the BZ process are necessary; otherwise, the black holes will be spinning maximally.

    \item We illustrated the effect of saturation on the evolution of the $M_{\bullet} (z) = K_{0}(z) \sigma^{p(z)}$ relation.
    \item By running the models backward in time, we retrodict the formation parameters of seed black holes. This will enable us to discriminate among models of black hole formation.
     \item Stellar capture can be considered as a viable process for formation of SMBH seeds, as this dominates the accretion process when $M_{\bullet} \leq 2 \times 10^{4} M_{\odot}$. 
    \item We expect our transparent and detailed formulation in a fully relativistic framework to be useful for future simulational studies.
\end{enumerate}
This model can be improved by incorporating a model for time variation of $\eta$ which is an uncertain input. The data from future surveys at high redshift, for example, from TMT, VLT, and ELT, along with measurements of $\sigma$ from SKA, can be used to probe the $M_{\bullet}$--$\sigma$ evolution to test our model. We also plan to work on the demographics of the black hole, based on a model of seed mass and spin distribution functions.

\acknowledgements
We would like to thank the anonymous referee for the constructive comments that helped us to improve the paper. We acknowledge DST SERB CRG grant No. 2018/003415 for financial support. We would like to thank Saikat Das for helping us with Figure \ref{diagram}. We acknowledge IIA for the use of the HPC facility for computations and Kavalur Observatory (VBO) for hospitality during our visits. We thank P. Natarajan for discussions during our visit to ICTS (supported by ICTS/smbh2019/12).

%\bibliography{references}
%\newpage

\appendix 

\section{Analytic Approximation to the Tidal Radius and Numerical Statistics to the Tidal Radius}\label{Appendix}
Taking $y = 1 / x_{t}$ and $\tilde{y} = y / y_{t0} = 1 + \delta$, we find the first-order approximation to $y$ defining $y_{t0}$ as the inverse of the dimensionless tidal radius to be
\[
\displaystyle y_{t0} = \frac{1}{x_{t0}} = M_{8}^{-\frac{1}{3}} \bigg(\frac{\rho_{*}}{\rho_{\odot}}\bigg) ^{-\frac{1}{3}} 10^{5} \cdot r_{g}~ {\rm pc^{-1}}.
\] 
The sixth-order equation for $\delta$ is
\be
2( 1 + \delta )^{3} - 3 (l^{2} + Q) y_{t0} (1 + \delta)^{4} + 12 (1 + \delta)^{5} [(j - l)^{2} + Q] y_{t0}^{2} + 10 j^{3} Q y_{t0}^{3} (1 + \delta)^{6} - 1 = 0. \label{sixthdelta}
\ee
Solving Equation (\ref{sixthdelta}) numerically, we obtain $\delta(j, Q)$.% which is shown in Figure \ref{delta}.

%\begin{figure}[H]
    %\centering
   % \includegraphics[scale =0.4]{delta.pdf}
    %\caption{A contour plot of $\delta$($j, Q$) defined by $y \equiv y_{t0} (1 + \delta)$ for $l = l_{c}$ and $k = -1$ is shown , which lies in the range of 0.32 $\pm$ 0.05 for $Q ~ \in ~[0, 4]$ and $j~ \in ~ [0,1]$.}
    %\label{delta}
%\end{figure}

\begin{enumerate}
\item $x_{t} (M_{8}, j, Q = 0)$ is shown in Figure \ref{figrta}, where we see that at a fixed $j$, $x_{t}$ decreases with an increase of $M_{\bullet}$. In the high-mass regime, the variation of $x_{t}$ is small with spin, but it shows more variation in the low-mass regime, which is also reflected in the calculation of the rate of star capture presented later in this \S 2.
%\item Fig \ref{figrtaQ} shows the dependence of $x_{t}(M_{8}, j, Q)$ on $Q$. 
\item We find that for a fixed value of $j$, $x_{t}$ has a small dependence on $Q$ as a function of $M_{8}$. However, $x_{t}(M_{8}, j, Q)$ decreases as a function of $Q$ for the retrograde case for a fixed value of $M_{\bullet}$. But for the prograde case, $x_{t}(M_{8}, j, Q)$ initially decreases for higher $Q$, but subsequently it shows an increase with increasing $Q$. $x_{\ell}(M_{8}, j, k, Q)$ for different $Q$ values, are shown in Figure \ref{figrcaQ}, for both prograde and retrograde cases.
\end{enumerate}

\begin{figure}[H]
\centering
\subfigure[]{\includegraphics[scale=0.2]{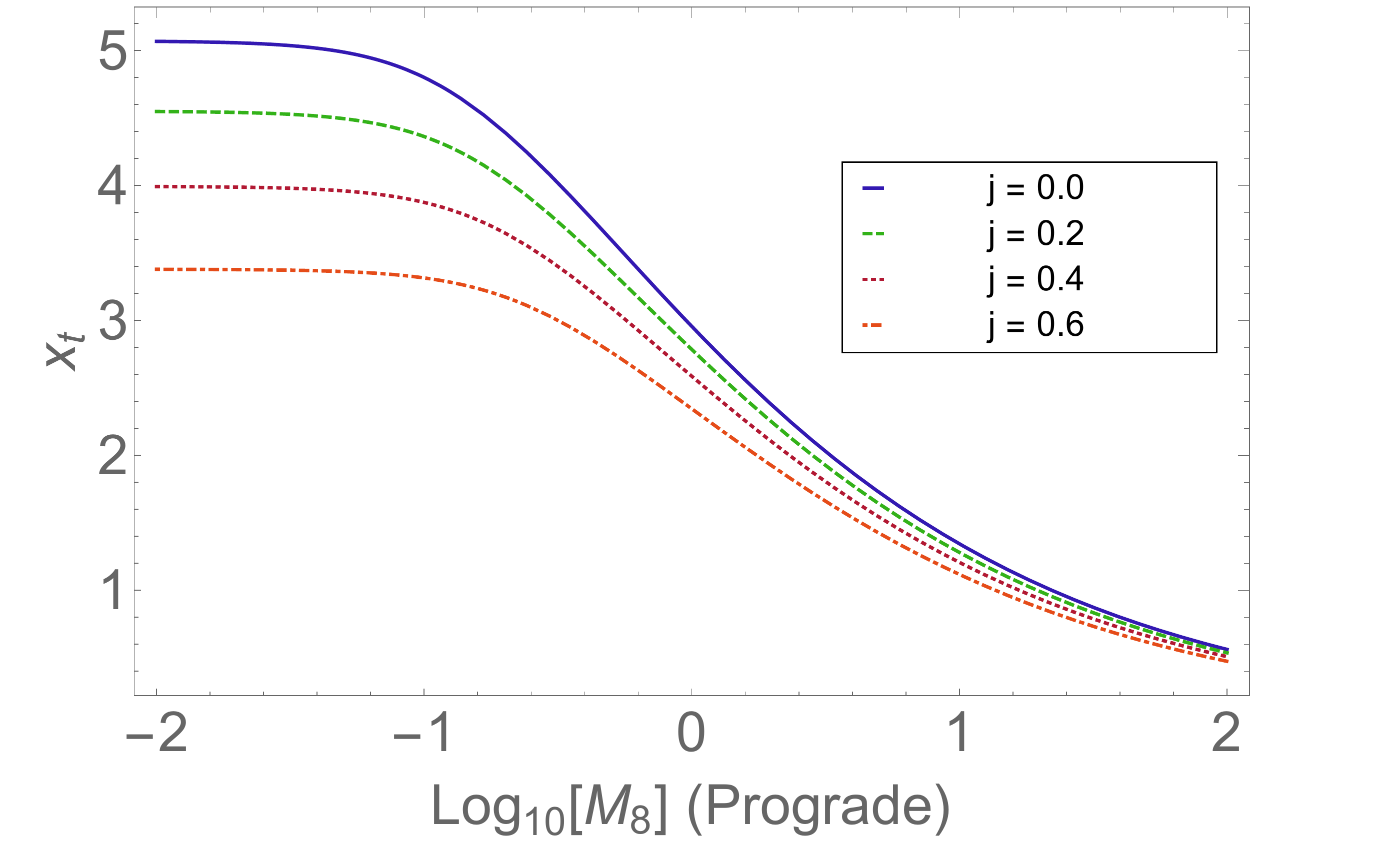}} \hspace{0.3 cm}
\subfigure[]{\includegraphics[scale=0.2]{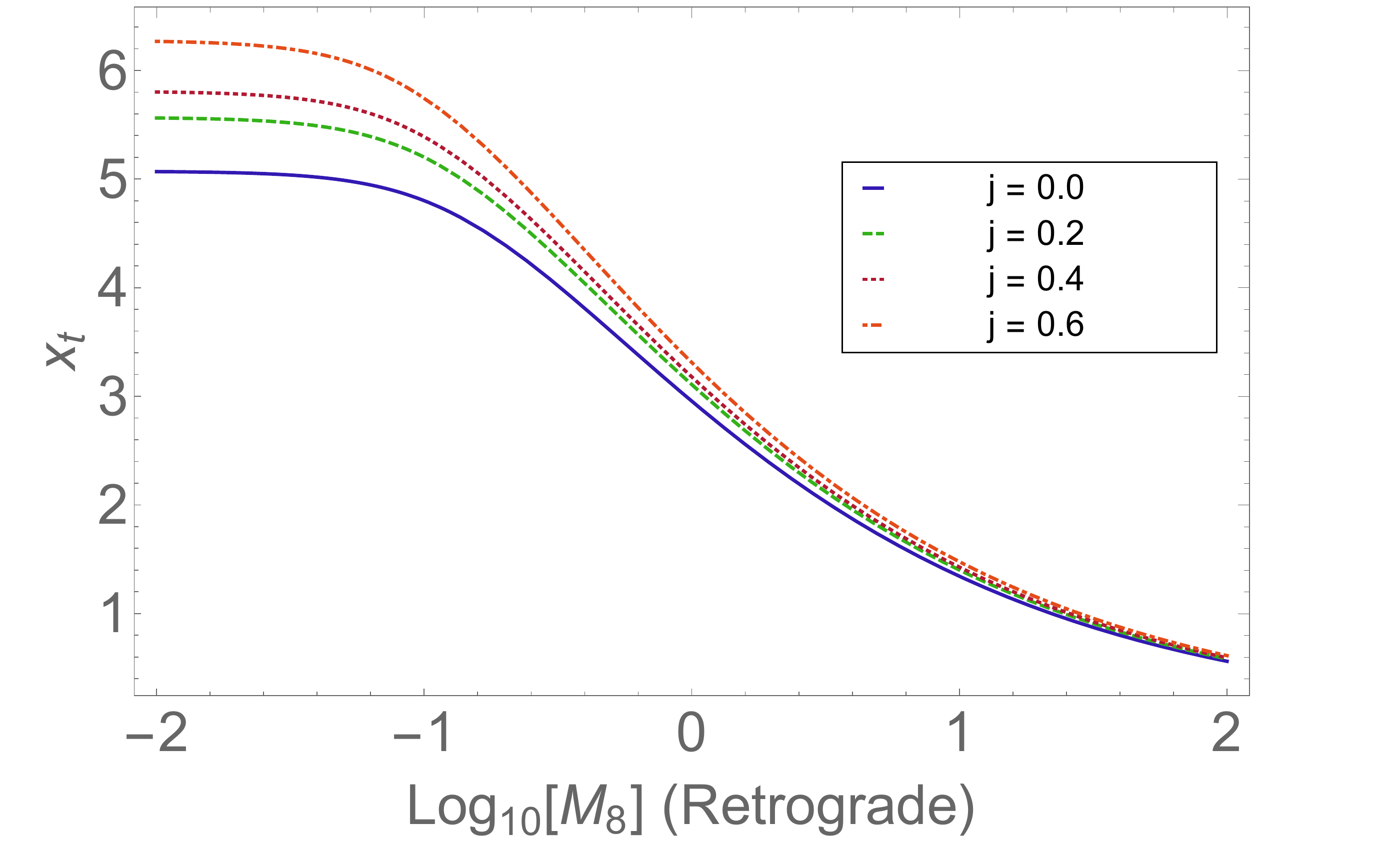}} \hspace{0.3 cm}\\
\subfigure[]{\includegraphics[scale=0.2]{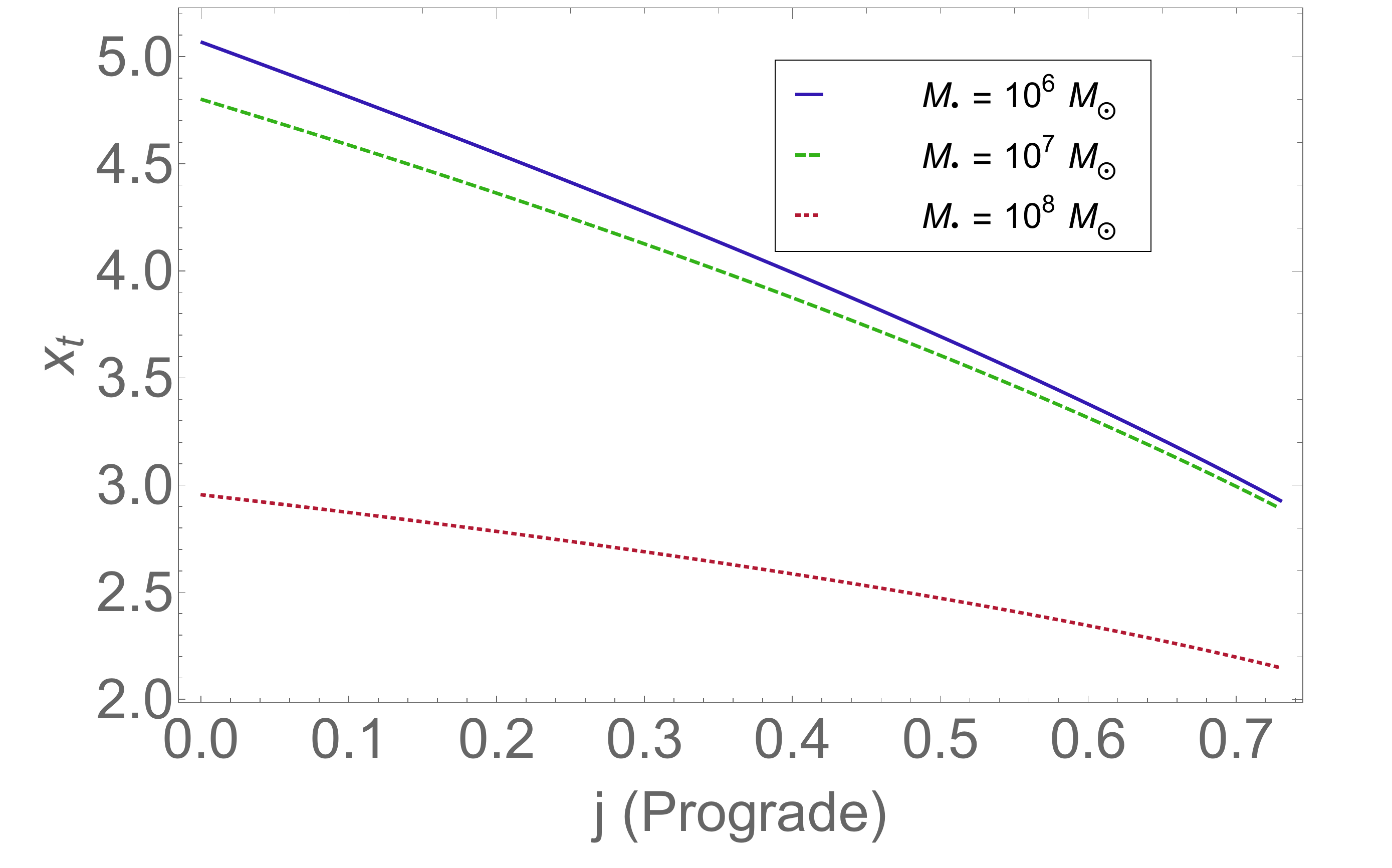}} \hspace{0.3 cm}
\subfigure[]{\includegraphics[scale=0.2]{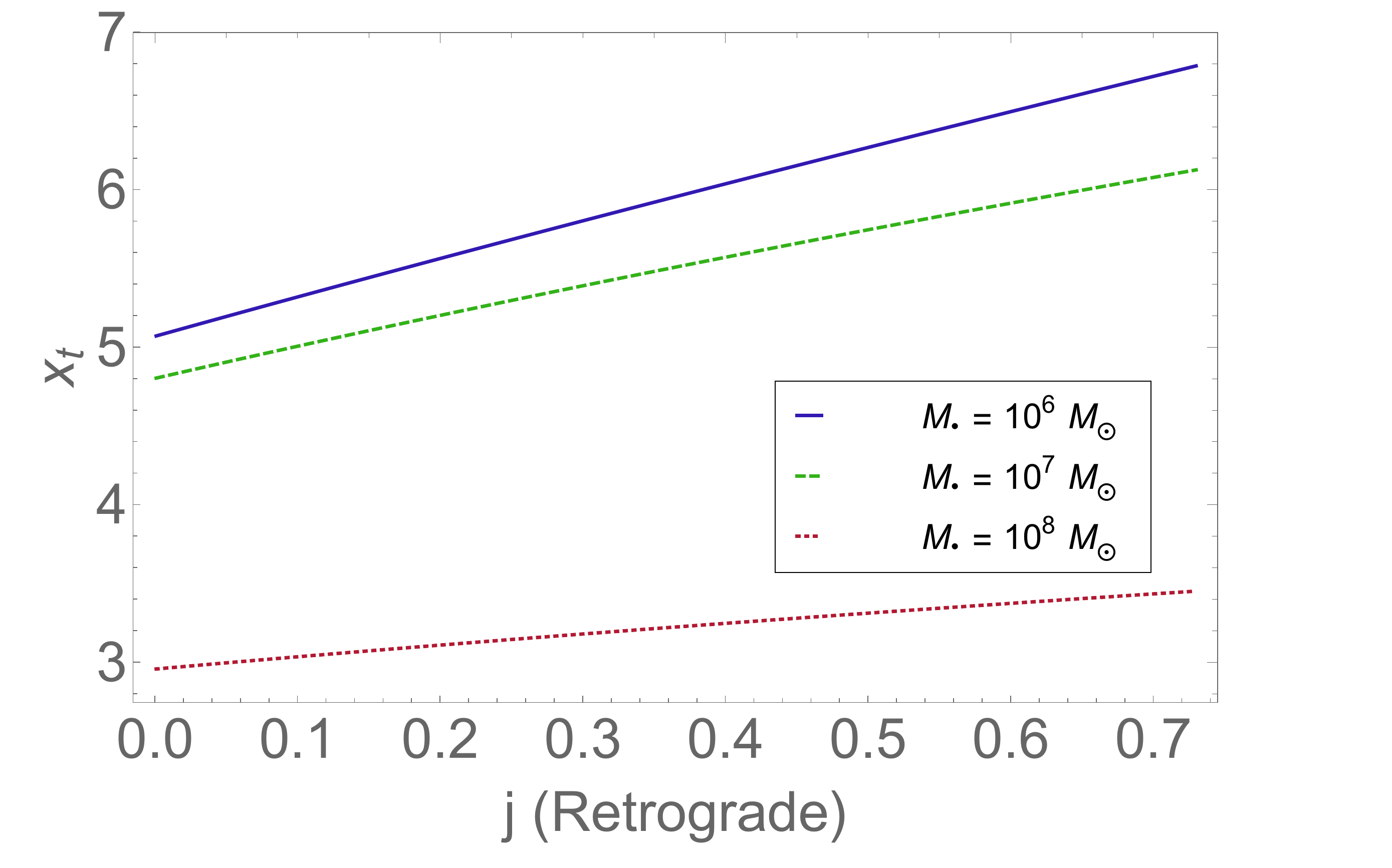}} 
\caption{Tidal radius [$\displaystyle x_{t} (M_{8}, j, Q) = r_{t} (M_{8}, j, Q) / r_{g}$] given by Equation (\ref{rtEquation}) is shown as a function of $M_{8}$ (panels (a) and (b)) and $j$ (panels (c) and (d)) for $Q =0$, or prograde (panels (a) and (c)) and retrograde (panels (b) and (d)) motion.} 
\label{figrta}
\end{figure}

%\begin{figure}[H]
%\centering
%\subfigure[Prograde]{\includegraphics[scale=0.22]{xtmk1Q.pdf}} \hspace{0.3 cm}
%\subfigure[Retrograde]{\includegraphics[scale=0.22]{xtmk0Q.pdf}} \hspace{0.3 cm}\\
%\subfigure[Prograde]{\includegraphics[scale=0.22]{xtjk1Q.pdf}} \hspace{0.3 cm}
%\subfigure[Retrograde]{\includegraphics[scale=0.22]{xtjk0Q.pdf}}
%\caption{The tidal radius ($\displaystyle x_{t} (M_{8}, j, Q)= r_{t} (M_{8}, j, Q)/ r_{g}$) given by Equation (\ref{rtEquation}) is shown as a function of $M_{8}$ for $j$ = 0.2 (top) and $j$ for $M_{8}$ = 0.1 (bottom).} 
%\label{figrtaQ}
%\end{figure}

\begin{figure}[H]
\centering
\subfigure[]{\includegraphics[scale=0.2]{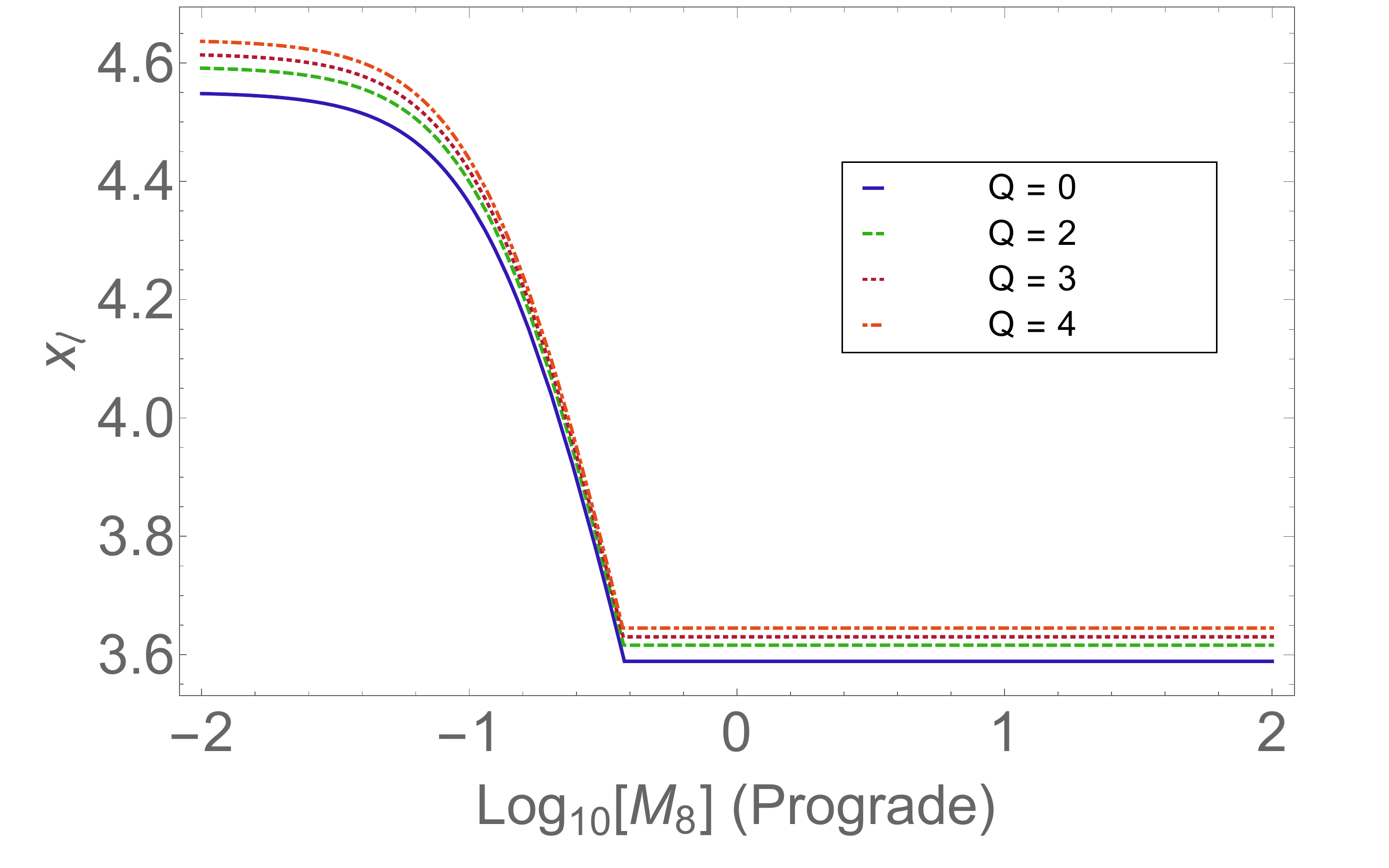}} \hspace{0.3 cm}
\subfigure[]{\includegraphics[scale=0.2]{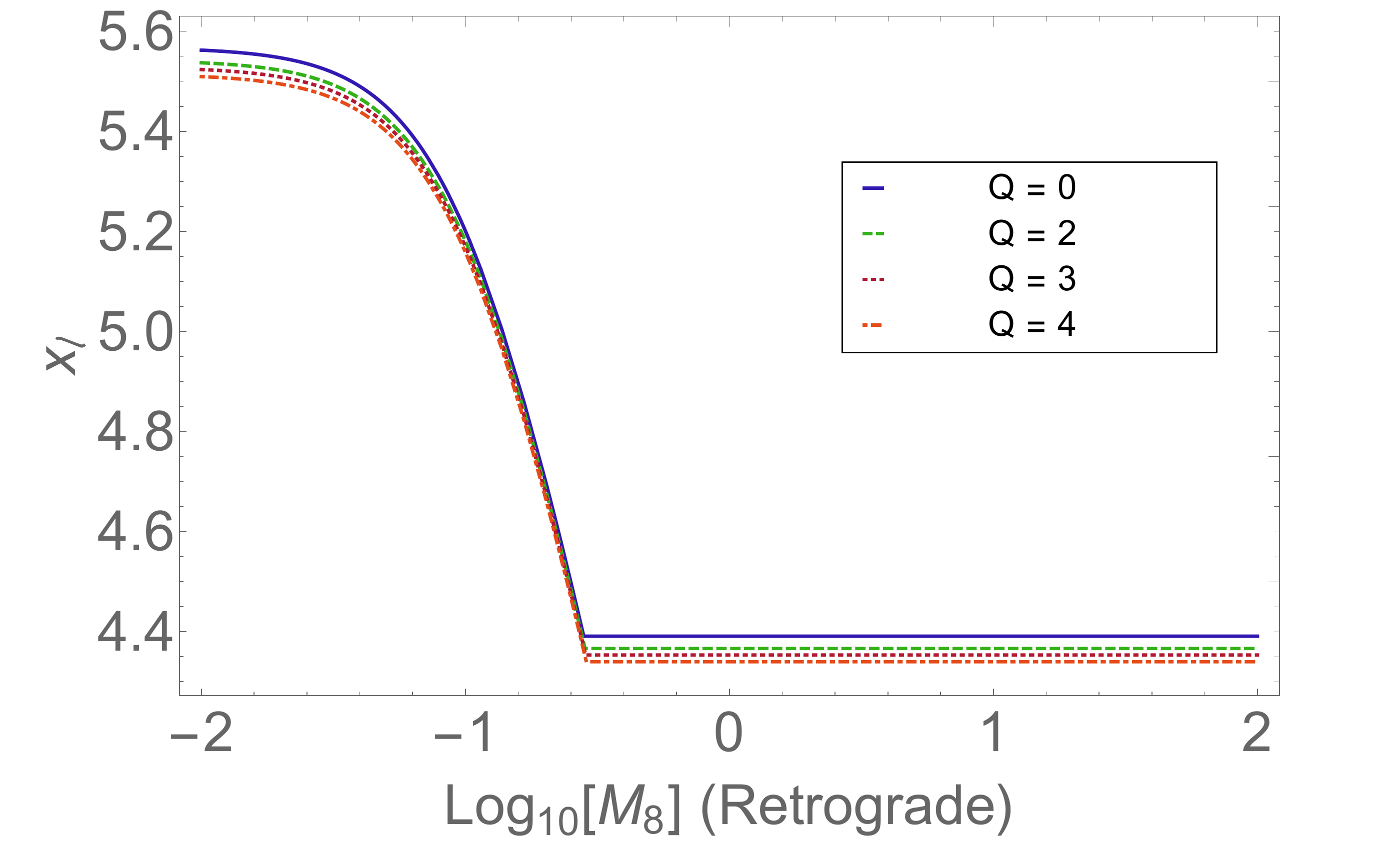}} \hspace{0.3 cm}\\
\subfigure[]{\includegraphics[scale=0.2]{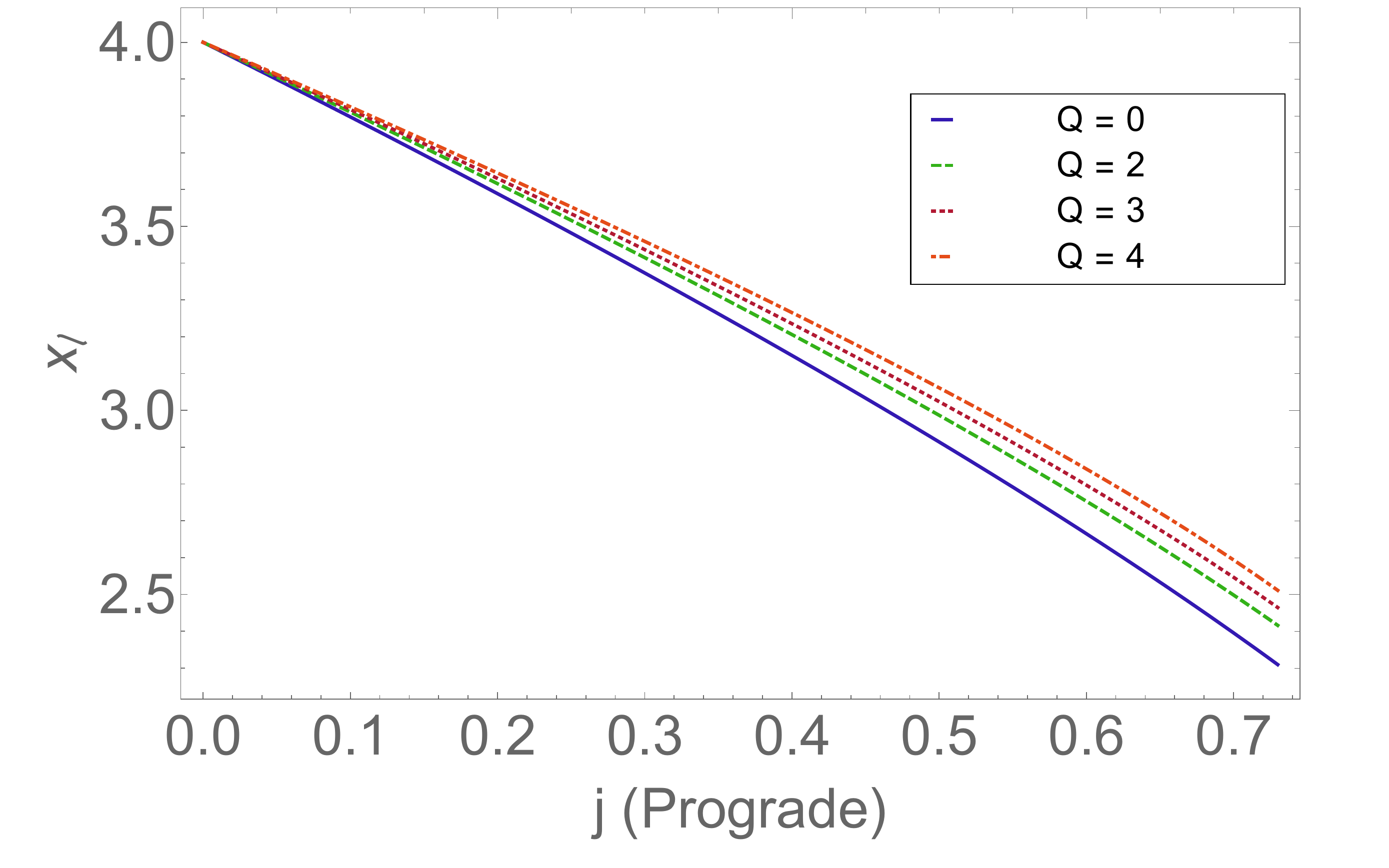}} \hspace{0.3 cm}
\subfigure[]{\includegraphics[scale=0.2]{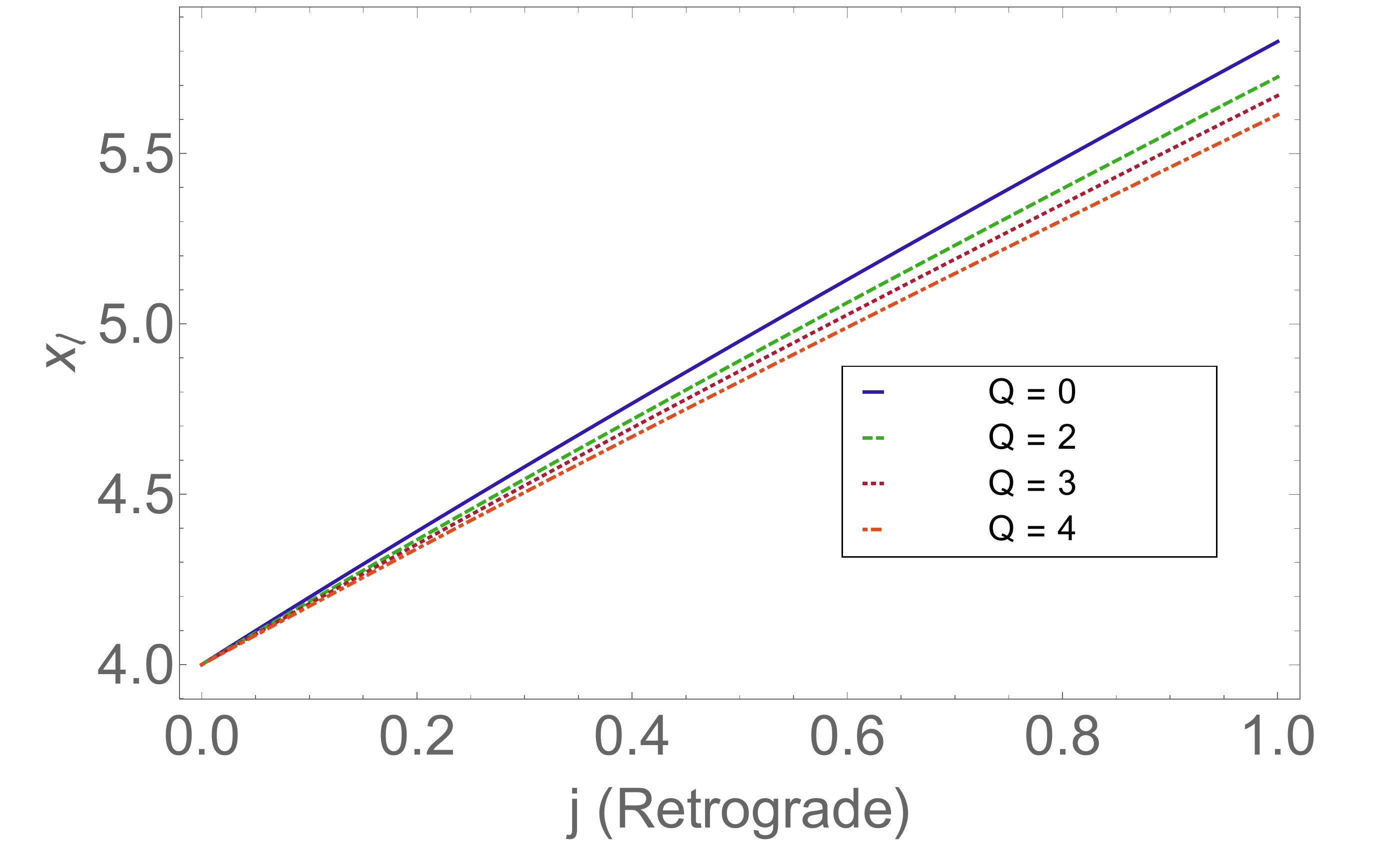}} 
\caption{The loss cone radius [$\displaystyle x_{\ell} (M_{8}, j, Q)= r_{\ell} (M_{8}, j, Q) / r_{g} = {\rm Max}[r_{t}(M_{8}, j, Q), r_{c}(M_{8}, j, Q)] / r_{g}$] is shown as a function of $M_{8}$ for $j$ = 0.2 (a, b) and $j$ for $M_{8}$ = 1 (c, d) for different $Q$ values, for prograde (a, c) and retrograde (b, d) motion.} 
\label{figrcaQ}
\end{figure}

\section{Justification for Using Steady Loss Cone Theory over Full Loss Cone Theory}\label{flcslc}
The stars captured populate a loss cone whose angular size is given by \citep{1976MNRAS.176..633F}
\be 
\theta_{\ell}^{2} (r) = \frac{r_{c}}{r^{2}}\frac{GM_{\bullet}}{\sigma^{2}},
\ee 
where $\theta_{\ell}$ is the half angle of the loss cone. The angle scattered in a dynamical time $\displaystyle t_{d} = r / \sigma$ is approximated by
\begin{eqnarray} 
\displaystyle
\theta_{d} = \sqrt{\frac{t_{d}}{t_{R}}} \\
\displaystyle t_{d} = \left\{\begin{array}{lr}
        \sqrt{\frac{r^{3}}{GM_{\bullet}}} & \text{\rm for~} r \leq r_{h} \\
       \frac{r}{\sigma} & \text{\rm for~} r \geq r_{h}
        \end{array}\right. \\
t_{R} = \frac{\sigma^{3}}{3 \ln \Lambda G^{2}m_{*} n_{c}};
\end{eqnarray} 
where $\Lambda$ is the Coulomb logarithm of the ratio of maximum and minimum values of the impact parameter and $n_{c}$ is the cluster mass density, with $m_{*}$ being the stellar mass \citep{1999MNRAS.306...35S}. In the diffusive regime, $\theta_{d} < \theta_{\ell}$, the loss cone is empty as the star is removed from the loss cone within a dynamical time scale. At the other extreme, $\theta_{d} > \theta_{\ell}$, the loss cone is always full.  Both the regimes are shown in Figure \ref{thetalcd} for $m_{*} = M_{\odot}$ and $n_{c} = 10^{4} M_{\odot} {\rm pc}^{-3}$ for the mass range $M_{\bullet} = 10^{4}$--$10^{8} M_{\odot}$ (we have assumed the $M_{\bullet}$--$\sigma$ relation with $p$ = 4).
\begin{figure}[H]
\begin{center}
\subfigure[]{\includegraphics[scale=0.24]{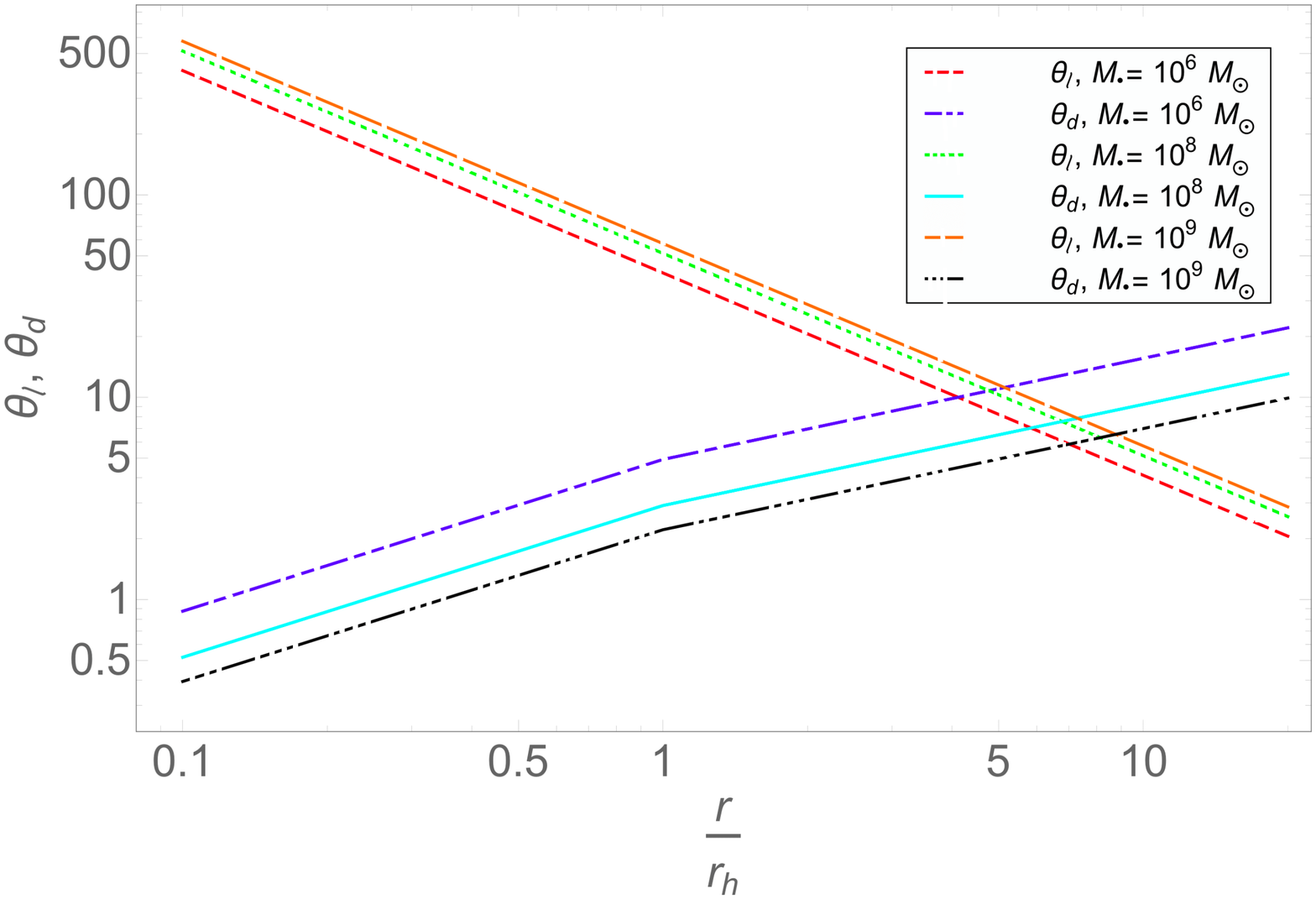}}
\subfigure[]{\includegraphics[scale=0.26]{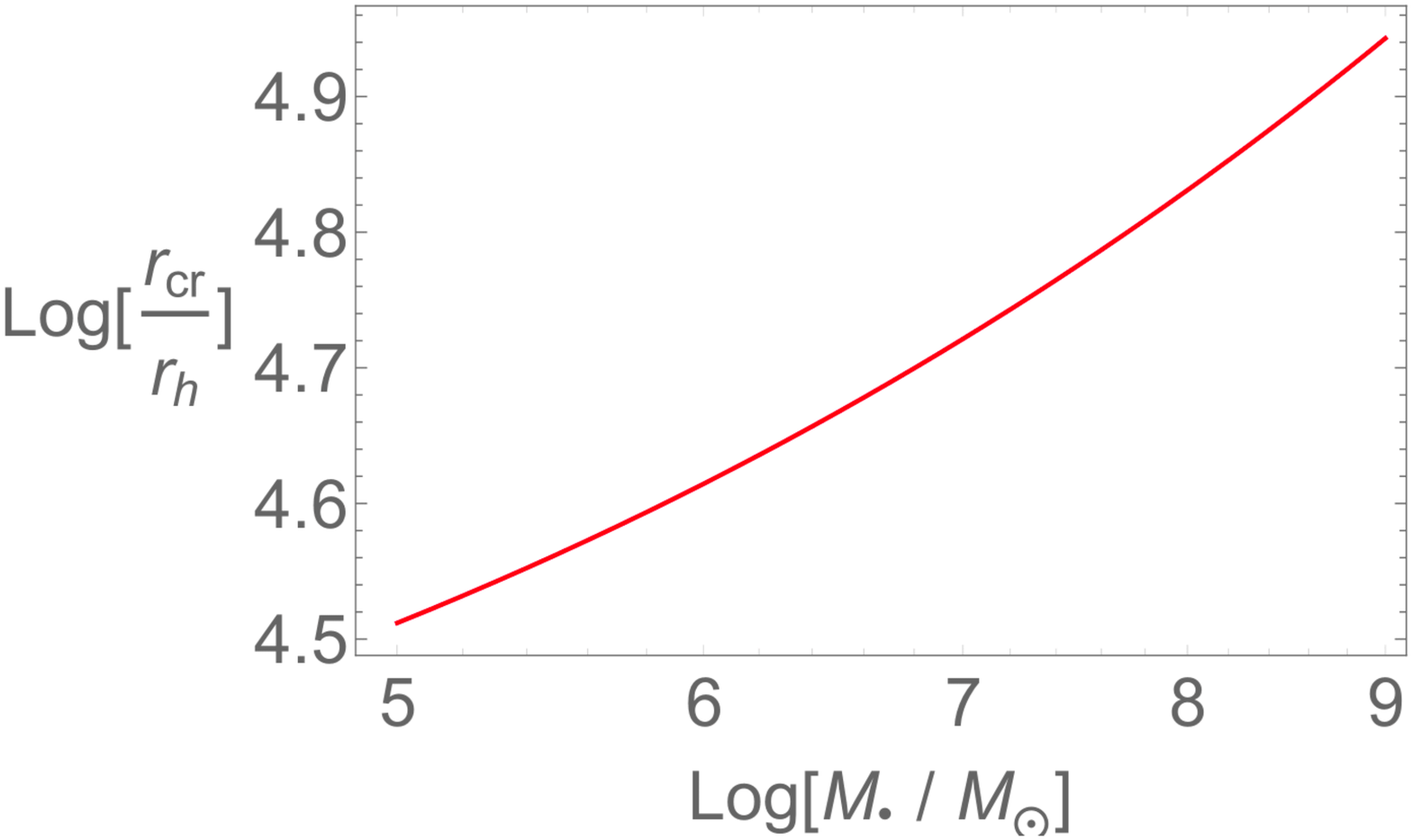}}
\caption{Variation of the angular size of loss cone, $\theta_{\ell}$ and angle scattered in dynamical time, $\theta_{d}$ with $r / r_{h}$ for $m_{*} = M_{\odot}$ and $n_{c} = 10^{4} M_{\odot} {\rm pc}^{-3}$, for the range $M_{\bullet} = 10^{4}$--$10^{8} M_{\odot}$ (a) and (b) the variation of the crossing point, $r_{cr}(M_{\bullet}$) defined by $\theta_{lc}(r_{cr}) = \theta_{d}(r_{cr})$, for $m_{*} = M_{\odot}$ and $n_{c} = 10^{4} M_{\odot} {\rm pc}^{-3}$ \citep{1999MNRAS.306...35S}.}
\label{thetalcd}
\end{center}
\end{figure}
From Figure \ref{thetalcd}, we see that as the mass of the black hole increases, the crossing point ($r_{cr} / r_{h}$) of the two curves shifts toward the right, which implies that the diffusive region expands with an increase in mass.

\section{Calculation of the stellar capture rate}\label{capturerate}
%\begin{figure}[H]
%\begin{center}
%\subfigure[Prograde]{\includegraphics[scale=0.25]{stdlcndk1.pdf}}
%\subfigure[Retrograde]{\includegraphics[scale=0.25]{stdlcndk0.pdf}}\\
%\subfigure[Prograde]{\includegraphics[scale=0.2]{stdlcndk1Q.pdf}}
%\subfigure[Retrograde]{\includegraphics[scale=0.2]{stdlcndk0Q.pdf}}
%\caption{The capture rate, $\dot{N}_{s}(M_{\bullet},j, k, Q, \epsilon_{s}, \sigma)$, is shown which reduces monotonically with $M_{8}$ and increases slightly with $j$ when $k$ = 1 (left), -1 (right) for %$Q$ = 0 (top) and (bottom) the same plots for 
%$Q$ = 4 where the lower limit of the $\epsilon_{s}$ integration is taken to be $\epsilon_{m}$ = -10, $\gamma$ = 1.1 and $\sigma$ = 200 km/sec.}
%\label{stdlcQ}
%\end{center}
%\end{figure}

\begin{figure}[H]
\begin{center}
\subfigure[Prograde]{\includegraphics[scale=0.2]{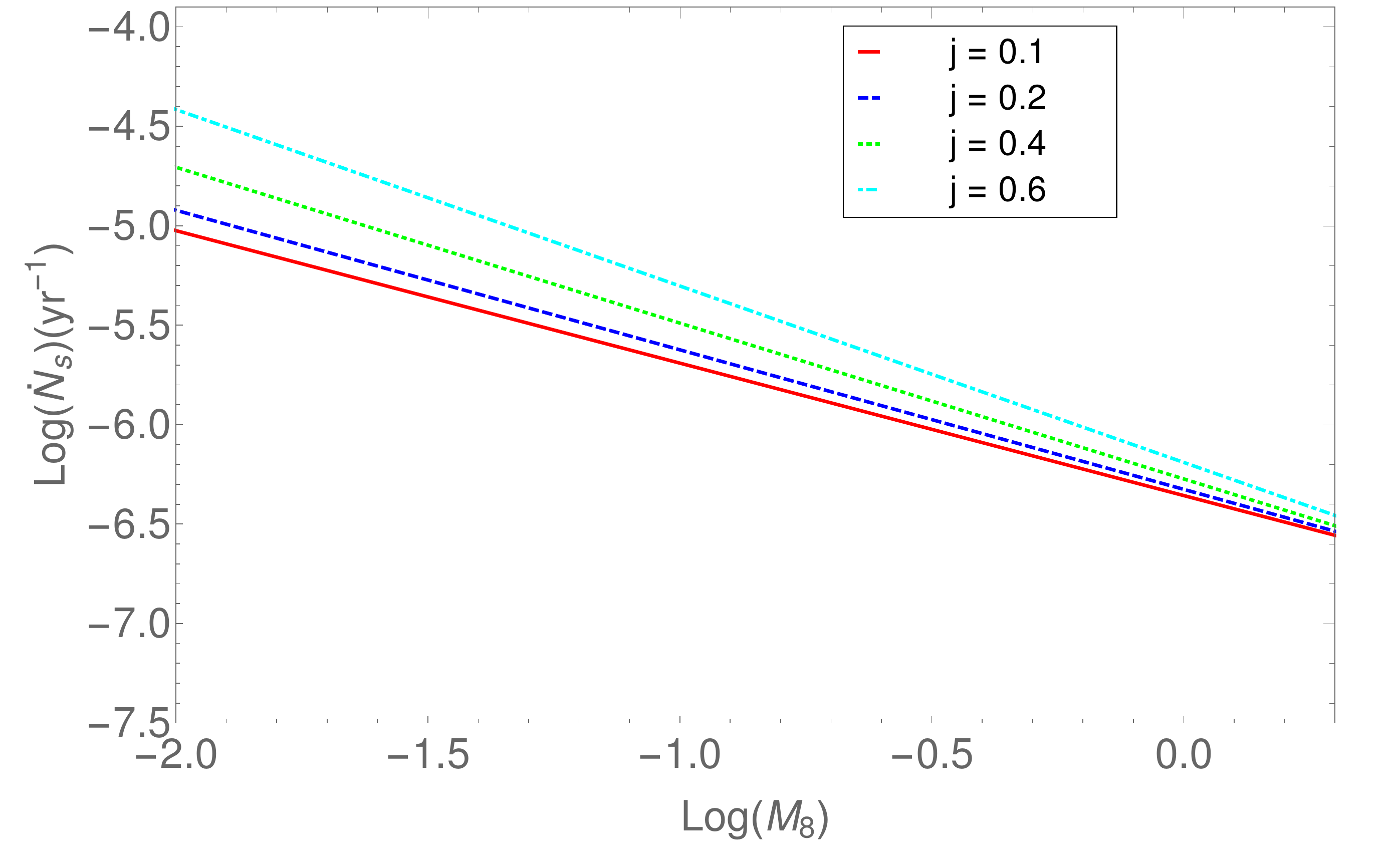}}
\subfigure[Retrograde]{\includegraphics[scale=0.2]{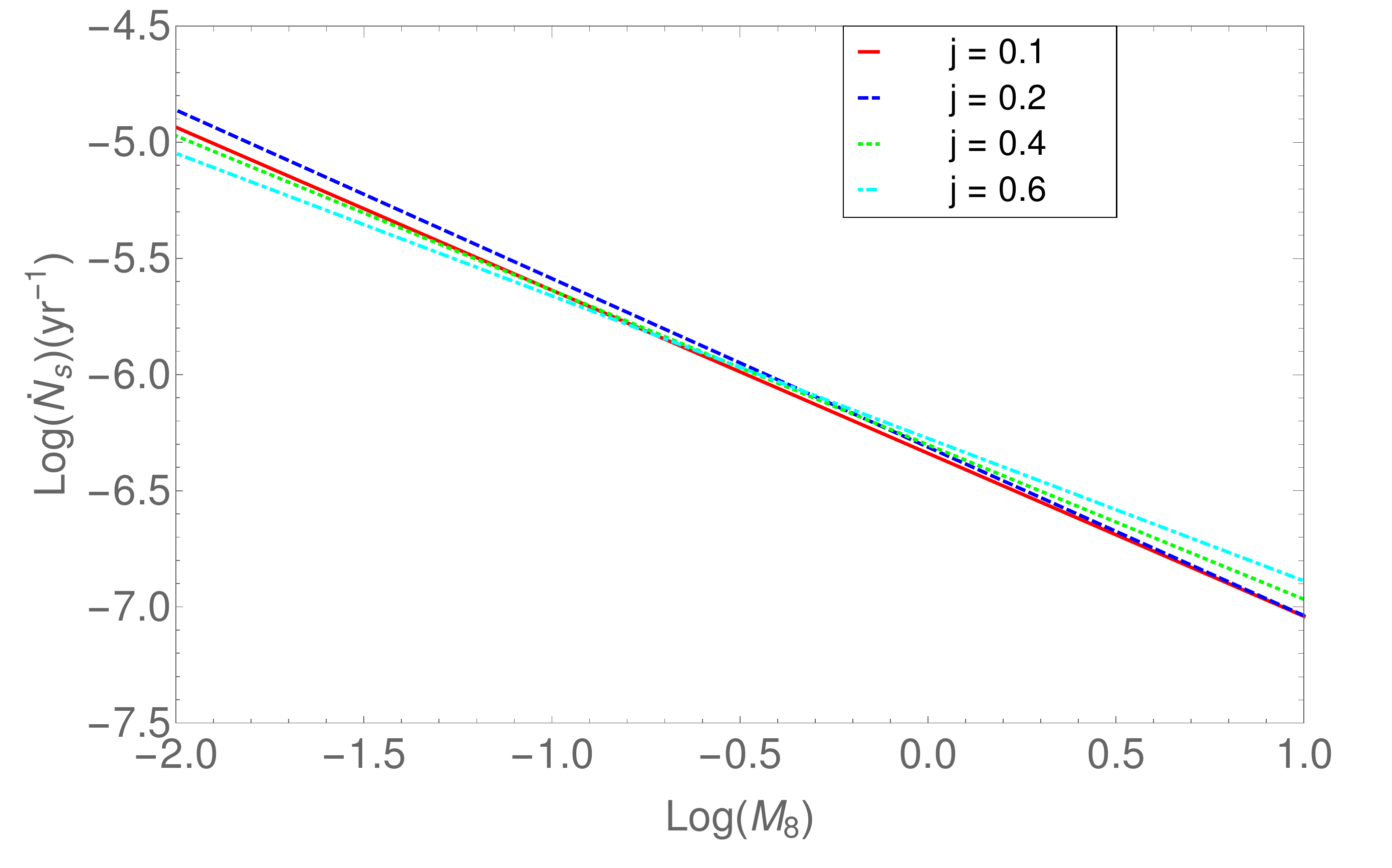}}\\
\subfigure[]{\includegraphics[scale=0.2]{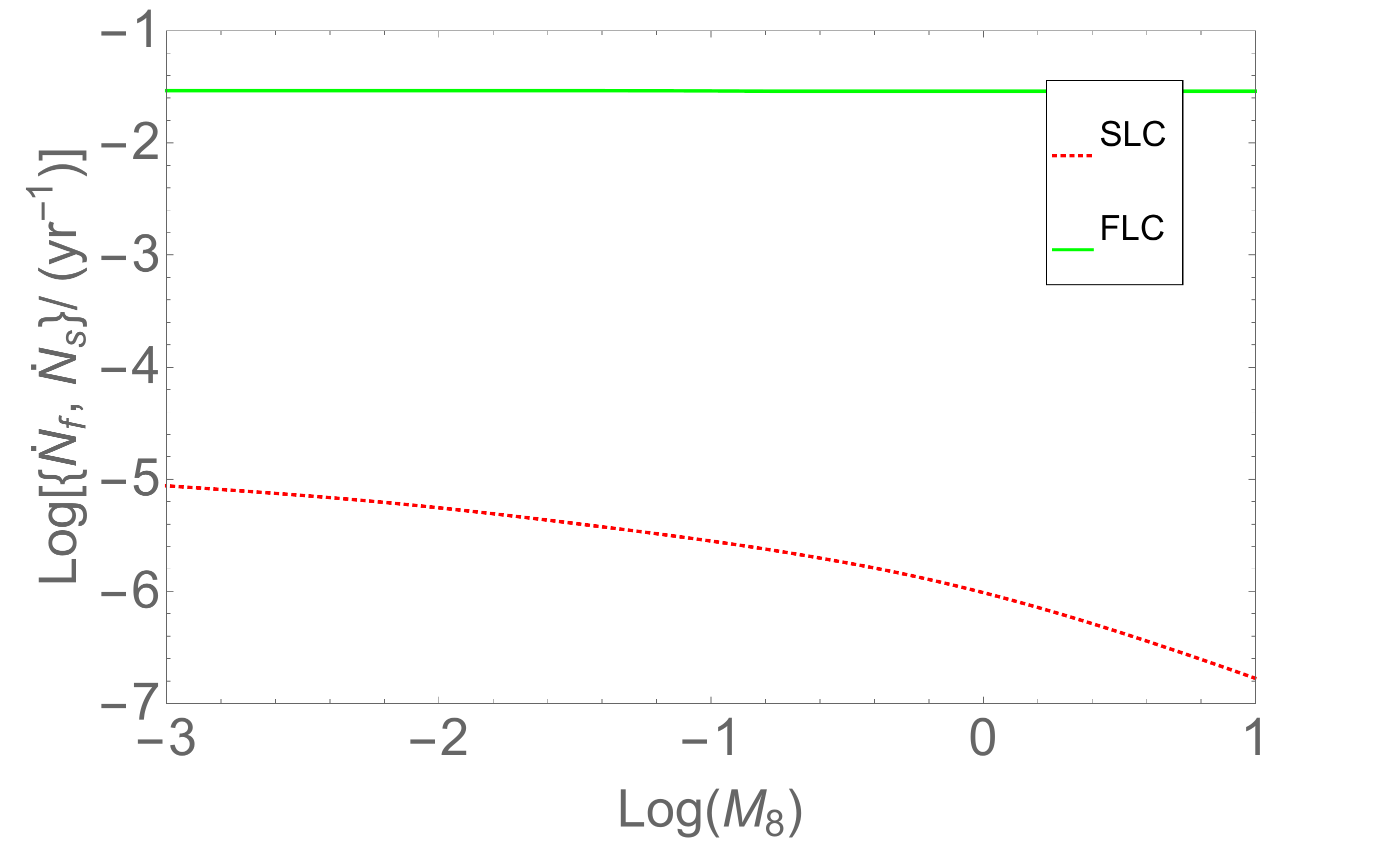}}
\subfigure[]{\includegraphics[scale=0.2]{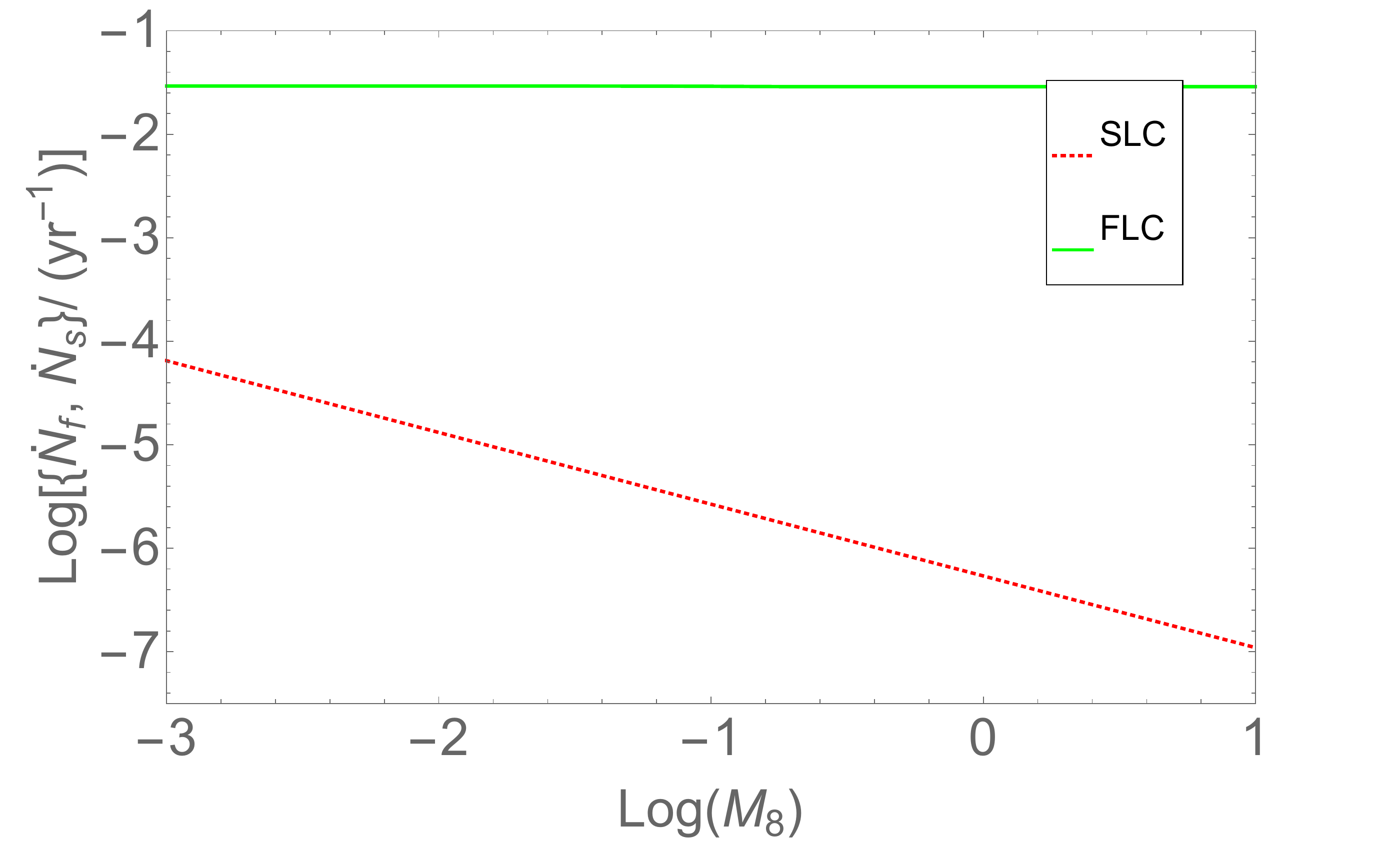}}\\
\caption{Capture rate, $\dot{N}_{s}(M_{\bullet},j, k, Q, \epsilon_{s})$, is shown for different values of $j$ using the $M_{\bullet}$--$\sigma$ relation ($p$ = 4.86), where (a) $k$ = 1 for the prograde case and (b) -1 for the retrograde case, with the lower limit of $\epsilon_{s}$ taken to be $\epsilon_{m}$ = -10, $\gamma$ = 1.1. (c, d) $\dot{N}_{f}$ and $\dot{N}_{s}(M_{\bullet},j, k, Q, \epsilon_{s}, \sigma)$  are shown for both the steady and the full loss cone theory with $j = 0$, $Q=0$, $k = -1$, $\gamma$ = 1.1, $\epsilon_{m}$ = -10 and $\sigma$ = 200 km s$^{-1}$ using the $M_{\bullet}$--$\sigma$ relation with $p$ = 4.86, $\gamma$ = 1.1, $k = -1$..}
\label{stdlcmsigma}
\end{center}
\end{figure}

%\begin{figure}[H]
%\centering
%\subfigure[ \label{slcflcwtms}]{\includegraphics[scale=0.23]{slc_flc_n1.pdf}}
%\subfigure[ \label{slcflcwms}]{\includegraphics[scale=0.23]{slc_flc_n2.pdf}}
%\caption{$\dot{N}_{f}$ and $\dot{N}_{s}(M_{\bullet},j, k, Q, \epsilon_{s}, \sigma)$  are shown for both the steady and the full loss cone theory with $j = 0$, $Q=0$, $k = -1$, $\gamma$ = 1.1, $\epsilon_{m}$ = -10 and $\sigma$ = 200 km/sec (Left) and (right) using the $M_{\bullet}$--$\sigma$ relation with $p$ = 4.86, $\gamma$ = 1.1, $k = -1$.}
%\label{slcflc}
%\end{figure}

From the bottom panels of Figure \ref{stdlcmsigma}, we see that the rate of the number of stars falling into the loss cone is higher in the case of full loss cone theory than in the case of steady loss cone by an order of magnitude. Also, the slope is positive at the lower-mass end, and it becomes negative as it reaches the higher mass. $\dot{N}_{f}$ is almost constant and mainly dependent on the $\sigma$ term, throughout the whole range, because its dependence on mass through $l_{\ell}^{2}$ is small, while the mass dependence of $\dot{N}_{s}$ is strong. We determine the slope of the steady loss cone rate for both the curves with and without using the $M_{\bullet}$--$\sigma$ relation. Without applying the $M_{\bullet}-\sigma$ relation, the slope is --0.3, and by applying the relation with $p$ = 4.86, the slope is --0.6. Therefore, we conclude that $\dot{N}_{s}$ using the $M_{\bullet}$--$\sigma$ relation is smaller than for the case assuming $\sigma$ independent of mass.

%\section{\underline{Effect of mergers and BZ torque}}\label{mergerBZ}
%\begin{figure}[H]
%\centering
%\includegraphics[scale=0.35]{mbh_merger.pdf}
%\subfigure[\label{mergerm}]{\includegraphics[scale=0.25]{mbh_merger.pdf}} 
%\subfigure[\label{BZ_j}]{\includegraphics[scale=0.25]{BZ_j.pdf}} 
%\caption{The evolution of black hole mass due to mergers is shown for different formation redshifts for $q$ = 0.1 and $f$ = $3 \times 10^{5}$ (left) and (right) the spin down, $j(t)$, due to BZ torque for $B_{4}$ = 5, $M_{\bullet}$ = $10^{6} M_{\odot}$ is shown with an initial spin, $j_{i}$ = 0.3.}
%\label{mergerm}
%\end{figure}

\section{Calculation of timescales of spin and mass evolution for all the effects}\label{timescales}
%\section[Calculation of timescales of spin and mass evolution for all effects]{\texorpdfstring{Calculation of timescales of spin and mass evolution \\ for all effects}}
\begin{enumerate}
    \item {\it{Gas accretion}}: From Equation (\ref{mdotst}) we see that \be \tau_{M, g} = \frac{M_{\bullet}}{\dot{M}_{\bullet g}} = \frac{1}{k_{1}} \simeq 1 ~{\rm Gyr},\ee
    and from  Equation (\ref{jtcacc})
    \be \tau_{j, g} = \frac{M_{\bullet}}{\dot{M}_{\bullet g}} \bigg(l_{I} (j) - 2 \epsilon(j)j\bigg) \simeq 1 ~{\rm Gyr}\ee
    \item {\it{Stellar Capture}}: We estimate from Equation (\ref{mdotst}) that \be \tau_{M, *} = \frac{M_{\bullet}}{\dot{M}_{\bullet *}} = \frac{M_{\bullet}}{M_{\odot}}10^{5.5} ~{\rm yr} \simeq 10 ~{\rm Gyr~for}~ M_{\bullet} = 10^{5} M_{\odot} \ee
    \item {\it{Merger}}: Also from Equation (\ref{mg_mt}) we find \be \tau_{M, m} = \frac{M_{\bullet}}{\dot{M}_{\bullet m}} = \frac{M_{\bullet}}{M_{5}^{1.15} 8.058 \times 10^{-3} \cdot (1 + z)^{2.2} 10^{5}M_{\odot}} ~{\rm Gyr} \simeq 10 ~{\rm Gyr~for}~ M_{\bullet} = 10^{5} M_{\odot}, z \simeq 3,\ee
    and from Equation (\ref{mg_jt})
    \be \tau_{j, m} = \frac{M_{\bullet}}{\dot{M}_{\bullet m}} \bigg(-\frac{7}{3} + \frac{9q}{\sqrt{2}j^{2}}\bigg) \simeq 10 ~{\rm Gyr}\ee
    \item{\it{BZ Torque}}: We see from Equation (\ref{bz_jt}) \be
\displaystyle
\tau_{j, BZ} = \frac{\mathcal{J}_{0}}{\mathcal{G}_{0}} \int_{j_{f}}^{j_{i}} \frac{\diff j}{r^{3}(j) j} = 7.0 \times 10^{8} yr \frac{(\kappa (j_{i}, j_{f}) / 0.1)}{B_{4}^{2} M_{9} f_{BZ}} \simeq 1 ~{\rm Gyr},
\ee
where $M_{9}$ is $M_{\bullet}$ in units of $10^{9} M_{\odot}$ and 
\be
\kappa (j_{i}, j_{f}) = \bigg[\bigg(\frac{1}{16}\bigg) \log \bigg(\frac{2 - w}{w}\bigg) + \bigg(\frac{3 w^{2} + 3 w -4}{24 w^{3}} \bigg)\bigg]_{w_{f}}^{w_{i}},
\ee
with $w_{i} = x_{+}(j_{i})$, $w_{f} = x_{+} (j_{f})$.
\end{enumerate}

\section{Experiment 1: Only gas accretion is present}\label{AppendixA}
In the presence of only accretion the spin and mass evolution equations (Equations \ref{jt}, \ref{mt}) take the form
\be \displaystyle
\frac{\diff j}{\diff \tau} = \frac{\dot{\mu}_{g}}{\mu_{\bullet}} \bigg(l_{I} (j) - 2 \epsilon_{I}(j)j\bigg),\label{jtcacc}
\ee
\be
\frac{\diff \mu_{\bullet}}{\diff \tau} = \epsilon_{I}(j) \dot{\mu}_{g}.\label{mtcacc}
\ee
where $ \displaystyle
e_{I} = \bigg( 1 - \frac{2}{3x_{I}}\bigg)^{\frac{1}{2}},~ l_{I} = \frac{2}{3 \sqrt{3}} r_{g} ( 1 + 2(3x_{+} - 2)^{\frac{1}{2}})$ \citep{1970Natur.226...64B}. The solution of $j$ as a function of black hole mass when there is only accretion present was derived by \cite{1970Natur.226...64B} using the solution of the geodesic equation for the Kerr metric found by \cite{1968NCimB..57..351F} and \cite{1968PhRv..174.1559C} to be
\be
j = \frac{1}{3} x_{+}^{\frac{1}{2}} ( 4 - (3x_{+} - 2)^{\frac{1}{2}} ),\label{bardeeneq}
\ee
where $x_{I} = r_{I} / r_{g}$, $x_{+} = r_{+} / r_{g}$, where $r_{+}$ is the horizon, $e_{I}$ is the energy per unit mass, and $l_{I}$ is the angular momentum per unit mass for the ISCO. The value of $x$ varies from 6 to 1 for $j$ varying from 0 to 1. Equation (\ref{bardeeneq}), is derived as follows.
%\be
%l_{I} - je_{I} = \pm \frac{1}{\sqrt{3}} x_{I},
%\ee
Using the expressions of $l_{I}$ and $e_{I}$ \citep{1972ApJ...178..347B},
\be
l_{I} = \frac{\sqrt{x_{I}}(x_{I}^{2} + j^{2} -2j\sqrt{x_{I}})}{x_{I} (x_{I}^{2} -3 x_{I} + 2j\sqrt{x_{I}})^{1/2}}, ~ e_{I} = \frac{(x_{I}^{2} -2 x_{I} + j\sqrt{x_{I}})}{x_{I}(x_{I}^{2} -3 x_{I} + 2j\sqrt{x_{I}})}, \label{bardeen72}
\ee
and after squaring both sides we finally arrive at a quadratic equation of $j$ given as
\be
3j^{2} + 6x_{I} - 8j\sqrt{x_{I}} - x_{I}^{2} = 0,
\ee
whose solution is
\be
j = \frac{\sqrt{x_{I}}}{3} (4 \pm \sqrt{3x_{I} -2)}.
\ee
Since $j < 1$, the negative sign is the correct choice, so that
\be 
j(x_{+}) = \frac{1}{3} x_{+}^{\frac{1}{2}} ( 4 - (3x_{+} - 2)^{\frac{1}{2}} ).
\ee
Now, the condition $\displaystyle V_{eff} = \frac{e_{I}^{2} - 1}{2}$ along with Equation (6) in RM19 with $Q = 0$, after some algebra, gives \be
l_{I} = je_{I} \pm \frac{x_{I}}{\sqrt{3}}. 
\ee
The final expression for $l_{I}$ becomes \citep{1972ApJ...178..347B},
\be
l_{I} = \frac{2}{3\sqrt{3}} [2(3 x_{I} - 2)^{1/2} + 1] + \frac{x_{I}}{\sqrt{3}}(-1 \pm 1). 
\ee
Here a positive sign is the correct choice since for $a = 1$, $l_{I} = 0$. Therefore, the final expression of $l_{I}$ is
\be
l_{I} (x_{I}, m_{d}) = \frac{2}{3 \sqrt{3}} m_{d} ( 1 + 2(3x_{I} - 2)^{\frac{1}{2}}),
\ee
where $m_{d}$ is the mass of the disk consumed by the hole. The analytic relation between $x_{+}$ and $r_{g}$ \citep{1970Natur.226...64B} is
\be 
\bigg(\frac{x_{+}}{x_{1}}\bigg) =  \bigg(\frac{r_{g1}}{r_{g}}\bigg)^{2},
\ee
where $x_{1}$ and $r_{g1}$ are the initial values when $j = 0$.
Using this, it is found  that \citep{1970Natur.226...64B},
\be
\frac{r_{g}}{r_{g1}} = \bigg(\frac{3x_{1}}{2} - 1\bigg)^{\frac{1}{2}} \sin\bigg[\bigg(\frac{2}{3x_{1}}\bigg)^{\frac{1}{2}} \frac{\Delta m_{0}}{r_{g1}}\bigg] + \cos\bigg[\bigg(\frac{2}{3x_{1}}\bigg)^{\frac{1}{2}} \frac{\Delta m_{0}}{r_{g1}}\bigg],
\ee
where $\Delta m_{0}$ is the accreted rest mass when the change in mass is from $r_{g}$ to $r_{g1}$. 
%\begin{figure}[H]
    %\centering
   % \includegraphics[scale=0.35]{bardeen_1970.pdf}
   % \caption{The \cite{1970Natur.226...64B} solution, $j(\displaystyle \mu_{\bullet})$ of the spin is shown when there is only accretion (run \#1.1).}
    %\label{bardeen}
%\end{figure}
We obtained $j(\displaystyle \mu_{\bullet})$ for the \cite{1970Natur.226...64B} solution using Equations (\ref{jtcacc}) and (\ref{mtcacc}) where there is only accretion, which is shown in Figure \ref{acc_bardeen}. After the black hole spin saturates, only the mass increases, leaving the spin parameter unchanged at the saturated value of 1.

\begin{figure}[H]
\centering
\subfigure[\label{acc_bz_j_bardeen}]{\includegraphics[scale=0.25]{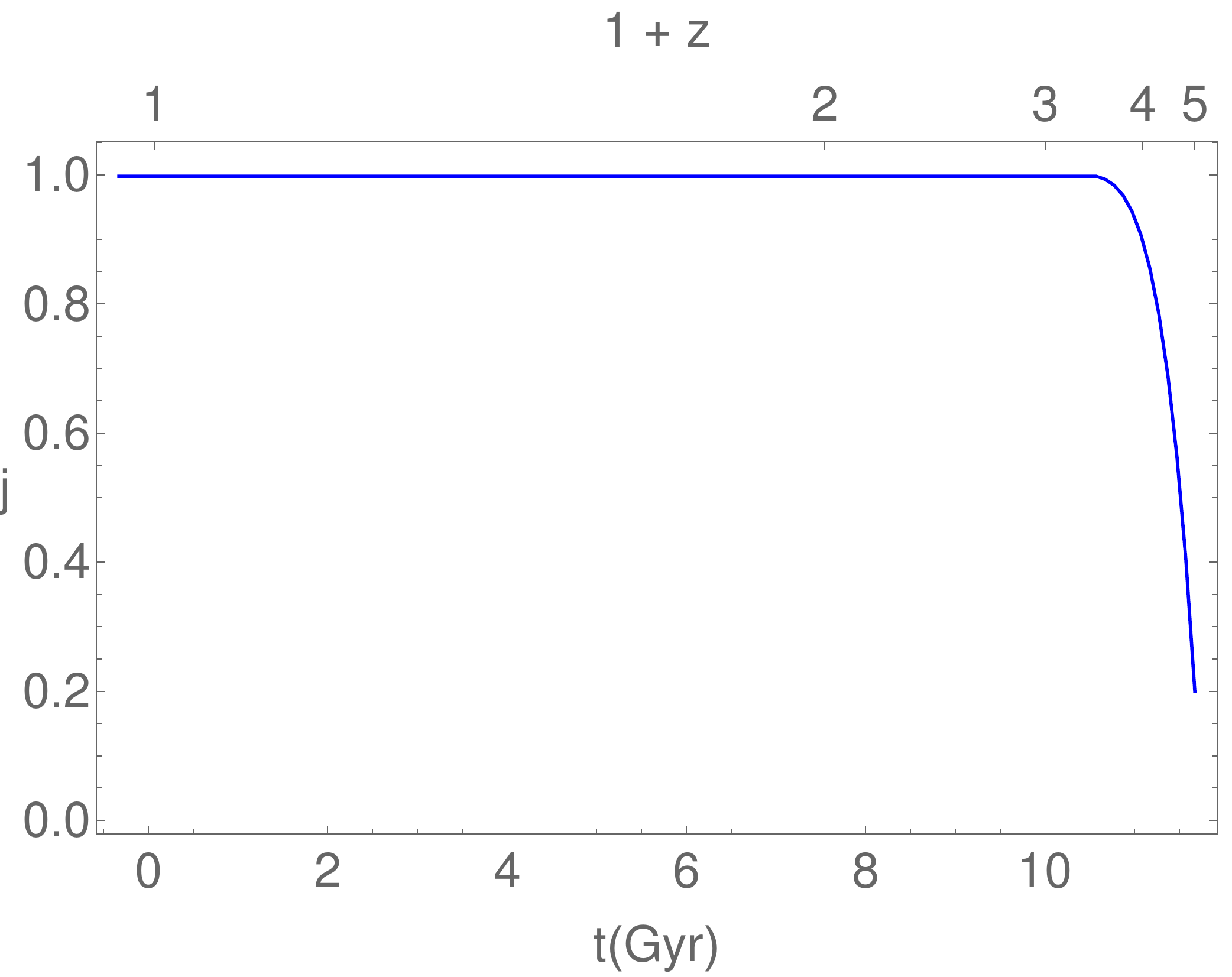}}  \hspace{0.1 cm}
\subfigure[\label{acc_bz_m_bardeen}]{\includegraphics[scale=0.27]{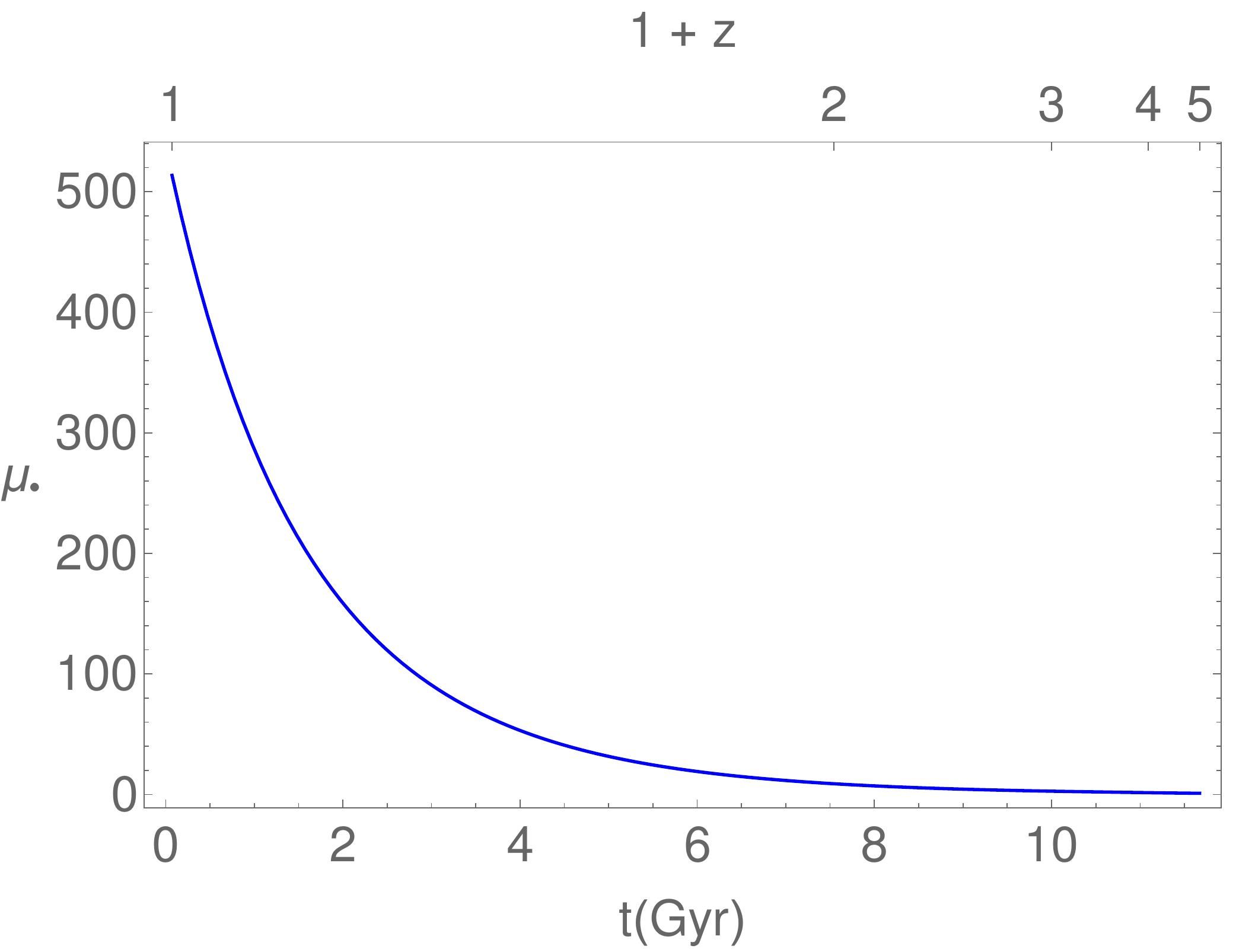}}  \hspace{0.1 cm}
\caption{(a) Spin evolution, $j(t)$, and (b) the mass evolution, $\displaystyle\mu_{\bullet}(t)$, for $B_{4}$ = 5, $z_{f}$ = 4, $\eta$ = 0.09, $M_{s} = 10^{5} M_{\odot}$ are shown for the case when only accretion is present (run \# 1.2).}
\label{acc_bardeen}
\end{figure}
Figure \ref{acc_bardeen} shows the solutions to Equations (\ref{jtcacc}) and (\ref{mtcacc}), $j(t)$ and $\displaystyle\mu_{\bullet}(t)$, when there is only gas accretion present. The mass continues to grow and the spin reaches a saturated value, and afterward it remains the same.

\section{Experiment 2: Non - relativistic accretion feedback and full loss cone theory}\label{AppendixB}
%\section[Experiment 2: Non - relativistic accretion feedback and full loss cone theory]{\texorpdfstring{Experiment 2: Non - relativistic accretion feedback\\ and full loss cone theory}}
{\it{Full loss cone theory:}}
Here we consider the case of the full loss cone ($\theta_{d} >> \theta_{\ell}$) where the mass density in the galaxy cusp follows a single power-law profile and the stars are able to quickly fill the loss cone on dynamical time scales. Therefore,
\be \rho = \rho_{0}r^{-\gamma},\ee 
where $\gamma$ is the power-law index. The distribution function of stars in a such a galaxy is given by \citep{2013CQGra..30x4005M}
\be f_{s}(E) = \frac{3-\gamma}{8}\sqrt{\frac{2}{\pi^{5}}} \frac{\Gamma(\gamma + 1)}{\Gamma(\gamma - \frac{1}{2})} \frac{M_{\bullet}}{m_{\star}} \frac{\phi_{0}^{\frac{3}{2}}}{(GM_{\bullet})^{3}} \bigg(\frac{|E|}{\phi_{0}}\bigg)^{\gamma - \frac{3}{2}},\ee 
where $\displaystyle \phi_{0} = \frac{GM_{\bullet}}{r_{m}}$, $E$ is the energy, $r_{m}$ is the gravitational influence radius of the black hole defined as $GM_{\bullet} / \sigma^{2}$ and $m_{*}$ is the stellar mass. The rate of capture of stars within the loss cone is
\be  F_{f}(E) = 4 \pi^{2}L_{\ell}^{2}(E)f_{s}(E),\ee
where $L_{\ell} (E)$ is the loss cone angular momentum of the star. An integration of this over all energies gives the total rate of capture in the loss cone,
\be  \dot{N}_{f} = \int_{-\infty}^{\phi_{0}} F_{f}(E) \diff E, \ee 
so that
\be
\dot{M}_{\bullet *f} = m_{\star}\dot{N}_{f} =   \frac{3-\gamma}{8}\sqrt{\frac{1}{2\pi}} \frac{\Gamma(\gamma + 1)}{\Gamma(\gamma - \frac{1}{2})} \frac{l_{\ell}^{2}}{GM_{\bullet}}\frac{1}{r_{m}}\bigg(\frac{GM_{\bullet}}{r_{m}^{3}}\bigg)^{\frac{1}{2}} M_{\bullet}
 =  \frac{3-\gamma}{8}\sqrt{\frac{1}{2\pi}} \frac{\Gamma(\gamma + 1)}{\Gamma(\gamma - \frac{1}{2})} \frac{l_{\ell}^{2}(M_{\bullet})}{G c^{2}} \sigma^{5}, \label{flceq}
\ee
where $L_{\ell} \equiv (GM_{\bullet} / c)~ l_{\ell}$. For the nonrelativistic case, $L_{\ell}$ is given by Equation (\ref{nonrell}). After simplification, it is seen that this expression for  $\dot{M}_{\bullet *} \propto \sigma^{5}$ for the nonrelativistic case does not depend on $M_{\bullet}$. However, for the relativistic case, $\dot{M}_{\bullet *}$ depends on both $\sigma$ and $M_{\bullet}$ through the capture radius. In the full loss cone regime, the depleted orbits are repopulated within orbital periods by the relaxation process; this is a reasonable assumption for $M_{\bullet} << 10^{5} M_{\odot}$. 

Now, we study the nonrelativistic case with no spin and full loss cone theory applied to stellar capture for which a fully analytic solution can be obtained. We solve the mass evolution equation (Equation (\ref{mtdim}) to find that
\be 
        \displaystyle
        t (M_{\bullet}) = \left\{\begin{array}{lr}
        \int_{M_{s}}^{M_{\bullet}} \frac{dM_{\bullet}}{\dot{M_{g}} + \dot{M_{*}}} & \text{\rm for~} t > t_{s} \\ \\
        t_{s} + \int_{M_{\bullet t}}^{M_{\bullet}} \frac{dM_{\bullet}}{\dot{M_{*}}}.  & \text{\rm for~} t \leq t_{s} , 
        \end{array}\right.
        \label{nras}
\ee 
where $t_{s}$ is time at which feedback has stopped accretion. Solving Equation (\ref{nras}) using Equation (\ref{macc}) for $\dot{M}{g}$ and Equation (\ref{flceq}) for $\dot{M}_{*}$ for $t\leq t_{s}$, we find $\tau_{s} = k_{1}t_{s}$ where
\be 
\tau_{s} = \log \bigg[ \frac{k_{1}M_{\bullet t} + k_{2}\sigma^{5}}{k_{1}M_{s} + k_{2}\sigma^{5}} \bigg] = k_{1}t_{z} (z_{s}) - k_{1} t_{z} (z_{f}), 
\ee
where $M_{*} = k_{2} \sigma^{5}$ for the nonrelativistic full loss cone theory and $k_{1}$ is defined by Equation (\ref{acceq}). Using Equation (\ref{tz}), it is found that
\be
\bigg(\tau_{s} + k_{1} t_{z} (z_{f}) - k_{1} \frac{2}{3} \frac{1}{\sqrt{1 - \Omega_{m}}} \log \sqrt{1 - \Omega_{m}} \bigg) \frac{3\sqrt{1 - \Omega_{m}}}{2k_{1}} = \log (\alpha_{1} + \sqrt{\Omega_{m} + \alpha_{1}^{2}}),\label{flceqn}
\ee
where \be \nonumber
\alpha_{1}^{2} = \frac{1 - \Omega_{m}}{(1 + z_{s})^{3}}.
\ee
Writing the left-hand side of Equation (\ref{flceqn}) as $\log \beta_{1}$, we derive
\be 
z_{s} = \bigg[ \frac{2\beta_{1}\sqrt{1 - \Omega_{m}}}{\beta_{1}^{2} - \Omega_{m}}\bigg]^{\frac{2}{3}} - 1.
\ee
After solving Equation (\ref{nras}) for $t > t_{s}$, the final equation for $M_{\bullet}$ as a function of redshift is given by
\be 
        %\displaystyle
       M_{\bullet} (\tau, M_{s}, \sigma_{100}) = \left\{\begin{array}{lr}
       \mu_{M}(\tau) M_{s} + M_{s} = M_{s} + (e^{\tau} + C \sigma_{100}^{5}(e^{\tau} - 1))M_{s}  & \text{\rm for~} z < z_{s} \\
        M_{\bullet t} + [\mu_{s} + (\tau - \tau_{s})C \sigma_{100}^{5}]M_{s}.  & \text{\rm for~} z \geq z_{s} , 
        \end{array}\right.
        %\label{nras}
\ee 
where $C = k_{2}(100 ~{\rm km~ sec^{-1}})^{5} / (k_{1}M_{s})$ and
\be \mu_{M} = \frac{M_{\bullet} - M_{s}}{M_{s}} = \mu_{\bullet} - 1, \ee 
where $M_{s} = f_{b}M_{b} \sigma^{5}$.

\begin{figure}[H]
\centering
\includegraphics[scale=0.25]{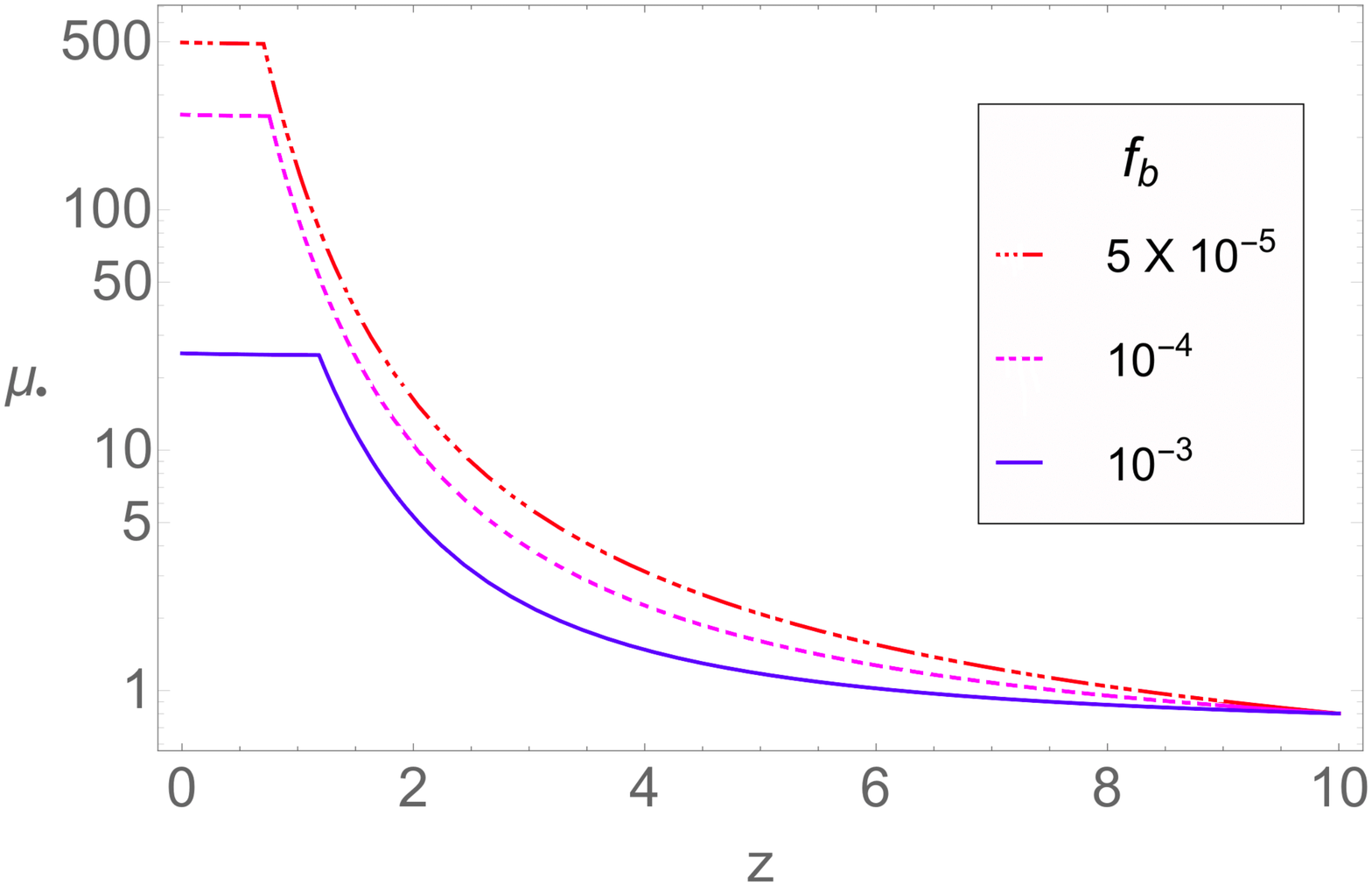}
\caption{\small {Mass evolution, $\mu_{\bullet}(z)$, is shown for different $f_{b}$ for $\sigma_{100} = 1$ (run \#2.1, \#2.2, \#2.3) with $j_{0}$ = 0, $B_{4}$ = 0}.}
\label{nrwtsp}
\end{figure}
In Figure \ref{nrwtsp}, the late evolution represents the black hole mass growth only by capture of stars, and the mass growth rate by accretion of gas dominates much earlier.

\begin{enumerate}
\item In the later experiments, we have scaled Equation (\ref{jt}) by Equation (\ref{mt}) to obtain an equation for $\displaystyle \frac{\diff j}{\diff M_{\bullet}}$ that is solved to find $j(M_{\bullet})$ and fed into Equation (\ref{mt}) to obtain $M_{\bullet}(t)$. 
\item We derive $j(t)$ similarly, using the solution of $j(M_{\bullet})$ in Equation (\ref{jt}). 
\item All the solutions are dependent on the value of $\sigma$ which we have considered to be constant throughout for a particular galaxy. 
\end{enumerate}
%Taking into account the saturation, we will now present the results for the more realistic evolution experiments (3 and 4) in Table \ref{expts} where we include the effects one at a time. 

\section{Experiment 3: Effect of Gas accretion and BZ torque}\label{3.1}
Here the spin and mass evolution equations (Equations \ref{jt}, \ref{mt}) take the form
\be \displaystyle
\frac{\diff j}{\diff \tau} = \frac{\dot{\mu}_{g}}{\mu_{\bullet}} \bigg(l_{I} (j) - 2 \epsilon_{I}(j)j\bigg)  + \frac{4}{9}\times 10^{-5} f_{BZ} B_{4} \mu_{\bullet} M_{s5}x_{+}^{3}(j)j.\label{jtc1}
\ee
\be
\frac{\diff \mu_{\bullet}}{\diff \tau} = \epsilon_{I}(j) \dot{\mu}_{g} .\label{mtc1}
\ee
First, we study the canonical case (run \# 3.1.1, experiment 3), and then we change the parameters one by one, keeping others constant. We now present the results for different runs listed in Table \ref{table_acc_bz_st_mg}.
%\begin{figure}[H]
%\centering
%\subfigure[\label{acc_bz_jan}]{\includegraphics[scale=0.3]{acc_bz_j_n.pdf}}  \hspace{0.1 cm}
%\subfigure[\label{acc_bz_mcn}]{\includegraphics[scale=0.32]{acc_bz_m_n.pdf}}  \hspace{0.1 cm}
%\caption{The spin evolution, $j$($t$), (Figure \ref{acc_bz_jan}) and the mass evolution $\displaystyle\mu_{\bullet}$($t$), (Figure \ref{acc_bz_mcn}) are shown for $B_{4}$ = 5, $z_{f}$ = 4, $\eta$ = 0.09, $M_{s} = 10^{5} M_{\odot}$ when there is only accretion and BZ torque present and accretion continues to occur (canonical case, run \# 3.1.1).}
%\label{acc_bz_jn}
%\end{figure}

\begin{figure}[H]
\centering
\subfigure[\label{acc_bz_ja}]{\includegraphics[scale=0.25]{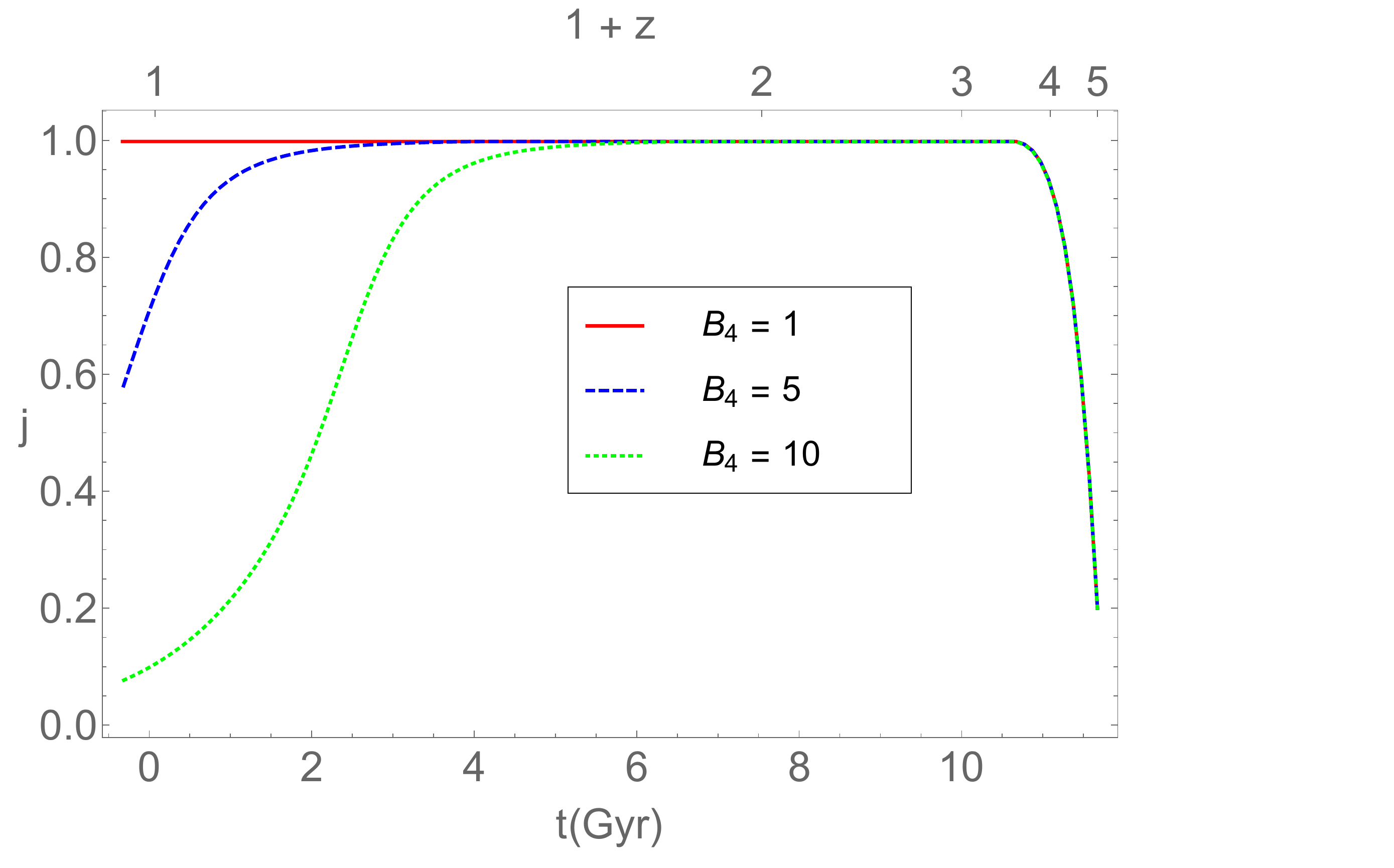}}  \hspace{0.1 cm}
\subfigure[\label{acc_bz_mc}]{\includegraphics[scale=0.25]{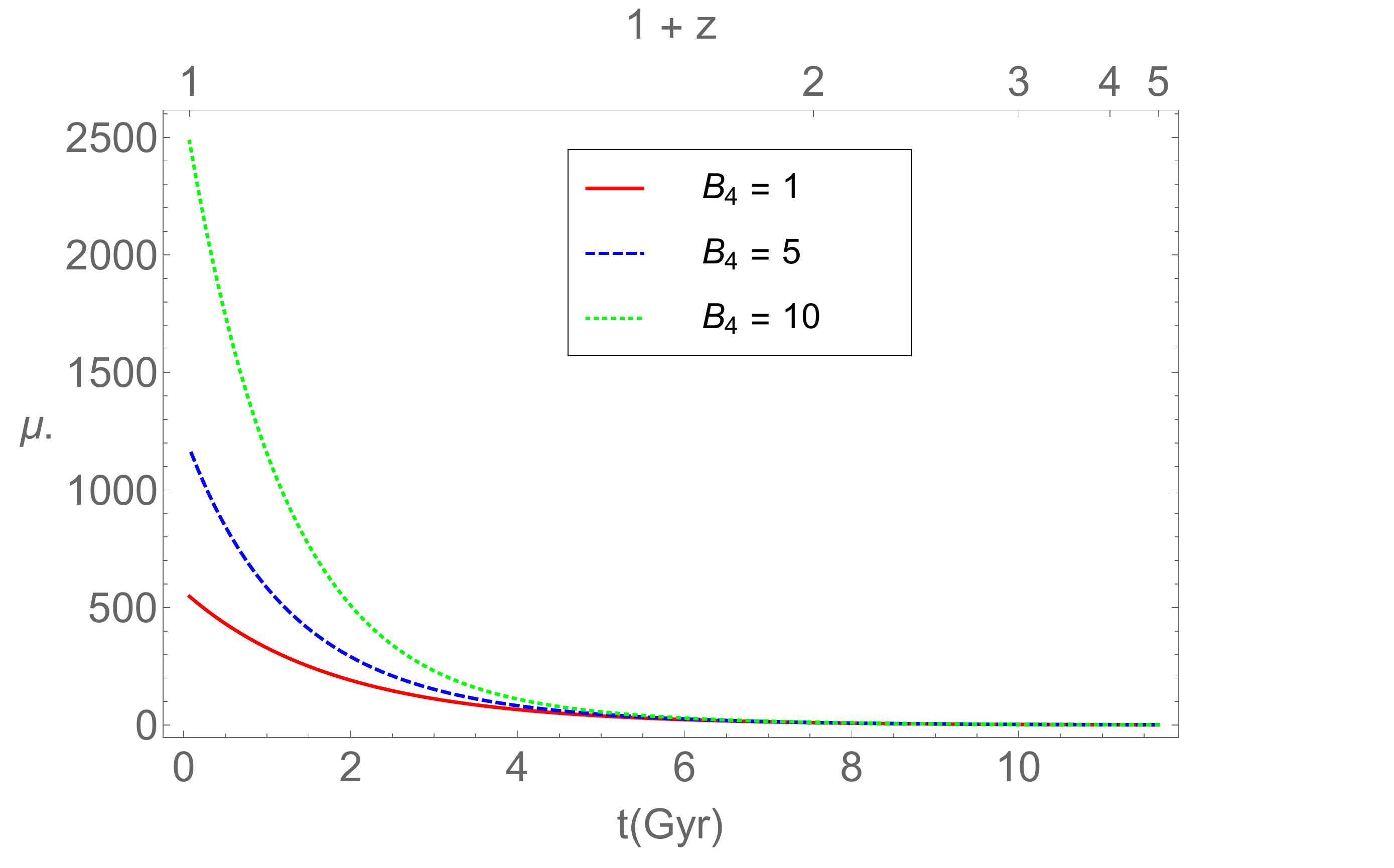}}  \hspace{0.1 cm}
\caption{(a) Spin evolution, $j$($t$), and (b) mass evolution, $\displaystyle\mu_{\bullet}$($t$), are shown for run\# 3.1.1 and run \# 3.1.2 in Table \ref{table_acc_bz_st_mg} when there is only accretion and BZ torque present.}
\label{acc_bz_j}
\end{figure}

\begin{enumerate}
%\item From Figures \ref{acc_bz_mcn} and \ref{acc_bz_jan}, we see that the saturates without feedback and the mass continues to grow while the spin saturates.
\item It can be seen from Figure \ref{acc_bz_ja} that the BZ torque causes the spin-down of the black hole reducing it from the highest saturated spin value. As the $B_{4}$ value is increased, the spin-down is more effective while the accretion is enhanced.
%\end{itemize}

\begin{figure}[H]
\centering
\subfigure[\label{acc_bz_j_la}]{\includegraphics[scale=0.25]{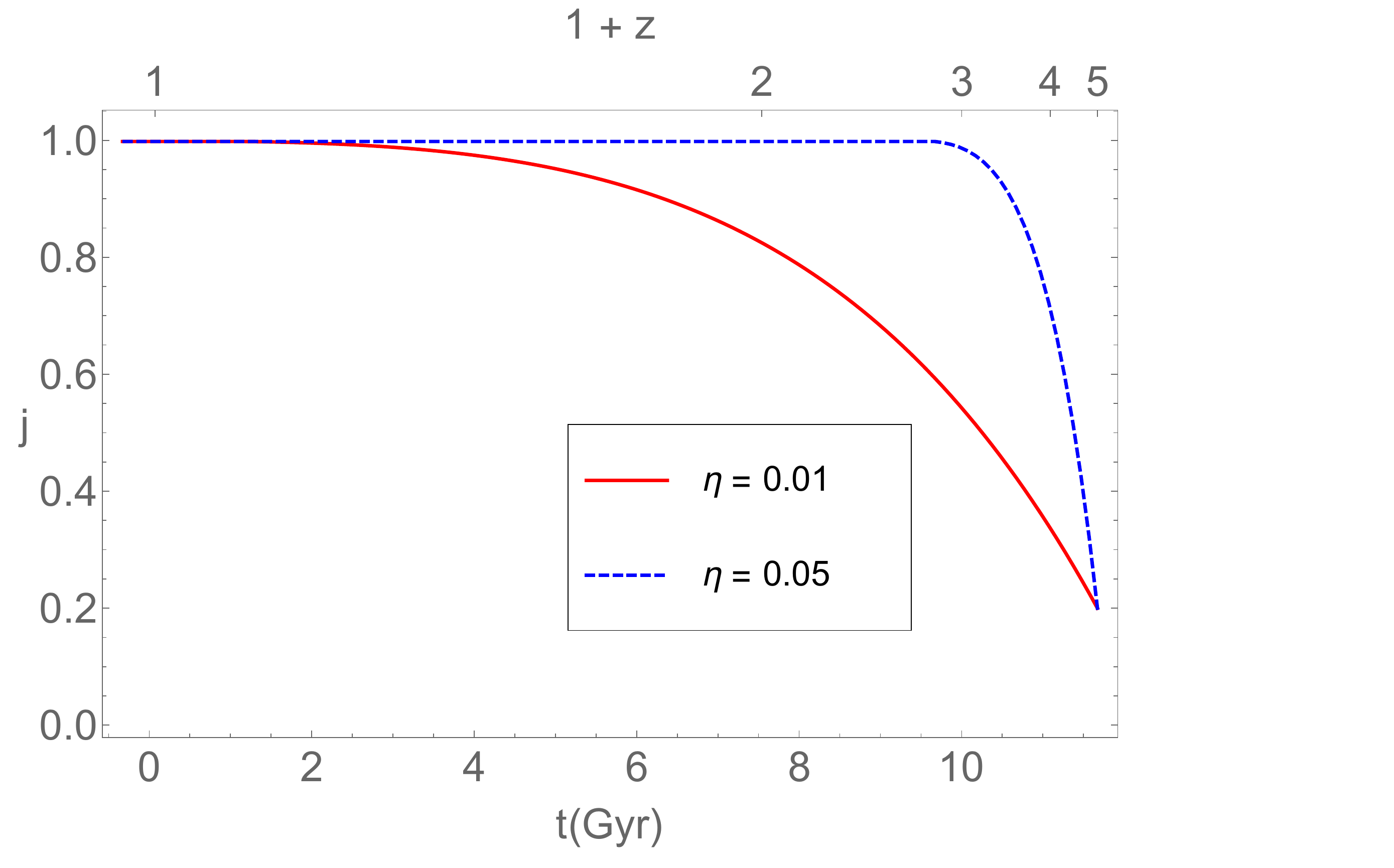}}  \hspace{0.1 cm}
\subfigure[\label{acc_bz_m_lb}]{\includegraphics[scale=0.25]{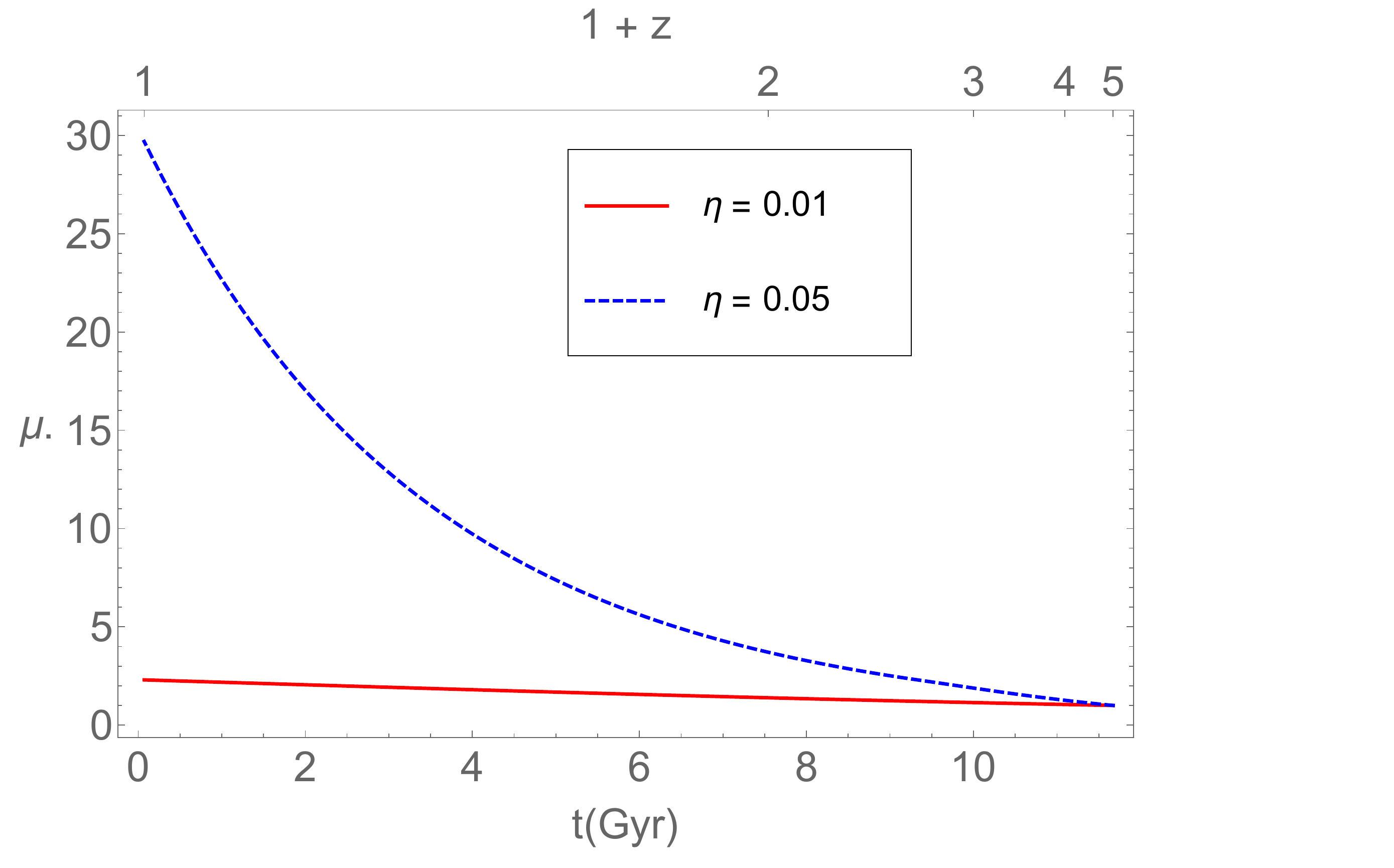}} 
\caption{(a) Spin evolution, $j$($t$), and (b) mass evolution, $\displaystyle\mu_{\bullet}$($t$), are shown for run \# 3.2.1 and \# 3.2.2 [see Table \ref{table_acc_bz_st_mg}] when there is only accretion and BZ torque present.}
\label{acc_bz_mj_l}
\end{figure}

%\begin{itemize}
\item From the plots (see Figure \ref{acc_bz_mj_l}) it can be seen that mass growth by accretion with an efficiency of $\eta$ = 0.01 or 0.05 is very small, which cannot generate high-mass black holes in the universe, indicating that $\eta \geq$ 0.05.
\end{enumerate}

\section{Experiment 4: Gas accretion, stellar capture, and BZ torque are present}\label{3.2}
In this experiment, the spin and mass evolution equations (Equations \ref{jt}, \ref{mt}) take the form
\be \displaystyle
\frac{\diff j}{\diff \tau} = \frac{\dot{\mu}_{g}}{\mu_{\bullet}} \bigg(l_{I} (j) - 2 \epsilon_{I}(j)j\bigg) + \frac{\dot{\mu}_{*}}{\mu_{\bullet}} \bigg(l_{*} (j) - 2 \epsilon(j)j\bigg) + \frac{4}{9}\times 10^{-5} f_{BZ} B_{4} \mu_{\bullet} M_{s5}x_{+}^{3}(j)j.\label{jtc2}
\ee
\be
\frac{\diff \mu_{\bullet}}{\diff \tau} = \epsilon_{I}(j) \dot{\mu}_{g}+ \epsilon(j) \dot{\mu}_{*}.\label{mtc2}
\ee
where accretion, BZ torque, and the stellar capture with steady loss cone theory are taken into account.% (Figure \ref{acc_bz_st_jn}).
%\begin{figure}[H]
%\centering
%\subfigure[\label{acc_bz_jans}]{\includegraphics[scale=0.3]{acc_bz_st_j_n.pdf}}  \hspace{0.1 cm}
%\subfigure[\label{acc_bz_mcns}]{\includegraphics[scale=0.32]{acc_bz_st_m_n.pdf}}  \hspace{0.1 cm}
%\caption{(a) Evolution of the spin, $j$($t$), and (b) mass, $\displaystyle\mu_{\bullet}$($t$), are shown for $B_{4}$ = 5, $z_{f}$ = 6, $\eta$ = 0.07, $\gamma$ = 1.1, $M_{s} = 10^{5} M_{\odot}$ when there is accretion, stellar capture and BZ torque present for the canonical case (run \# 4.1.4).}
%\label{acc_bz_st_jn}
%\end{figure}

\begin{enumerate}
\item  By studying the canonical case (run \# 2.1), we find that the mass evolution does not show any significant variation with changes in parameters ($k$, $\gamma$, $j_{0}$) and the spin evolution does not show variation for changes in ($k$, $\gamma$). This is because the accretion is a dominant process and the factors that control the stellar capture do not make a significant impact on the range of parameters considered. 
\end{enumerate}
We present the results and discuss the runs (\# 4.1 to \# 4.6) given in Table \ref{table_acc_bz_st_mg}.

\begin{figure}[H]
\centering
\subfigure[\label{acc_bz_st_ma}]{\includegraphics[scale=0.2]{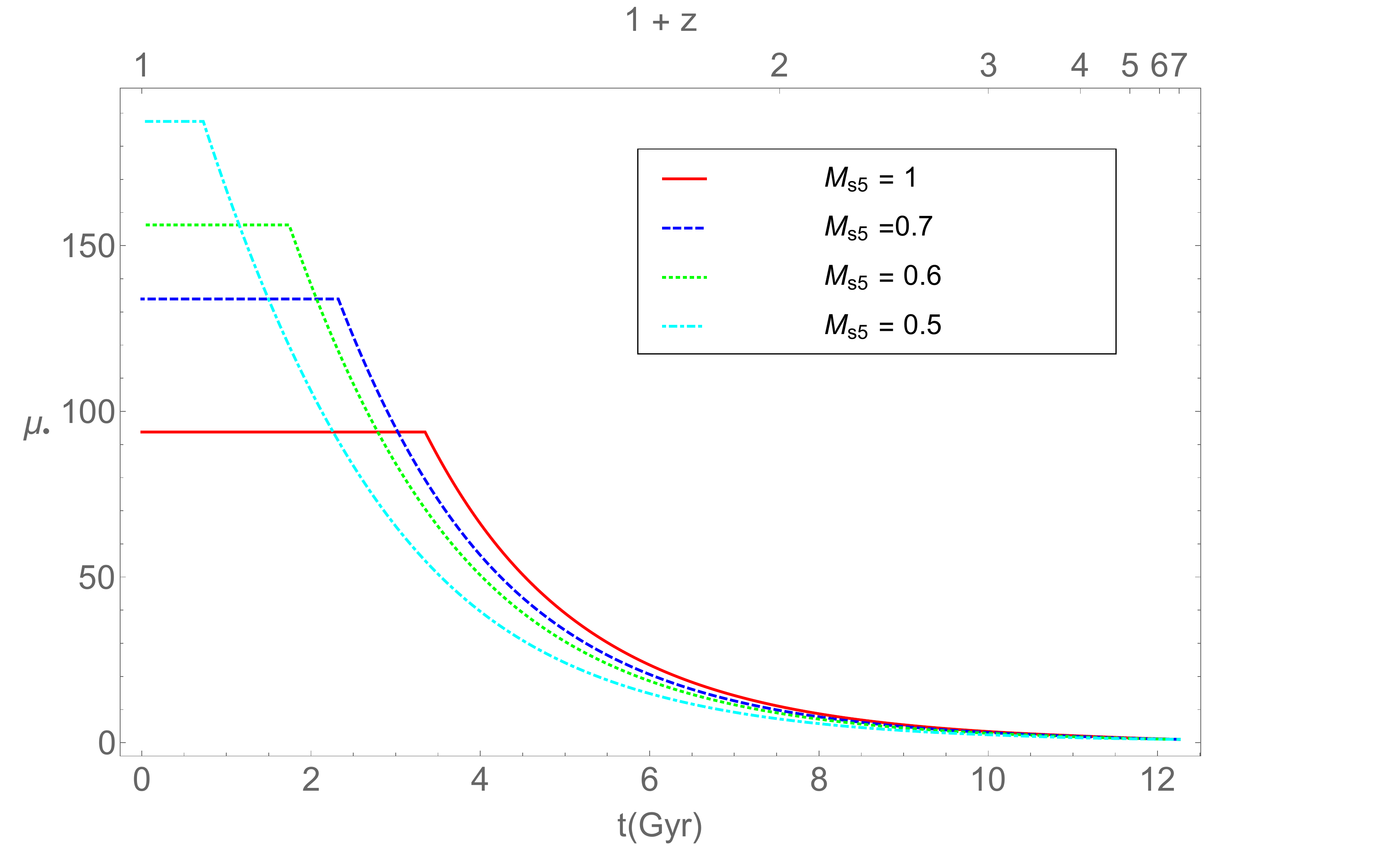}} \hspace{0.1 cm}
%\subfigure[\label{acc_bz_st_mb}]{\includegraphics[scale=0.25]{acc_bz_st_m_bz10_rel.pdf}} 
%\\
\subfigure[\label{acc_bz_st_md}]{\includegraphics[scale=0.2]{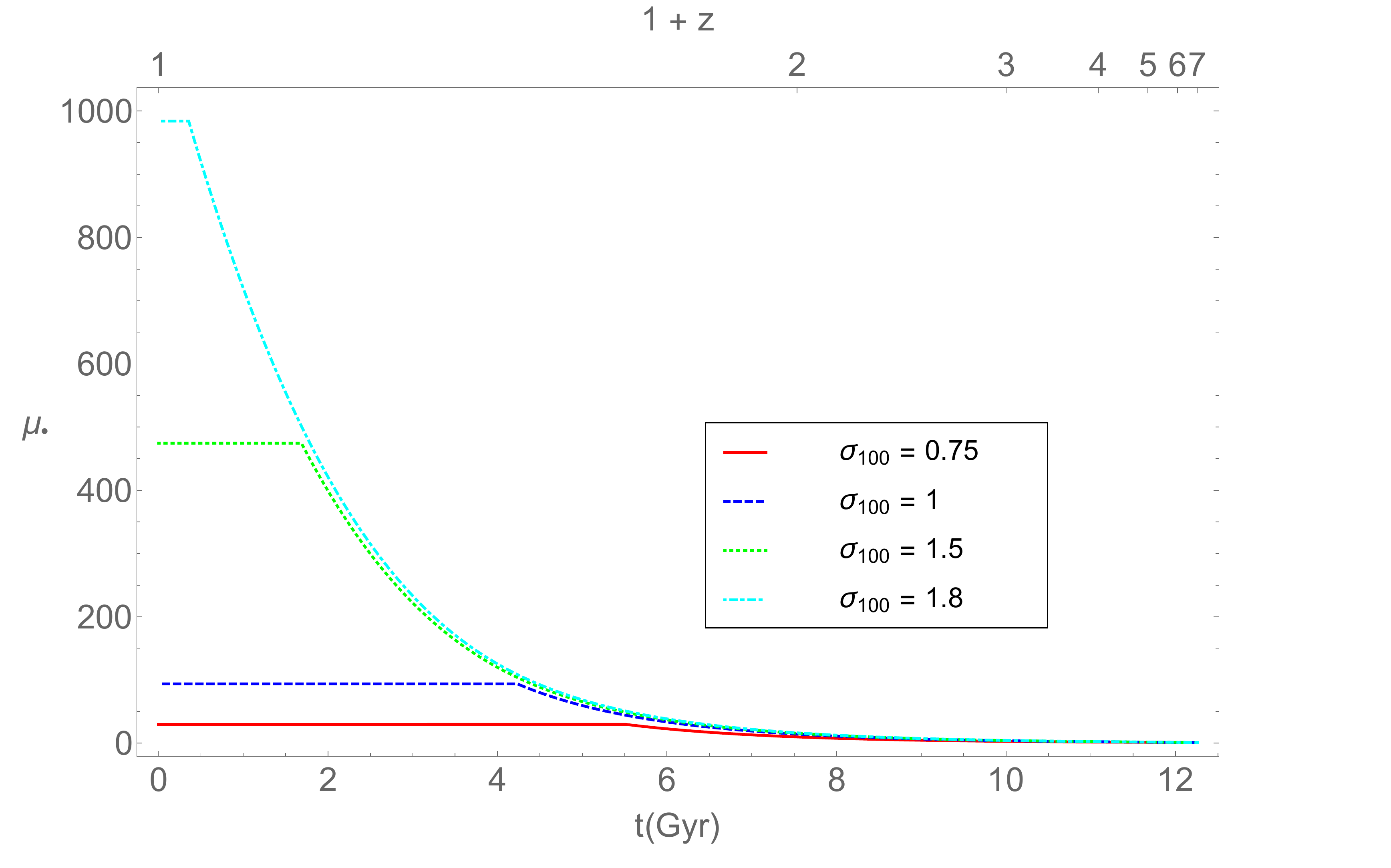}} \\ 
\subfigure[\label{acc_bz_st_me}]{\includegraphics[scale=0.2]{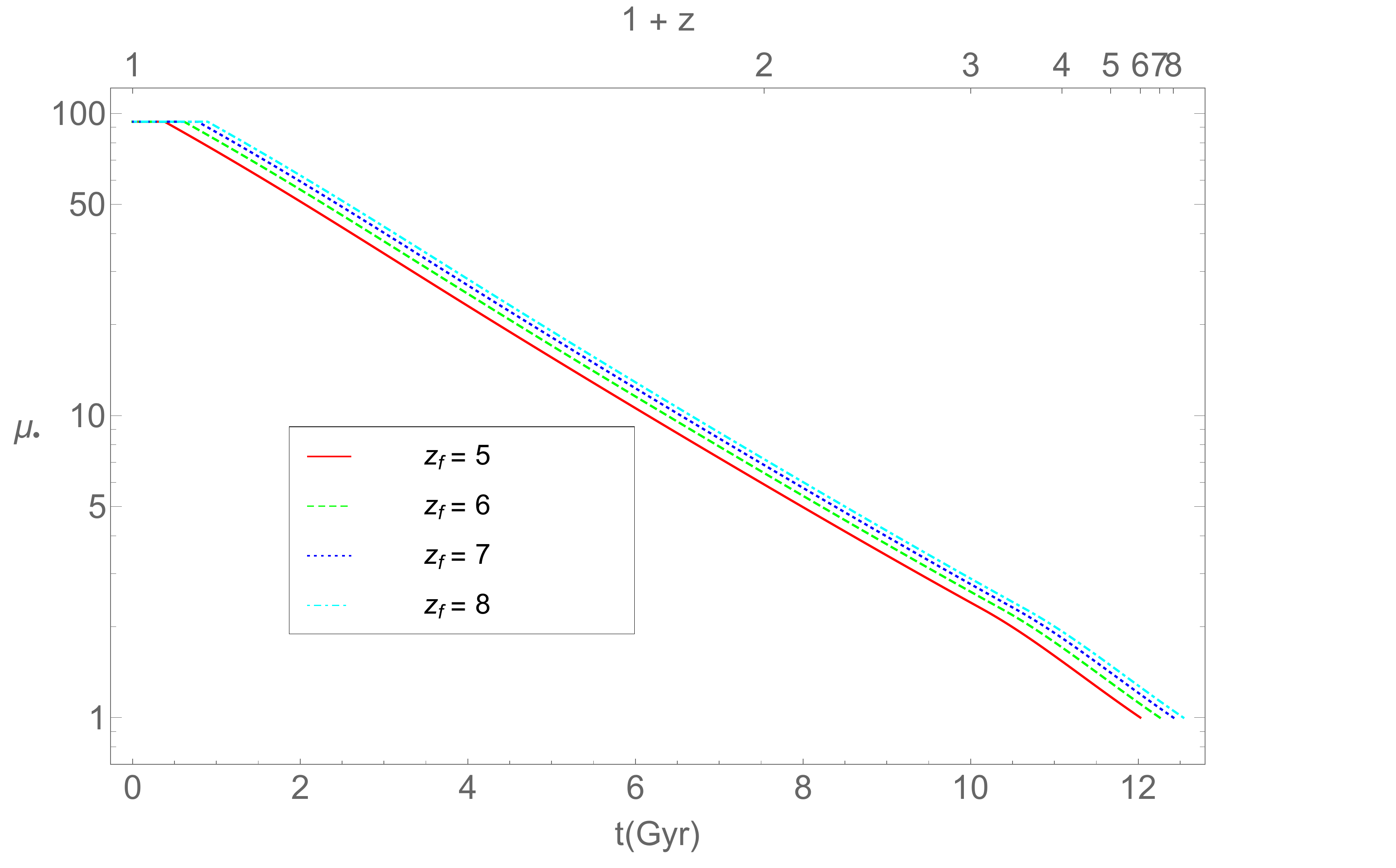}} \hspace{0.1 cm}
\subfigure[\label{acc_bz_st_mf}]{\includegraphics[scale=0.20]{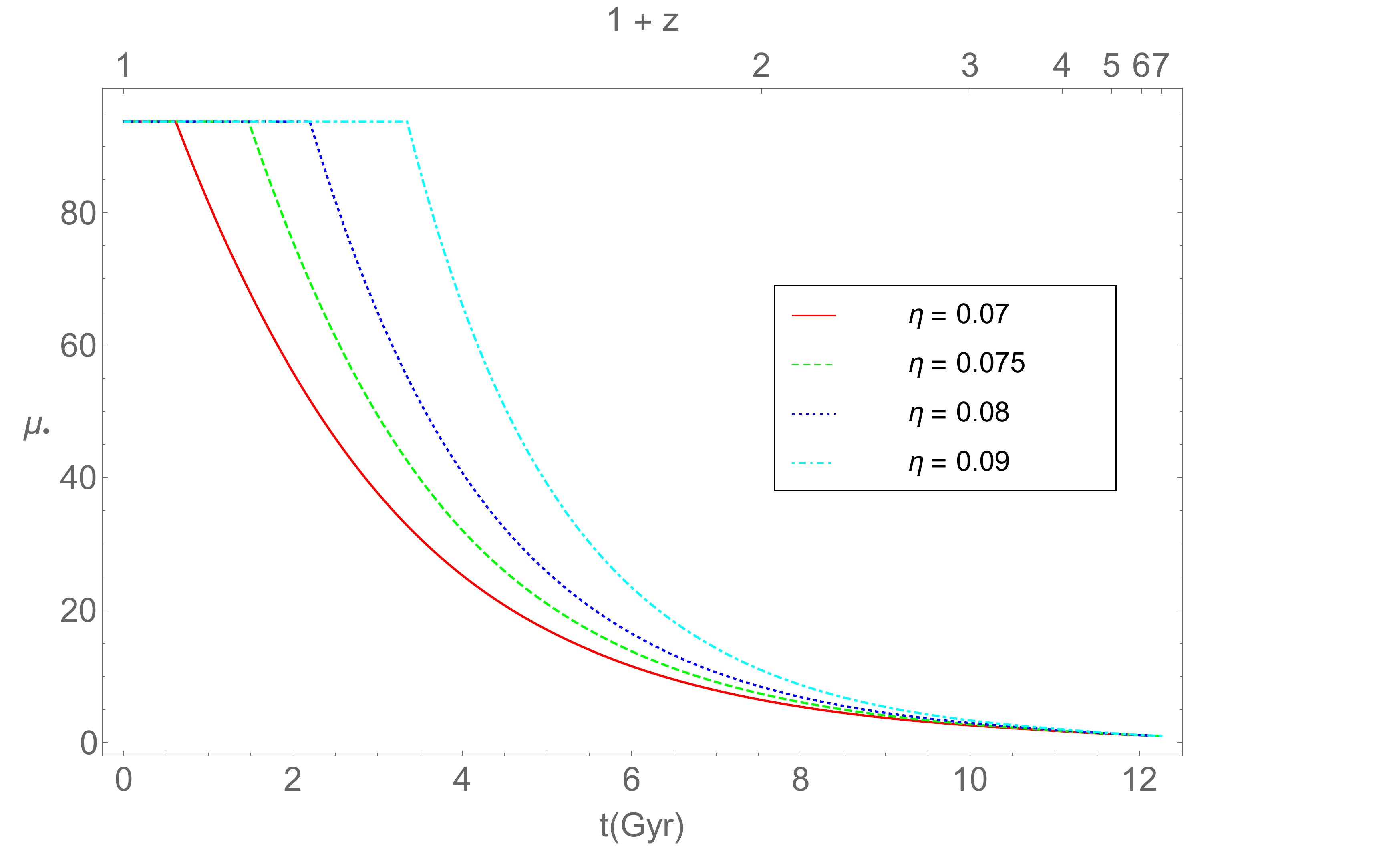}} \\

\caption{Mass evolution, $\displaystyle\mu_{\bullet}$($t$), for run \# 4.1, \# 4.3, \# 4.4, \# 4.5 [see Table \ref{table_acc_bz_st_mg}] (a -- d) are shown, for the case when there is accretion, stellar capture and BZ torque present.}
\label{acc_bz_st_m}
\end{figure}

\begin{enumerate}
\item Figure \ref{acc_bz_st_m} show the evolution of black hole mass for run \# 4.1 to run \# 4.5 given in Table \ref{table_acc_bz_st_mg}. We see by studying its deviation from the canonical set that the evolution has a small dependence on the parameters $\{k$, $\gamma$, $j_{0}, B_{4}\}$, so we do not show those cases here.

\item If the $\sigma$ is the same, then the final mass will be almost the same, irrespective of their initial masses [see Figure \ref{acc_bz_st_ma}]. 
\item Change in $z_{f}$ (run \# 4.4) makes little impact on the evolution and does not affect the final mass much [see Figure \ref{acc_bz_st_me}]. 
\item Variation of $\sigma$ (run \# 4.3) shifts the saturation point owing to the dependence on $\sigma$ [see Equation (\ref{satm})]. The higher the $\sigma$, the higher the saturation mass, and longer the time taken to reach the saturation point [see Figure \ref{acc_bz_st_md}]. 
\item Increase of $\eta$ (run \# 4.5) increases the accretion rate, which is the main source of mass growth. Hence, for higher $\eta$, the system reaches the saturation point earlier [see Figure \ref{acc_bz_st_mf}].
%\end{itemize}

\begin{figure}[H]
\centering
\subfigure[\label{acc_bz_st_ja}]{\includegraphics[scale=0.2]{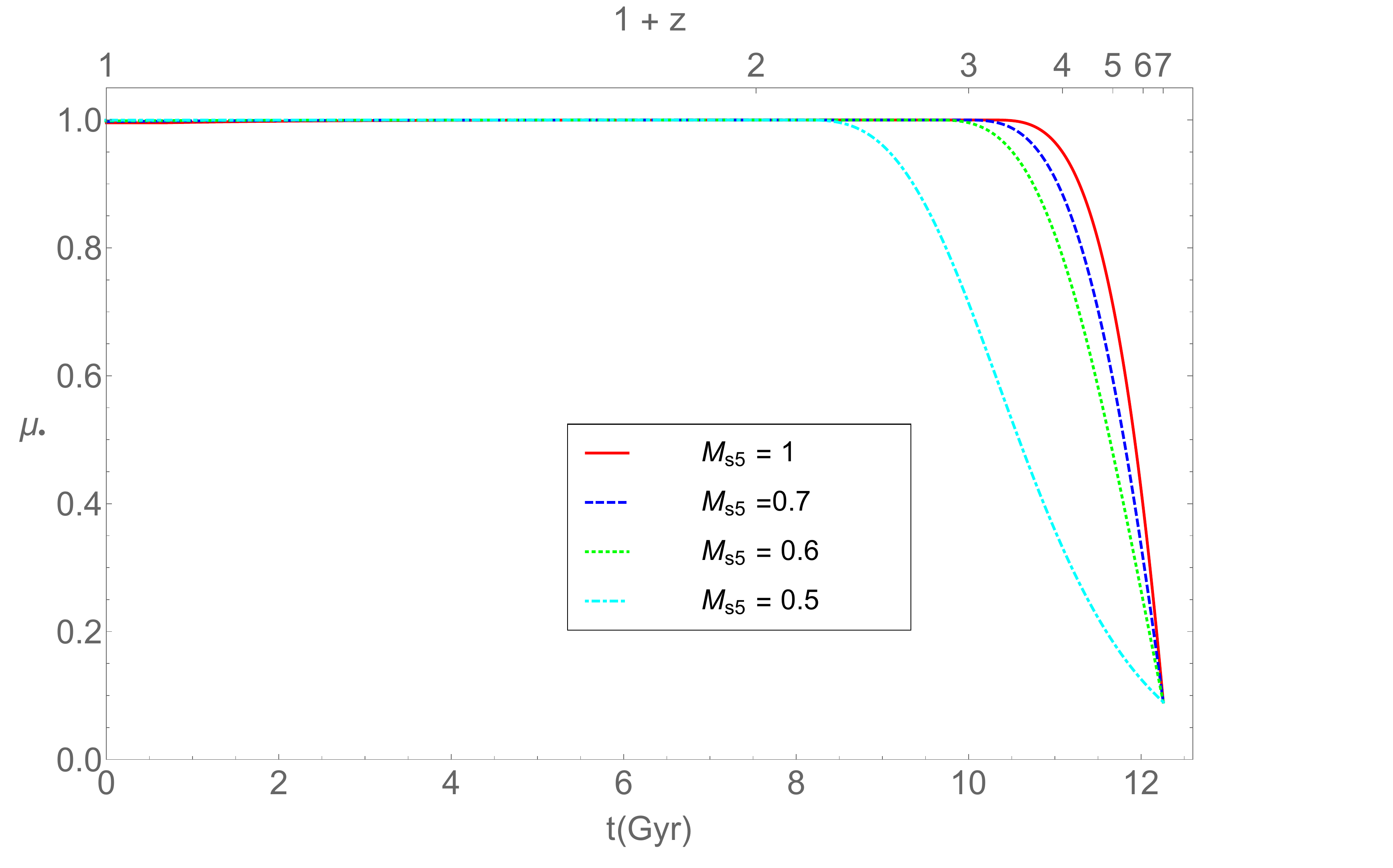}} \hspace{0.1 cm}
\subfigure[\label{acc_bz_st_jb}]{\includegraphics[scale=0.2]{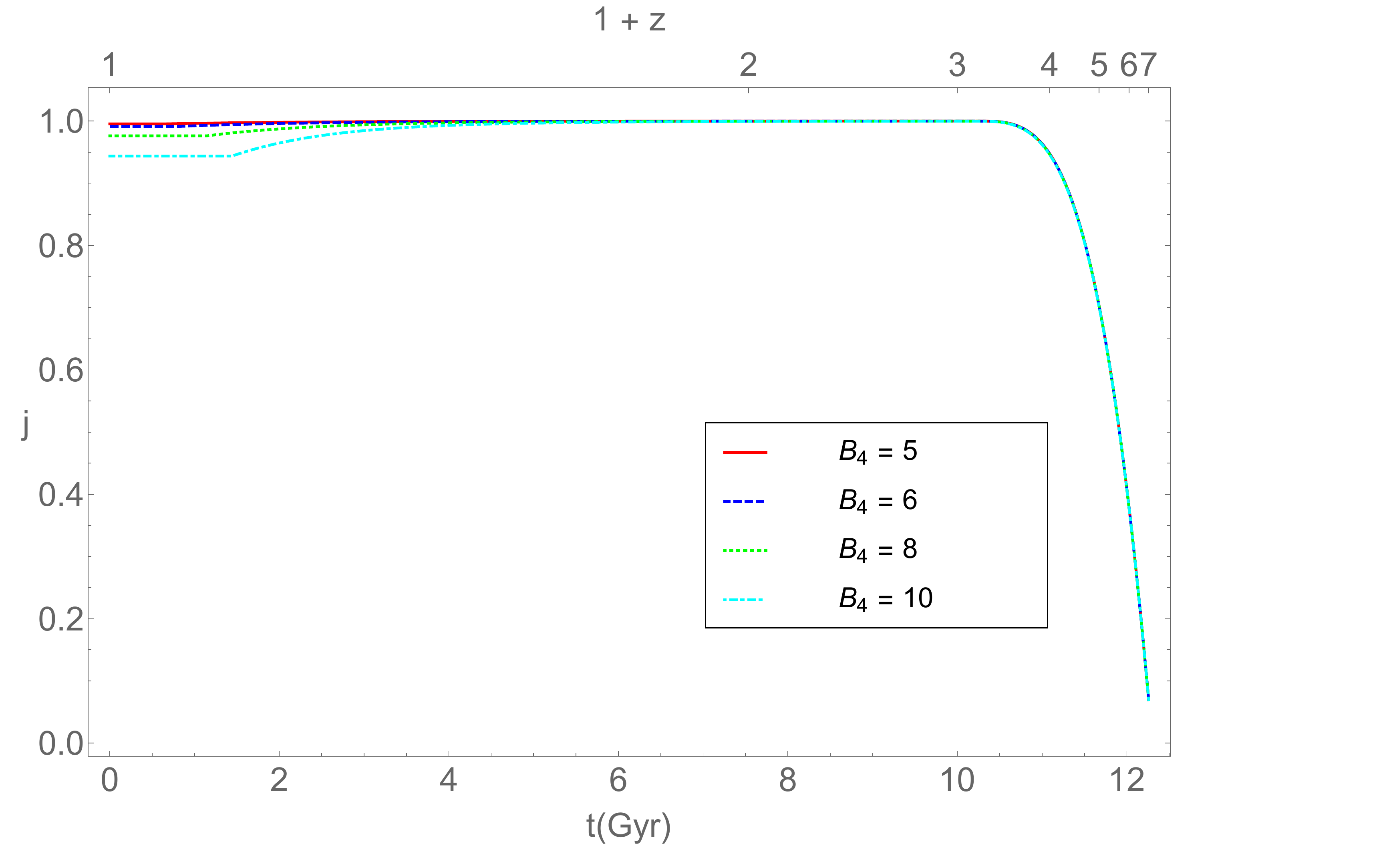}} 
\\
\subfigure[\label{acc_bz_st_jd}]{\includegraphics[scale=0.2]{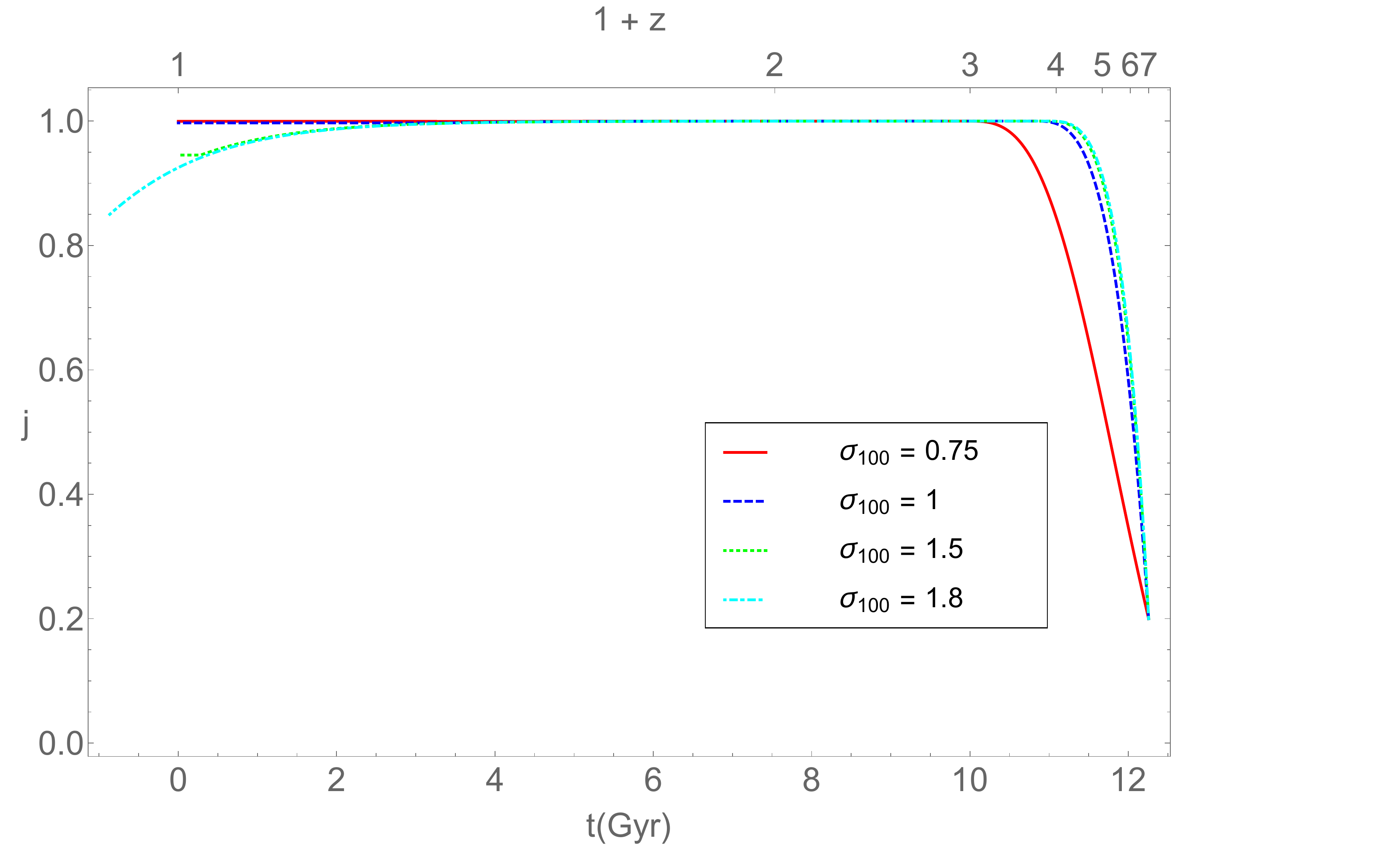}}  
\subfigure[\label{acc_bz_st_je}]{\includegraphics[scale=0.2]{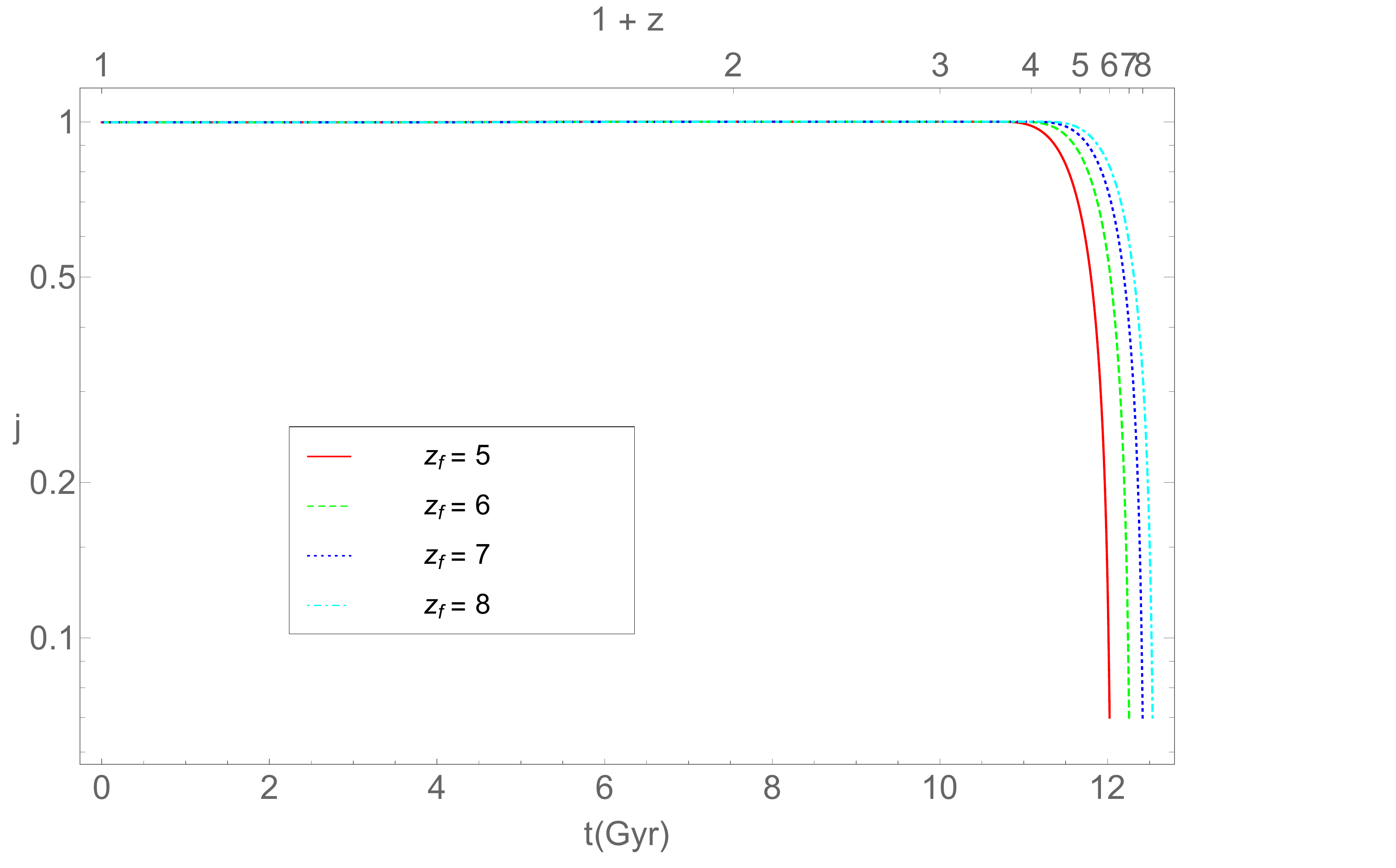}} \hspace{0.1 cm}\\
\subfigure[\label{acc_bz_st_jf}]{\includegraphics[scale=0.2]{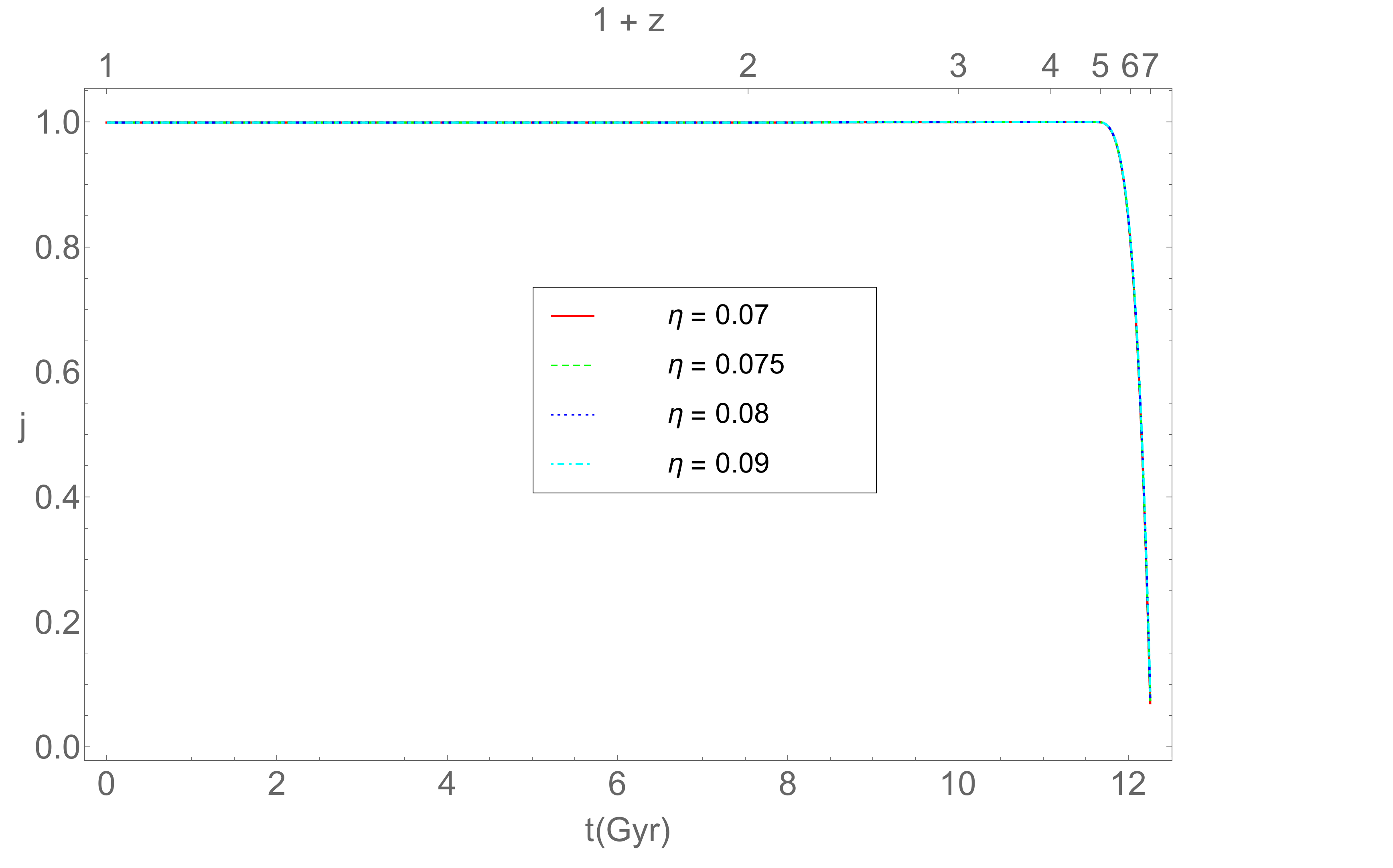}} 
\subfigure[\label{acc_bz_st_jh}]{\includegraphics[scale=0.2]{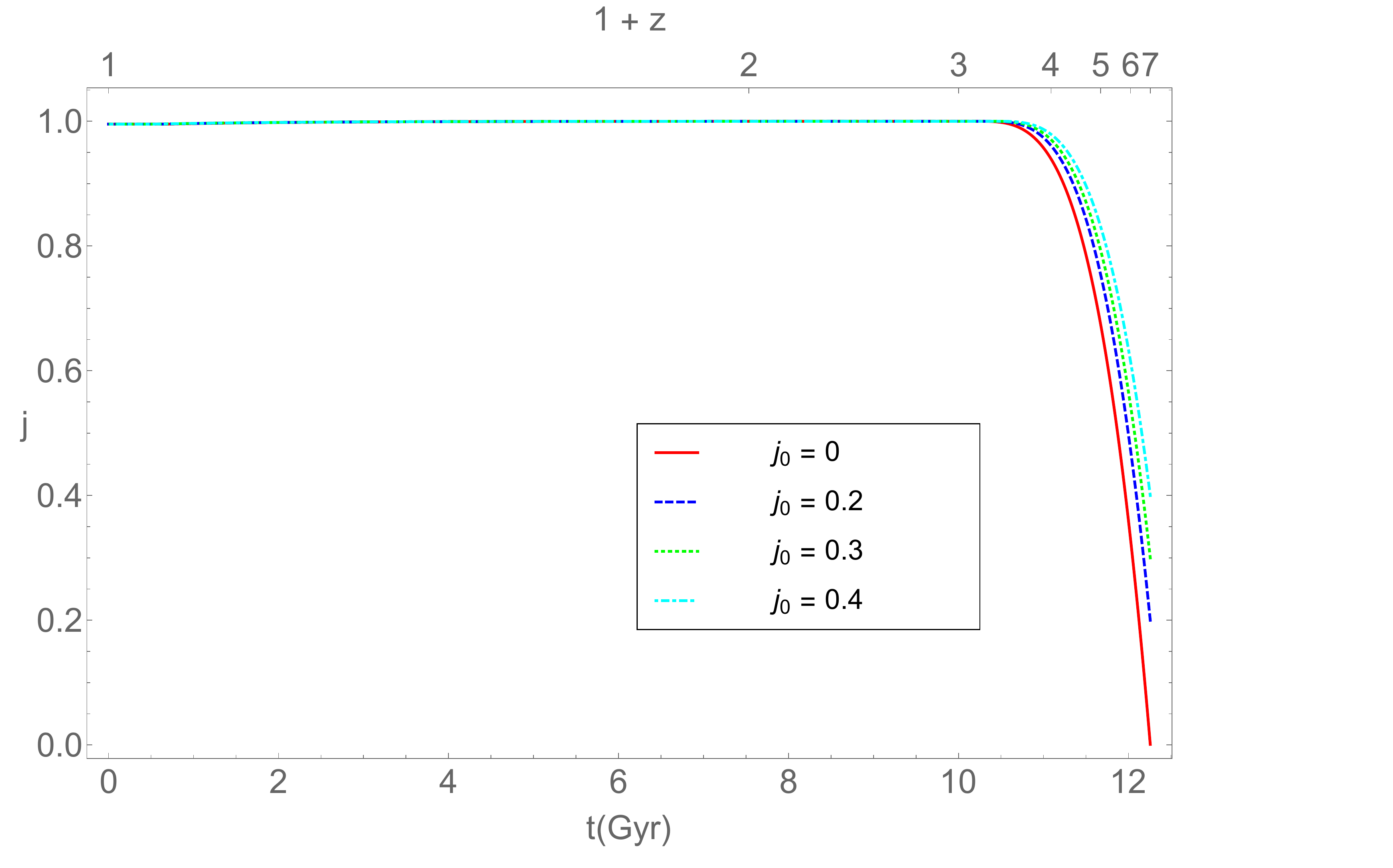}} 
\caption{Spin evolution, $j$($t$), for run \# 4.1 to run \# 4.6 (a -- f) [see Table \ref{table_acc_bz_st_mg}] are shown, for the cases when there is accretion, stellar capture, and BZ torque present.}
\label{acc_bz_st_j}
\end{figure}

%\begin{itemize}
\item The spin $j$ has a smaller dependence on the initial parameters $\{M_{s}$, $z_{f}$, $\eta$, $j_{0}\}$ [run \# 4.1, \# 4.4, \# 4.5, \# 4.6, respectively; see Figures \ref{acc_bz_st_ja}, \ref{acc_bz_st_je}, \ref{acc_bz_st_jf}, \ref{acc_bz_st_jh}], where a variation is seen at the starting points because of different initial values, but the final values attained are nearly the same. This result is different from that of experiment 3, where only accretion is present; we incorporate the concept of saturated mass here, which causes the accretion to stop, thereby reducing the final mass attained. 
\item The decrease in $j$ occurs at the high-mass end because of the BZ effect, which is small compared to the run \# 3.1.1 [see Figure \ref{acc_bz_st_jb}]. This is because, in experiment 3, we did not incorporate the saturation of black hole mass, which resulted in a high final mass reducing $j$ owing to the BZ effect. It is also seen that an increase in $B_{4}$ value (run \# 4.2) decreases the final spin, as expected.
%\item Figure \ref{acc_bz_st_jb} shows that an increase in $B_{4}$ value (run \# 4.2), decreases the final spin, as expected.
\item A higher $\sigma$ (run \# 4.3) causes a higher final mass of the black hole; hence, the final spin value decreases with an increase in $\sigma$, (see Figure \ref{acc_bz_st_jd}) while keeping $M_{s}$ constant.
\end{enumerate}

\bibliography{references}

\end{document}